\documentclass{aa}  

\usepackage{graphicx}
\usepackage{txfonts}
\usepackage{color}

\begin{document}

  \title{The Close AGN Reference Survey (CARS)}

  \subtitle{IFU survey data and the BH mass dependence of long-term AGN variability}

  \author{
  B.~Husemann\inst{1}
  \and
  M.~Singha\inst{2}
  \and
  J.~Scharw\"achter\inst{3}
  \and 
  R.~McElroy\inst{4}
  \and 
  J.~Neumann\inst{5}
  \and
  I.~Smirnova-Pinchukova\inst{1}
  \and 
  T.~Urrutia\inst{6}
  \and 
  S.~A.~Baum\inst{2}
  \and
  V.~N.~Bennert\inst{7}
  \and 
  F.~Combes\inst{8}
  \and
  S.~M.~Croom\inst{4}
  \and
  T.~A.~Davis\inst{9}
  \and
  Y.~Fournier\inst{6}
  \and 
  A.~Galkin\inst{6}
  \and
  M.~Gaspari\inst{10,11}  
  \and 
  H.~Enke\inst{6}
  \and
  M.~Krumpe\inst{6}
  \and 
  C.~P.~O'Dea\inst{2}
  \and M.~Pérez-Torres \inst{12,13}
  \and
  T.~Rose\inst{14}
  \and
  G.~R.~Tremblay\inst{15}
  \and
  C.~J.~Walcher\inst{6}
  }

  \institute{
  Max-Planck-Institut f\"ur Astronomie, K\"onigstuhl 17, D-69117 Heidelberg, Germany, \email{husemann@mpia.de}
  \and
  Department of Physics \& Astronomy, University of Manitoba, Winnipeg, MB R3T 2N2, Canada
  \and
  Gemini Observatory/NSF’s NOIRLab, 670 N. A’ohoku Place, Hilo, Hawai’i, 96720, USA
  \and
  Sydney Institute for Astronomy, School of Physics, University of Sydney, NSW 2006, Australia
  \and
  Institute of Cosmology and Gravitation, University of Portsmouth, Burnaby Road, Portsmouth, PO1 3FX, UK
  \and
  Leibniz-Institut f\"ur Astrophysik Potsdam, An der Sternwarte 16, 14482 Potsdam, Germany
   \and
  Physics Department, California Polytechnic State University, San Luis Obispo, CA 93407, USA
  \and
  LERMA, Observatoire de Paris, PSL Univ., Coll\`ege de France, CNRS, Sorbonne Univ., Paris, France
  \and
  School of Physics \& Astronomy, Cardiff University, Queens Buildings, The Parade, Cardiff, CF24 3AA, UK
  \and
  INAF - Osservatorio di Astrofisica e Scienza dello Spazio, via P. Gobetti 93/3, I-40129 Bologna, Italy
  \and 
  Department of Astrophysical Sciences, Princeton University, 4 Ivy Lane, Princeton, NJ 08544-1001, USA
  \and
  Instituto de Astrofísica de Andaluc\'{i}a, Glorieta de las Astronom\'{i}a s/n, 18008 Granada, Spain
  \and 
  Departamento de F\'{\i}sica Te\'orica, Facultad de Ciencias, Universidad de Zaragoza, E-50009 Zaragoza, Spain
  \and
  Centre for Extragalactic Astronomy, Durham University, DH1 3LE, UK
  \and
  Center for Astrophysics $|$ Harvard \& Smithsonian, 60 Gardent St., Cambridge, MA 02138, USA
  }

   \date{14 May 2021 / 31 October 2021}

  \abstract
  % context heading (optional)
   {Active galactic nuclei (AGN) are thought to be intimately connected with their host galaxies through feeding and feedback processes. A strong coupling is predicted and supported by cosmological simulations of galaxy formation, but the details of the physical mechanisms are still observationally unconstrained.} %leave it empty if necessary  
  % aims heading (mandatory)
   {Galaxies are complex systems of stars and a multiphase interstellar medium (ISM). A spatially resolved multiwavelength survey is required to map the interaction of AGN with their host galaxies on different spatial scales and different phases of the ISM. The  goal of the Close AGN Reference Survey (CARS) is to obtain the necessary spatially resolved multiwavelength observations for an unbiased sample of local unobscured luminous AGN.}
  % methods heading (mandatory)
   {We present the overall CARS survey design and the associated wide-field optical integral-field unit (IFU) spectroscopy for all 41 CARS targets at $z<0.06$ randomly selected from the Hamburg/ESO survey of luminous unobscured AGN. This data set provides the backbone of the CARS survey and allows us to characterize host galaxy morphologies, AGN parameters, precise systemic redshifts, and ionized gas distributions including excitation conditions, kinematics, and metallicities in unprecedented detail. }
  % results heading (mandatory)
   {We focus our study on the size of the extended narrow-line region (ENLR) which has been traditionally connected to AGN luminosity. Given the large scatter in the ENLR size--luminosity relation, we performed a large parameter search to identify potentially more fundamental relations. Remarkably, we identified the strongest correlation between the maximum projected ENLR size and the black hole mass, consistent with an $R_\mathrm{ENLR,max}\sim M_\mathrm{BH}^{0.5}$ relationship. We interpret the maximum ENLR size as a timescale indicator of a single black hole (BH) radiative-efficient accretion episode for which we inferred $\left\langle\log(t_\mathrm{AGN}/[\mathrm{yr}])\right\rangle = (0.45\pm0.08)\log(M_\mathrm{BH}/[\mathrm{M}_\odot]) + 1.78^{+0.54}_{-0.67}$ using forward modeling. The extrapolation of our inferred relation toward higher BH masses is consistent with an independent lifetime estimate from the \ion{He}{ii} proximity zones around luminous AGN at $z\sim3$.}
  % conclusions heading (optional), leave it empty if necessary 
   {While our proposed link between the BH mass and AGN lifetime might be a secondary correlation itself or impacted by unknown biases, it has a few relevant implications if confirmed. For example, the famous AGN Eigenvector 1 parameter space may be partially explained by the range in AGN lifetimes. Also, the lack of observational evidence for negative AGN feedback on star formation can be explained by such timescale effects. Further observational tests are required to confirm or rule out our BH mass dependent AGN lifetime hypothesis.}


   \maketitle
%@arxiver{ENLR_size_MBH.pdf,}
%________________________________________________________________

\section{Introduction}
Active galactic nuclei (AGN) have drawn a lot of attention over the last decades because they have been beacons for the existence and demographics of super-massive black holes (BHs) throughout the history of the Universe \citep[e.g.,][]{Soltan:1982,Kollmeier:2006,Greene:2007,Schulze:2010, Kelly:2013, Kormendy:2013, Schulze:2015, Banados:2018}. The release of gravitational binding energy via accretion of matter onto these BHs is expected to have a profound impact on the evolution of their host galaxies \citep[e.g.,][]{Silk:1998, Granato:2004, Matteo:2005, Springel:2005, Hopkins:2008a, Somerville:2008, Fabian:2012, Harrison:2017, Harrison:2018, Gaspari:2019, Nelson:2019}. Large spectroscopic surveys such as the Sloan Digital Sky Survey \citep[SDSS,][]{York:2000,Abazajian:2009}, the 2df QSO redshift survey \citep[2QZ,][]{Croom:2004}, the VIMOS VLT Deep Survey \citep[VVDS,][]{LeFevre:2013} or the VIMOS Public Extragalactic Redshift Survey \citep[VIPERS,][]{Scodeggio:2018} in combination with several deep X-ray surveys taken with \textit{ROSAT} \citep{Voges:1999}, \textit{Chandra} \citep{Elvis:2009,Xue:2011}, \textit{XMM-Newton} \citep{Pierre:2016}, and \textit{eROSITA} \citep{Predehl:2021} as well as large radio surveys \citep[e.g.,][]{Becker:1995,Condon:1998,Smolcic:2017,Shimwell:2019,Lacy:2020,Gordon:2021} have provided an enormous data set to characterize the AGN population and its evolution with redshift in great detail. While the standard model for the AGN central engine has been successful in unifying the various classes of AGN appearance \citep{Antonucci:1993,Urry:1995, Padovani:2017}, some aspects such as changing-look AGN \citep[CLAGN, e.g.][]{MacLeod:2016,
Ruan:2016,Graham:2017,Noda:2018} and tidal-disruption events \citep[e.g.,][]{Gezari:2009,Merloni:2015,Auchettl:2017} are just being explored more extensively in the time domain. 

It remains challenging to understand the impact of AGN on their host galaxies as a function of AGN luminosity and dominant mode of accretion, that is radiatively efficient or inefficient, across cosmic time. Modern cosmological galaxy evolution simulations implement energy coupling between the AGN and its host galaxy to prevent excessive star formation in massive halos \citep[e.g.,][]{Vogelsberger:2014,Crain:2015,Steinborn:2015,Angles-Alcazar:2017,Weinberger:2017}. However, the different physical subgrid prescriptions of AGN feedback in those simulations yield similar results for the galaxy population despite different assumptions. This suggests that the necessity of the impact is well established, but not the accurate physical process at play. Observations of AGN and their host galaxies are therefore required to reveal the various physical processes operating on different spatial and temporal scales. 

AGN can directly affect the surrounding interstellar and intergalactic medium (ISM and IGM) in various ways. The ISM-IGM can be ionized from kiloparsec to megaparcsec scales given the hard ionizing photon flux of AGN, which has been re\-cognized as extended emission-line regions \citep[EELRs,][]{Stockton:1983,Baum:1988,Heckman:1991,Villar-Martin:1997,Christensen:2006,Fu:2008,Sun:2018,Villar-Martin:2018,Balmverde:2021} or extended narrow-line regions \citep[ENLR][]{Unger:1987,Bennert:2002,Husemann:2013a,Hainline:2013,Husemann:2014,Chen:2019} at low redshifts and as the AGN proximity zones at high redshifts \citep[e.g.,][]{Bajtlik:1988,Bolton:2007,Eilers:2017,Schmidt:2017,Worseck:2021}. Radio jets and their associated mechanical power can heat the ISM/IGM and the surrounding halos and potentially quench the condensation of gas onto the galaxies \citep[e.g.,][]{McNamara:2000,Brueggen:2002,Fabian:2006b,Forman:2007,Fabian:2012,Gaspari:2012,Barai:2016,Gaspari:2020}. Gas and dust outflows from the circumnuclear region out to the galaxy and into the IGM can also be driven by the radiation and/or jets from the AGN \citep[e.g.,][]{Morganti:2005,Holt:2008,Nesvadba:2008,Feruglio:2010,Greene:2011,Combes:2013,Liu:2013b,Harrison:2014,Carniani:2015,Harrison:2015,Fiore:2017,Jarvis:2019,Husemann:2019b,Yang:2019,Santoro:2020}.  All these processes are known to exist, but how they play together, what relative impact they have on galaxy evolution, and how frequent various process occur, remains elusive.

A major complication in understanding the impact of AGN on galaxies lies in the complexity of the galaxy itself. Galaxies exhibit a multiphase ISM with various different temperatures that cannot be easily observed together \citep[e.g.,][]{Cicone:2018}; their evolution is not only driven by galaxy mass but also linked with their specific environment \citep[e.g.,][]{Peng:2010b}; and the dynamical time of galaxies (several hundred million years) is probably much longer than the short outbursts of bright AGN phases \citep[e.g.,][]{Hickox:2014}. Hence, a comprehensive study of AGN host galaxies probing all the different aspects of their evolution across the underlying galaxy population is needed to quantify the impact of the various coupling mechanisms.  

Numerous AGN surveys have already been conducted focusing on the ionized gas \citep[e.g.,][]{Stockton:1987,Schmitt:2003b,Husemann:2013a, Fu:2009, Liu:2013, Liu:2014, McElroy:2015, Harrison:2016b, Villar-Martin:2016, Woo:2016, Bischetti:2017, Circosta:2018, Chen:2019, Mingozzi:2019}, the cold gas \citep{Scoville:2003,Evans:2006, Bertram:2007,Villar-Martin:2013, Husemann:2017,Kakkad:2017, Rosario:2018,Shangguan:2018,Rose:2019,Shangguan:2020,Koss:2021,Zhuang:2021}, the hot gas \citep{Birzan:2004, Fabian:2006, Greene:2014, Lansbury:2018} and the star formation rates \citep[e.g.,][]{Lutz:2008, Shao:2010, Diamond-Stanic:2012, Harrison:2012b, Mullaney:2012b, Santini:2012, Chen:2013, Rosario:2013, Azadi:2015,  Shimizu:2015, Shimizu:2017} of their host galaxies. Despite various efforts, it remains observationally controversial whether AGN enhance, suppress or do not alter star formation in their host. The role of AGN-driven winds and gas heating in the required dissipation of energy remains unclear and controversial. In order to provide such a comprehensive and representative characterization, our team initiated the Close AGN Reference Survey \citep[CARS,][]{Husemann:2017b} which is a spatially resolved multiwavelength survey of luminous AGN in the nearby Universe. 

In this paper, we present the first set of optical IFU observations taken as part of CARS, which represents the backbone of our first public data release. We describe the reduction and processing of the data including the adopted AGN--host galaxy deblending method, the combined stellar continuum and emission line fitting procedure, and a morphological characterization of the host galaxies. This paper is one in a series of three core papers from the CARS survey and focuses on the ENLR properties and its scaling relations. As a key result we interpret the size of the ENLR as a long-term AGN variability indicator and report a BH mass dependence of the AGN variability on $10^{4}$--$10^{6}$\,yr timescale.

The paper is organized as follows. In Sect.~\ref{sect:CARS} we present the survey aims and basic sample characteristics for CARS. The details of the IFU observations and data reduction are described in Sect.~\ref{sect:IFU_data} which is followed by the data analysis in Sect.~\ref{sect:analysis} including AGN-host galaxy deblending, host galaxy characterization, emission-line mapping and AGN parameter determination. We present and discuss the results in Sect.~\ref{sect:results} and close the paper with our conclusions in Sect.~\ref{sect:conclusions}. Ancillary information on our AGN variability model and figures for the entire sample can be found in the Appendix. In this paper, we assume a concordance cosmology with $H_0 = 70$\,km s$^{-1}$ Mpc$^{-1}$, $\Omega_M = 0.3$, and $\Omega_{\Lambda} = 0.7$.

\section{The close AGN reference survey }\label{sect:CARS}
\subsection{Survey strategy and main aims}
Galaxies are composed of stars and a complex multiphase interstellar medium that is also linked to the larger gas reservoir of the intergalactic medium. It has been argued that investigating only a single gas-phase provides a very limited or biased picture about AGN outflows, feedback and feeding \citep[e.g.,][]{Cicone:2018}. In addition, the gas-phases often need to be spatially resolved to properly differentiate between possible AGN feeding and feedback processes \citep[e.g.,][]{Husemann:2018b}. While large statistical studies of AGN with integrated properties can address some of the questions from a statistical point-of-view, it is impossible to discern all internal galaxy processes based solely on integrated measurements.

Various surveys have started to obtain detailed spatially resolved multiphase observations of AGN host galaxies to complement large statistical investigations. For example, the KMOS AGN Survey at High-$z$ \citep[KASHz,][]{Harrison:2016b,Scholtz:2020}, Survey for Unveiling the Physics and the Effect of Radiative feedback \citep[SUPER,][]{Circosta:2018, Kakkad:2020, Circosta:2021} and the WISE/SDSS-selected hyper-luminous quasar survey \citep[WISSH,][]{Bischetti:2017,Vietri:2018,Bischetti:2021} map the properties of outflows and host galaxies of luminous high-redshift AGN by combining NIR spectroscopy with molecular gas and dust mapping with ALMA. At low redshifts, the BAT AGN Spectroscopic Survey \citep[BASS,][]{Koss:2017} provides a multiwavelength spectroscopic characterization for a large AGN sample, while the Nuclei of Galaxies survey \citep[NUGA,][]{Garcia-Burillo:2005,Haan:2009}, the Galactic Activity, Torus and Outflow Survey \citep[GATOS,][]{Alonso-Herrero:2019}, the Measuring Active Galactic Nuclei Under MUSE Microscope survey \citep[MAGNUM,][]{Cresci:2015,Mingozzi:2019}, the Local Luminous AGN with Matched Analogs survey \citep[LLAMA,][]{Rosario:2018, Caglar:2020}, and the Quasar Feedback Survey \citep{Jarvis:2021} provide extensive spatially resolved data  for smaller samples. Such low-redshift samples achieve much better spatial resolution and sensitivity than possible at high redshifts, but are often limited to low-luminosity AGN or selecting a specific AGN parameter space. 

The aim of CARS is to provide a representative reference data set for a significant sample of the most luminous AGN in the nearby Universe that bridges the gap in AGN luminosity between the low-redshift and high-redshift surveys. This requires a wide area AGN survey as a parent sample which contains such rare AGN in the local Universe with sufficient numbers. The CARS AGN survey is limited to $z<0.06$ to ensure a subkiloparsec spatial resolution across the host galaxies in most wavelength regimes. While the backbone of the survey is optical IFU spectroscopy of the ionized gas phase, we are combining the IFU data with ALMA, VLA and SOFIA observations to map the cold gas, \textit{Chandra} for the hot gas, and deep panchromatic imaging from various resources. With such a comprehensive data set, CARS aims to tackle a few major open questions about the connection between AGN and their host galaxies in terms of feeding and feedback cycle such as the incidence and properties of multiphase gas outflows, relative role of radiation and radio jet-driven outflows, suppression or enhancement of star formation, timescale of AGN accretion and outflows, localized versus. global impact of AGN feedback, and signature for fueling mechanisms on host galaxy scales.

The value of this multiwavelength approach for addressing those questions has already been explored for the CARS galaxy HE~1353$-$1917 \citep{Husemann:2019a,Smirnova-Pinchukova:2019}. Furthermore, limits on detecting the hot gas in X-rays with current facilities has been highlighted in \citep{Powell:2018} for two objects (Mrk~1044 and HE~1353-1917). In addition to our discovery of a prominent changing look (CLAGN) event in Mrk~1018 \citep{McElroy:2016,Husemann:2016b,Krumpe:2017} within CARS, we performed a systematic study of star formation within the bars of the galaxies presented by \citet{Neumann:2019}, which indicates differences in star formation rates that are likely related to secular evolution rather than AGN feedback.

\subsection{The unobscured AGN parent sample for CARS}
The CARS sample is entirely focused on unobscured (type 1) AGN for which black hole masses and accretion rates can be easily estimated from the accretion disk brightness and broad line characteristics from a single spectrum \citep[e.g.,][]{Peterson:2000,Kaspi:2000}. In order to select the rare bright AGN in the nearby Universe we use the Hamburg/ESO survey of bright UV-excess sources \citep[HES][]{Wisotzki:2000} which covers a large area of $\sim$9500$\,\mathrm{deg}^{2}$ in the Southern sky. The limiting magnitude of HES is $B_J < 17.3\,\mathrm{mag}$ with a dispersion of $0.5\,\mathrm{mag}$ across fields and allows a selection of bright type 1 AGN up to a redshift of $z\approx 3.2$. Follow-up spectroscopy as part of HES is used to confirm AGN and determine accurate redshifts \citep{Wisotzki:2000,Schulze:2009}. 

Applying a redshift cut of $z \leq 0.06$ to the HES catalog leads to a sample of 99 AGN, which defines the CARS parent sample. The chosen redshift cut ensures a subkiloparsec resolution in seeing-limited optical and near-infrared observations. This is necessary for a detailed structural decomposition and in particular enables us to separate the AGN point-source emission from the host galaxy components \citep[e.g.,][]{Busch:2014}. The exact value was chosen such that the CO(2-0) band-heads are still observable in NIR $K$-band in order to map the stellar kinematics and populations from high-angular NIR observations \citep[e.g.,][]{Fischer:2006,Busch:2016}.

Bolometric luminosities, derived from the X-ray luminosity, are lower than those of high-$z$ QSOs by about one order of magnitude, but the Eddington ratios of CARS targets are similar to more high-$z$ AGN \citep{Laha:2018}. Molecular gas masses ranging from $0.4 \times 10^9\,M_\odot$ to $9.7 \times 10^9\,M_\odot$ \citep{Bertram:2007} and neutral atomic gas masses ranging from $1.1 \times 10^9\,M_\odot$ to $3.8 \times 10^{10}\,M_\odot$ \citep{Koenig:2009} also fall between those of nearby low-luminosity AGN and high-$z$ luminous QSOs \citep[e.g.,][]{Kakkad:2017,Circosta:2018}. Our sources can therefore serve as an important bridge between these AGN populations \citep{Moser:2012}.

\subsection{The CARS AGN sample and data release 1}
\begin{figure*}
 \includegraphics[width=\textwidth]{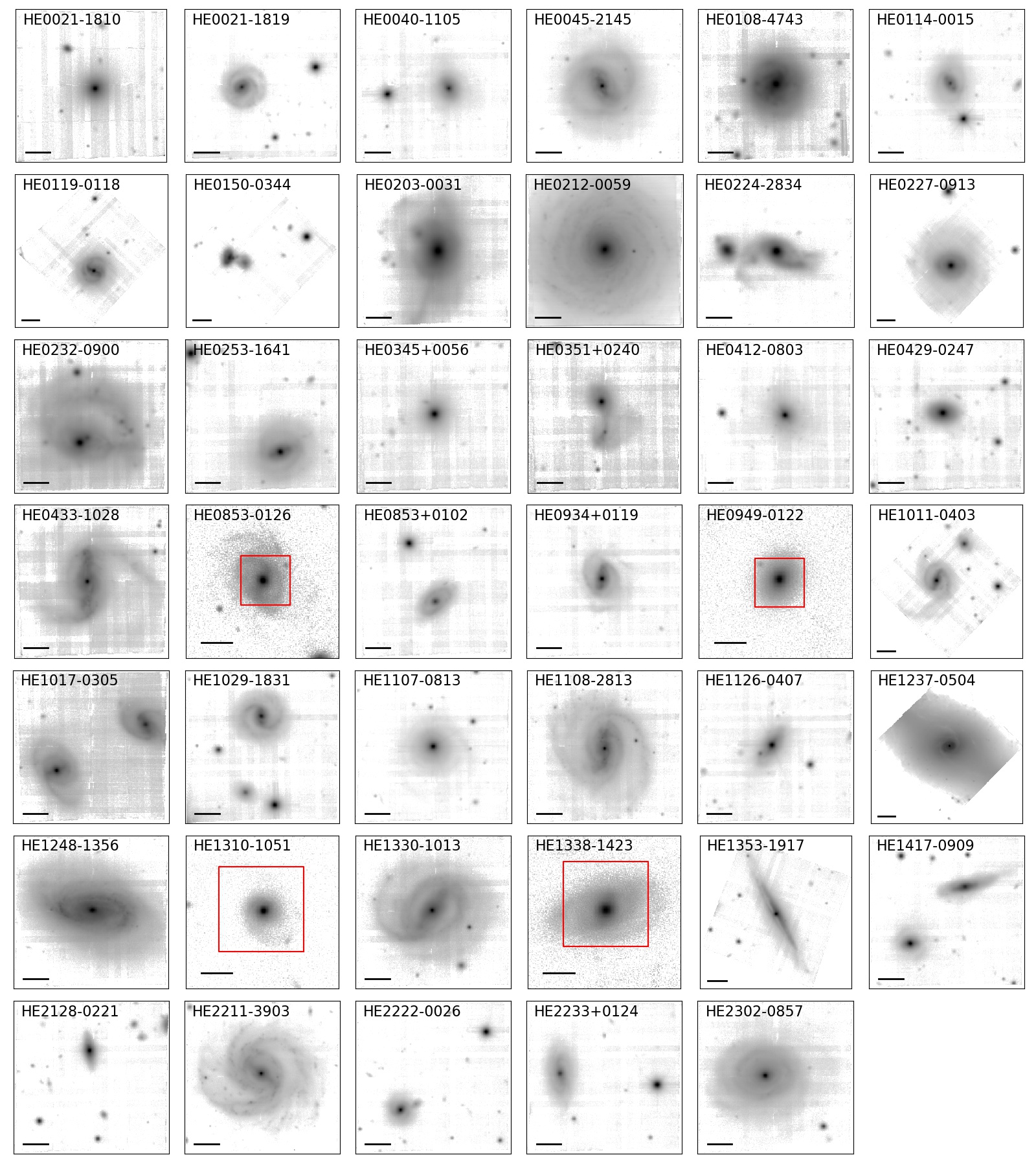}
 \caption{Gallery of reconstructed $i$ band images from the MUSE IFU observations of CARS targets. For the CARS targets observed with PMAS and VIMOS we show Pan-STARRS image instead and highlight the respect IFU FoV as red rectangles. The scale bar corresponds to 10\arcsec\ and north is up and east is to the left. }\label{fig:overview}
\end{figure*}

From the parent sample of 99 objects described above, the CARS sample is a representative subset of 41 galaxies that were randomly chosen from the parent sample for follow-up single-dish CO(1-0) observations as presented by \citet{Bertram:2007}. A gallery of broad-band images for the entire sample is shown in Fig.~\ref{fig:overview}. The availability of single-dish submillimeter observations for this sample was considered as an advantage for future follow-up spatially resolved observations with ALMA and the VLA. All CARS targets are listed in Table~\ref{tab:sample} together with the absolute $B_J$-band magnitude from HES, the continuum radio flux from the FIRST or NVSS survey, the soft X-ray flux from ROSAT as well as the integrated $H_2$ \citep{Bertram:2007} and \ion{H}{i} \citep{Koenig:2009} line fluxes. 

\begin{table*}
\label{tab:sample}
\caption{Basic CARS sample characteristics from the literature}
\centering
\begin{small}
\begin{tabular}{llccccccccccc}\hline\hline
Target & Other Name\tablefootmark{a} & $\alpha$ (J2000) & $\delta$ (J2000) & $z$\tablefootmark{b} & $M_{B_J}$ & $f_{\mathrm{1.4GHz}}$\tablefootmark{c} & $f_{\mathrm{0.1-2.4keV}}$\tablefootmark{d} & $I_\mathrm{CO(1-0)}$\tablefootmark{e} & $I_\mathrm{\ion{H}{i}}$\tablefootmark{f} \\
&  & [h:m:s] & [\degr:\arcmin:\arcsec] &  & [mag] & [mJy] & [$10^{-12}$\,erg\,s$^{-1}$\,cm$^{-2}$] & [K\,km\,s$^{-1}$] & [Jy\,km\,s$^{-1}$] \\\hline
HE0021-1810 &   & 00:23:39.4 & -17:53:54 & 0.0537 & -20.44 & $40.0\pm1.3$ & $1.2\pm0.2$ & $<$0.1 & ...\\
HE0021-1819 &  & 00:23:55.4 & -18:02:51 & 0.0533 & -20.18 & $<2.5$ & $1.0\pm0.1$ & $0.4\pm0.0$ & $<$2.0\\
HE0040-1105 & RBS101 & 00:42:36.9 & -10:49:22 & 0.0419 & -19.38 & $<2.5$ & $7.4\pm0.8$ & $1.0\pm0.1$ & $<$1.6\\
HE0045-2145 & MCG-04-03-014 & 00:47:41.2 & -21:29:28 & 0.0214 & -20.09 & $8.0\pm0.5$ & ... & $8.2\pm0.2$ & $3.3\pm0.1$\\
HE0108-4743 & RBS162 & 01:11:09.7 & -47:27:37 & 0.0239 & -19.92 & $<2.5$ & $6.2\pm0.6$ & $2.2\pm0.1$ & ...\\
HE0114-0015 & RBS175 & 01:17:03.6 & 00:00:27 & 0.0458 & -20.29 & $1.3\pm0.1$ & $3.7\pm0.4$ & $0.9\pm0.1$ & $<$1.8\\
HE0119-0118 & Mrk1503 & 01:21:59.8 & -01:02:24 & 0.0548 & -21.35 & $4.2\pm0.5$ & $7.8\pm1.1$ & $2.4\pm0.1$ & $<$1.4\\
HE0150-0344 &  & 01:53:01.5 & -03:29:23 & 0.0480 & -20.02 & $3.2\pm0.5$ & ... & $0.9\pm0.1$ & $2.7\pm0.3$\\
HE0203-0031 & Mrk1018 & 02:06:16.0 & -00:17:29 & 0.0425 & -21.59 & $4.2\pm0.4$ & $3.6\pm0.3$ & $<$0.2 & ...\\
HE0212-0059 & Mrk590 & 02:14:33.6 & -00:46:00 & 0.0264 & -21.17 & $16.2\pm0.6$ & $25.5\pm0.8$ & $3.5\pm0.1$ & $2.0\pm0.2$\\
HE0224-2834 & AM0224-283 & 02:26:25.9 & -28:21:01 & 0.0602 & -20.84 & $2.6\pm0.5$ & $0.8\pm0.3$ & $0.7\pm0.1$ & $2.6\pm0.3$\\
HE0227-0913 & Mrk1044 & 02:30:05.5 & -08:59:53 & 0.0165 & -19.86 & $2.4\pm0.5$ & $83.4\pm3.4$ & $1.9\pm0.2$ & $2.6\pm0.2$\\
HE0232-0900 & NGC985 & 02:34:37.8 & -08:47:15 & 0.0427 & -22.04 & $15.3\pm1.0$ & $38.7\pm2.1$ & $6.9\pm0.3$ & $3.3\pm0.4$\\
HE0253-1641 &  & 02:56:02.6 & -16:29:15 & 0.0319 & -20.09 & $14.2\pm1.0$ & $2.6\pm0.6$ & $2.9\pm0.1$ & $<$2.6\\
HE0345+0056 &  & 03:47:40.2 & 01:05:14 & 0.0310 & -21.26 & $32.0\pm1.0$ & $1.9\pm0.8$ & $<$0.1 & ...\\
HE0351+0240 & RBS489 & 03:54:09.5 & 02:49:31 & 0.0354 & -19.09 & $3.7\pm0.5$ & $4.1\pm0.4$ & $<$0.2 & ...\\
HE0412-0803 &  & 04:14:52.7 & -07:55:40 & 0.0380 & -20.69 & $8.9\pm0.5$ & $2.7\pm0.2$ & $<$0.1 & ...\\
HE0429-0247 & RBS550 & 04:31:37.1 & -02:41:24 & 0.0423 & -19.49 & $<2.5$ & $29.0\pm3.1$ & $<$0.3 & ...\\
HE0433-1028 & Mrk618 & 04:36:22.2 & -10:22:34 & 0.0355 & -20.76 & $17.0\pm0.7$ & $61.9\pm4.2$ & $9.4\pm0.3$ & $<$2.2\\
HE0853-0126 &  & 08:56:17.8 & -01:38:08 & 0.0596 & -20.85 & $2.5\pm0.5$ & $18.4\pm2.9$ & $1.3\pm0.1$ & $<$1.0\\
HE0853+0102 &  & 08:55:54.2 & 00:51:11 & 0.0527 & -20.32 & $0.9\pm0.2$ & ... & $<$0.2 & ...\\
HE0934+0119 & Mrk707 & 09:37:01.0 & 01:05:43 & 0.0507 & -20.70 & $<2.5$ & $14.5\pm1.1$ & $<$0.1 & ...\\
HE0949-0122 & Mrk1239 & 09:52:19.2 & -01:36:43 & 0.0197 & -20.12 & $62.2\pm1.9$ & $1.5\pm0.4$ & $1.2\pm0.1$ & $5.9\pm0.5$\\
HE1011-0403 & PG1011-040 & 10:14:20.6 & -04:18:40 & 0.0587 & -22.26 & $<2.5$ & $0.1\pm0.0$ & $1.4\pm0.1$ & $<$0.6\\
HE1017-0305 & Mrk1253 & 10:19:32.9 & -03:20:14 & 0.0491 & -20.80 & $<2.5$ & $5.4\pm0.6$ & $1.2\pm0.1$ & $<$1.2\\
HE1029-1831 &  & 10:31:57.4 & -18:46:34 & 0.0405 & -20.99 & $10.3\pm0.6$ & $5.6\pm0.9$ & $4.2\pm0.2$ & $1.1\pm0.1$\\
HE1107-0813 &  & 11:09:48.5 & -08:30:15 & 0.0585 & -21.44 & $7.3\pm1.2$ & ... & $0.8\pm0.1$ & $<$1.0\\
HE1108-2813 & ESO438-9 & 11:10:48.0 & -28:30:04 & 0.0240 & -20.88 & $15.2\pm0.6$ & $2.9\pm0.3$ & $8.3\pm0.3$ & $<$0.5\\
HE1126-0407 & PG1126-041 & 11:29:16.6 & -04:24:08 & 0.0605 & -22.69 & $1.2\pm0.1$ & ... & $1.5\pm0.1$ & $<$1.8\\
HE1237-0504 & NGC4593 & 12:39:39.4 & -05:20:39 & 0.0083 & -19.16 & $4.4\pm0.5$ & $40.5\pm2.0$ & $7.5\pm0.6$ & $7.5\pm0.4$\\
HE1248-1356 & IC3834 & 12:51:32.4 & -14:13:17 & 0.0145 & -16.99 & $2.9\pm0.6$ & ... & $3.0\pm0.2$ & $0.8\pm0.0$\\
HE1310-1051 & PG1310-109 & 13:13:05.8 & -11:07:42 & 0.0343 & -20.29 & $<2.5$ & $4.1\pm0.4$ & $<$0.2 & ...\\
HE1330-1013 & MCG-02-35-001 & 13:32:39.1 & -10:28:53 & 0.0225 & -19.30 & $<2.5$ & $119.0\pm10.7$ & $1.8\pm0.2$ & $1.8\pm0.3$\\
HE1338-1423 & RBS1303 & 13:41:13.0 & -14:38:41 & 0.0413 & -21.32 & $5.3\pm0.6$ & $33.2\pm3.1$ & $<$0.1 & ...\\
HE1353-1917 & ES O578-9 & 13:56:36.7 & -19:31:45 & 0.0348 & -19.36 & $12.6\pm0.6$ & $2.9\pm0.7$ & $3.2\pm0.2$ & $<$2.4\\
HE1417-0909 &  & 14:20:06.2 & -09:23:14 & 0.0437 & -20.06 & $<2.5$ & $2.2\pm0.3$ & $<$0.1 & ...\\
HE2128-0221 &   & 21:30:49.9 & -02:08:15 & 0.0527 & -19.76 & $<2.5$ & $7.2\pm1.7$ & $<$0.1 & ...\\
HE2211-3903 & ESO344-16 & 22:14:42.0 & -38:48:23 & 0.0397 & -20.62 & $3.5\pm0.6$ & $16.9\pm1.4$ & $1.9\pm0.1$ & ...\\
HE2222-0026 &  & 22:24:35.3 & -00:11:04 & 0.0581 & -19.97 & $<2.5$ & ... & $0.2\pm0.0$ & $<$1.2\\
HE2233+0124 &  & 22:35:42.0 & 01:39:33 & 0.0567 & -20.67 & $<2.5$ & $1.6\pm0.5$ & $0.9\pm0.1$ & $<$2.9\\
HE2302-0857 & Mrk926 & 23:04:43.4 & -08:41:09 & 0.0470 & -20.91 & $32.6\pm1.1$ & $30.9\pm1.3$ & $2.3\pm0.2$ & $1.9\pm0.2$\\
\noalign{\smallskip}\hline
\end{tabular}
\tablefoot{
\tablefoottext{a}{Most common alternative target identifier for targets if available.} 
\tablefoottext{b}{Accurate systemic redshift of the host galaxies based on the stellar continuum in our observed IFU data.} 
\tablefoottext{c}{Flux density at 1.4GHz as measured by the NRAO VLA sky survey (NVSS) and the VLA FIRST survey if undetected in NVSS. For non-detection we assume the completeness limit of NVSS at 2.5\,mJy as the upper limit for the source.} 
\tablefoottext{d}{Soft X-ray flux as seen by ROSAT.} 
\tablefoottext{e}{CO(1-0) line fluxes and upper limits for non-detection as reported in \citet{Bertram:2007}.} 
\tablefoottext{f}{\ion{H}{i} line fluxes and upper limits for non-detection as reported in \citet{Koenig:2009}.} 
}

\end{small}
\end{table*}

\begin{figure}
 \resizebox{\hsize}{!}{\includegraphics{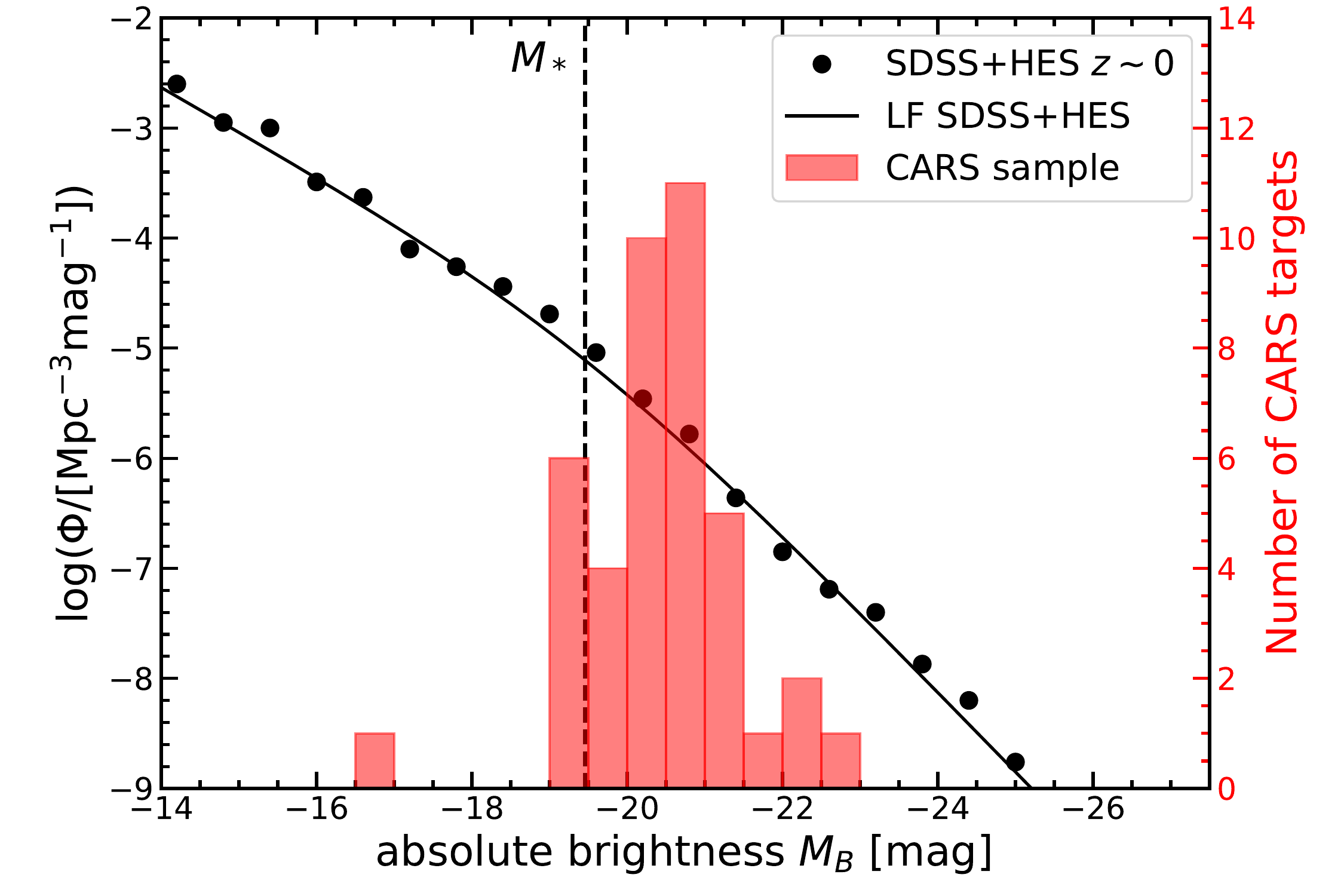}}
 \caption{Nuclear $B_J$-band absolute magnitude distribution of the CARS sample in comparison to the local ($z\sim0$) type 1 AGN luminosity function. The data points represent measurements of the combined SDSS+HES luminosity function as inferred by \citet{Schulze:2009} and the solid line is the corresponding best-fit double Schechter function as reported by \citet{Schulze:2009}. The vertical dashed line highlights the break luminosity $M_{B,*}$ of the Schechter function. The red histogram shows the nuclear $B_J$-band absolute magnitude distribution of the CARS sample.}\label{fig:AGN_LF_CARS}
\end{figure}

The overall AGN luminosity function at low redshifts was determined by \citet{Schulze:2009}. They combined the large SDSS and HES type 1 AGN samples and derived a double Schechter function to describe the observed luminosity function as shown in Fig.~\ref{fig:AGN_LF_CARS}. The distribution in AGN luminosity for the CARS sample peaks around an absolute magnitude  of $M_B=-20.5$\,mag which is one magnitude brighter than the break luminosity $M_{B,*}=-19.46$\,mag of the best-fit Schechter function \citep{Schulze:2009}. Hence, the CARS sample is confirmed to probe the bright tail of the AGN luminosity function in the local Universe, but still lacks the bright QSOs due the limited volume imposed by the redshift cut. An exception in this regard is the faint AGN HE~1248$-$1356 with an absolute magnitude of $M_{B_J}=-16.99$\,mag. It has only been detected by HES due to its low redshift of $z=0.0145$.

As a first step in the CARS project we investigated optical IFU spectroscopy as well as panchromatic images collected from various resources. We release all those observations and high-level data products to the community as part of the CARS data release 1 (DR1) that can be accessed at  \url{http://cars.aip.de}. The details of the processing of the distributed data products are described in this paper for the IFU data and ionized gas measurements, in \citet{Singha:2021} for the circumnuclear [\ion{O}{iii}] kinematics and in \citet{Smirnova-Pinchukova:2021} for the overall spectral energy distributions and star formation properties. One important result from the CARS IFU follow-up observations related to the sample is that two of the CARS targets, HE~0045$-$2145 and HE~0150$-$0344, turned out to be starburst galaxies which were initially identified as AGN by HES.  The exceptionally blue continuum and broad line components from starburst-driven outflows could not be distinguished from type 1 AGN signatures in the low-resolution HES spectra. We can exclude changing-look AGN as a cause for this discrepancy after comparing with the original HES spectra (Schulze priv. communications). This misclassification in the parent sample is reducing the actual number of AGN targets from 41 to 39, but we keep the other two galaxies in the DR for completeness and determine all relevant parameters.

\begin{table*}
\label{tab:observations}
\caption{Observational parameters}\label{tab:observations}
\centering
\begin{tabular}{lcccccccc}\hline\hline
Target & Instrument\tablefootmark{a} & Date\tablefootmark{b} & $t_\mathrm{exp}$\tablefootmark{c} & $\theta_\mathrm{FWHM}$\tablefootmark{d} & $q_\mathrm{phot}$\tablefootmark{e} & ID\tablefootmark{f} & comment\tablefootmark{g}\\
&  & [yyyy-mm--dd] & [s] & [\arcsec] &  & &  \\\hline
HE0021$-$1810 & MUSE/WFM & 2015-01-12 & 700 & 0.77 & 0.85 & 094.B-0345(A) & twilight observations\\
HE0021$-$1819 & MUSE/WFM & 2015-01-12 & 1400 & 0.61 & 1.05 & 094.B-0345(A) & \\
HE0040$-$1105 & MUSE/WFM & 2015-01-12 & 800 & 0.62 & 1.10 & 094.B-0345(A) & \\
HE0045$-$2145 & MUSE/WFM & 2015-07-18 & 900 & 0.66 & 0.97 & 095.B-0015(A) & starburst galaxy\\
HE0108$-$4743 & MUSE/WFM & 2015-01-11 & 600 & 1.32 & 1.59 & 094.B-0345(A) & \\
HE0114$-$0015 & MUSE/WFM & 2015-08-24 & 900 & 0.50 & 1.10 & 095.B-0015(A) & \\
HE0119$-$0118 & MUSE/WFM & 2015-01-12 & 600 & 0.65 & 1.03 & 094.B-0345(A) & \\
HE0150$-$0344 & MUSE/WFM & 2015-01-10 & 1200 & 1.07 & 1.03 & 094.B-0345(A) & starburst galaxy\\
HE0203$-$0031 & MUSE/WFM & 2015-07-18 & 800 & 1.12 & 0.91 & 095.B-0015(A) & changing-look AGN\\
HE0212$-$0059 & MUSE/WFM & 2017-11-14 & 7700 & 0.58 & 0.97 & 099.B-0249(A) & changing-look AGN\\
HE0224$-$2834 & MUSE/WFM & 2015-08-11 & 900 & 1.47 & 0.95 & 095.B-0015(A) & \\
HE0227$-$0913 & MUSE/WFM & 2015-01-10 & 1200 & 1.03 & 1.01 & 094.B-0345(A) & \\
HE0232$-$0900 & MUSE/WFM & 2015-01-11 & 1200 & 1.03 & 1.99 & 094.B-0345(A) & \\
HE0253$-$1641 & MUSE/WFM & 2014-12-26 & 800 & 0.84 & 1.63 & 094.B-0345(A) & \\
HE0345$+$0056 & MUSE/WFM & 2014-12-26 & 1600 & 0.80 & 1.02 & 094.B-0345(A) & \\
HE0351$+$0240 & MUSE/WFM & 2014-12-26 & 1600 & 0.65 & 1.66 & 094.B-0345(A) & \\
HE0412$-$0803 & MUSE/WFM & 2014-11-30 & 1600 & 0.67 & 1.09 & 094.B-0345(A) & \\
HE0429$-$0247 & MUSE/WFM & 2015-01-10 & 3200 & 0.85 & 1.85 & 094.B-0345(A) & \\
HE0433$-$1028 & MUSE/WFM & 2014-12-28 & 1800 & 0.51 & 1.59 & 094.B-0345(A) & \\
HE0853$-$0126 & PMAS/Larr & 2018-12-10 & 2400 & 1.15 & 1.47 & H18-3.5-010 & \\
HE0853$+$0102 & MUSE/WFM & 2015-01-12 & 2400 & 0.59 & 1.04 & 094.B-0345(A) & \\
HE0934$+$0119 & MUSE/WFM & 2015-04-13 & 1350 & 0.66 & 1.01 & 095.B-0015(A) & \\
HE0949$-$0122 & PMAS/Larr & 2018-12-10 & 3300 & 1.71 & 1.65 & H18-3.5-010 & \\
HE1011$-$0403 & MUSE/WFM & 2015-01-14 & 600 & 0.72 & 1.07 & 094.B-0345(A) & \\
HE1017$-$0305 & MUSE/WFM & 2015-06-25 & 900 & 0.64 & 1.11 & 095.B-0015(A) & \\
HE1029$-$1831 & MUSE/WFM & 2015-01-14 & 600 & 0.62 & 1.08 & 094.B-0345(A) & \\
HE1107$-$0813 & MUSE/WFM & 2015-07-11 & 2700 & 0.61 & 1.09 & 095.B-0015(A) & \\
HE1108$-$2813 & MUSE/WFM & 2015-01-15 & 400 & 0.43 & 0.98 & 094.B-0345(A) & \\
HE1126$-$0407 & MUSE/WFM & 2015-07-11 & 1800 & 0.70 & 1.08 & 095.B-0015(A) & \\
HE1237$-$0504 & MUSE/WFM & 2017-04-24 & 2700 & 0.46 & 0.98 & 099.B-0242(B) & no sky frames taken\\
HE1248$-$1356 & MUSE/WFM & 2015-07-11 & 600 & 0.47 & 0.96 & 095.B-0015(A) & \\
HE1310$-$1051 & VIMOS/HR-blue-old & 2009-04-22 & 2000 & 1.69 & 0.73 & 083.B-0801(A) & \\
 & VIMOS/HR-orange & 2009-04-25 & 3000 & 1.39 & 1.04 & 083.B-0801(A) & \\
HE1330$-$1013 & MUSE/WFM & 2015-06-20 & 600 & 0.67 & 0.82 & 095.B-0015(A) & \\
HE1338$-$1423 & VIMOS/HR-blue-old & 2009-04-27 & 2000 & 1.44 & 0.77 & 083.B-0801(A) & \\
 & VIMOS/HR-orange & 2009-04-27 & 3000 & 1.35 & 0.97 & 083.B-0801(A) & \\
HE1353$-$1917 & MUSE/WFM & 2015-06-20 & 900 & 0.69 & 0.85 & 095.B-0015(A) & \\
HE1417$-$0909 & MUSE/WFM & 2015-06-20 & 1350 & 0.65 & 0.61 & 095.B-0015(A) & \\
HE2128$-$0221 & MUSE/WFM & 2015-05-21 & 1350 & 0.70 & 0.90 & 095.B-0015(A) & \\
HE2211$-$3903 & MUSE/WFM & 2015-05-28 & 900 & 0.42 & 0.80 & 095.B-0015(A) & \\
HE2222$-$0026 & MUSE/WFM & 2015-06-19 & 1350 & 0.60 & 0.91 & 095.B-0015(A) & \\
HE2233$+$0124 & MUSE/WFM & 2015-06-25 & 1350 & 0.77 & 0.57 & 095.B-0015(A) & \\
HE2302$-$0857 & MUSE/WFM & 2015-05-29 & 600 & 0.63 & 1.13 & 095.B-0015(A) & \\
\noalign{\smallskip}\hline
\end{tabular}
\tablefoot{
\tablefoottext{a}{IFU instrument used to conduct the observations for a given target. The no-AO mode with the nominal wavelength range is used for MUSE, the Lens-Array with 1\arcsec\ sampling for PMAS, and the 0\farcs66 sampling for the high-resolution grating observations with VIMOS.} 
\tablefoottext{b}{Date of the observing night for the observations.} 
\tablefoottext{c}{Total on source exposure time combined for the final cube after rejecting low-quality individual exposures.} 
\tablefoottext{d}{Seeing in the final combined cubes inferred from 2D Moffat modeling of the broad H$\alpha$ (or H$\beta$ for the VIMOS/HR-blue setup) intensity maps. In case of Mrk1018 and the two starburst galaxies, we use the median FWHM of the guide star instead which is less precise. } 
\tablefoottext{e}{Empirically estimated photometric scale factor $q_\mathrm{phot}=f_\mathrm{IFU}/f_\mathrm{ref}$ between the MUSE data and PANSTARRS or SKYMAPPER broad band images.}
\tablefoottext{f}{Proposal ID of the data set under which the program was executed at the respective observatories.} 
\tablefoottext{g}{Additional important comment about the observation or the target itself.} 
}

\end{table*}

\section{Optical IFU observation and data reduction}\label{sect:IFU_data}
\subsection{Multi-unit spectroscopic explorer}
Follow-up optical IFU spectroscopy of the CARS sample was mainly taken with the Multi-Unit Spectroscopic Explorer \citep[MUSE,][]{Bacon:2010,Bacon:2014a} at the Very Large Telescope (VLT) as part of two filler programs 094.B-0345(A) and 095.B-0015(A) (PI: B.~Husemann). MUSE maps a large $1\arcmin\times1\arcmin$ field-of-view (FoV) at a sampling of $0\farcs2$ with a spectral coverage of 4750--9300\,\AA\ and a spectral resolution of $1800<R<3500$. 35 targets from the CARS sample have been obtained with our program. The total integration times range from 600\,s up to 2400\,s split into several exposures as listed in Table~\ref{tab:observations}. Some observations were repeated due to nonoptimal observing conditions, but some of those exposures are added if data appeared usable after inspection. Because all those CARS targets are significantly smaller than the MUSE FoV, no dedicated sky exposures were taken and sky background is estimated from the blank areas within the science exposures.

\begin{figure*}
 \includegraphics[width=\textwidth]{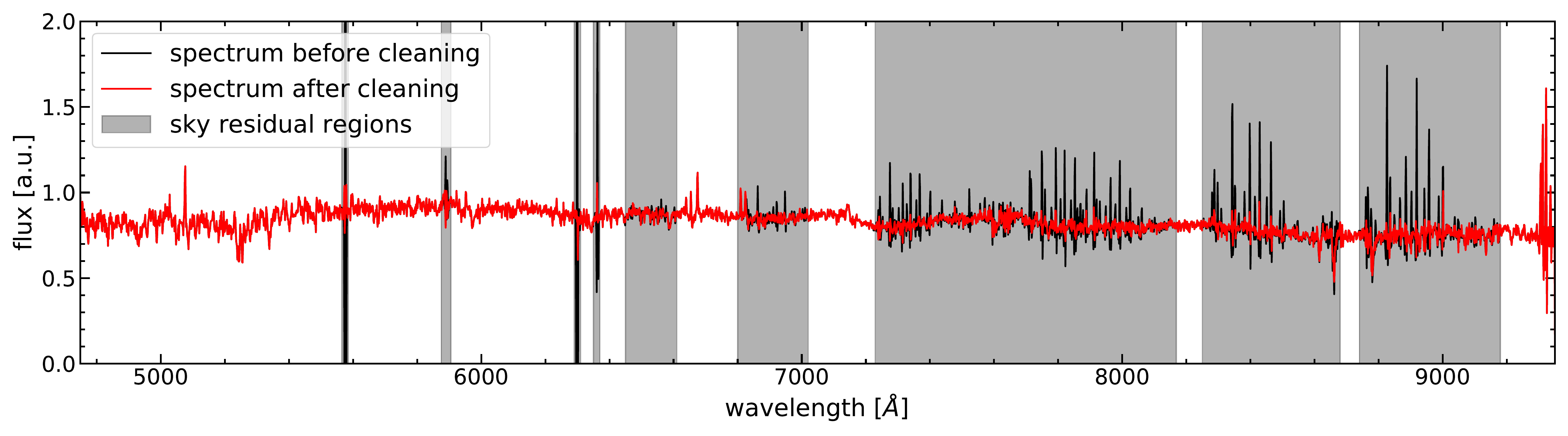}
 \caption{Example for the PCA-driven sky-line residual suppression with CubePCA. We show a coadded spectrum within a $4\arcsec\times4\arcsec$ region before (black line) and after cleaning (red line) the sky line residuals using CubePCA. Only residuals in the wavelength-range highlighted by the shaded gray area are cleaned while the other wavelengths remain untouched.}\label{fig:CubePCA}
\end{figure*}

In addition to our two dedicated programs, 2 CARS targets were observed by other programs and are publicly available in the ESO archives. HE~0212$-$0059 (Mrk~590) is a well-known changing-look AGN and was targeted as part of program 099.B-0249(A) (PI: Raimundo) for $8\times1100$\,s with interleaved dedicated sky frame exposures due to the large size of the galaxy \citep{Raimundo:2019}. HE~1237$-$0504 was observed as part of the MUSE Atlas of Disc (MAD) project \citep{Erroz-Ferrer:2019} under program 099.B-0242(B) (PI: Carollo)  with $4\times900$\,s exposures. Unfortunately, no dedicated sky frames were obtained for this galaxy although it is significantly more extended than the MUSE FoV.

We process all data with the official MUSE pipeline v2.8.1 \citep{Weilbacher:2012,Weilbacher:2014,Weilbacher:2020} following standard procedures for reducing the bias frames, the continuum lamp exposures to trace the slit, the arc lamp exposure to establish the wavelength solution, the standard star exposures for flux calibration, and the twilight flat frames. Those calibrations are applied to each individual exposure. A simple mean sky spectrum is created from the area within the FoV selected from the lower 10\% of the flux distribution in the white light images. This mean sky spectrum is then subtracted from the full cube which leaves significant sky line residuals which we further suppress as described in more detail in the next section. Telluric absorption bands are approximately corrected by dividing with the normalized transmission from the standard star exposures closest in time. Given that standard stars are usually not taken close in time, some significant residuals can still be present. For some targets a bright star is in the field which we used to calibrate and subtract the residual telluric features whenever possible. 

\subsubsection*{Suppression of sky line residuals}
Due to significant variation of the line spread function (LSF) in MUSE it is very challenging to accurately subtract the sky line across the entire FoV. A principal component analysis (PCA) scheme has shown to be very efficient in those cases to characterize the spectral residuals \citep[e.g.,][]{Kurtz:2000, Sharp:2010, Bai:2017}. A minimal set of orthogonal basis spectra are created that can describe any sky line residual as a linear super-position. Fitting those linear contributions to individual spectra allows us to remove most of the systematic residual sky line emission patterns. While a general and flexible software tool has been designed for MUSE, called Zurich Atmosphere Purge \citep[ZAP,][]{Soto:2016}, the parameters still need to be optimized depending on specific content of the observations. For CARS we created a simplified sky residual suppression algorithm based on the PCA approach which requires less free parameters to be set and is therefore more robust with respect to the actual science content of the observations. The code termed CubePCA\footnote{\url{https://git.io/cubepca}} is publicly available and we briefly describe the algorithm in the following. 

First, a 2D image with the same spatial dimension of the MUSE cube is created to flag empty sky regions (1 for sky and 0 otherwise). The corresponding Eigenspectra are computed from this selected region of empty sky spectra which were limited to a maximum subset of 20,000 spectra. Any remaining continuum signal is subtracted beforehand by running a median filter of 25 pixel width. The derived orthogonal Eigenspectra are then used to clean the sky line residuals across the entire science cube. The science cube is temporarily cleaned from any continuum signal using a running median filter of the same width. A linear $\chi^2$ fitting of the first 50 Eigenspectra is independently performed for each spaxel in the cube to find the best linear coefficients of the Eigenspectra to match the sky line residuals. Here, the $\chi^2$ is only computed in predefined wavelength regions covering bright sky lines and the best-fit linear combination is only subtracted in those wavelength regions. Afterwards the previously subtracted continuum signal from the median filtering is added back to the data. We highlight the resulting significant improvements of the sky line residual suppression with CubePCA in Fig.~\ref{fig:CubePCA}.

\subsection{VIsible Multi-Object Spectrograph}
Archival observations with the VIsible Multi-Object Spectrograph \citep[VIMOS,][]{LeFevre:2003} were available for HE~1310$-$1041 and HE~1338$-$1423 under program 083.B-0801(A) (PI: Jahnke). The VIMOS data cover $27\arcsec\times27\arcsec$ on the sky with $0\farcs66$ sampling. The high-resolution blue ($R\sim2550$) and orange ($R\sim2650$) gratings of VIMOS were used to cover all important emission lines from H$\beta$ to [\ion{S}{ii}].  The information content of the VLT-VIMOS data is nearly equal compared to the VLT-MUSE data, but has much larger spaxels ($0\farcs66$ instead of $0\farcs2$) and does not cover the CaT stellar absorption lines.

The VIMOS data were taken in April 2009 with three 500\,s and 750\,s dithered exposures in the blue and orange gratings, respectively, for each target. Continuum and arc lamp calibrations were attached to each observation for accurate tracing of fibers and wavelength calibration given the significant flexure  of VIMOS. The data were reduced with the \texttt{Py3D} data reduction package, which was initially developed for the CALIFA survey \citep{Sanchez:2012a,Husemann:2013b} and successfully applied to this VIMOS data set as described in \citet{Husemann:2014}, where more details on the data reduction can be found. Briefly, the reduction process includes basic steps such as bias subtraction, fiber tracing, flexure correction, wavelength calibration, flat fielding, and flux calibration. Since the spectral resolution significantly  varies in dispersion and cross-dispersion direction across the detector, we first characterized the variation based on the arc lamp exposure and then adaptively smooth the data to a common spectral resolution of 3\,\AA\ (FWHM). 

Since no dedicated sky frames were taken, a mean sky spectrum was created from the object free regions and subtracted from the individually calibrated exposures. Given the size of HE~1338$-$1423 this leads to a some over-subtraction of the stellar continuum but not of the emission lines. Each observation is mapped to a cube so that the position of the AGN is traced as a function of wavelength. All exposures are then drizzled \citep{Fruchter:2002} to a common grid using the AGN position as a reference for each wavelength to align the different exposures.

\subsection{Potsdam multi-aperture spectrophotometer}
As the MUSE data was only a filler program and not fully completed, we observed the two missing sources, HE~0853$-$0126 and HE~0949$-$0122, from the CARS sample with the Potsdam Multi-Aperture Spectrophotometer \citep[PMAS,][]{Roth:2005,Kelz:2006} at the 3.5m telescope of the Calar Alto Observatory under program H18-3.5-010 in December 2018. We used the PMAS lens-array with a $16\arcsec\times16\arcsec$ FoV and $1\arcsec\times1\arcsec$ spaxel size sufficient to map nearly the entire galaxies. The employed V600 grating covers the rest-frame optical wavelength range from H$\beta$ to [\ion{S}{ii}] at a spectral resolution of $R\sim1500$. 

Total integration times were $2400$\,s split-up into two exposures bracketed by three  300\,s long dedicated sky exposures for both HE~0853$-$0126 and and HE~0949$-$0122. Continuum and arc lamps calibrations were taken together after or before the target observations sky flats and standard star were taken in twilight for the matching spectral setups. We also used \texttt{Py3D} to reduce this data set following the same basic reduction steps as for the VIMOS data. Here, we adaptively smooth the spectral resolution also to a common value of 3\,\AA\ across the entire field and along wavelengths.

For background subtraction, we measure the sky background as the mean of all spectra in the dedicated sky exposures and subtracted it from the science exposures closest in time. The sky-subtracted cubes were then resampled to a cube in order to characterize the  differential atmospheric refraction by tracing the position of the AGN position as a function of wavelength. Based on the positions measured for each observation at each wavelength, we resampled the data into one cube following the drizzle algorithm \citep{Fruchter:2002} similar to the VIMOS data.

\subsection{Galactic extinction and absolute photometry}
Most of the observations are not conducted under photometric conditions and standard stars are not always obtained close in time, in particular for the ESO service mode observation. Hence, it is important to infer the absolute photometric correction factor for a given observations. We use calibrated Pan-STARRS DR2 \citep{Chambers:2016, Magnier:2020} and SkyMapper DR2 \citep{Onken:2019} cut-out $r$ and $i$ band images as a reference for our absolute photometric reference system. We construct corresponding broad-band images from the IFU cubes and compare the AB magnitude of stars in the IFU FoV with the reference images. In cases where no bright stars are captured in the IFU FoV, we use bright off-nuclear galaxy components or the integrated galaxy to minimize the impact of AGN variability. In particular for the changing-look AGN HE~0203$-$0031 and HE~0212$-$0059 it is essential to exclude the nucleus in the comparison apertures. The derived photometric scale factor $q_\mathrm{phot}=f_\mathrm{IFU}/f_\mathrm{ref}$ is listed in Table~\ref{tab:observations}. All reported fluxes in this paper are therefore divided by this factor to ensure a common photometric scaling across the survey with an absolute photometric uncertainty of $<$10 per cent as inherited from the reference surveys.

Foreground Galactic extinction can significantly reduce the observed flux and alter the overall shape of the spectra recorded for our extra-galactic targets. We therefore correct all MUSE, VIMOS and PMAS data cubes from Galactic extinction by dividing with the \citet{Cardelli:1989} Milky Way optical extinction curve as a final step. The extinction curve is scaled to the line-of-sight $V$-band extinction as reported by the NASA/IPAC Extragalactic Database (NED) which is based on the far-IR maps presented by \citet{Schlegel:1998} or SDSS stars \citep{Schlafly:2011}.

\begin{figure*}
 \includegraphics[width=\textwidth]{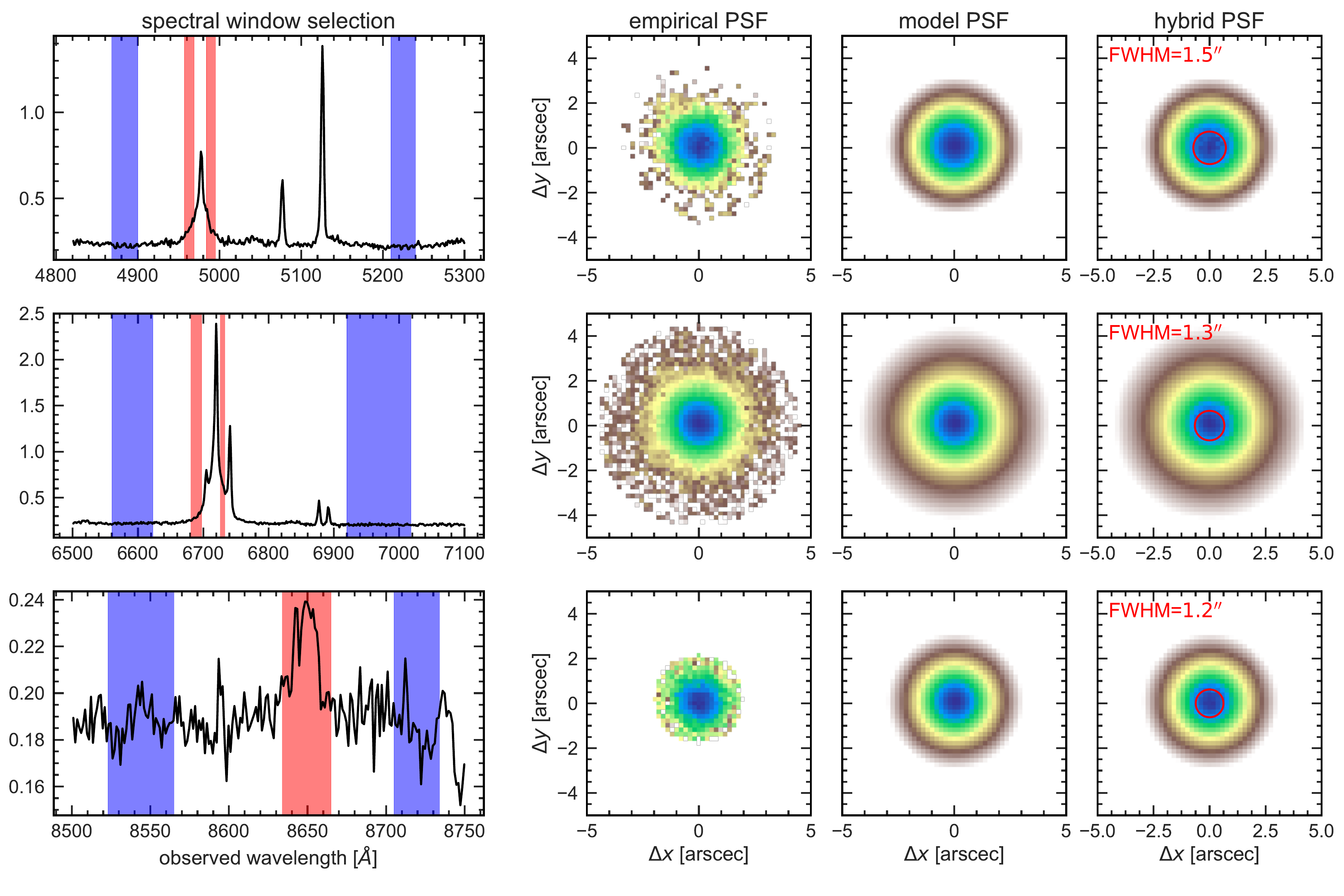}
 \caption{Example of PSF measurements for HE~0108$-$4743 based on the MUSE data. The continuum and broad line spectral wavelength regions from which the broad line intensity is mapped across the field are shown, as blue and red areas respectively, in the left panels for the spectral regions around H$\beta$, H$\alpha$ and \ion{O}{i}\,$\lambda$8446+\ion{Ca}{ii}\,$\lambda8498$  (from top to bottom). The resulting empirical PSF are shown in the 2nd column with a logarithmic intensity scaling and their best-fit 2D Moffat models are shown in the 3rd column. The hybrid PSF in the right column are created by replacing the empirical PSFs with the modeled PSFs after a certain radius which depends on the S/N of the empirical PSFs. The FWHM of PSF is indicated as the red circle on the hybrid PSFs.}\label{fig:PSF_modelling}
\end{figure*}

\section{Data analysis}\label{sect:analysis}
\subsection{AGN-host deblending}
The bright point-like emission from the AGN is able to dominate the light out to a few arcsec in ground-based seeing-limited observation. It is therefore important to subtract its contribution from each spaxel before the host galaxy extended emission can be properly analyzed. Here, we employ the AGN-host deblending method described in \citet{Husemann:2013a} and \citet{Husemann:2014}, but expand the algorithm to deal with the wavelength-dependent PSF. 

The AGN-host deblending process consists of five steps, 1) estimation of the PSF from broad lines at different wavelengths using \texttt{QDeblend$^\mathrm{3D}$} \citep[][available at \url{https://git.io/qdeblend3d}]{Husemann:2013a}, 2) modeling of the PSFs with a 2D Moffat profile to suppress noise at large radii, 3) interpolating the PSF as a function of wavelength, 4) reconstructing the intrinsic host galaxy surface brightness profile from 2D image modeling, and 5) applying an iterative AGN-host deblending scheme combining the wavelength-dependent PSF with the host galaxy surface brightness profile. In the following subsections, we describe each of the steps in more details.

\begin{table*}
 \caption{Best-fit 2D single and double Sersic profiles for the $i$ band images}\label{tab:host_model}
 \small
 \begin{tabular}{lcccccccccccccccc}\hline\hline
& \multicolumn{6}{c}{single Sersic} & & \multicolumn{6}{c}{double Sersic}\\\cline{2-7}\cline{9-14}
Object & $m_\mathrm{AGN}$ & $m_\mathrm{host}$ & $n$ & $r_\mathrm{e}$ & $b/a$ & PA & &$m_\mathrm{AGN}$ & $m_\mathrm{host}$ & $n$ & $r_\mathrm{e}$ & $b/a$ & PA\\
 & [mag] & [mag] &  & [\arcsec] & & [\degr] & & [mag] & [mag] & & [\arcsec] & & [\degr]\\\hline
HE0021-1810 & 17.0 & 15.0 & 4.7 & 4.4 & 0.92 & -13 & & ... & ... & ... & ... & ... & ... \\
HE0021-1819 & 18.1 & 15.6 & 2.6 & 3.7 & 0.76 & -52 & & 18.0 & (16.7,16.4) & (0.4,1.4) & (6.2,1.9) & (0.87,0.66) & (+36,-52) \\
HE0040-1105 & 16.9 & 15.4 & 2.6 & 2.4 & 0.74 & +23 & & 17.3 & (16.1,16.1) & (1.0,3.5) & (3.5,0.9) & (0.74,0.74) & (+19,+27) \\
HE0108-4743 & 16.0 & 13.5 & 1.2 & 6.6 & 0.83 & -32 & & 16.2 & (13.9,15.1) & (0.5,0.8) & (7.4,3.0) & (0.95,0.57) & (-23,-34) \\
HE0114-0015 & 17.7 & 15.0 & 1.8 & 3.3 & 0.62 & +32 & & 17.7 & (16.7,15.2) & (0.3,2.8) & (2.8,4.3) & (0.38,0.77) & (+37,+20) \\
HE0119-0118 & 15.9 & 14.4 & 1.7 & 5.9 & 0.68 & +67 & & 15.9 & (14.9,15.9) & (0.5,0.9) & (5.8,2.7) & (0.93,0.33) & (-29,+66) \\
HE0212-0059 & 18.5 & 11.5 & 11.0 & 60.5 & 0.97 & +25 & & 16.2 & (15.0,12.2) & (0.7,4.0) & (1.2,18.8) & (0.84,0.96) & (-33,+34) \\
HE0224-2834 & 15.9 & 15.5 & 1.9 & 4.8 & 0.64 & +61 & & 16.1 & (16.5,15.8) & (0.2,2.6) & (10.6,2.6) & (0.33,0.70) & (+69,+50) \\
HE0227-0913 & 14.8 & 13.7 & 2.2 & 6.5 & 0.77 & +87 & & 15.2 & (14.0,15.4) & (1.1,1.0) & (6.8,0.7) & (0.73,0.95) & (+87,-2) \\
HE0232-0900 & 14.4 & 14.2 & 0.8 & 6.0 & 0.78 & +87 & & 14.5 & (14.8,14.5) & (0.2,1.4) & (13.2,3.7) & (0.52,0.76) & (+61,-71) \\
HE0253-1641 & 15.2 & 14.8 & 2.4 & 5.9 & 0.83 & -65 & & 15.2 & (15.3,15.9) & (0.9,0.9) & (8.1,3.3) & (0.88,0.36) & (-61,-73) \\
HE0345+0056 & 14.2 & 15.1 & 3.5 & 2.1 & 0.81 & -17 & & 14.2 & (15.5,16.7) & (2.0,0.4) & (1.3,5.9) & (0.82,0.81) & (-17,-8) \\
HE0351+0240 & 16.2 & 15.5 & 3.5 & 3.9 & 0.73 & +15 & & 16.5 & (16.9,15.9) & (0.3,5.2) & (4.9,1.5) & (0.59,0.77) & (+31,+5) \\
HE0412-0803 & 14.7 & 15.3 & 3.0 & 3.4 & 0.75 & +33 & & 14.7 & (16.0,16.3) & (1.8,0.8) & (1.7,5.7) & (0.69,0.88) & (+36,+19) \\
HE0429-0247 & 16.3 & 15.9 & 1.9 & 2.5 & 0.66 & +86 & & 16.4 & (16.3,17.1) & (0.7,0.7) & (3.5,0.9) & (0.64,0.70) & (+87,+82) \\
HE0433-1028 & 15.0 & 13.7 & 0.8 & 9.7 & 0.35 & -1 & & 15.0 & (14.6,13.9) & (0.2,2.2) & (10.1,12.2) & (0.23,0.66) & (+0,-5) \\
HE0853+0102 & 17.8 & 15.3 & 3.8 & 4.3 & 0.83 & -55 & & 17.7 & (16.0,16.8) & (0.3,0.8) & (5.4,0.9) & (0.61,0.83) & (-44,-81) \\
HE0853-0126\tablefootmark{a} & 16.9 & 15.1 & 1.4 & 8.0 & 0.73 & +3 & & 17.3 & (15.5,16.3) & (0.5,2.8) & (8.9,3.3) & (0.72,0.51) & (+31,-15) \\
HE0934+0119 & 15.8 & 15.6 & 1.2 & 4.5 & 0.46 & -8 & & 15.8 & (16.3,16.4) & (0.5,0.4) & (3.4,5.9) & (0.36,0.87) & (-8,+23) \\
HE0949-0122\tablefootmark{a} & 14.8 & 14.1 & 3.0 & 3.0 & 0.73 & -27 & & 15.1 & (15.6,14.3) & (0.4,3.5) & (5.8,1.5) & (0.82,0.71) & (-23,-27) \\
HE1011-0403 & 15.4 & 14.9 & 2.7 & 5.8 & 0.52 & -20 & & 15.3 & (15.8,15.8) & (0.3,0.8) & (6.8,3.1) & (0.89,0.39) & (+50,-21) \\
HE1017-0305 & 15.8 & 14.5 & 2.2 & 5.7 & 0.78 & -78 & & 15.9 & (15.5,15.1) & (0.3,2.2) & (9.9,3.8) & (0.50,0.65) & (+21,-74) \\
HE1029-1831 & 15.9 & 14.4 & 4.1 & 3.8 & 0.65 & +13 & & 15.5 & (15.4,15.4) & (0.4,0.7) & (7.0,2.1) & (0.83,0.53) & (+73,+7) \\
HE1107-0813 & 14.5 & 15.0 & 3.6 & 3.7 & 0.96 & +17 & & 14.5 & (15.7,15.9) & (3.0,0.9) & (1.4,5.9) & (0.96,0.93) & (+0,+39) \\
HE1108-2813 & 15.2 & 13.4 & 1.7 & 8.3 & 0.50 & -6 & & 15.2 & (13.5,15.6) & (1.8,0.2) & (9.7,5.8) & (0.60,0.23) & (-10,-1) \\
HE1126-0407 & 14.3 & 15.2 & 2.6 & 3.6 & 0.42 & -30 & & 14.3 & (15.7,16.5) & (0.8,1.1) & (5.1,1.2) & (0.44,0.32) & (-36,-24) \\
HE1237-0504\tablefootmark{a} & 16.1 & 10.3 & 5.1 & 64.6 & 0.63 & +69 & & 15.5 & (12.5,11.0) & (1.5,1.1) & (5.5,51.1) & (0.72,0.41) & (-78,+60) \\
HE1248-1356 & 16.9 & 12.6 & 2.3 & 15.6 & 0.51 & +85 & & 19.3 & (13.0,15.2) & (0.8,1.5) & (12.8,1.1) & (0.50,0.63) & (+85,+82) \\
HE1310-1051\tablefootmark{a} & 15.7 & 15.5 & 1.6 & 3.5 & 0.86 & +89 & & 15.6 & (15.8,17.6) & (0.7,0.1) & (3.9,1.3) & (0.87,0.66) & (+79,-69) \\
HE1330-1013 & 16.6 & 13.5 & 2.3 & 16.3 & 0.53 & -42 & & 16.6 & (14.6,14.5) & (0.2,1.8) & (14.6,8.3) & (0.76,0.44) & (-81,-39) \\
HE1338-1423\tablefootmark{a} & 15.4 & 13.8 & 2.9 & 8.1 & 0.73 & -53 & & 15.7 & (14.4,14.9) & (3.6,0.3) & (3.3,12.9) & (0.70,0.48) & (-30,-76) \\
HE1353-1917 & 15.4 & 14.4 & 2.4 & 12.4 & 0.17 & +28 & & 15.4 & (16.7,14.6) & (0.1,2.7) & (13.8,11.4) & (0.16,0.17) & (+29,+28) \\
HE1417-0909 & 16.4 & 15.6 & 4.9 & 2.7 & 0.67 & -87 & & 16.4 & (18.1,15.5) & (0.6,8.3) & (2.2,4.4) & (0.27,0.84) & (-82,+83) \\
HE2128-0221 & 17.1 & 16.3 & 1.8 & 2.1 & 0.45 & +8 & & 17.1 & (16.5,18.5) & (1.7,1.8) & (2.1,2.6) & (0.40,0.70) & (+8,-79) \\
HE2211-3903 & 14.9 & 13.4 & 3.7 & 15.1 & 0.65 & +52 & & 15.0 & (15.0,13.8) & (0.2,5.6) & (11.0,13.1) & (0.69,0.57) & (-0,+56) \\
HE2222-0026 & 16.8 & 16.0 & 3.2 & 2.3 & 0.78 & -47 & & 16.7 & (16.9,16.9) & (0.5,1.1) & (3.9,1.1) & (0.88,0.66) & (+57,-44) \\
HE2233+0124 & 16.6 & 14.9 & 2.3 & 6.9 & 0.42 & +7 & & 17.3 & (15.7,15.6) & (0.6,8.3) & (7.3,2.5) & (0.42,0.42) & (+2,+13) \\
HE2302-0857 & 14.6 & 13.5 & 2.9 & 11.2 & 0.80 & -79 & & 14.9 & (14.5,14.3) & (0.3,5.5) & (10.5,3.9) & (0.80,0.82) & (-82,-79) \\
\noalign{\smallskip}\hline
\end{tabular}
\tablefoot{
\tablefoottext{a} 2D image modelling are performed on PANSTARRS images as those targets were not observed with MUSE or in the case of HE1237-0504 are only covered partially with MUSE.
}

\end{table*}

\subsubsection{Creation of PSFs from broad emission lines}
The unobscured AGN selection for CARS has the great advantage that the PSF for a given observations can be empirically determined from the intensity distribution of broad emission lines emitted by the AGN broad-line region (BLR). The size of the BLR is known to be a few light months at most based on reverberation mapping studies \citep[e.g.,][]{Kaspi:2000,Peterson:2004,Bentz:2009c}. Considering the distance to the galaxies, the BLR is inherently spatially unresolved in seeing-limited optical data and represents a perfect point source for our purposes. As initially outlined by \citet{Jahnke:2004}, IFU observations are well suited to obtain the PSFs for unobscured AGN by constructing narrow-band images centered on the broad line wings from which the continuum emission is subtracted based on narrow-band images from the adjacent continuum. Similar approaches have been widely used to create PSFs from IFU data to properly recover extended emission around luminous AGN \citep[e.g.,][]{Christensen:2006,Husemann:2008,Herenz:2015,Borisova:2016,Cantalupo:2019, Drake:2019}. 

Based on the wavelength coverage of MUSE and the redshift range of our targets, we determined empirical PSFs from the broad H$\beta$, H$\alpha$ and \ion{O}{i}\,$\lambda 8446$+\ion{Ca}{ii}\,$\lambda8498$ \citep[see][]{Matsuoka:2007} emission lines using the \texttt{QDeblend$^\mathrm{3D}$} \citep{Husemann:2013a,Husemann:2014}. The advantage of \texttt{QDeblend$^\mathrm{3D}$} is the visual definition of broad-line spectral windows that are uncontaminated by narrow emission-line from the host galaxy across the PSF size. An example of the PSF determination for HE~0108$-$4743 is shown in Fig.~\ref{fig:PSF_modelling}. The method cannot be applied to HE~0021$-$1810, for which the poor data quality does not allow to construct the PSF from the broad \ion{O}{i}+\ion{Ca}{ii} emission lines, and HE~0203$-$0031, which is a changing-look AGN \citep{McElroy:2016,Husemann:2016b,Krumpe:2017}, for which the broad emission lines are nearly absent at the time of observations. HE~0045$-$2145 and HE~0150$-$0344 do not host an AGN, so that broad lines are naturally absent and the AGN-host deblending is unnecessary. 

The wavelength coverage of the PMAS observation only allows us to construct the empirical PSFs from the H$\beta$ and H$\alpha$ lines. The VIMOS observations are taken with two spectral setups, of which the HR blue grism covers the broad H$\gamma$, H$\beta$ and \ion{He}{i} lines while the HR orange grism only covers the broad \ion{He}{i} and H$\alpha$ lines. 

\subsubsection{Modelling each PSF with a 2D Moffat}
Although estimating empirical PSFs is easy in IFU observations of unobscured AGN, they are not free from measurement uncertainties. The confusion with the host galaxy spectrum dominates at radii a few times the FWHM of the seeing disk, depending on the brightness of the AGN with respect to the host galaxy. Applying the pure empirical PSF would therefore significantly degrade the S/N in the host galaxy due to the high uncertainties in the light profiles in the wings of the PSF. To suppress the degradation of S/N in the empirical PSF as a function of radius, we model the PSF with a 2D Moffat function which is a good representation of the PSF profile in seeing-limited observations \citep[e.g.,][]{Moffat:1969,Racine:1996,Trujillo:2001,Jahnke:2004b,Gadotti:2008}. 

While we could in principle use the best-fit 2D Moffat as our PSF profile, the empirical PSF is superior close to the AGN position where the most accurate PSF determination is required. Hence, we create a hybrid PSF as a combination of the modeled and empirical PSF, where the modeled PSF is replaced by the empirical PSF within the a certain radius of the peak position. The replacing radius is individually set for each PSF at the point where noise starts to dominate. 

\subsubsection{Interpolating the PSFs with wavelengths}
The PSF is wavelength dependent as can be seen from Fig.~\ref{fig:PSF_modelling}. Therefore, we need to interpolate the 2D PSFs across the entire wavelength range. Here, we are limited in precision by the sparse sampling of PSFs in wavelength space for all IFU data sets. However, we obtained PSFs close to the blue and red end of the covered wavelength range, so that no significant extrapolation of the measured PSFs is necessary. We use a very simple but effective approach to interpolate the PSFs along the wavelengths. After normalizing each hybrid PSF to a peak flux of 1, we describe the wavelength-dependence of the normalized PSF flux for each spaxel with a simple polynomial function. The order of the polynomial is set such that it passes through the provided data points, so that we use we use a 2nd order polynomial for three PSFs. This ensures that the PSF cube is equal to the actual measurements at the wavelength where the hybrid PSF are provided. 

For all MUSE targets with broad lines we used the nominal three PSFs with the only exception of HE~0021$-$1810. For this target the broad \ion{O}{i}+\ion{Ca}{ii} blended line is too weak to allow a PSF determination, so that we only used the PSF from H$\beta$ and H$\alpha$ PSF to interpolated them with a 1st order polynomial. This limits the usable wavelength range for the AGN-host galaxy separation.  For the VIMOS data we can use three PSFs for the HR blue grism, but only two PSFs for the HR orange grism. The PMAS data allows us to use three PSFs distributed across the wavelength range for the interpolation. 

\subsubsection{2D surface brightness modeling of the host}\label{sect:2Dhost}
\begin{figure}
 \resizebox{\hsize}{!}{\includegraphics{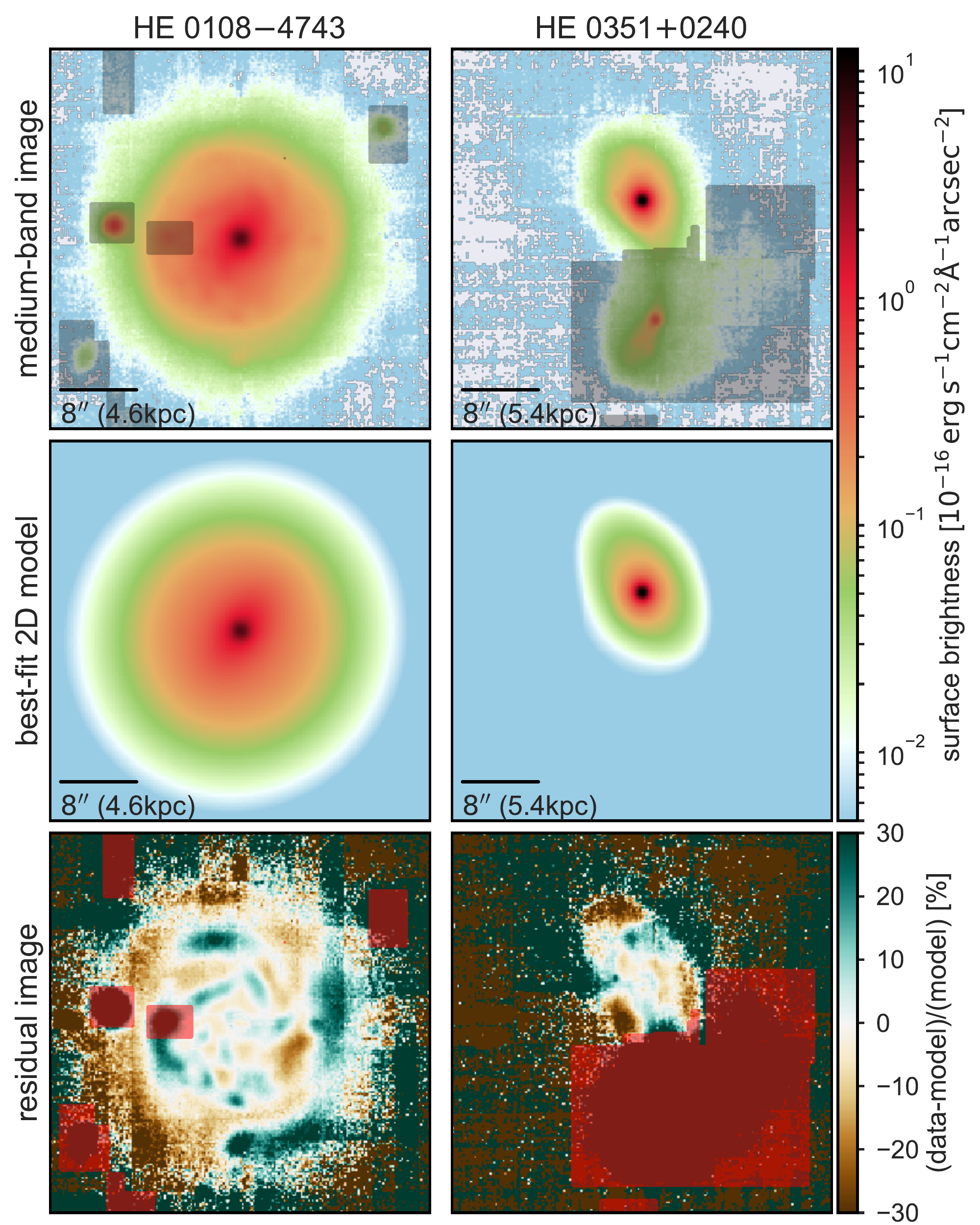}}
 \caption{Two examples for the 2D surface brightness modeling of $i$-band images reconstructed from MUSE. \textit{Top panels:} Broad-band image extracted from the MUSE cubes covering for the $i$-band. \textit{Middle panels:} Best-fit 2D surface brightness model determined with \texttt{galfit}, which consists of a PSF for the AGN and two Sersi\'c components for the host galaxy. \textit{Bottom panels:} Residuals of the best-fit model which highlights the substructure of the galaxies such a spiral arms, but also foreground stars, companions and background galaxies are visible. Those features are masked out during the fitting as shown by the shaded areas.}\label{fig:2Dhost_model}
 \end{figure}

For the next step in the iterative IFU AGN-host deblending, we need to create a surface brightness model of the host galaxy to avoid over-subtraction of the host galaxy. We therefore created $i$-band images from the MUSE cube itself. A corresponding PSF image is created from the interpolated PSF cube. We show two examples of the re-constructed broad-band images  for HE~0108$-$4743 and HE~0351$+$0240 in Fig.~\ref{fig:2Dhost_model}. For the PMAS and VIMOS targets we use the Pan-STARRS $i$ band images as the sensitivity, sampling and FoV is limiting the quality of the re-constructed images for those instruments. Similarly, we prefer to use the Pan-STARRS image also for HE~1237$-$0505 as the galaxy extends significantly beyond the MUSE FoV. 

We model the re-constructed broad-band image for all galaxies with a PSF-convolved 2D surface brightness model using \texttt{galfit} (v3.0.5) \citep{Peng:2010}. Generally, two Sersic components plus a point source for the AGN leads to a sufficient model of the host galaxy for our purpose. While a more sophisticated modeling has been performed for some of the CARS galaxies to investigate their bar properties in more detail \citep{Neumann:2019}, we only need to infer the central surface brightness profile of the host galaxy within the central 1\arcsec\ for the IFU AGN-host galaxy deblending. Here, we manually masked out nearby stars, background galaxies and interacting companions to keep the input model as simple as possible. The best-fit model and residuals are shown for HE~0108$-$4743 and HE~0351$+$0240 in Fig.~\ref{fig:2Dhost_model} and we list the inferred model parameters for a single and double Sersi\'c model for all galaxies in Table~\ref{tab:host_model}. 

\begin{figure*}
 \includegraphics[width=\textwidth]{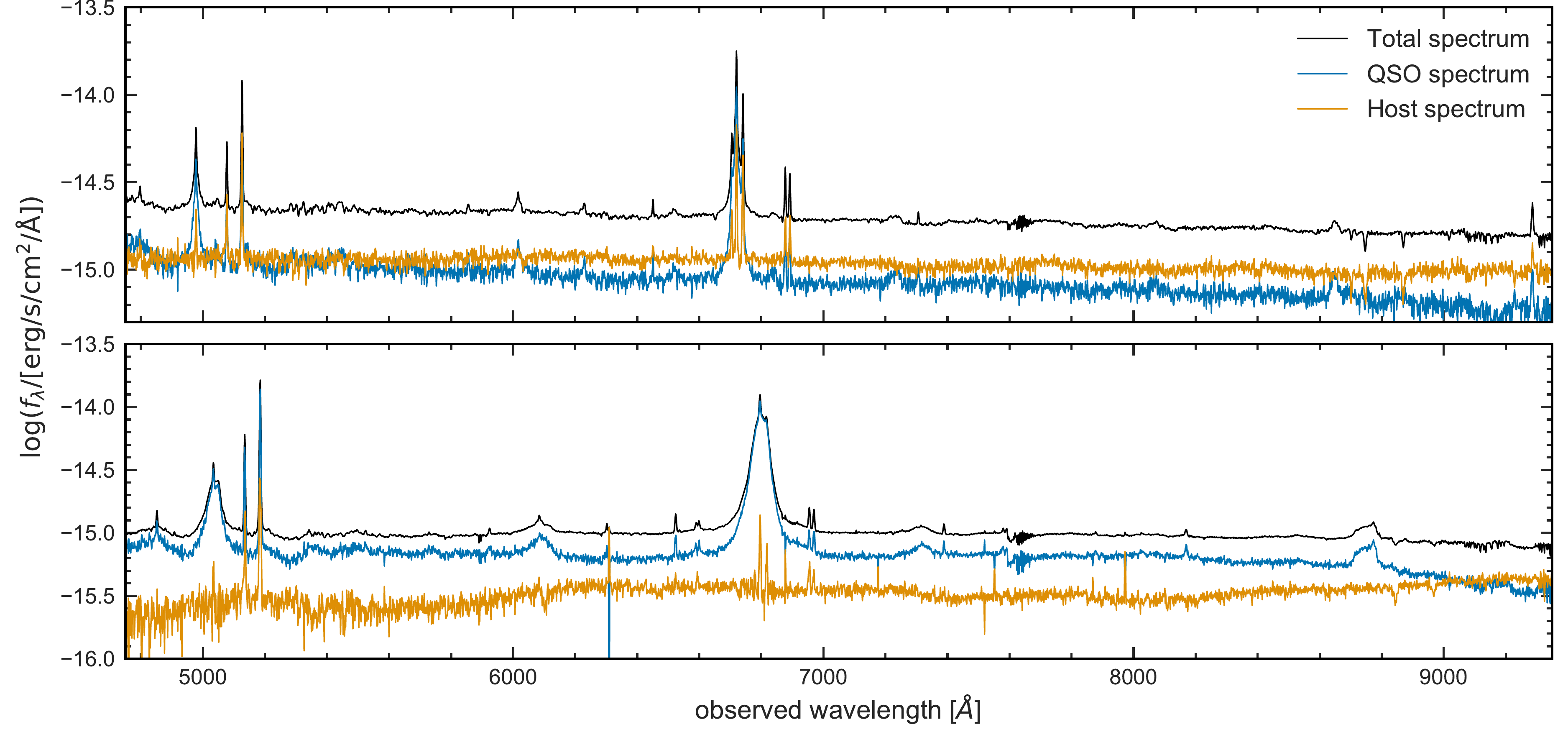}
 \caption{Result of the iterative AGN-host deblending for HE~0108$-$4743 and HE~0351$+$0240 as two examples. $3\farcs2$ aperture spectra centered on the QSO position are shown in both cases extracted from the original data (black line), the AGN contribution (blue line), the host galaxy contribution (orange line). }\label{fig:example_decomp}
\end{figure*}
\subsubsection{Iterative AGN-host deblending of the IFU data}
The last step is the actual AGN-host deblending which creates an AGN cube and a host galaxy cube. In principle, a host galaxy cube can already be created by subtracting  the previously determined wavelength-dependent PSF cube scaled to match the actual AGN spectrum at the brightest, highest S/N, spaxel. However, even the brightest spaxel at the position of the AGN will include host galaxy emission so that some over-subtraction will occurs. As outlined in \citet{Husemann:2014}, we therefore perform an iterative AGN subtraction where the AGN spectrum is corrected for host galaxy contribution. We briefly recap the iterative part of the algorithm in the following.

The first iteration starts with a pure PSF subtraction as outlined above. Here, the PSF cube is scaled to the average of the central $3\times3$ spaxels in the MUSE cubes.  After subtracting the scaled PSF cube from the original data a host galaxy spectrum is created from a rectangular shell surrounding the previous AGN spectrum region with a 2 spaxels width. For the PMAS and VIMOS data we use the brightest spaxel for the AGN region and the surrounding spaxels as the host galaxy rectangular shell region to be consistent given the coarser spatial sampling of those data sets. The AGN-subtracted host galaxy spectrum is then scaled up in surface brightness toward the central AGN spectrum extraction region based on the 2D surface brightness of the host as determined from the MUSE data (see Sect.~\ref{sect:2Dhost}). Thereby, we implicitly assume that the circumnuclear host galaxy spectrum is the same as the center of the galaxy and can be subtracted from the initial AGN spectrum used for scaling the PSF cube. The underlying assumption here is that the central part of the galaxy is more extended than a point source and its spectral signature will be visible after point-source subtraction.  The iterative process is repeated until the AGN and host galaxy spectrum have converged to a stable solution. We have consistently chosen 6 iterations for all IFU data sets after we confirmed a convergence to better than 1 per cent toward the final solution. Of course, the quality of the iterative process depends on the physical sampling and resolution. Any stellar population feature that appears point-like as the AGN will still be associated with the point-like AGN, but also does not lead to further over-subtraction across the resolved galaxy which is the main point of the iterative approach.

The result of the iterative algorithm is shown in Fig.~\ref{fig:example_decomp} for two representative targets observed with MUSE. The central 3\arcsec\ aperture spectra are shown for the total light and separated into the deblended host galaxy (extended) and AGN (point-like) contribution. Only the AGN spectrum contains broad emission lines, while the absorption lines from the stellar continuum are associated with the host galaxy spectrum. Although narrow emission lines are certainly extended, they are visible in both spectra, because part of the NLR originates from pc scales and therefore appear point-like in our seeing-limited observations. The narrow emission line contribution is thus shared between both components depending on the flux ratio of the unresolved versus resolved emission. Furthermore, the 2D surface brightness distribution in the iterative algorithm has been determined from the stellar continuum which may not necessarily apply also for the 2D surface brightness distribution of emission line flux. The iterative AGN-host galaxy deblending provides only a first order subtraction of point-like emission of narrow emission lines and a more detailed analysis is required to map the line properties in the central $<$1\arcsec\  as performed by \citet{Singha:2021}. Given the angular size of tens of arcsec for the CARS AGN host galaxies, our AGN-host deblending algorithm is capable of subtracting any compact emission. It allows us a detailed 2D investigation of the stellar and ionized gas properties in the AGN host galaxy down to 100--600\,pc from the nucleus across the CARS target redshift range.

\begin{table*}
 \small
 \caption{Visual classifications and isophotal host galaxy sizes}\label{table:visual_morph}
 \begin{tabular}{lccccccccccccc}\hline\hline
Object & \multicolumn{3}{c}{morphology} & &\multicolumn{3}{c}{barredness} & &\multicolumn{2}{c}{environment} & \multicolumn{2}{c}{$R_\mathrm{host}$\tablefootmark{a}} & $b/a$\tablefootmark{b} \\\cline{2-4}\cline{6-8}\cline{10-11}
& disc & bulge & irrgular & & barred & unbarred & unclear & & interacting & isolated & [arcsec] & [kpc] & \\\hline
HE0021-1810 & 0 & 100 & 0 & & 0 & 80 & 20 & & 20 & 80 & $12.7\pm0.5$ & $13.2\pm0.5$ & $0.97\pm0.01$ \\
HE0021-1819 & 90 & 10 & 0 & & 90 & 0 & 10 & & 60 & 40 & $10.1\pm0.1$ & $10.5\pm0.1$ & $0.96\pm0.01$ \\
HE0040-1105 & 20 & 70 & 10 & & 20 & 50 & 30 & & 20 & 80 & $12.8\pm0.5$ & $10.6\pm0.4$ & $0.88\pm0.02$ \\
HE0045-2145 & 90 & 0 & 10 & & 90 & 0 & 10 & & 20 & 80 & $24.0\pm0.4$ & $10.4\pm0.2$ & $0.92\pm0.02$ \\
HE0108-4743 & 90 & 0 & 10 & & 90 & 0 & 10 & & 20 & 80 & $18.3\pm0.3$ & $8.8\pm0.2$ & $0.94\pm0.01$ \\
HE0114-0015 & 100 & 0 & 0 & & 80 & 0 & 20 & & 80 & 20 & $15.4\pm0.4$ & $13.9\pm0.3$ & $0.69\pm0.02$ \\
HE0119-0118 & 100 & 0 & 0 & & 100 & 0 & 0 & & 20 & 80 & $14.8\pm0.2$ & $15.8\pm0.2$ & $0.98\pm0.01$ \\
HE0150-0344 & 0 & 0 & 100 & & 10 & 10 & 80 & & 100 & 0 & $7.0\pm0.1$ & $6.6\pm0.1$ & $0.91\pm0.03$ \\
HE0203-0031 & 0 & 20 & 80 & & 0 & 40 & 60 & & 100 & 0 & $32.5\pm0.4$ & $27.2\pm0.3$ & $0.71\pm0.07$ \\
HE0212-0059 & 90 & 10 & 0 & & 10 & 80 & 10 & & 10 & 90 & $44.0\pm0.6$ & $23.4\pm0.3$ & $0.88\pm0.01$ \\
HE0224-2834 & 30 & 10 & 60 & & 10 & 20 & 70 & & 100 & 0 & $24.4\pm0.5$ & $28.4\pm0.5$ & $0.34\pm0.28$ \\
HE0227-0913 & 90 & 10 & 0 & & 30 & 20 & 50 & & 20 & 80 & $29.6\pm0.8$ & $9.9\pm0.3$ & $0.86\pm0.01$ \\
HE0232-0900 & 20 & 10 & 70 & & 0 & 10 & 90 & & 100 & 0 & $28.8\pm0.5$ & $24.3\pm0.4$ & $0.98\pm0.01$ \\
HE0253-1641 & 100 & 0 & 0 & & 100 & 0 & 0 & & 11 & 89 & $17.8\pm0.2$ & $11.4\pm0.1$ & $0.92\pm0.05$ \\
HE0345+0056 & 0 & 100 & 0 & & 0 & 89 & 11 & & 44 & 56 & $11.3\pm0.4$ & $7.0\pm0.2$ & $0.98\pm0.01$ \\
HE0351+0240 & 12 & 38 & 50 & & 0 & 62 & 38 & & 100 & 0 & $11.9\pm0.3$ & $8.4\pm0.2$ & $0.64\pm0.01$ \\
HE0412-0803 & 10 & 90 & 0 & & 0 & 80 & 20 & & 20 & 80 & $12.4\pm0.3$ & $9.3\pm0.2$ & $0.91\pm0.01$ \\
HE0429-0247 & 50 & 50 & 0 & & 0 & 100 & 0 & & 20 & 80 & $9.0\pm0.1$ & $7.5\pm0.1$ & $0.74\pm0.02$ \\
HE0433-1028 & 90 & 0 & 10 & & 100 & 0 & 0 & & 10 & 90 & $30.5\pm0.4$ & $21.6\pm0.3$ & $0.80\pm0.01$ \\
HE0853+0102 & 100 & 0 & 0 & & 10 & 80 & 10 & & 20 & 80 & $11.7\pm0.1$ & $12.0\pm0.1$ & $0.64\pm0.02$ \\
HE0853-0126 & 100 & 0 & 0 & & 60 & 30 & 10 & & 10 & 90 & $18.4\pm0.6$ & $21.1\pm0.7$ & $0.67\pm0.04$ \\
HE0934+0119 & 100 & 0 & 0 & & 100 & 0 & 0 & & 0 & 100 & $14.6\pm0.4$ & $14.5\pm0.4$ & $0.68\pm0.01$ \\
HE0949-0122 & 0 & 100 & 0 & & 10 & 90 & 0 & & 44 & 56 & $14.6\pm0.2$ & $5.8\pm0.1$ & $0.82\pm0.01$ \\
HE1011-0403 & 100 & 0 & 0 & & 100 & 0 & 0 & & 90 & 10 & $16.5\pm0.4$ & $18.8\pm0.5$ & $0.72\pm0.01$ \\
HE1017-0305 & 90 & 10 & 0 & & 100 & 0 & 0 & & 100 & 0 & $20.4\pm0.2$ & $19.5\pm0.2$ & $0.69\pm0.05$ \\
HE1029-1831 & 100 & 0 & 0 & & 100 & 0 & 0 & & 60 & 40 & $12.9\pm0.2$ & $10.3\pm0.2$ & $0.99\pm0.04$ \\
HE1107-0813 & 60 & 30 & 10 & & 10 & 60 & 30 & & 60 & 40 & $15.1\pm0.4$ & $17.1\pm0.4$ & $0.91\pm0.01$ \\
HE1108-2813 & 90 & 0 & 10 & & 100 & 0 & 0 & & 20 & 80 & $27.7\pm0.6$ & $13.4\pm0.3$ & $0.89\pm0.04$ \\
HE1126-0407 & 90 & 0 & 10 & & 10 & 20 & 70 & & 60 & 40 & $14.0\pm0.9$ & $16.4\pm1.0$ & $0.67\pm0.05$ \\
HE1237-0504 & 100 & 0 & 0 & & 100 & 0 & 0 & & 0 & 100 & $114.3\pm2.1$ & $19.4\pm0.4$ & $0.64\pm0.03$ \\
HE1248-1356 & 100 & 0 & 0 & & 20 & 50 & 30 & & 20 & 80 & $35.5\pm0.4$ & $10.5\pm0.1$ & $0.68\pm0.01$ \\
HE1310-1051 & 90 & 0 & 10 & & 20 & 60 & 20 & & 90 & 10 & $13.8\pm0.4$ & $9.4\pm0.3$ & $0.74\pm0.01$ \\
HE1330-1013 & 90 & 0 & 10 & & 100 & 0 & 0 & & 70 & 30 & $28.1\pm0.3$ & $12.7\pm0.1$ & $0.92\pm0.03$ \\
HE1338-1423 & 100 & 0 & 0 & & 50 & 10 & 40 & & 0 & 100 & $25.4\pm0.3$ & $20.7\pm0.2$ & $0.57\pm0.01$ \\
HE1353-1917 & 100 & 0 & 0 & & 20 & 10 & 70 & & 0 & 100 & $27.9\pm0.4$ & $19.3\pm0.3$ & $0.29\pm0.01$ \\
HE1417-0909 & 100 & 0 & 0 & & 70 & 10 & 20 & & 90 & 10 & $10.9\pm0.3$ & $9.4\pm0.3$ & $0.89\pm0.01$ \\
HE2128-0221 & 80 & 20 & 0 & & 0 & 10 & 90 & & 60 & 40 & $9.2\pm0.3$ & $9.4\pm0.3$ & $0.60\pm0.01$ \\
HE2211-3903 & 100 & 0 & 0 & & 100 & 0 & 0 & & 10 & 90 & $26.7\pm0.5$ & $21.0\pm0.4$ & $0.85\pm0.01$ \\
HE2222-0026 & 100 & 0 & 0 & & 60 & 0 & 40 & & 80 & 20 & $8.7\pm0.1$ & $9.8\pm0.1$ & $0.88\pm0.04$ \\
HE2233+0124 & 100 & 0 & 0 & & 90 & 0 & 10 & & 20 & 80 & $17.5\pm0.4$ & $19.3\pm0.4$ & $0.51\pm0.01$ \\
HE2302-0857 & 90 & 0 & 10 & & 30 & 60 & 10 & & 10 & 90 & $28.9\pm0.5$ & $26.6\pm0.4$ & $0.94\pm0.08$ \\
\noalign{\smallskip}\hline
\end{tabular}
\tablefoot{The table lists the visual classifications as percentage of votes for each choice in the distintinc categories morphology, baredness and environment. 
\tablefoottext{a}{Major axis of the elliptical isophotes at an intrinsic $i$ band surface brightness of 24.5\,mag\,arcsec${}^{-2}$.}
\tablefoottext{b}{Axis ratio of the isophote at an intrinsic $i$ band surface brightness of 24.5\,mag\,arcsec${}^{-2}$.}
}

\end{table*}

\subsection{Host galaxy properties}
\subsubsection{Visual classification of galaxies}
Due to the selection of our sample based solely on the presence of broad emission lines, the host galaxies of our targets exhibit a variety in morphologies from disks to ongoing major mergers. Visual classifications of the AGN host galaxy morphologies were determined independently by 9--10 CARS team members based on re-constructed MUSE and archival broad-band images. A visual morphology can be easily obtained even without AGN subtraction in ground-based seeing-limited observations due to the low-redshift of the targets. Here, we limit the visual classification to very basic parameters and do not try to apply the full Hubble sequence classification. We focused mainly on basic morphology, bar presence and environment of the galaxies. The classifiers were asked to decide whether a galaxy was 1) bulge-dominated, disk-dominated or irregular, 2) unbarred, barred or uncertain bar presence, as well as whether they showed 3) the presence of tidal tails, nearby companions or if the galaxy appears isolated. The third category of classification are not necessarily mutually exclusive because a galaxy may appear isolated but exhibit tidal tails from a past interaction. We therefore combined the classifications into the two classes, interacting and isolated galaxy, where a decision for either tidal tails or companions led to an ``interacting galaxy'' classification and galaxies determined as isolated were classified as an ``isolated galaxy''. Since the detection of faint tidal features is strongly dependent on image depth, our classifications need to be considered relative across the sample. Because the classification images have similar depth and the CARS sample covers only a narrow redshift range, our classifications are not significantly affected by surface brightness dimming. The fraction of classifications in percent are listed in Table~\ref{table:visual_morph} for each galaxy.

\begin{figure}
 \resizebox{\hsize}{!}{\includegraphics{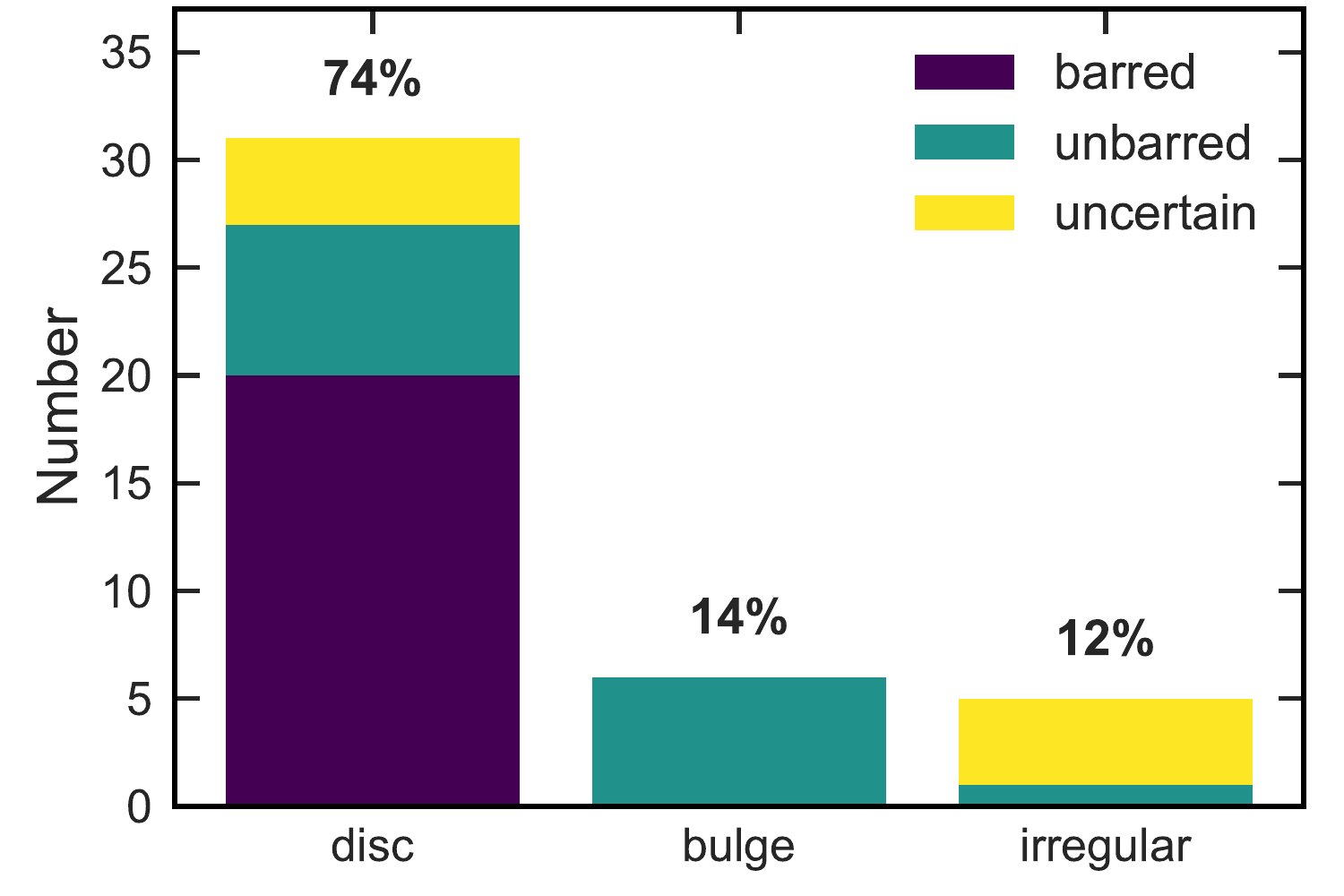}}
 \caption{Statistics of the visual morphological classification from the CARS team members. Galaxies were classified into disk-, bulge-, or irregular-like and whether they exhibit a bar or not with the option of an uncertain bar classification. Majority decision is highlighted in boldface fonts.}\label{fig:morph_stat}
\end{figure}

In all cases, except for HE~0429$-$0247, a classification was obtained by absolute majority of classifications. The resulting distribution of majority classifications is shown in Fig.~\ref{fig:morph_stat} which reveals that the CARS sample is clearly dominated by disk-dominated AGN host galaxies with a fraction of 74\%. A minority of CARS targets is found in bulge-dominated or irregular host galaxies. Strongly disturbed systems are clearly not prevalent in the sample, which is unsurprising given that no excess of strongly disturbed AGN host galaxies have been reported in the literature \citep[e.g.,][]{Cisternas:2011, Schawinski:2012, Mechtley:2016, Villforth:2017,Marian:2019} at least in the AGN luminosity range of the CARS sample. About 40\% of the targets show some signs of interactions including major mergers and galaxy pairs, but these are limited to close companions due to the limiting size of the images. A full environmental study with 1\,Mpc around the CARS galaxies is in preparation, so we cannot identify potential wide pairs and loose group environments for which the host would still be classified as isolated by the current metric.

The overall bar fraction is $\sim$50\% within the entire CARS sample and  64\% among the disk-dominated galaxy subgroup. This fraction could even be slightly higher as a couple of disk-dominated galaxies are too much inclined to identify bars and are counted as uncertain, in addition to most of the irregular galaxies. A similarly high bar fraction was reported for X-ray-selected low-redshift AGN by \citet{Cisternas:2015}, while a bar fraction of 28.5\% was reported by \citet{Alonso:2013} for local galaxies selected from SDSS hosting obscured AGN. It is clear that sample selections effects, intrinsic strength of bars, and their classification method have a significant influence on the observed bar fraction and so this fraction needs to be interpreted with caution.  The bar fraction of the overall low-redshift galaxy population is a strong function of color, stellar mass and bulge-to-disk ratio \citep{Masters:2011} ranging from $\sim$10\% to $\sim$60\% with a mean fraction of 29.4\%$\pm$0.5\%. A much higher bar fraction reaching $\sim70$\% is typically found in IR imaging observations \citep[e.g.,][]{Buta:2015,Erwin:2018}, which implies that the recovered bar fraction may be wavelength dependent.  

It is not the purpose of CARS to investigate the role of bars or interactions in triggering or fostering BH accretion in a statistical sense, but the presence of bars or strong gravitational interactions with companions can have a strong impact on the galaxy dynamics and distribution of gas on host galaxy scales. This needs to be carefully taken into account in the interpretation of the stellar and gas kinematics and distribution of star formation across the galaxies in terms of AGN outflows and AGN feedback. Hence, the visual classifications of the CARS sample presented here will be essential for all investigations performed with this sample.

\begin{figure*}
 \centering
 \includegraphics[width=\textwidth]{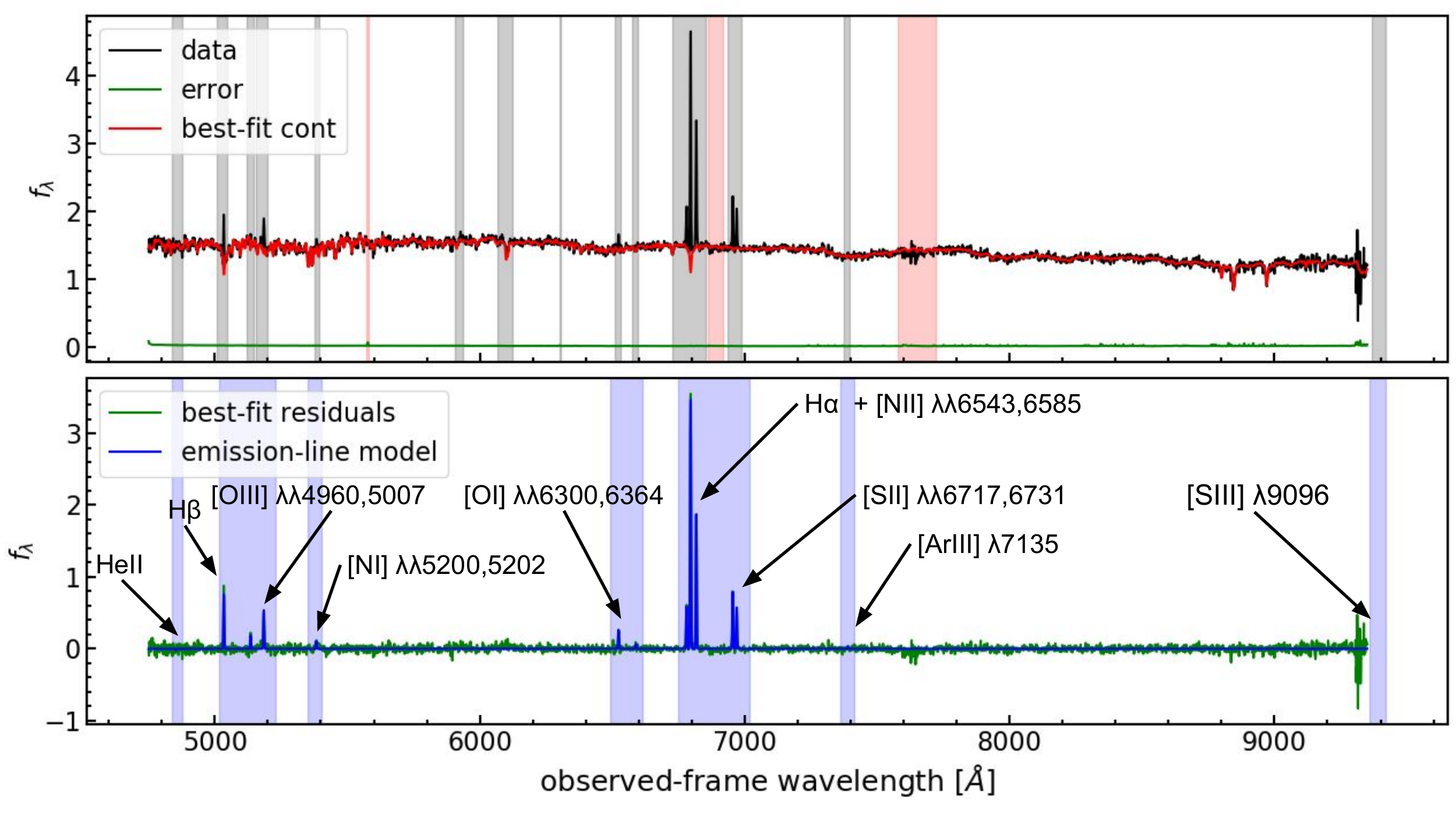}
 \caption{Example of the full spectral modeling using PyParadise for one binned spectrum of HE~0433$-$1028. The input spectrum and error spectrum are shown in the top panels as a black and green line, respectively. Emission-line contaminated regions and prominent sky or telluric absorption features are indicated by the gray and red shaded areas. The corresponding  best-fit continuum model is shown as the red line. The residual spectrum is highlighted in the lower panel with best-fit single Gaussian emission-line models represented by the blue lines. The wavelength range modeled for the emission-line detection are represented by the blue shaded areas. All fitted emission lines are annotated for reference and identification.}\label{fig:PyParadise_fit_example}
\end{figure*}

\subsubsection{Host galaxy sizes and axis ratios}
The size and inclination of a galaxy with respect to our line of sight are important basic parameters for the CARS AGN hosts. For example, the impact of the AGN radiation on the host galaxy can be significantly affected by the orientation of central AGN engine with respect to the stellar disk \citep[e.g.,][]{Husemann:2019a}. We estimated the effective radius $R_e$ and axis ratios through the 2D surface brightness modeling as part of the QSO-host galaxy deblending process, but those measurements can be influenced by a strong bar structure and are more reliable for the central part of the galaxy. We therefore performed an isophotal ellipse fitting additionally on the archival Pan-STARRS $i$-band images. Exceptions are HE~1108$-$2813, for which we use the $r$-band due to insufficient data quality in the $i$-band, HE~0108$-$4743 and HE~2211$-$3903, for which we use our dedicated wide-field imager (WFI) $B$-band images as presented in \citet{Smirnova-Pinchukova:2021} given the lack of Pan-STARRS coverage. 

The fitting is performed using the python package \texttt{Photutils}, an \texttt{Astropy} package for detection and photometry of astronomical sources \citep{Bradley:2019}. We use the tools in \texttt{photutils.isophote} to fit ellipses to the isophotes in the images, which are measured using an iterative method described in \citet{Jedrzejewski:1987}. The procedure determines for each ellipse the coordinates of its center, the semi-major axis, the ellipticity and the position angle; based on visually informed initial parameter guesses. It iteratively increases the sizes of the ellipses up to a predetermined surface brightness limit. In case of significant contamination of nonelliptical features, we use sigma-clipping, which especially improves our ability to fit in low signal-to-noise regions.

From the ellipse fitting we determine the semi-major axis and axis ratio at a limiting $i$-band surface brightness limit $\Sigma_\mathrm{lim} = 24.5\,\mathrm{mag\,arcsec}^{-2}+2.5\log((1+z)^4)$ to correct for the surface brightness dimming across the redshift range of our sample. We apply an average color correction for the limiting brightness of $r-i\sim0.25$\,mag and $B-r\sim0.65$\,mag for star-forming galaxies in the three cases without $i$ band images. This ensures that the host galaxy sizes of these three galaxies are comparable to the other ones. The corresponding measurements are listed in Table~\ref{table:visual_morph} for the entire sample.

\subsubsection{IFU continuum and emission-line fitting}
From the IFU data we want to map the stellar and ionized gas properties across the host galaxies. The process to subtract the point-like QSO emission is described above so that we can work with QSO-subtracted cubes at this stage. Here, we model the stellar continuum and ISM emission lines subsequently using \texttt{PyParadise} \citep[see][]{Walcher:2015,Weaver:2018,Husemann:2019a}. The subsequent fitting of stellar continuum and line emission ensures that complex line shapes do not impact the continuum modeling. While \texttt{pPXF} \citep{Cappellari:2004} has become the standard for the stellar continuum modeling in the IFU analysis of nearby galaxies \citep[e.g][]{Westfall:2019,Bittner:2019,Croom:2021}, the PSF interpolation along wavelength for the QSO subtraction leads to significant low-frequency continuum variation close to the nucleus that cannot be easily described with low-order polynomials used by \texttt{pPXF}. 

\texttt{PyParadise} instead normalizes the stellar continuum and template spectral library before fitting and is therefore much less sensitive to global unphysical continuum variations. While some parameters, such as line-of-sight extinction for the stellar continuum, cannot be determined this way, it allows us to obtain robust stellar kinematics and absorption line equivalent width much closer to the nucleus. The emission lines are fitted to the continuum residuals and we choose single Gaussians coupled in radial velocity and velocity dispersion for all emission lines as a zero-order approximation. A coupling of the lines enhances the robustness of the flux measurements for fainter lines and increases the precision on global gas kinematics. Follow-up analysis in specific regions with complex line kinematics can be performed later on with dedicated methods \citep{Husemann:2019a}.  PyParadise is publicly available at \url{https://git.io/pyparadise} and more details on the algorithm can be found in the user manual. The S/N of individual spaxels in the IFU data can be too low to properly model the stellar continuum and low surface-brightness emission line regions. We therefore employ two different binning strategies as described in the following.

\begin{figure*}
 \centering
 \includegraphics[width=0.85\textwidth]{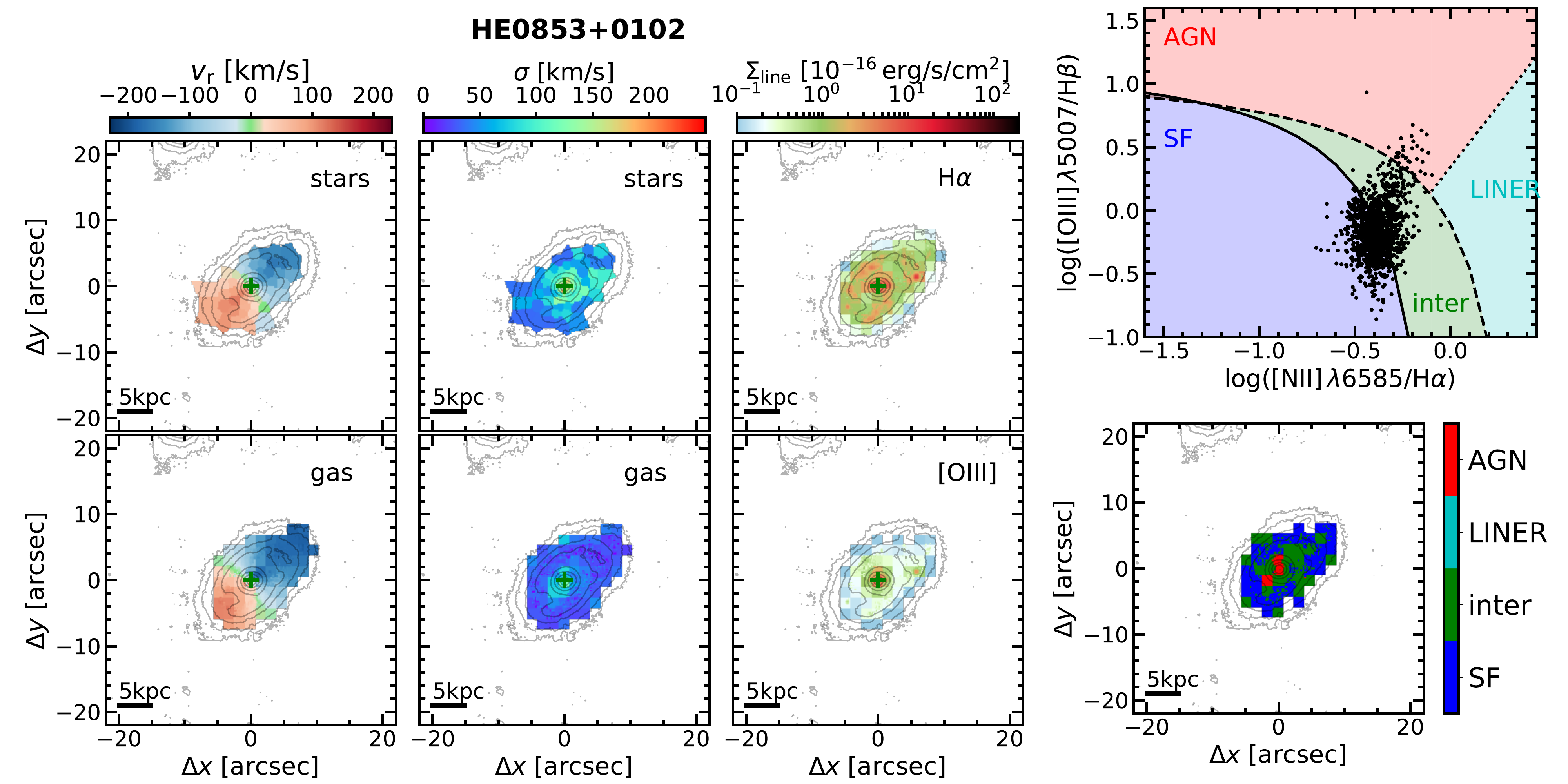}
 \caption{Results of the 2D continuum and emission line modeling of the AGN-subtracted data of HE~0853$+$0102. The radial velocity and velocity dispersion of the stars and gas are shown in the first two columns of the left panel, and the surface brightness maps of the H$\alpha$ and [\ion{O}{iii}] are shown in the right column of the left panel. The classical BPT diagram for all spaxels with S/N$>$3 in all lines are presented in the upper right panel with empirical demarcations curves from \citet{Kauffmann:2003}, \citet{Kewley:2001} and \citet{Stasinska:2008} overplotted as solid, dashed and dotted lines, respectively, defining classifications into star forming (SF), intermediate (inter), Low-Ionization Nuclear Emission Region (LINER) and AGN as shaded areas. The corresponding 2D maps of the classification are shown below the BPT to visualize the ionization distribution.}\label{fig:PyParadise_results}
\end{figure*}

For bright emission line regions we want to retain the native MUSE spatial sampling of 0\farcs2 for which the S/N in the stellar continuum may not be sufficient for the continuum modeling. We therefore perform an initial Voronoi binning to achieve a minimum continuum S/N of 20 per Voronoi cell per spectral pixel at around 6000\AA. We then model the binned stellar continuum spectra with PyParadise using stellar spectra from the INDO-US spectra library \citep{Valdes:2004} and remap the inferred stellar kinematics back to the native MUSE sampling grid. Afterwards we model the stellar continuum in the full cube while keeping the stellar kinematics fixed, so that the emission lines can be fitted to the residual continuum spectra for each individual MUSE spectra. Additionally, we bin the QSO-subtracted datacube by 8$\times$8 pixels for MUSE, 2$\times$2 pixels for PMAS, and 4$\times$4 pixels for VIMOS corresponding to a binned sampling of 1\farcs6, 2\farcs0, and 2\farcs64, respectively.  A full continuum and emission-line modeling with \texttt{PyParadise} is performed on the binned cubes without the intermediate Voronoi binning step for the stellar continuum. The analysis of the binned data allows us to trace the emission lines nearly an order of magnitude fainter in surface brightness than in the unbinned data. 

An example of a single spaxel fitting with PyParadise is shown in Fig.~\ref{fig:PyParadise_fit_example}. All masked regions for the emission-line modeling are indicated as gray shaded areas, whereas regions of strong atmospheric absorption and sky line residuals are highlighted as red shaded areas. We also label all emission lines that we incorporate in the automatic modeling. An important part of this process is to infer proper errors on all measured parameters to identify robust measurements and to set upper limits of undetected emission lines for diagnostics purposes. Errors on the stellar kinematics are inferred using an MCMC approach as part of \texttt{PyParadise} iterative continuum modeling. The errors on all emission-line parameters are inferred combining a bootstrapping and Monte-Carlo approach. The entire modeling of the stellar continuum (at fixed stellar kinematics) and emission-line modeling are repeated 50 times after randomly modifying each spectral pixel within its noise and randomly reducing the number of input template stellar library spectra to 60\%. This way we incorporate some systematic uncertainty of the stellar continuum modeling also in the derived emission line parameters. This is most important for emission lines that are superimposed on stellar absorption lines, mainly H$\beta$ and H$\alpha$.

An example of the inferred 2D kinematics and emission-line distribution is shown in Fig.~\ref{fig:PyParadise_results}. For the emission-line surface brightness maps we combined the results from the unbinned and binned QSO-subtracted data. Assuming a minimum line detection of S/N$>$5, we replace undetected regions in the unbinned data with the surface brightness of the binned data if the S/N limits were achieved for a given binned spaxel.

\begin{table}
\caption{ENLR size measurements}\label{tab:ENLR_sizes}
 \begin{tabular}{lcccccccc}\hline\hline
Object & $R_\mathrm{ENLR,max}$\tablefootmark{(a)} & $R_{\mathrm{ENLR},10^{-16}}$\tablefootmark{(b)} & $R_{\mathrm{ENLR},10^{-15}}$\tablefootmark{(c)}\\
 & [$\arcsec$] & [$\arcsec$] & [$\arcsec$] \\\hline
 HE0021-1819 & $5.3\pm0.6$  & $2.2\pm0.6$ & $1.0\pm0.6$\\
 HE0040-1105 & $11.1\pm0.6$  & $5.1\pm0.6$ & $2.3\pm0.6$\\
 HE0108-4743 & $2.1\pm1.3$  & $2.1\pm1.3$ & $2.1\pm1.3$\\
 HE0114-0015 & $7.4\pm0.5$  & $<0.8$ & <0.8\\
 HE0119-0118 & $5.8\pm0.7$  & $4.7\pm0.7$ & <0.8\\
 HE0203-0031 & $20.7\pm1.1$  & $8.2\pm1.1$ & $3.2\pm1.1$\\
 HE0212-0059 & $29.4\pm0.6$  & $0.9\pm0.6$ & $0.9\pm0.6$\\
 HE0224-2834 & $24.9\pm1.5$  & $23.1\pm1.5$ & $4.6\pm1.5$\\
 HE0227-0913 & $21.2\pm1.0$  & $9.3\pm1.0$ & <0.8\\
 HE0232-0900 & $12.1\pm1.0$  & $8.0\pm1.0$ & $5.1\pm1.0$\\
 HE0253-1641 & $4.6\pm0.8$  & $3.4\pm0.8$ & $3.1\pm0.8$\\
 HE0345+0056 & <0.8  & $<0.8$ & <0.8\\
 HE0351+0240 & $14.4\pm0.7$  & $13.7\pm0.7$ & $10.5\pm0.7$\\
 HE0412-0803 & $28.0\pm0.7$  & $19.0\pm0.7$ & $7.4\pm0.7$\\
 HE0429-0247 & $9.6\pm0.8$  & $6.7\pm0.8$ & $2.2\pm0.8$\\
 HE0433-1028 & $11.1\pm0.5$  & $5.2\pm0.5$ & $1.8\pm0.5$\\
 HE0853+0102 & $3.4\pm0.6$  & $2.0\pm0.6$ & $0.6\pm0.6$\\
 HE0853-0126 & <1.0  & $<1.0$ & <1.0\\
 HE0934+0119 & $7.6\pm0.7$  & $1.0\pm0.7$ & $1.0\pm0.7$\\
 HE0949-0122 & $8.1\pm1.7$  & $8.1\pm1.7$ & $4.0\pm1.7$\\
 HE1011-0403 & $4.5\pm0.7$  & $1.4\pm0.7$ & <0.8\\
 HE1017-0305 & $7.0\pm0.6$  & $3.5\pm0.6$ & <0.8\\
 HE1029-1831 & $9.1\pm0.6$  & $3.3\pm0.6$ & <0.8\\
 HE1107-0813 & $25.0\pm0.6$  & $15.7\pm0.6$ & $1.6\pm0.6$\\
 HE1108-2813 & $4.9\pm0.4$  & $2.3\pm0.4$ & $1.6\pm0.4$\\
 HE1126-0407 & $9.2\pm0.7$  & $3.7\pm0.7$ & $2.3\pm0.7$\\
 HE1237-0504 & $12.7\pm0.5$  & $12.3\pm0.5$ & $1.7\pm0.5$\\
 HE1248-1356 & $12.3\pm0.5$  & $7.7\pm0.5$ & $3.2\pm0.5$\\
 HE1310-1051 & <0.7  & $<0.7$ & <0.7\\
 HE1330-1013 & $6.8\pm0.7$  & $2.4\pm0.7$ & $0.6\pm0.7$\\
 HE1338-1423 & $10.0\pm1.4$  & $8.7\pm1.4$ & $5.1\pm1.4$\\
 HE1353-1917 & $35.0\pm0.7$  & $35.0\pm0.7$ & $5.3\pm0.7$\\
 HE1417-0909 & $6.6\pm0.7$  & $4.7\pm0.7$ & $0.6\pm0.7$\\
 HE2128-0221 & $1.3\pm0.7$  & $1.3\pm0.7$ & $1.0\pm0.7$\\
 HE2211-3903 & $6.7\pm0.4$  & $3.4\pm0.4$ & $1.0\pm0.4$\\
 HE2222-0026 & $3.9\pm0.6$  & $<0.8$ & <0.8\\
 HE2233+0124 & $22.2\pm0.8$  & $2.2\pm0.8$ & <0.8\\
 HE2302-0857 & $30.0\pm0.6$  & $8.3\pm0.6$ & $6.1\pm0.6$\\
\noalign{\smallskip}\hline
\end{tabular}
\tablefoot{
\tablefoottext{a}{Maximum projected distance of all robust AGN-ionized regions with respect to the AGN location as seen in each IFU observation.}
\tablefoottext{b}{Maximum projected distance of robust AGN-ionized regions with respect to the AGN location down to an intrinsic [\ion{O}{iii}] surface brightness limit of $\Sigma_{16}=10^{-16}\mathrm{erg\,s}^{-1}\mathrm{cm}^{-2}\,\mathrm{arcsec}^{-2}(1+z)^{-4}$.}
\tablefoottext{c}{Maximum projected distance of robust AGN-ionized regions with respect to the AGN location down to an intrinsic [\ion{O}{iii}] surface brightness limit of $\Sigma_{15}=10^{-15}\mathrm{erg\,s}^{-1}\mathrm{cm}^{-2}\,\mathrm{arcsec}^{-2}(1+z)^{-4}$.}
}

\end{table}

\subsubsection{Ionized gas excitation conditions and ENLR properties}
It is a common approach to map the excitation conditions of the ionized gas across the AGN host galaxies. The Baldwin-Phillips-Terlevich \citep[BPT,][]{Baldwin:1981} diagram comparing the [\ion{O}{iii}]/H$\beta$ versus [\ion{N}{ii}]/H$\alpha$ line ratios has been extensively used at optical wavelengths for this task \citep[e.g.,][]{Veilleux:1987,Kewley:2001,Kauffmann:2003,Kewley:2006,CidFernandes:2011}. We show the overall PyParadise results on stellar kinematics and emission line properties  for the CARS data of HE~0853$+$0102 in Fig.~\ref{fig:PyParadise_results} together with the derived BPT diagram and associated classification. The full gallery of figures for all CARS targets can be found in the Appendix~\ref{apx:maps_all}. The BPT diagram is one way to distinguish between different sources of ionized gas excitation, such as ionization by young stars in star-forming sides often referred to as \ion{H}{ii} regions,  the photoionization by the hard radiation field of an AGN and low-ionization nuclear emission-line regions (LINERs). Clearly distinguishing between such mechanisms is not trivial, but several demarcations lines have been proposed to discriminate between those sources. 

Given the redshift of our targets, the spatial resolution is limited to a few 100\,pc, which does not allow to see individual line emitting clouds. Mixing of different ionization mechanisms along the line-of-sight is an additional complication. This is evident in Fig.~\ref{fig:PyParadise_results} where the majority of spaxels are consistent with ionization by star forming regions, which has a smooth  extension toward the AGN location in the BPT. The 2D spatial mapping of the excitation confirms that the higher ionization is found closer to the AGN. This is known as the SF-AGN mixing-sequence \citep[e.g.,][]{Davies:2014b, Richardson:2014, Davies:2016} and is important to take into account when inferring the current star formation using the H$\alpha$ line powered solely by \ion{H}{ii} regions. Such a detailed analysis will be presented in the companion paper \citep{Smirnova-Pinchukova:2021} and has been applied for a subsample of galaxies \citep{Neumann:2019}. Here, we focus our investigation on the characterization of clouds significantly influenced by the AGN photoionization, which can be unambiguously associated with the ENLR.

From the excitation maps as shown in Fig.~\ref{fig:PyParadise_results}, we determine the maximum distance of AGN-ionized regions ($R_\mathrm{ENLR,max}$) with respect to the AGN location that we can cover with our depth and area.  To suppress the impact of noise on misclassification close to the borders of the demarcation lines, we require that at least six surrounding unbinned spaxels share the same AGN-ionization classification or the classification is present in the spatially binned data with its intrinsically higher S/N. We need to exclude the LINER and the intermediate BPT regions, because those are not necessarily associated with AGN photoionization and can originate from post-AGN stars \citep[e.g.,][]{Binette:1994,Singh:2013}, shocked gas from stellar winds \citep[e.g.,][]{Ho:2016,Lopez-Coba:2019}, or the diffuse ionized gas inbetween star forming regions \citep[e.g.,][]{Lacerda:2018,Levy:2019}.  We define the maximum ENLR size, $R_\mathrm{ENLR,max}$, as the maximum projected distance of a robust AGN region to the AGN location detectable within the IFU FoV at the observational depth. In general, the maximum ENLR extent can be biased due to surface brightness dimming and observational depth. This is not a concern for the large FoV of MUSE and the narrow redshift distribution of the CARS sample. Nevertheless, we also determine two additional ENLR radii, $R_\mathrm{ENLR,15}$ and $R_\mathrm{ENLR,16}$, corresponding to the ENLR size out to a surface brightness limit of $\Sigma_{15}=10^{-15}\mathrm{erg\,s}^{-1}\mathrm{cm}^{-2}\,\mathrm{arcsec}^{-2}(1+z)^{-4}$ \citep[e.g][]{Liu:2013,Hainline:2013,Liu:2014,Hainline:2014}, or $\Sigma_{16}=10^{-16}\mathrm{erg\,s}^{-1}\mathrm{cm}^{-2}\,\mathrm{arcsec}^{-2}(1+z)^{-4}$ \citep{Chen:2019b}, respectively. All these measurements are listed in Table~\ref{tab:ENLR_sizes}. The ENLR size can become unresolved for the CARS targets, in particular, at high intrinsic ENLR surface brightnesses. For such nondetections we set the upper limits for the ENLR size to be the angular size of the binned spaxels. The error on the sizes are generally driven by the spatial resolution of a given observations, so that we assume the error on all sizes to be the FWHM of seeing for each observation as reported in Table~\ref{tab:observations}.

\subsection{AGN characteristics}
\subsubsection{AGN spectral fitting}
The IFU data also provide high S/N unobscured AGN spectra which allow us to characterize various AGN parameters. We specifically model the AGN spectra in the wavelength range covering the broad H$\beta$ line for which many import calibrations have been established in the literature \citep[e.g.,][]{Kaspi:2000,Peterson:2004,Greene:2005,Vestergaard:2006,Denney:2009,Bentz:2013,Woo:2015,Vietri:2020}. As a first step we subtract the best-fit stellar continuum model datacube, determined via \texttt{PyParadise} after QSO-host deblending, from the original observations to remove any continuum signal from the stars without introducing additional noise. Afterwards we extract integrated AGN spectra within an aperture of 3\arcsec\ diameter centered on the AGN position. 

As the blue part of the optical spectrum below $<$4700\AA\ is not covered by the MUSE data, a full spectrum fitting of the optical AGN spectra usually performed for large AGN surveys \citep[e.g.,][]{Shen:2008b, Park:2012, Coffey:2019} is not ideal in our case. In particular, the use of \ion{Fe}{ii} templates \citep[e.g.,][]{Boroson:1992, Veron-Cetty:2004,Kovacevic:2010} often leads to bad fits for strong \ion{Fe}{ii} emitters in our high S/N MUSE spectra and quality was deemed insufficient for several individual cases. Furthermore, the \ion{Fe}{ii} blends on the blue side of H$\beta$ are not covered in most of our IFU observations which reduces the robustness of the template fitting. Therefore, we decided to model only the rest-frame wavelength from 4750\AA\ to 5100\AA\ and adopt a linear pseudo-continuum for this narrow wavelength range as described in \citet{Singha:2021}. For the broad and narrow emission lines we adopted super-positions of Gaussian line profiles. More specifically we adopted one or two Gaussians for the broad H$\beta$ line and the two prominent isolated \ion{Fe}{ii} $\lambda\lambda4923,5018$ lines originating both from the broad-line region of the AGN. The narrow lines of [\ion{O}{iii}]\,$\lambda\lambda4960,5007$ and H$\beta$ are commonly modeled as two Gaussian components to account for the typical asymmetry of those lines in AGN \citep[e.g.,][]{Greene:2005b, Mullaney:2013,Harrison:2016b,Bennert:2018}. To keep the modeling robust and reduce the number of free parameters, all broad and narrow line components are coupled in redshift and kinematics across line transitions. Additionally, the [\ion{O}{iii}] doublet line ratio is constrained to be 3 \citep{Storey:2000} and we fix the flux ratio \ion{Fe}{ii} $\lambda4923$/\ion{Fe}{ii}\,$\lambda5018=0.81$, which is the mean of the empirically measured ratios when leaving the fluxes unconstrained.

\begin{figure}
 \resizebox{\hsize}{!}{\includegraphics{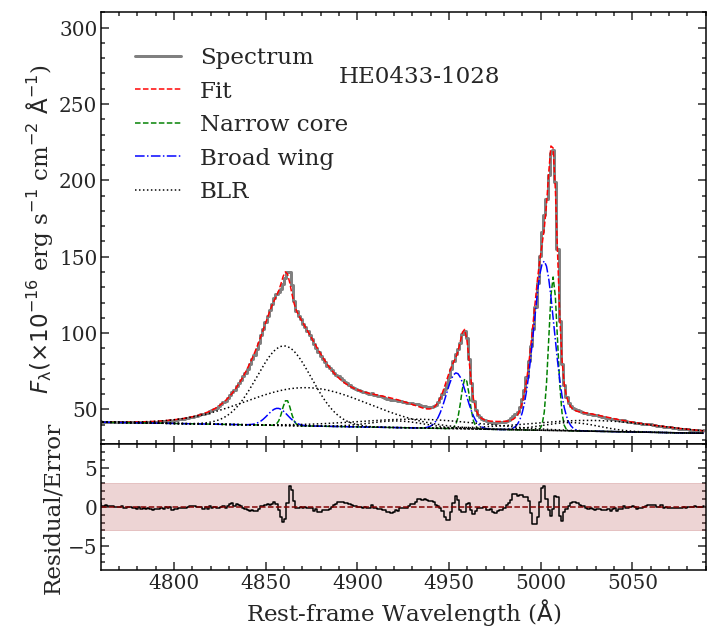}}
 \caption{Example of the AGN spectral modeling for the case of HE~0433$-$1028. The modeled wavelength range is limited to the rest-frame wavelength range 4750\AA--5100\AA\ covering the prominent H$\beta$ and [\ion{O}{iii}]\,$\lambda\lambda4960,5007$ emission lines. The spectrum with the full best-fit model and various line components for the BLR and the narrow and core component for the NLR of H$\beta$ and [\ion{O}{iii}] are individually shown with different line styles and colors. The residual spectrum and the 5 $\sigma$ limiting band are shown in the lower panel.}\label{fig:AGN_spectra_fit}
\end{figure}

An example of the modeling is shown in Fig.~\ref{fig:AGN_spectra_fit} and entire sample fits are presented in \citet{Singha:2021}, where we focus on the line shape of [\ion{O}{iii}]. For the purpose of this paper, we are only interested in measurements that are relevant to estimate bolometric luminosities, BH mass, etc., so we list only the total line fluxes of the broad H$\beta$ and \ion{Fe}{ii} lines together with their line widths as well as the total [\ion{O}{iii}] flux in Table~\ref{tab:QSO_par}. Errors are determined through a Monte Carlo approach as described in \citet{Singha:2021}, but a 10\% systematic uncertainty is added on the line fluxes to take into account the absolute photometric calibration uncertainty of the IFU observations. 

\begin{figure}
\resizebox{\hsize}{!}{\includegraphics{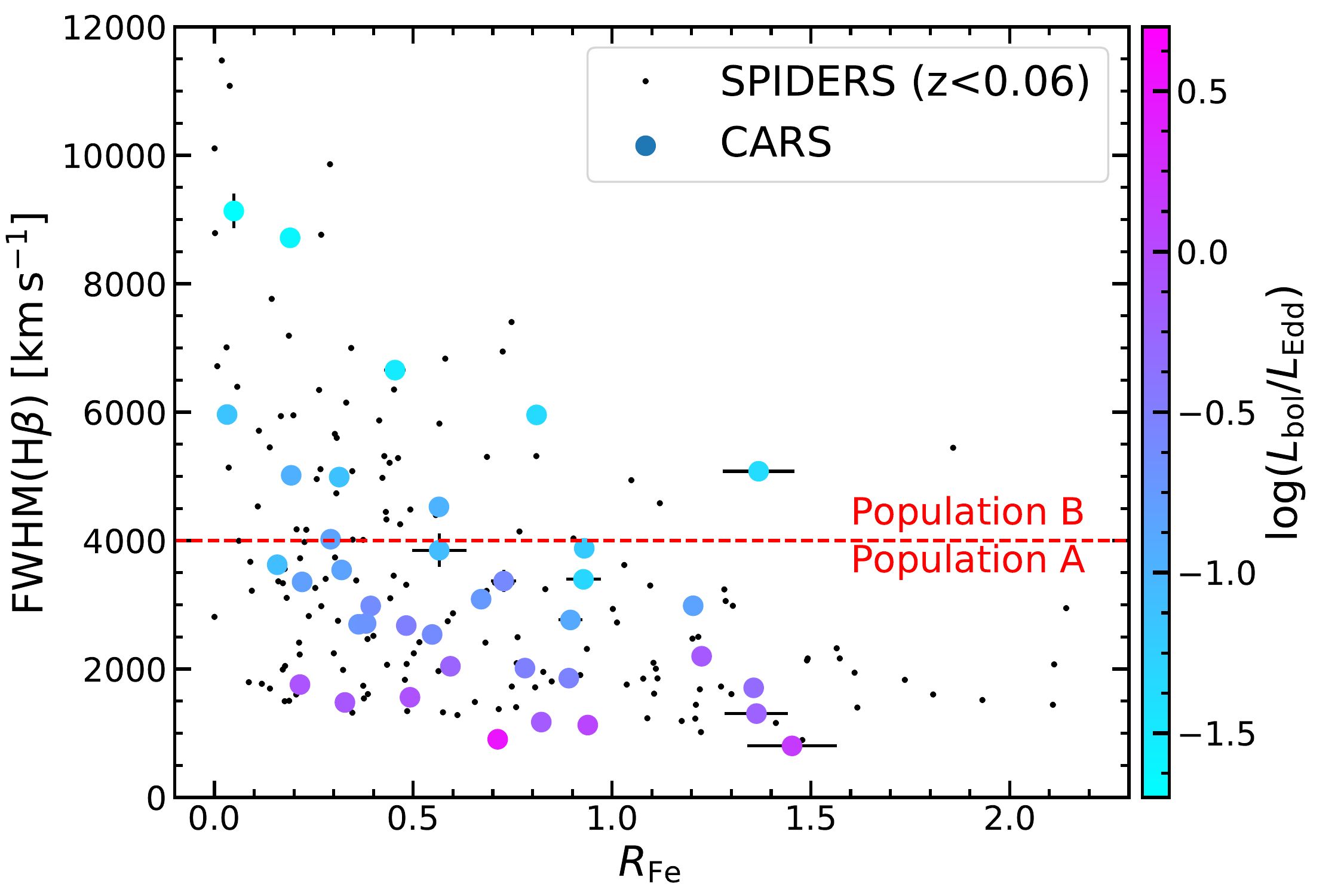}}\\
 \resizebox{\hsize}{!}{\includegraphics{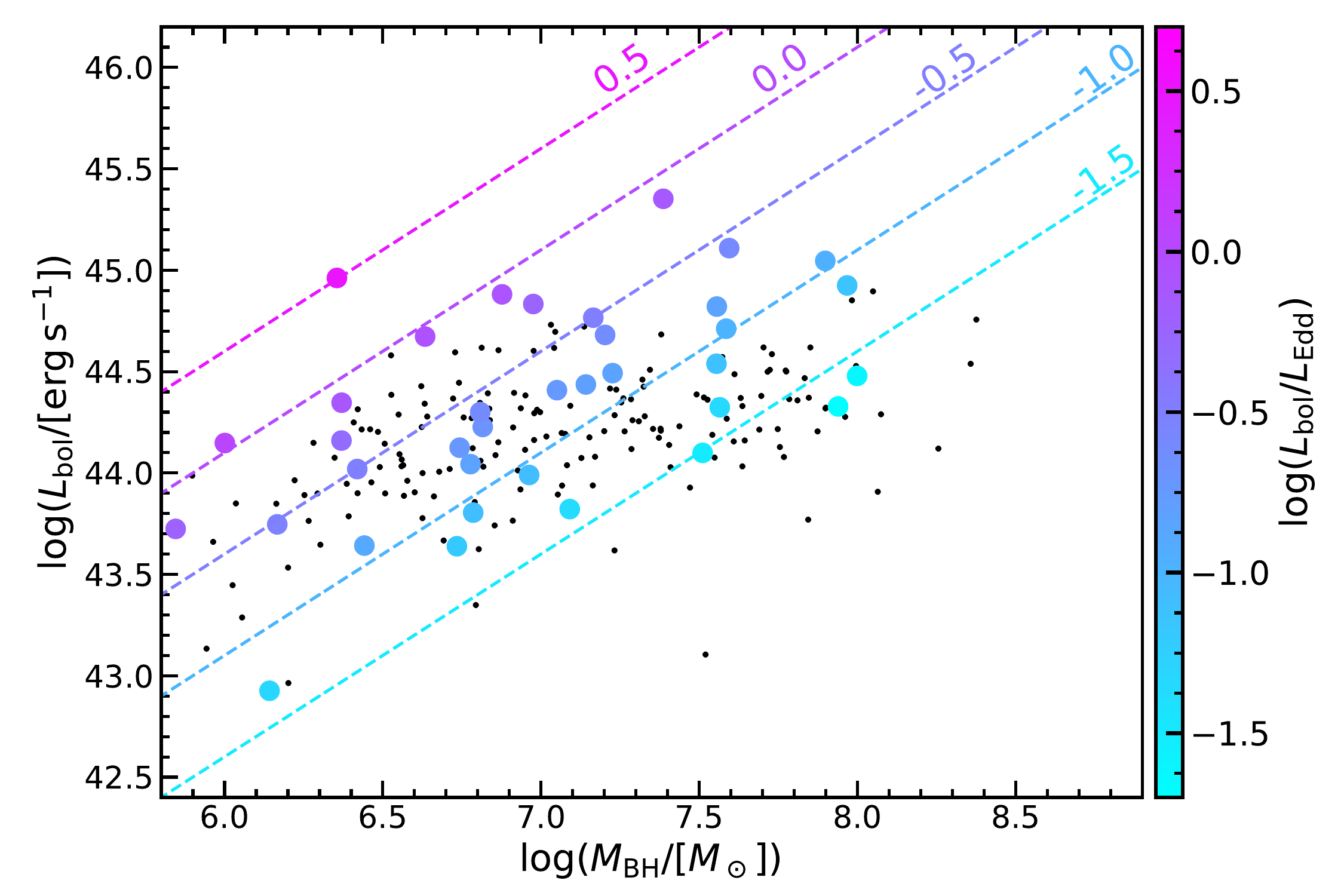}}
 \caption{Primary AGN parameters and Eigenvector 1 population distribution of CARS in comparison to 170 AGN at $z<0.06$ from the X-ray selected SPIDERS AGN sample \citep{Coffey:2019}. \textit{Upper panel: } Broad H$\beta$ FWHM against the \ion{Fe}{ii} strength as $R_{\ion{Fe}{ii}} = f(\ion{Fe}{ii}(4434-4685))/f(\mathrm{BLR\ H}\beta)$. The CARS sample is further color-coded with the Eddington ratio. The formal division between population A and B at H$\beta$ FWHM of 4000\,$\mathrm{km\,s}^{-1}$ is indicated by the dashed line. \textit{Lower panel:} Bolometric luminosity $L_\mathrm{bol}$ against $M_\mathrm{BH}$. Lines of constant Eddington ratios are indicated by the dashed line to guide the eye.} \label{fig:CARS_population}
\end{figure}

\begin{table*}
 \caption{AGN properties from the modeling of the H$\beta$, [\ion{O}{iii}] and \ion{Fe}{ii} spectral region}\label{tab:QSO_par}
 \begin{tabular}{lccccccccccc}\hline\hline
Object & $f(\mathrm{H}\beta)$\tablefootmark{(a)} & $f(\mathrm{FeII})$\tablefootmark{(b)} & $f(\mathrm{OIII})$\tablefootmark{(c)} & $\mathrm{FWHM}_{\mathrm{H}\beta}$\tablefootmark{(d)} & $\sigma_{\mathrm{H}\beta}$\tablefootmark{(e)}   & $M_\mathrm{BH}$\tablefootmark{(f)} & $L_\mathrm{bol}$\tablefootmark{(g)} & $\lambda_\mathrm{Edd}$\tablefootmark{(h)}\smallskip\\\cline{2-4}
 & \multicolumn{3}{c}{$10^{-16}\mathrm{erg}\,\mathrm{s}^{-1}\mathrm{cm}^{-2}$} & [$\mathrm{km}\,\mathrm{s}^{-1}$] & [$\mathrm{km}\,\mathrm{s}^{-1}$] & [$10^6M_\sun$] & [$10^{43}\mathrm{erg}\,\mathrm{s}^{-1}$] & \smallskip\\\hline
HE0021-1810 & $95\pm5$ &  $90\pm3$ & $185\pm1$ & $5079\pm65$ &  $2194\pm28$  & $12.3\pm3.2$ &  $6.6\pm0.7$ &  $0.04\pm0.01$\\
HE0021-1819 & $92\pm0$ &  $10\pm1$ & $289\pm0$ & $3624\pm22$ &  $1539\pm9$  & $6.1\pm1.3$ &  $6.4\pm0.6$ &  $0.08\pm0.02$\\
HE0040-1105 & $349\pm1$ &  $88\pm2$ & $501\pm0$ & $2696\pm6$ &  $1144\pm2$  & $5.5\pm0.8$ &  $13.4\pm1.3$ &  $0.19\pm0.04$\\
HE0108-4743 & $307\pm2$ &  $174\pm4$ & $283\pm1$ & $1174\pm19$ &  $934\pm24$  & $0.5\pm0.1$ &  $4.3\pm0.4$ &  $0.69\pm0.15$\\
HE0114-0015 & $83\pm1$ &  $52\pm2$ & $61\pm0$ & $2764\pm39$ &  $1174\pm17$  & $2.8\pm0.6$ &  $4.4\pm0.4$ &  $0.13\pm0.03$\\
HE0119-0118 & $863\pm3$ &  $235\pm4$ & $746\pm1$ & $2983\pm31$ &  $1861\pm11$  & $15.9\pm2.4$ &  $48.0\pm4.8$ &  $0.24\pm0.04$\\
HE0212-0059 & $1524\pm83$ &  $52\pm1$ & $560\pm4$ & $9133\pm228$ &  $4329\pm120$  & $86.8\pm17.5$ &  $21.3\pm2.1$ &  $0.02\pm0.00$\\
HE0224-2834 & $488\pm2$ &  $106\pm2$ & $386\pm1$ & $4989\pm20$ &  $2119\pm9$  & $35.8\pm3.7$ &  $34.6\pm3.5$ &  $0.08\pm0.01$\\
HE0227-0913 & $2658\pm5$ &  $1719\pm4$ & $442\pm1$ & $1127\pm5$ &  $1025\pm4$  & $1.0\pm0.2$ &  $14.2\pm1.4$ &  $1.12\pm0.27$\\
HE0232-0900 & $3663\pm5$ &  $491\pm5$ & $2820\pm2$ & $5017\pm7$ &  $2131\pm3$  & $79.1\pm9.8$ &  $111.2\pm11.1$ &  $0.11\pm0.02$\\
HE0253-1641 & $673\pm4$ &  $632\pm4$ & $890\pm2$ & $1707\pm10$ &  $1376\pm13$  & $2.3\pm0.3$ &  $14.5\pm1.5$ &  $0.49\pm0.08$\\
HE0345+0056 & $5929\pm31$ &  $2901\pm7$ & $2140\pm7$ & $906\pm8$ &  $1141\pm8$  & $2.3\pm0.2$ &  $92.5\pm9.2$ &  $3.22\pm0.31$\\
HE0351+0240 & $643\pm1$ &  $170\pm4$ & $576\pm1$ & $2710\pm7$ &  $1228\pm4$  & $6.6\pm1.1$ &  $16.9\pm1.7$ &  $0.20\pm0.05$\\
HE0412-0803 & $2579\pm836$ &  $523\pm3$ & $6880\pm200$ & $4022\pm1258$ &  $1709\pm36$  & $35.9\pm16.9$ &  $66.2\pm6.6$ &  $0.15\pm0.15$\\
HE0429-0247 & $616\pm2$ &  $140\pm2$ & $331\pm0$ & $1479\pm23$ &  $1390\pm7$  & $2.3\pm0.4$ &  $22.3\pm2.2$ &  $0.75\pm0.15$\\
HE0433-1028 & $2640\pm7$ &  $883\pm4$ & $1389\pm2$ & $2676\pm14$ &  $1667\pm7$  & $14.6\pm1.7$ &  $58.4\pm5.8$ &  $0.32\pm0.06$\\
HE0853+0102 & $154\pm26$ &  $61\pm2$ & $124\pm5$ & $3850\pm1557$ &  $1571\pm493$  & $9.2\pm4.3$ &  $9.8\pm1.0$ &  $0.08\pm0.10$\\
HE0853-0126 & $40\pm1$ &  $40\pm2$ & $15\pm1$ & $794\pm32$ &  $1079\pm105$  & $0.2\pm0.0$ &  $3.8\pm0.4$ &  $1.45\pm0.32$\\
HE0934+0119 & $985\pm1$ &  $335\pm2$ & $363\pm0$ & $1559\pm6$ &  $1053\pm5$  & $4.3\pm0.6$ &  $47.3\pm4.7$ &  $0.87\pm0.13$\\
HE0949-0122 & $582\pm23$ &  $550\pm5$ & $735\pm3$ & $1300\pm22$ &  $1959\pm17$  & $0.7\pm0.2$ &  $5.3\pm0.5$ &  $0.61\pm0.17$\\
HE1011-0403 & $1111\pm2$ &  $457\pm4$ & $325\pm1$ & $2043\pm6$ &  $1104\pm6$  & $9.5\pm1.1$ &  $68.3\pm6.8$ &  $0.57\pm0.11$\\
HE1017-0305 & $1167\pm5$ &  $457\pm5$ & $206\pm1$ & $4525\pm20$ &  $1925\pm8$  & $38.5\pm4.1$ &  $51.5\pm5.2$ &  $0.11\pm0.01$\\
HE1029-1831 & $594\pm19$ &  $225\pm4$ & $377\pm1$ & $2538\pm20$ &  $1293\pm38$  & $6.4\pm0.8$ &  $20.0\pm2.0$ &  $0.25\pm0.04$\\
HE1107-0813 & $2273\pm124$ &  $1147\pm4$ & $108\pm6$ & $3369\pm219$ &  $1817\pm26$  & $39.3\pm7.6$ &  $128.6\pm12.9$ &  $0.26\pm0.05$\\
HE1108-2813 & $890\pm2$ &  $743\pm8$ & $480\pm1$ & $2985\pm9$ &  $1267\pm4$  & $6.0\pm1.1$ &  $11.1\pm1.1$ &  $0.15\pm0.03$\\
HE1126-0407 & $4056\pm7$ &  $3443\pm5$ & $668\pm2$ & $2199\pm8$ &  $1404\pm4$  & $24.4\pm4.2$ &  $225.5\pm22.6$ &  $0.73\pm0.16$\\
HE1237-0504 & $2625\pm4$ &  $1694\pm3$ & $1697\pm2$ & $3880\pm7$ &  $1651\pm3$  & $5.4\pm1.0$ &  $4.4\pm0.4$ &  $0.06\pm0.01$\\
HE1248-1356 & $136\pm2$ &  $88\pm4$ & $433\pm1$ & $3397\pm54$ &  $1443\pm23$  & $1.4\pm0.4$ &  $0.8\pm0.1$ &  $0.05\pm0.02$\\
HE1310-1051 & $1363\pm3$ &  $302\pm3$ & $1185\pm1$ & $3540\pm8$ &  $1503\pm4$  & $16.8\pm1.6$ &  $31.1\pm3.1$ &  $0.15\pm0.02$\\
HE1330-1013 & $464\pm3$ &  $286\pm5$ & $136\pm1$ & $1858\pm20$ &  $1357\pm24$  & $1.5\pm0.4$ &  $5.6\pm0.6$ &  $0.30\pm0.09$\\
HE1338-1423 & $2583\pm20$ &  $385\pm2$ & $1535\pm4$ & $1751\pm35$ &  $1167\pm11$  & $7.5\pm1.0$ &  $76.2\pm7.6$ &  $0.81\pm0.15$\\
HE1353-1917 & $472\pm22$ &  $149\pm3$ & $392\pm1$ & $6655\pm41$ &  $2491\pm106$  & $32.4\pm4.7$ &  $12.6\pm1.3$ &  $0.03\pm0.01$\\
HE1417-0909 & $727\pm29$ &  $111\pm3$ & $611\pm42$ & $3356\pm13$ &  $1425\pm5$  & $13.9\pm2.2$ &  $27.4\pm2.7$ &  $0.16\pm0.03$\\
HE2128-0221 & $166\pm1$ &  $90\pm2$ & $194\pm0$ & $2014\pm18$ &  $856\pm8$  & $2.6\pm0.4$ &  $10.5\pm1.0$ &  $0.32\pm0.06$\\
HE2211-3903 & $659\pm8$ &  $371\pm6$ & $298\pm1$ & $5958\pm59$ &  $2533\pm25$  & $36.7\pm5.7$ &  $21.1\pm2.1$ &  $0.05\pm0.01$\\
HE2222-0026 & $372\pm2$ &  $173\pm2$ & $85\pm1$ & $3087\pm20$ &  $1542\pm7$  & $11.2\pm1.7$ &  $25.7\pm2.6$ &  $0.18\pm0.03$\\
HE2233+0124 & $470\pm17$ &  $62\pm3$ & $105\pm2$ & $8715\pm152$ &  $3975\pm101$  & $99.7\pm12.9$ &  $30.1\pm3.0$ &  $0.02\pm0.00$\\
HE2302-0857 & $2214\pm35$ &  $49\pm4$ & $2701\pm134$ & $5963\pm112$ &  $2172\pm26$  & $92.8\pm13.0$ &  $84.2\pm8.4$ &  $0.07\pm0.02$\\
\noalign{\smallskip}\hline
\end{tabular}
\tablefoot{
\tablefoottext{a}{Integrated BLR component flux of H$\beta$ excluding the NLR contribution.} 
\tablefoottext{b}{Integrated flux of the \ion{Fe}{ii}(5100--5550\AA) band.} 
\tablefoottext{c}{Integrated NLR component flux of [\ion{O}{iii}]}.
\tablefoottext{d}{FWHM of the H$\beta$ BLR component.} 
\tablefoottext{e}{Line dispersion of the H$\beta$ BLR component.} 
\tablefoottext{f}{BH mass estimated from the BLR H$\beta$  luminosity and FWHM following \citet{Greene:2005}, see Eq.~\ref{eq:BH_mass}. Errors are based on the propagation of measurement errors and uncertainties of the scaling relation. A systematic error of 0.3dex is typically assumed for the precision of single-epoch BH mass estimate for a single object.}
\tablefoottext{g}{Bolometric luminosity estimated from the H$\beta$ luminosity (see text for details). Errors are dominated by the systematic uncertainty of the absolute flux calibration which is assumed to be 10 per cent.} 
\tablefoottext{h}{Eddington ratio defined as $L_\mathrm{bol}/L_\mathrm{Edd}$.} 
}

\end{table*}

\subsubsection{AGN parameter estimation}
The main AGN parameters we are interested in are the bolometric luminosity, the BH mass and the Eddington ratio which are easy to obtain from the unobscured AGN spectra. Bolometric correction factors have been estimated for the AGN continuum light at 5100\AA\, such as $L_\mathrm{bol} \sim (8$--$12)\times \lambda L_{5100}$ \citep[e.g.,][]{Kaspi:2000,Richards:2006,Runnoe:2012b,Netzer:2019} or the broad emission lines \citep[e.g.,][]{Greene:2005,Stern:2012a,Shen:2012,Jun:2015} as those optical features should  directly respond to the AGN accretion disk luminosity. Here, we use the broad H$\beta$ line luminosity to estimate the bolometric luminosity, because the QSO-host deblending adds a systematic uncertainty to the continuum luminosity at 5100\,\AA. \citet{Greene:2005} empirically calibrated the following relation between  $L_{5100}$ and $L_{H\beta}$ as
\begin{equation}
 \frac{L_{H\beta}}{\mathrm{erg\,s}^{-1}} = (1.425\pm0.007)\times10^{42}\left(\frac{L_{5100}}{10^{44}\mathrm{erg\,s}^{-1}}\right)^{1.133\pm0.005}
\end{equation}
from which we computed $L_\mathrm{bol}$, listed in Table~\ref{tab:QSO_par}, adopting a bolometric correction factor of 10 for $L_{5100}$ consistent with \citet{Richards:2006}.

Single-epoch BH masses are based on the empirical BLR size-luminosity relation and a virial factor $f$ that captures assumptions on the detailed  BLR geometry in order to convert the observed broad-line width into a gravitational potential. Different size-luminosity relations \citep[e.g.,][]{Kaspi:2000,Bentz:2009b,Bentz:2013}, virial factors \citep[e.g.,][]{Onken:2004, Woo:2013, Pancoast:2014}, and BLR line width definitions, such as FWHM versus line dispersion \citep[e.g.,][]{Collin:2006}, have been used in the literature which makes it hard to compare different samples. We adopt the relation of \citet{Greene:2005} to compute BH mass 
\begin{equation}
 \frac{M_\mathrm{BH}}{M_\sun} = (3.6\pm0.2)\times 10^6 \left(\frac{L_{\mathrm{H}\beta}}{10^{42}\mathrm{erg\,s}^{-1}}\right)^{0.56\pm0.02}\left(\frac{\mathrm{FWHM}}{10^3\mathrm{km\,s}^{-1}}\right)^2\quad ,\label{eq:BH_mass}
\end{equation}
which provides one specific single-epoch BH mass estimate. In Table~\ref{tab:QSO_par}, we provide all measurements necessary to re-calculate the BH mass using different calibrations when needed for comparison purposes. Beside the measurement errors, a systematic uncertainty of 0.3\,dex is typically assumed for a single-epoch BH mass estimate of a single object. We also compute the corresponding Eddington luminosity as $L_\mathrm{Edd}/[\mathrm{erg\,s}^{-1}] = 1.26\times10^{38}M_\mathrm{BH}/[M_\sun]$ and define the Eddington ratio $\lambda=L_\mathrm{bol}/L_\mathrm{Edd}$.

\begin{figure*}
 \includegraphics[width=\textwidth]{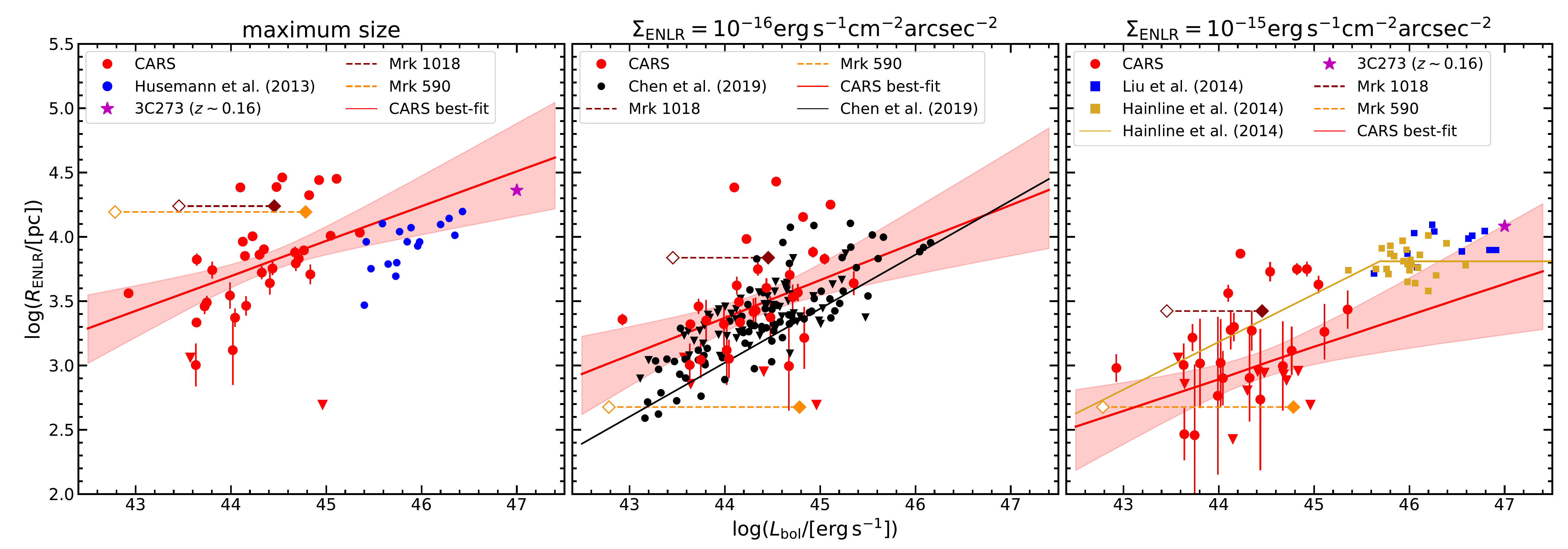}
 \caption{ENLR size against AGN bolometric luminosity for three different ENLR size definition. The maximum size (left panel), the ENLR out to an intrinsic limiting surface brightness of $\Sigma_\mathrm{16}=10^{-16}/(1+z)^4\,\mathrm{erg\,s}^{-1}\mathrm{cm}^{-2}\mathrm{arcsec}^{-2}$ (middle panel) and to $\Sigma_\mathrm{15}=10^{-15}/(1+z)^4\,\mathrm{erg\,s}^{-1}\mathrm{cm}^{-2}\mathrm{arcsec}^{-2}$ (right panel). The measurements for the famous changing-look AGN Mrk~590 \citep{Denney:2014} and Mrk~1018 \citep{McElroy:2016, Krumpe:2017} are included as horizontal dashed lines to highlight the range in luminosity with filled and open symbols marking the maximum and minimum brightness. In the left panel we add measurements on maximum ENLR sizes on unobscured QSOs from \citet{Husemann:2013a} and 3C~273 \citet{Husemann:2019b}. In the middle panel we include ENLR size measurements at $\Sigma_\mathrm{16}$ for a large sample of obscured AGN from the MaNGA survey presented by \citet{Chen:2019b}. For the ENLR size at $\Sigma_\mathrm{15}$, we included data for unobscured QSO from \citet{Liu:2014}, obscured QSOs from \citet{Hainline:2014} and the additional data point for 3C~273 \citep{Husemann:2019b}. Literature ENLR scaling relations as well as the best-fit linear relation to the CARS data are shown as solid lines. The red shaded band indicates the 1$\sigma$ confidence band for the ENLR relation of CARS.}\label{fig:ENLR_size_overview}
\end{figure*}

\subsubsection{AGN and Eigenvector 1 parameter space}
One of the major reasons to focus on type 1 AGN for CARS is that the primary AGN parameters such as $M_\mathrm{BH}$, $L_\mathrm{bol}$  and $\lambda$ can be directly inferred from the AGN spectra, which are often difficult to constrain for type 2 AGN without additional assumptions. One intrinsic correlation of AGN parameters has been the famous Eigenvector 1 parameter space \citep[e.g.,][]{Boroson:1992,Sulentic:2000}, which shows that AGN spectra exhibit a striking correlation between [\ion{O}{iii}] peak height and H$\beta$ FWHM as well as an anticorrelation between the \ion{Fe}{ii} band strength and H$\beta$ FWHM. These trends have been confirmed to be tightly correlated with the Eddington ratio of the AGN as the FWHM has a much stronger impact in the BH mass determination than the luminosity. Therefore, we explore whether the CARS sample is representative also across this important parameter space.

Usually the \ion{Fe}{ii} band flux is measured from the blue side of H$\beta$ between 4434--4685\AA, but is not covered by MUSE. We measure the \ion{Fe}{ii} flux by fitting the wavelength region between 5070--5520\AA  with the \ion{Fe}{ii} templates from \citet{Kovacevic:2010}. A first-order polynomial continuum sampled at both sides of the \ion{Fe}{ii} blend is subtracted beforehand and the emission lines [\ion{N}{i}]\,$\lambda$5199, [\ion{Fe}{xiv}] \,$\lambda$5302 and \ion{He}{i}\,5411 are masked to avoid a potential contamination. During the fit, the individual groups defined in the \ion{Fe}{ii} templates are scaled in flux per group, while all groups are forced to share the same line-of-sight velocity and line width. The total \ion{Fe}{ii} flux was derived by integrating the resulting  model between 5100\AA\ and 5500\AA. The measurement was repeated 100 times after randomly modulating the input spectrum with the error spectrum. The \ion{Fe}{ii} flux values and errors are reported in Table~\ref{tab:QSO_par} which correspond to the mean and standard deviation of the 100 measurements. This needs to be converted to the \ion{Fe}{ii}(4434--4685\AA) flux in order to compare with literature data.  The mean ratio between the bands was measured to be $\frac{\ion{Fe}{ii}(4434-4685\AA)}{\ion{Fe}{ii}(5100-5550\AA)}=1.44\pm0.55$ by \citet{Kovacevic:2010}. We scale our flux accordingly and present the FWHM of H$\beta$ against $R_\mathrm{Fe}=\frac{\ion{Fe}{ii}(4434-4685\AA)}{\mathrm{H}\beta}$ for the CARS sample in the upper panel of Fig.~\ref{fig:CARS_population}. A classification of the AGN into a population A (H$\beta$ FWHM $<$ 4000\,$\,\mathrm{km\,s}^{-1}$)  and B (H$\beta$ FWHM $>$ 4000\,$\,\mathrm{km\,s}^{-1}$)  has been introduced by \citet{Sulentic:2000} based on radio properties and further refined by \citet{Sulentic:2002} into several subcategories.

We compare the distribution of CARS with AGN from the SPIDERS \citep[SPectroscopic IDentification of eROSITA Sources][]{Dwelly:2017} survey, which is an SDSS-IV wide-area spectroscopic follow-up program of X-ray selected  source from \textit{ROSAT} and \textit{XMM-Newton}. The modeling of the spectra and resulting catalog of parameters are reported in \citet{Coffey:2019}. About $\sim170$ AGN from their catalog are at $z<0.06$ consistent with our CARS sample selection. According to Fig.~\ref{fig:CARS_population}, the CARS sample is lacking extreme iron emitters at $R_\mathrm{Fe}>1.5$. Nevertheless, the trend of increasing Eddington ratio as FWHM is decreasing and $R_\mathrm{Fe}$ is increasing can be recovered. The bolometric luminosity against BH mass is presented in the lower panel of Fig.~\ref{fig:CARS_population} for CARS and SPIDERS. Again, the distributions are comparable, but CARS contains a few more higher luminosity AGN due to the larger effective area and volume covered by HES compared to the X-ray surveys. Both samples contain BH masses, AGN luminosities and Eddington ratios in the range of  $6.0<\log(M_\mathrm{BH}/[M_\odot])<8.3$, $43<\log(L_\mathrm{bol}/[\mathrm{erg\,s}^{-1}])<45.5$ and $-1.5<\log(\lambda)<0.5$, respectively. Hence, the CARS subsample of local AGN is representative for the population in all important AGN parameters independent of the optical versus X-ray selection of type 1 AGN.

\section{Results and discussion}\label{sect:results}
\subsection{The ENLR size--luminosity relation}
A relation between the ENLR size and the AGN luminosity has been observed in many studies based on imaging \citep[e.g.,][]{Bennert:2002,Fischer:2018,Sun:2018}, long-slit spectroscopy \citep[e.g.,][]{Greene:2012,Hainline:2013,Hainline:2014,Sun:2017} and IFU observations  \citep[e.g.,][]{Liu:2013,Husemann:2014,Kang:2018}. While \textit{HST} narrow-band imaging offers the highest angular resolution, the limited sensitivity of \textit{HST} for low surface brightness features can lead to an underestimation of the ENLR extent \citep[e.g][]{Greene:2011,Husemann:2013a}. This leads to the primary issue of defining an objective ENLR size criterion based on the [\ion{O}{iii}] light distribution. Like expensive multi narrow-band imaging, spectroscopic studies also have the  advantage that the ionization mechanism of the gas can be verified to be AGN ionization as required for the ENLR. In Fig.~\ref{fig:ENLR_size_overview}, we compare the maximum ENLR size $R_\mathrm{ENLR,max}$, as well as $R_\mathrm{ENLR,16}$ and $R_\mathrm{ENLR,15}$ with the bolometric luminosity of the AGN in our CARS sample.  Here, we excluded the targets observed with PMAS and VIMOS for the comparison as their depth and FoV coverage is not matching that of the MUSE IFU. This would introduce additional biases as discussed below, which we want to avoid.  While  $R_\mathrm{ENLR,max}$ was measured for all targets, upper limits are more frequent for $R_\mathrm{ENLR,16}$ and $R_\mathrm{ENLR,15}$ where the ENLR can be too faint for an extended detection. It is important to recall here that we subtracted the bright point-like emission of the compact NLR.  This could otherwise artificially increase the ENLR size specifically for $R_\mathrm{ENLR,15}$ and $R_\mathrm{ENLR,16}$ due to beam smearing effects.

\begin{figure}
  \resizebox{\hsize}{!}{\includegraphics{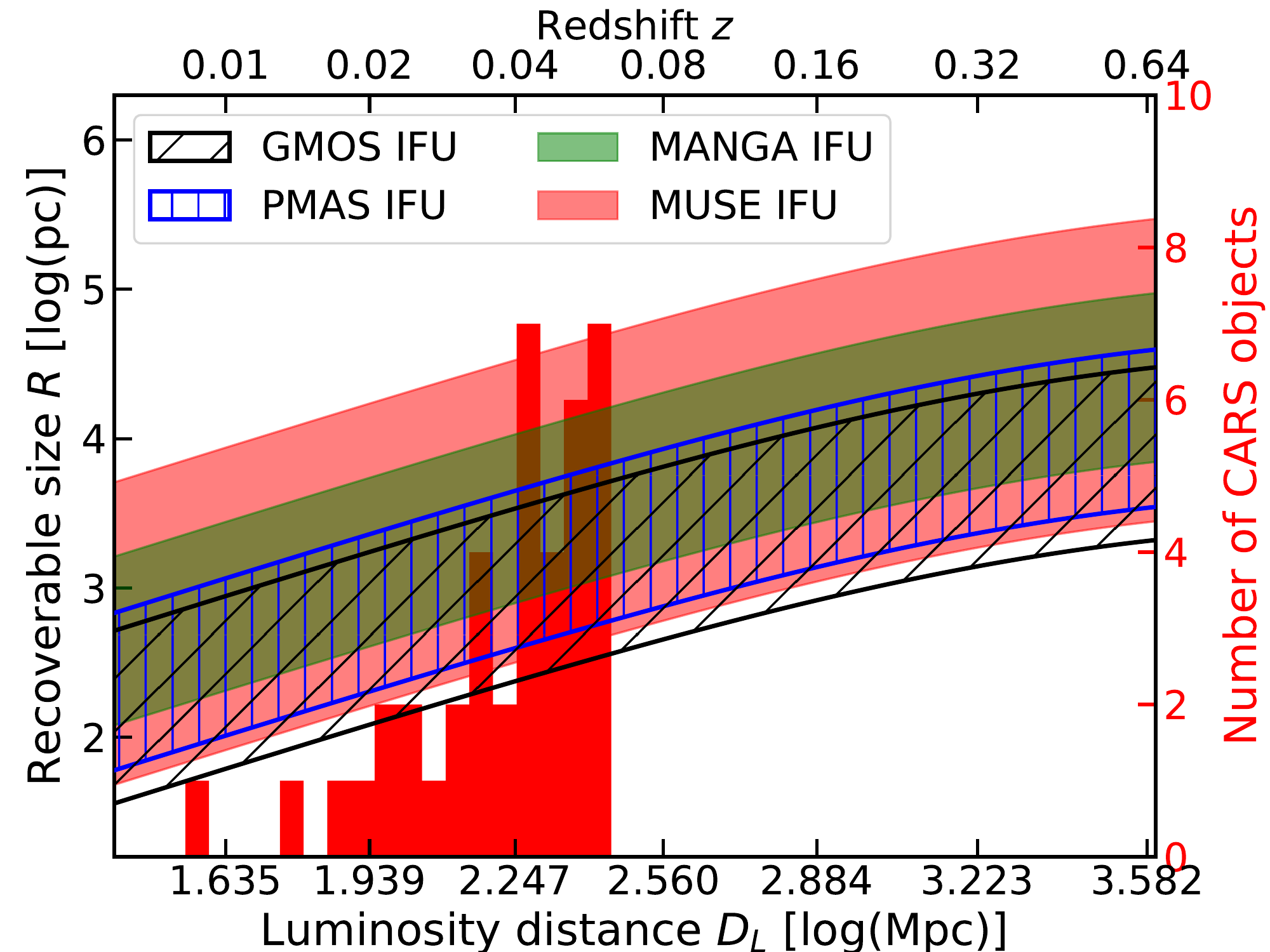}}
  \caption{Upper and lower measurement limits of ENLR sizes as a function of luminosity distance (redshift) for different commonly used IFU. Here we assume the following observing modes for the IFUs: Wide-field mode for MUSE, 2-slit mode for GMOS, $8\arcsec\times8\arcsec$ for the PMAS lens array, and the 27\arcsec\ diameter MaNGA IFU head. We defined the upper limit as the maximum distance to the corners from the center and the lower limit depends on the sampling and seeing which we set to 0\farcs4, 0\farcs3, 0\farcs5, and 1\farcs0 for MUSE, GMOS, PMAS, MaNGA respectively. The redshift distribution of CARS is highlighted as the red shaded histogram.}\label{fig:IFU_sizes}
\end{figure}

The ENLR sizes are successively decreasing from the $R_\mathrm{ENLR,max}$ definition toward the most brightest inner parts of the ENLR as measured by $R_\mathrm{ENLR,15}$. A strong trend with $L_\mathrm{bol}$ is only in all three ENLR definition when combining the CARS observations with more luminous AGN from the literature. Although previous studies reported relatively tight relations, the scatter within the CARS sample is enormous with more than an order of magnitude in the ENLR size at $L_\mathrm{bol}\sim10^{44}\,\mathrm{erg\,s}^{-1}$. This might be partially attributed to AGN variability as discussed in more depth later on. It is important to note that often the integrated [\ion{O}{iii}] line luminosity is used as bolometric indicator. This luminosity is implicitly adding the   ionizing photons emitted by the AGN over time, so that more extended ENLRs naturally have a higher integrated luminosity. It is therefore natural to introduce a tighter correlation between ENLR size and integrated [\ion{O}{iii}] luminosity. The Pearson correlation coefficients based on the CARS data alone are not high with $r_\mathrm{ENLR,max}=0.4$, $r_\mathrm{ENLR,16}=0.35$ and $r_\mathrm{ENLR,15}=0.39$. To deal with the upper limits in the ENLR sizes, we use a Bayesian approach to infer the best-fit scaling relations. We assume that the ENLR sizes follow a log-Normal probability distribution function at a given luminosity:
\begin{equation}
 p(R_\mathrm{ENLR}) \sim \exp\left[-\frac{-(\log(R_\mathrm{ENLR}/[\mathrm{pc}])-\mu)^2}{2\sigma_\mathrm{ENLR}^2}\right] \quad ,
\end{equation}
where $\mu = \alpha\,\log(L_\mathrm{bol}/10^{44}[\mathrm{erg}\,\mathrm{s}^{-1}])+\beta$ and $\sigma_\mathrm{ENLR}$ is the intrinsic scatter of $\log(R_\mathrm{ENLR})$ independent of luminosity.  We use the Monte-Carlo Markov Chain (MCMC) sampling algorithm \texttt{emcee} \citep{Foreman-Mackey:2013} to infer the posterior distribution functions for the linear relation parameters $\alpha, \beta$ and the intrinsic scatter $\sigma_\mathrm{ENLR}$. The corresponding best-fit ENLR size--luminosity relations for the three size definitions are the following: 
\begin{eqnarray}
 \frac{R_\mathrm{ENLR,max}}{\mathrm{pc}} & = & 10^{(3.70\pm0.09)}\left(\frac{L_\mathrm{bol}}{10^{44}\mathrm{erg\,s}^{-1}}\right)^{(0.27\pm0.13)} \\
 \frac{R_\mathrm{ENLR,16}}{\mathrm{pc}} & = & 10^{(3.37\pm0.11)}\left(\frac{L_\mathrm{bol}}{10^{44}\mathrm{erg\,s}^{-1}}\right)^{(0.29\pm0.15)} \\
 \frac{R_\mathrm{ENLR,15}}{\mathrm{pc}} & = & 10^{(2.89\pm0.12)}\left(\frac{L_\mathrm{bol}}{10^{44}\mathrm{erg\,s}^{-1}}\right)^{(0.25\pm0.16)} \quad .
\end{eqnarray}

The intrinsic scatter of the relations is quite large with $\sigma_\mathrm{ENLR}$ of $0.42\pm0.06$\,dex, $0.47\pm0.07$\,dex, $0.51\pm0.08$\,dex, for $R_\mathrm{ENLR,max}$, $R_\mathrm{ENLR,16}$, and $R_\mathrm{ENLR,15}$, respectively. In particular, we find numerous AGN host galaxies which display rather large ENLRs exceeding 10\,kpc below $L_\mathrm{AGN}<10^{45}\,\mathrm{erg\,s}^{-1}$ which are similar to the sizes of significantly more luminous QSOs in the literature even for $R_\mathrm{ENLR,15}$. One important aspect, which has not been discussed in the literature so far, is the implications of limited area and spatial resolution for IFU observations. This leads to the fact that the maximum and minimum ENLR size is significantly changing with target distance and redshift as shown in Fig.~\ref{fig:IFU_sizes}. These ranges also vary substantially between IFU instruments and observing site characteristics. The WFM of MUSE has by far the largest dynamical range for recovering the ENLR by area and likely also by depth compared to all  other current IFU instruments.  This might introduce tighter correlations and biased slopes as more luminous AGN are naturally targeted at higher redshifts than the lower redshift counter parts. 

The systematic offset toward smaller ENLR size is evident in the detected ENLRs for unobscured QSOs observed with the PMAS IFU instrument \citep{Husemann:2013a} shown in the left panel of Fig.~\ref{fig:ENLR_size_overview}. This is expected for the small PMAS FoV compared to the big MUSE FoV when trying to recover $R_\mathrm{ENLR,max}$. Using the intrinsic high surface brightness limits for the ENLR size definition makes the ENLR size generally smaller so that the resolution limit of the observations becomes more important than the recoverable maximum size. This leads to the issue that the sizes will cluster close to the resolution limit and an increased fraction of ENLRs become actually unresolved as seen in our CARS data and the MaNGA data of \citet{Chen:2019}. 

Combining different samples from various instruments at different redshifts therefore inevitably introduces ENLR size--luminosity relations with different slopes $\alpha$ depending on the details of target selection and analysis approaches. Slopes ranging from $\alpha=0.22\pm0.04$ \citep{Greene:2012}, $\alpha=0.25\pm0.02$ \citep{Liu:2013}, $\alpha\sim0.3$--0.4 \citep{Hainline:2013,Chen:2019}, to $\alpha\sim0.5$ \citep{Bennert:2002, Husemann:2014} are reported in the literature.  The slopes solely inferred from the CARS data are consistent with those reported by \citet{Greene:2012} and \citet{Liu:2013} and are therefore on the shallower side of previous estimates. Nevertheless, the scatter in the observed relation is significant and measured slope variations might be entirely attributed to the observationally induced biases as discussed above. A slope of $\alpha=0.5$ is reminiscent of the BLR size-luminosity relation, but would require a constant ionization parameter $U$ that demands a constant density with radius. This is not observed for the ENLR on kiloparsec scales \citep[e.g.,][]{Bennert:2006a,Kakkad:2018} and more detailed photoionization calculations are required to predict the shallower slopes inferred for most studies \citep{Dempsey:2018}. We cannot study the radial variations of $U$ as our snapshot MUSE observations are not deep enough to map the electron density given the too low S/N of the [\ion{S}{ii}] doublet on kpc scales. However, the photoionization calculations do not take into account variable ionizing flux from AGN on $10^5$\,yr time scales \citep{Schawinski:2015} and the various geometrical intersections of the ionizing radiation field with the gas distribution of the galaxies. The CARS survey is least biased with regard to $R_\mathrm{ENLR}$ given the narrow redshift range and large dynamic range offered by MUSE (see Fig.~\ref{fig:IFU_sizes}). Therefore, the CARS survey is one of the best data set to explore the origin of the significant scatter in ENLR size--luminosity relation and search for additional factors or more fundamental parameters controlling the ENLR size. 

\begin{figure*}
\sidecaption
 \includegraphics[width=12cm]{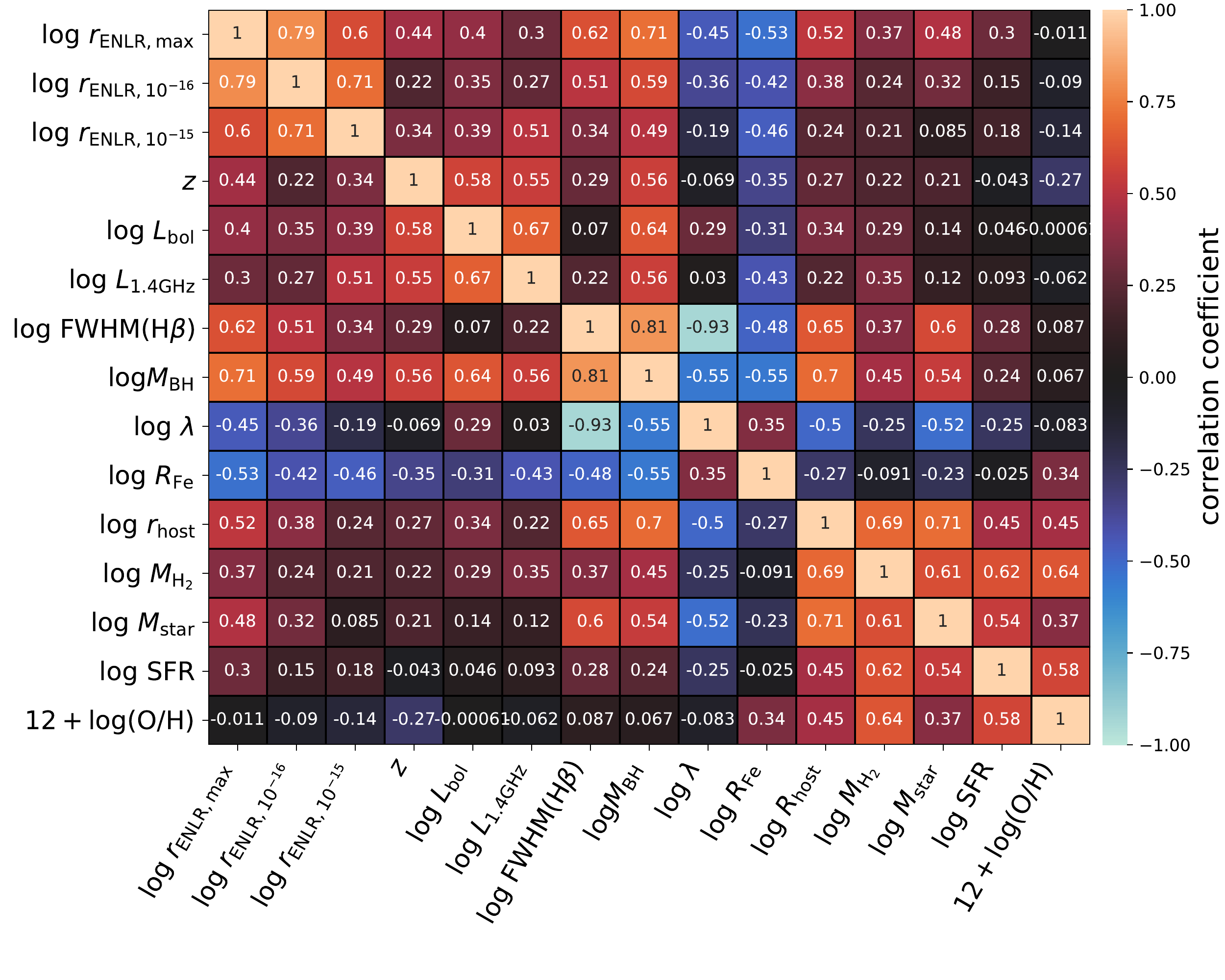}
 \caption{Full Pearson correlation coefficient matrix between the three ENLR size measurements, AGN and host galaxy parameters. Significant positive correlations appear in orange/red and negative correlation appear in blue/cyan colors, while dark blue/brown colors suggest no correlations between parameters. Each cell is labeled with the specific linear correlation coefficients of the matrix.}\label{fig:corr_matrix}
\end{figure*}

\subsection{BH mass as a more fundamental driver for the ENLR size}
\citet{Husemann:2008} already reported that the AGN luminosity cannot be the only parameter setting the ENLR size or its radial surface brightness distribution on kpc scales. They found that kpc-size ENLRs were preferentially present around AGN with H$\beta$ FWHM larger than 4000\,km/s and low Eddington ratios at a given AGN luminosity. Furthermore, large ENLR were preferentially detected around radio-loud AGN rather than radio-quiet AGN \citep[e.g.,][]{Stockton:1987}. We can perform a more extensive study with CARS by looking at a variety of host galaxy and AGN parameters that may be important to control the ENLR size and represent more fundamental correlations.

In Fig.~\ref{fig:corr_matrix}, we show the Pearson correlation matrix between the three ENLR sizes, redshift, AGN parameters, and various host galaxy parameters as determined by \citet{Smirnova-Pinchukova:2021}. We find that the ENLR sizes are strongly correlated with each other ($r>0.5$ nd $p<2\times 10^{-4}$ in all cases) as naturally expected. Similar internal correlations are found between the AGN parameters and the host galaxy parameters. We find only weak correlations between the ENLR sizes and redshift or AGN luminosity with $r<0.45$ and $p>0.02$ in all cases. Strikingly, we find the strongest correlation between the maximum ENLR size $R_\mathrm{ENLR,max}$ and the BH mass $M_\mathrm{BH}$ with a correlation coefficient of $r_\mathrm{ENLR,max}=0.71$ and $p=5\times10^{-6}$. Because single-epoch BH mass estimates are computed based on AGN luminosity and the FWHM of the H$\beta$ line \citep[e.g.,][]{Kaspi:2000,Vestergaard:2002,Peterson:2004}, it explains that 1) the ENLR shows a weak correlation with AGN luminosity and 2) the detection of the ENLR also depends on the FWHM of H$\beta$  at fixed luminosity \citep{Husemann:2008}. 

\begin{figure*}
 \includegraphics[width=\textwidth]{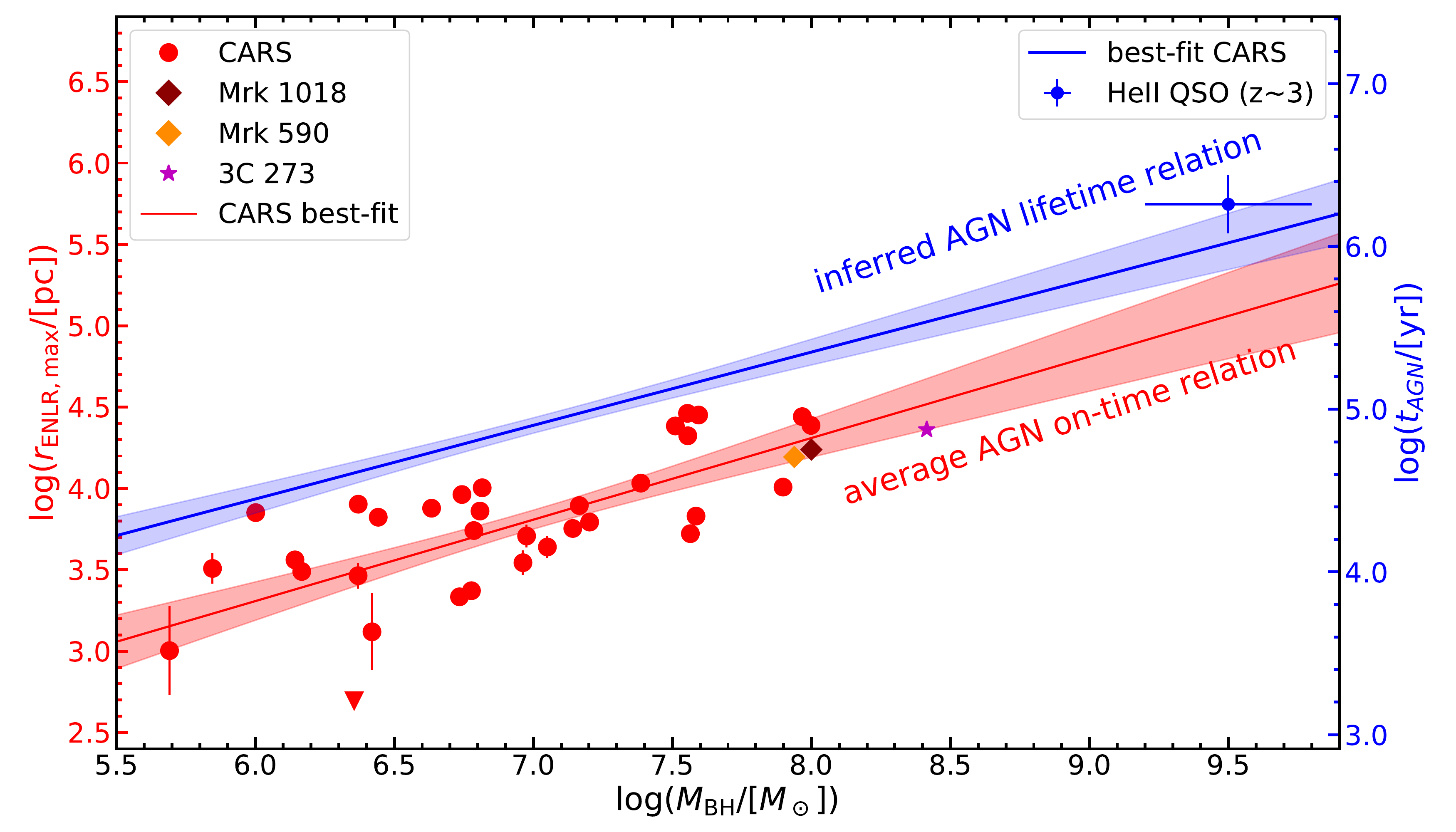}
 \caption{Maximum ENLR size $R_\mathrm{ENLR,max}$ as a function of BH mass $M_\mathrm{BH}$ for the CARS sample. A linear best-fit relation to the CARS data is shown as the red line with corresponding confidence area highlight as shaded area. The changing-look AGN Mrk~590 and Mrk~1018 are denoted by the colored diamond symbols and the local hyper-luminous QSOs 3C~273 from \citep{Husemann:2019b} as a star symbol. The right axis shows the AGN life time $t_\mathrm{AGN}$ in blue colors. Taking into account the effect of inclination and random sampling of AGN during their life time, our inferred AGN life time relation from MCMC sampling is shown as the blue line. An independent measurement of the AGN life time based on the \ion{He}{ii} proximity zones around $z\sim3$ QSOs from \citet{Khrykin:2021} is shown as a blue point with uncertainties.} \label{fig:ENLR_MBH_relation}
\end{figure*}

We plot $R_\mathrm{ENLR,max}$ as a function of $M_\mathrm{BH}$ in Fig.~\ref{fig:ENLR_MBH_relation} for the CARS sample excluding targets observed with VIMOS and PMAS to avoid observational biases due to their smaller FoV. Following the same Bayesian approach as for the ENLR size--luminosity relation we infer a scaling relation of 
\begin{equation}
\frac{R_\mathrm{ENLR,max}}{\mathrm{pc}}  = 10^{(3.81\pm0.06)}\left(\frac{M_\mathrm{BH}}{10^{7}\mathrm{M}_\sun}\right)^{(0.50\pm0.10)}
\end{equation}
with an intrinsic scatter of $\sigma_\mathrm{ENLR}=0.32\pm0.04$\,dex. This intrinsic scatter is significantly smaller compared to a scatter of $0.42\pm0.06$\,dex for the linear relation with $L_\mathrm{bol}$. Hence, $M_\mathrm{BH}$ can be considered a more fundamental driver for the maximum ENLR size. As $M_\mathrm{BH}$ is an inferred quantity derived from AGN luminosity and the FWHM of broad lines, we also infer a bivariate relation based directly on the actual observables which are the width and luminosity of the broad H$\beta$ component:  
\begin{equation}
  \frac{R_\mathrm{ENLR,max}}{\mathrm{pc}} = 10^{3.6\pm0.2}\left(\frac{L_{\mathrm{H}\beta}}{10^{43}\mathrm{erg\,s}^{-1}}\right)^{(0.24\pm0.09)}\left(\frac{\mathrm{FWHM}_{\mathrm{H}\beta}}{10^{3}\mathrm{km\,s}^{-1}}\right)^{(1.12\pm0.24)}
\end{equation}
with an intrinsic scatter of $\sigma_\mathrm{ENLR}=0.31\pm0.04$\,dex similar to the one using the $M_\mathrm{BH}$. As FWHM and H$\beta$ luminosity are directly observed quantities, they are independent of the various prescriptions used to compute BH masses. It is also possible that the apparent correlation with BH mass is emerging as a result of the correlation with AGN luminosity and FWHM which depends on the actual physical cause of such relations.

It is important to note that the relations of ENLR size as a function of BH mass and the BLR parameters are only applicable to confirmed AGN with radiative-efficient accretion and type 1 AGN host galaxies, respectively. One cannot predict an ENLR size around non-AGN galaxies or radio AGN. The CARS sample is specifically selected to host BH with ongoing radiative-efficient accretion starting roughly at 1\,\% Eddington except for the changing-look AGN in the sample. It is also possible that the absolute normalization of the $R_\mathrm{ENLR}$--$M_\mathrm{BH}$ relation will have a dependence on additional AGN or host galaxy parameters, but these dependencies are not detectable with the still relatively small sample size of CARS. While the relation may suggest that the BH mass could be estimated from the ENLR size for obscured type 2 AGN, the individual AGN lifetimes as discussed below introduce significant systematic uncertainties for individual galaxies. Further studies are needed to quantify such systematics and test additional applications.

\subsection{ENLR sizes of changing-look AGN}
The time variability of AGN luminosity becomes dramatically visible in changing-look AGN, which change their luminosity by orders of magnitude within a couple of years. They highlight that BH accretion is not necessarily preserved over long time periods and current observations provide only a current snapshot of AGN properties. However, the BH mass is a nearly constant quantity with time even for changing-look AGN and smoothly increases with time. Luckily, the CARS sample contains 2 well-known changing-look AGN, namely HE~0212$-$0059 aka Mrk~590 \citep{Osterbrock:1977,Denney:2014} and HE~0203$-$0019 aka Mrk~1018 \citep{Cohen:1986,McElroy:2016,Krumpe:2017}, which we use here to test the impact on the ENLR size scaling. 

We include both changing-look AGN as stripes in Fig.~\ref{fig:ENLR_size_overview} representing the range in AGN luminosity from the observed maximum to 1/100 and 1/10 of the luminosity as observed for Mrk~590 \citep{Denney:2014} and Mrk~1018 \citep{McElroy:2016,Krumpe:2017}, respectively, during their minimum state. It is clear that both sources can be significantly offset from the mean ENLR size--luminosity relation depending on the actual phase (minimum/maximum) of accretion rates. Such drastic changes might happen in any AGN and would only be noticed with a continuous monitoring program over decades. Therefore, we are not aware of the long-term variations and current accretion rate status of all other AGN in our sample, which might play a role in the scatter of the ENLR size--luminosity relation.

As shown in Fig.~\ref{fig:ENLR_MBH_relation}, both changing-look AGN Mrk~1018 and Mrk~590 are in agreement with the ENLR size--BH mass relation. For Mrk~590 we use the broad H$\beta$ parameters as measured from the MUSE IFU data during a recent rebrightening phase, and for Mrk~1018 we used the BH mass estimated during its bright phase \citep{Bennert:2015} as the broad H$\beta$ line remained too faint after the last changing-look event. The position of both changing-look AGN supports the notion that the $M_\mathrm{BH}$ is a more fundamental parameter in predicting the ENLR size and insensitive to short term AGN variability. 

\subsection{Is the ENLR size an AGN episode lifetime indicator?}
A new key result of this study is that the ENLR size is more strongly linked with the BH mass than with AGN luminosity. A major question is whether this relation has a physical origin directly linked to the BH mass or if it is a secondary correlation given that the BH mass is known to be linked with host galaxy properties. Indeed, we also observe a correlation between BH mass and the host galaxy size even for our predominately disk-dominated systems (see Fig.~\ref{fig:corr_matrix}). Such a link between BH mass and host size ($R_\mathrm{host}$) has been reported in the literature \citep[e.g.,][]{Bosch:2016} even for disk-dominated galaxies. Although $R_\mathrm{host}$ and $R_\mathrm{ENLR}$ are correlated, $R_\mathrm{ENLR}$ shows a much larger spread than $R_\mathrm{host}$. The ENLR sizes are often smaller than the host galaxies, which suggests that the ENLRs are likely to be ionization-bounded and the availability of gas clouds is not a major limiting factor for the overall ENLR size within the limited BH mass and AGN luminosity range probed by the CARS sample. Also we do not see the break-down of the ENLR size at $\sim$10\,kpc that was previously reported \citep{Hainline:2013,Hainline:2014,Dempsey:2018}.  While the complex and unknown gas distribution and ionization conditions will certainly have an impact on the ENLR size and vary from object to object, they should be marginalized statistically by the CARS sample size and comparable host galaxy properties.

Another mechanism predicted to scale with the BH mass is the condensation radius of gas cooling out of the turbulent hot halo of galaxies and groups via the Chaotic Cold Accretion (CCA) scenario \citep[e.g.,][]{Gaspari:2018,Gaspari:2019}. Indeed, more massive BHs are hosted in more extended and more massive halos \citep[e.g.,][]{Krumpe:2015}, which implies a larger rain of warm and cold gas. Cool gas with $\leq10^4$\,K is a requirement to detect AGN-ionized gas clouds via optical emission lines, so this mechanism can set an upper bound for $R_\mathrm{ENLR}$. Furthermore, the projected ENLR size and FWHM of H$\beta$ could both be similarly affected  by the inclination of the AGN central engine and thereby introduce a correlation between the two quantities. A $\sin i$ factor can impact the observed H$\beta$ FWHM if the BLR clouds have a disk-like distribution as reported from velocity-resolved reverberation mapping \citep[e.g.,][]{Grier:2013b,Pancoast:2014,Grier:2017,Williams:2018}. However, the inclination distribution of a random type 1 AGN sample has a well-described form when a simple cone geometry is assumed with a fixed half-opening angle $\theta_\mathrm{ENLR}$ (see Appendix~\ref{apx:pdfs}). The fraction of unobscured  X-ray selected AGN is $\sim$50\% at low redshift for AGN luminosities matching CARS sample \citep{Merloni:2014}, which corresponds to a half-opening angle of $\theta_\mathrm{ENLR}=60$\degr\ for the AGN ionization cones. The average inclination of a randomly selected unobscured AGN sample would then be $\langle i_\mathrm{ENLR}\rangle=39$\degr. The corresponding 1$\sigma$ confidence interval of the inclination distribution is $[22.4\degr,54.5\degr]$ where inclinations toward 0\degr\ become more and more unlikely. Such a narrow inclination range is consistent with the BLR inclination distribution directly determined from forward modeling of velocity-resolved reverberation mapping data of a random type 1 AGN sample \citep{Williams:2018}.  The inclination of the central engine alone might therefore lead to a $\log(R_\mathrm{ENLR})\propto \log(\mathrm{FWHM}_{\mathrm{H}\beta})$ relation and could potentially explain the link between a kpc-scale region and the kinematics of clouds at $<$pc distance. However, the projection of the ENLR size should be invariant to the small inclination range of type 1 AGN due to the much larger ionization cone angle, so that a correlation between $R_\mathrm{ENLR}$ and $\mathrm{FWHM}_{\mathrm{H}\beta}$ would not be introduced by the central engine inclination. Our large ionization cone angle assumption is supported by the fact that we do not find a systematic difference in $R_\mathrm{ENLR}$ size between our type 1 AGN sample of CARS and the type 2 AGN sample of \citet{Chen:2019} (see middle panel of Fig.~\ref{fig:ENLR_size_overview}). Based on that, we assume that  $R_\mathrm{ENLR}$ is insensitive to the central AGN engine inclination and does not introduce a direct correlation with H$\beta$ FWHM. This argument would break down if the ionizing radiation is not isotropic, but would lead to other major implications because an intrinsically isotropic AGN radiation field is a major assumption in many applications.

It is instructive to consider another possible physical parameter that can link the properties of the central AGN with the radiation field on kpc scales, which is AGN variability. A common implicit assumption is a constant ionizing flux of the AGN over time which may break down for a given AGN luminosity at various timescales \citep[e.g.,][]{Keel:2017}. The distance of an AGN-ionized gas cloud to the AGN engine directly translates into a light travel time. In this way, the ENLR size can be directly converted into a lower limit for the AGN lifetime if the BH is still accreting or an upper limit if the BH is not accreting anymore. Light echos of shutdown AGN \citep[e.g.,][]{Lintott:2009, Keel:2012, Schweizer:2013, Keel:2017} suggest an AGN life time for a single accretion episode to be $\sim10^{5}$yrs, which is consistent with the fraction of young AGN without strong NLR emission \citep{Schawinski:2015}. In addition, the ionization of neutral hydrogen or helium around high-redshift QSOs, the so-called proximity zones, can be used to determine current QSO ``on time'' ($t_\mathrm{on}$) for an individual source which is in the range of $10^{5}$ to $10^{7}$yr \citep[e.g.,][]{Eilers:2017,Worseck:2021}. Inferring the AGN lifetime ($t_\mathrm{AGN}$) from the ENLR or proximity zone size requires a statistical analysis to account for additional effects, such as neutral gas fraction, or random sampling of $t_\mathrm{on}$ for a specific AGN from the AGN lifetime distribution. One issue here is that AGN lifetime is not a well defined quantity as AGN brightness is known to vary on different timescales for different luminosities and accretion levels. In the picture we develop here, the best representation of AGN lifetime would be the stability time-scale of a radiative-efficient accretion disk with $L_\mathrm{bol}/L_\mathrm{Edd}\gtrsim0.01$.

Such a statistical framework has recently been developed and applied for the \ion{He}{ii} proximity zones around luminous $z\sim3$ QSOs \citep{Khrykin:2021}. They assume a log-normal distribution of AGN lifetimes with a mean $\langle t_\mathrm{AGN}\rangle$ and dispersion $\sigma_\mathrm{AGN}$ as well as a light-bulb light curve (top-hat function in time) for the AGN luminosity. Taking into account also random fluctuations in the initial \ion{He}{ii} fraction they create probability density functions (PDFs) for the proximity zone size across a grid of $\langle t_\mathrm{AGN}\rangle$ and $\sigma_\mathrm{AGN}$ values. Those PDFs are determined by applying a simple kernel density estimation for a large simulated population. In this way, the PDFs are effectively marginalized over all important stochastic relationships. Using the estimated PDFs, the likelihood function for the proximity zone measurements can be computed as a function of $\langle t_\mathrm{AGN}\rangle$ and $\sigma_\mathrm{AGN}$ and the posterior distribution function can be sampled with a MCMC sampler.  

\begin{figure}
 \resizebox{\hsize}{!}{\includegraphics{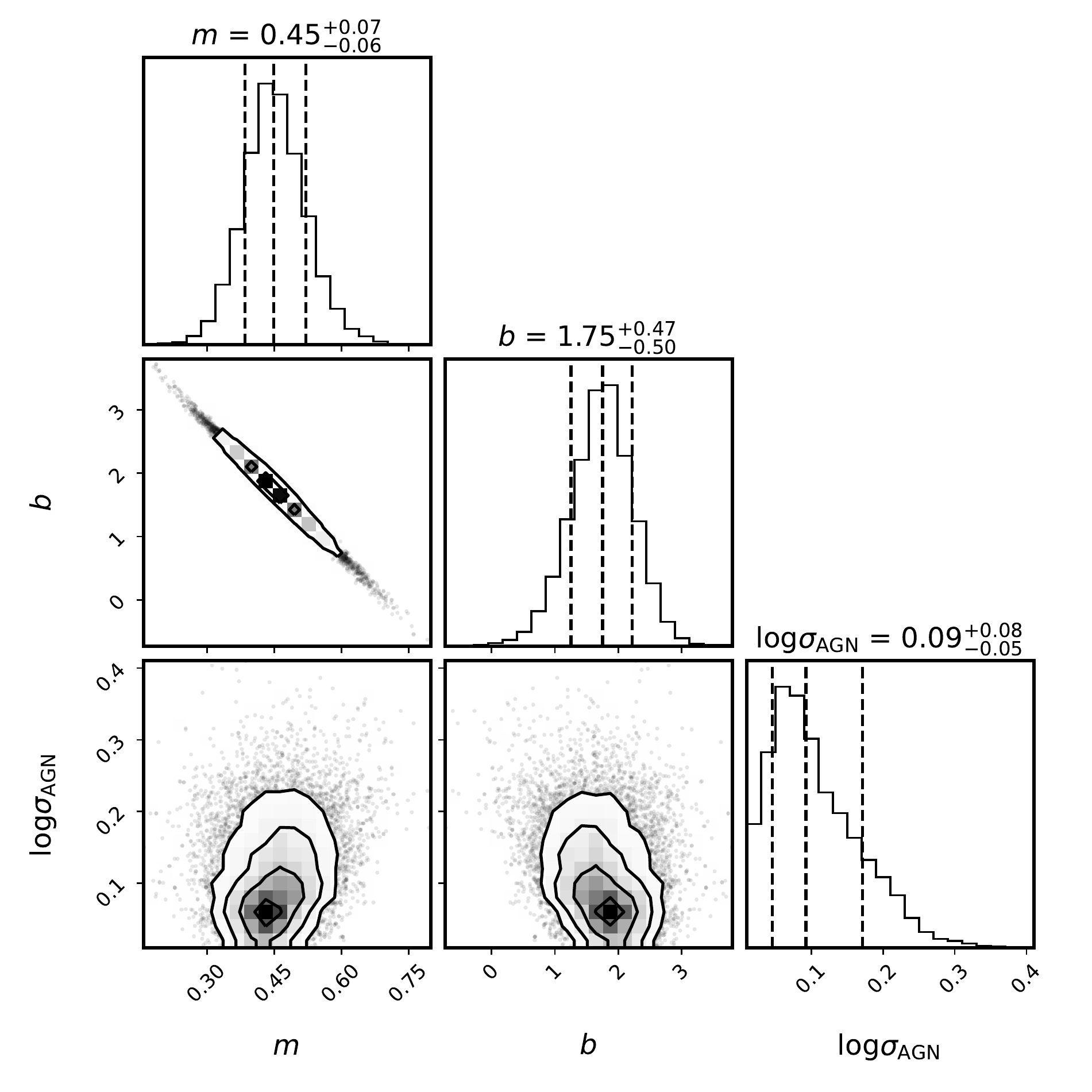}}
 \caption{Corner plot of the posterior distribution function for the parameters $m$, $b$ and $\log\sigma_\mathrm{AGN}$ of our AGN lifetime model. }\label{fig:MCMCcorner}
\end{figure}

In particular, the light-bulb approximation is a strong simplification of AGN variability and more complex light curves will be more realistic. In order to be consistent with the definition of \citet{Khrykin:2021} for $t_\mathrm{AGN}$ and considering our sample size, we follow their simple light-bulb assumption to avoid further complexities. It should be noted that the absolute normalization of $t_\mathrm{AGN}$ is dependent on this assumption. Instead of the variable \ion{He}{ii} fraction we need to incorporate some average inclination of our type 1 AGN sample with respect to our line-of-sight to account for the ionization cone projection effect. Here, we adopt an average inclination $\langle i_\mathrm{ENLR}\rangle=39$\degr\ as discussed above. Details on the PDF generation and the construction of the likelihood function are provided in Appendix~\ref{apx:pdfs}. Motivated by the strong positive correlation between $R_\mathrm{ENLR,max}$ and $M_\mathrm{BH}$, we assume a BH mass dependent mean AGN life time of the form $\langle\log(t_\mathrm{AGN}(M_\mathrm{BH})/[\mathrm{yr}]\rangle = m \cdot \log(M_\mathrm{BH}/[\mathrm{M}_\odot]) + b$. The inferred posterior distribution functions for $m$, $b$ and $\sigma_\mathrm{AGN}$ using \texttt{emcee} \citep{Foreman-Mackey:2013} are presented in Fig.~\ref{fig:MCMCcorner}.

We find a relation for the mean AGN lifetime of 
\begin{equation}
 \left\langle\log(t_\mathrm{AGN}/[\mathrm{yr}])\right\rangle = (0.45\pm0.07)\log(M_\mathrm{BH}/[\mathrm{M}_\odot]) + 1.75^{+0.47}_{-0.50}
\end{equation}
with a dispersion of $\sigma_\mathrm{AGN}=0.09_{-0.05}^{+0.08}$. The relation is shown in Fig.~\ref{fig:ENLR_MBH_relation} and represents an upper envelope to the actual ENLR sizes  while retaining the slope. Strikingly, our extrapolated relation to $M_\mathrm{BH}>10^{9}\mathrm{M}_\odot$ is consistent with the inferred AGN lifetime of $\left\langle \log(t_\mathrm{AGN}/[\mathrm{yr}])\right\rangle=6.22_{-0.25}^{+0.22}$ for high-mass BHs at $z\sim3$ \citep{Khrykin:2021} based on their \ion{He}{ii} proximity zones. Our  BH mass-dependent model for the duration of a single AGN lifetime episode can simultaneously predict the $10^{5}$\,yr lifetime of local shutdown AGN as well as the $10^{6}$\,yr lifetimes of more massive BH at higher redshifts. It is unclear if this is just a coincidence or a signature of an underlying physical mechanisms that is independent of redshift and mainly dependent on the gravitational potential of the central BH. It is interesting to note that the CCA framework mentioned above predicts a similar relation between AGN lifetime and BH mass.  The CCA-driven AGN activity roughly scales with the hot gas cooling time, $t_{\rm cool} \propto T_{\rm X}$, and $M_{\rm BH} \propto T_{\rm X}^2$ (\citealt{Gaspari:2019}), hence $t_{\rm AGN, CCA} \propto M_{\rm BH}^{0.5}$.

At low radiative AGN accretion efficiencies, powerful radio jets can emerge from an AGN from which a much more precise AGN lifetime can be estimated for individual systems \citep[e.g.,][]{Biava:2021}. The implied AGN lifetimes are typically on the order of tens of Myr given the observed jet lengths of several hundreds to thousands of kpc \citep[e.g.,][]{Hardcastle:1998} moving close to the speed of light ($>$0.1c), which likely depends on the ram pressure balance with the ambient medium. While this is much longer than the $t_\mathrm{AGN}\sim10^5$\,yr lifetime for the radiative-efficient accretion in the CARS AGN, which might partially related to the very different accretion mode of the central engine,  the discrepancy in time scales is actually not that large when considering the BH mass dependence. The fraction of galaxies hosting powerful radio-jets seems to be a strong function of stellar mass \citep[e.g.,][]{Best:2005} which is presumably tightly correlated with BH mass for bulge-dominated galaxies \citep{Marconi:2003,Haering:2004} reaching beyond $M_\mathrm{BH}\sim10^{10}M_\sun$ \citep[e.g.,][]{Magorrian:1998,McConnell:2011,Mehrgan:2019}. The extrapolation of our relation to such high BH masses would predict radiative-efficient AGN accretion lifetimes of several tens of Myr. In addition, Gigahertz Peak Spectrum (GPS) radio sources exhibit exceptionally compact jets with estimated expansion times of $10^4$\,yr and are therefore interpreted as young radio source \citep[e.g.,][]{Vink:2006, Czerny:2009}. Hence, there is also a big range in observed jet length that can only be explained in a time evolution picture. 

Interpreting the ENLR size to be induced predominantly by long-term AGN variability, as proposed by \citet{Schawinski:2015}, is therefore a possible scenario. Indeed, variability studies of AGN provide direct evidence that short-term and long-term X-ray AGN variability decreases with BH mass \citep[e.g.,][]{Bian:2003,O'Neill:2005,Lanzuisi:2014} and UV/optical variations on months to years timescales have a steeper slope with increasing BH \citep[e.g.,][]{Caplar:2017,Suberlak:2021}. However, none of those studies can provide evidence for a BH mass dependent AGN variability on a timescale of several thousand years. Of course, the overall AGN duty cycle must be much longer, on Gyr timescales, in order to explain the necessary BH mass growth. AGN therefore need to flicker on and off many times during their evolution in concert with the evolution of the host galaxies \citep[e.g.,][]{Hickox:2014,Schawinski:2015,Gaspari:2017}. Based on our morphological classification we know that the majority of CARS galaxies are isolated disk-dominated galaxies for which we do not expect any significant impact on the flicker cycles by external factors such as mergers or interactions. In addition, the ENLR are clearly associated with a single galaxy and its central BH in the vast majority of cases. This ensures that we cannot systematically mix-up the life-cycles of two independent BHs.  This might be an unknown source of confusion for the proximity zones at $z\sim3$ which exceed the sizes of galaxy groups that can host several independent BH being potentially active simultaneously \citep[e.g.,][]{Husemann:2019b}. Furthermore, the time a BH spends nearly in a quiescent state ($t_\mathrm{off}$) is crucial to understand the full AGN flicker cycle, which we cannot constrain from our observations. If $t_\mathrm{off}$ is significantly shorter than our measured $t_\mathrm{AGN}$, we would already measure a superposition of several cycles, which we cannot rule out  based on our observations and is a limitation of our model and data interpretation.

\section{Conclusion}\label{sect:conclusions}
In this paper we describe the full optical IFU data set for CARS with detailed information on the data reduction, QSO subtraction scheme and modeling of the stellar and emission lines. We focus on the characterization of the AGN-ionized region, which is referred to as the NLR or ENLR ranging from subkiloparsec to tens of kiloparsec in size. The main advantage of the type 1 AGN sample of CARS is the possibility of determining primary AGN parameters, in particular BH mass and Eddington ratios, which are difficult to determine for type 2 AGN more commonly targeted for these types of studies. As the CARS sample is a representative subsample of the luminous AGN population at low redshift, we are able to determine the  following key results:
\begin{itemize}
 \item We measure ENLR sizes ranging from a few 100\,pc to a few tens of kpc for three different ENLR size definitions, which are all not strongly correlated with the AGN luminosity.
 \item Comparing the ENLR sizes with all accessible AGN parameters, redshift and host galaxy properties, we find the strongest correlation between the maximum ENLR size  $\log(R_\mathrm{ENLR,max})$ and BH mass $\log(M_\mathrm{BH})$. 
 \item Interpreting the maximum ENLR size as time scale indicator, we recover a scaling relation for the lifetime of a single AGN episode of $t_\mathrm{AGN} \propto M_\mathrm{BH}^{0.45\pm0.07}$ taking into account projection effects and random AGN lifetime sampling applying a generative model to the data. Alternative interpretations are possible if secondary correlation are driving the observed BH mass dependence.
 \item Our BH mass dependent scaling relation for the AGN episode lifetime is consistent with the $\sim$3\,Myr lifetime estimate of high-mass BH at $z\sim3$ based on the \ion{He}{ii} proximity zone sizes \citep{Khrykin:2021}.
\end{itemize}

The apparent BH mass dependence of the ENLR size is the key result of our study. Nevertheless, we cannot fully rule out that the apparent correlation is a derivative of an even more fundamental hidden relation or through unknown selection effects considering the complexity of the ISM in galaxies, radiative transfer, and AGN variability. However, if the ENLR sizes are correctly interpreted as a proxy for the lifetime of a single AGN episode, it may provide possible explanations for various phenomena. First the famous Eigenvector 1 parameter space \citep[e.g.,][]{Boroson:1992, Sulentic:2000} could be interpreted as a time difference in the AGN phase. AGN with more massive BHs would be statistically observed at much later times in their episodic phase than lighter BHs due to the difference in AGN lifetime despite similar luminosity. In this picture, the so-called narrow-line Seyfert 1 (NLS1) galaxies would be younger AGN as previously proposed \citep[e.g.,][]{Mathur:2000, Grupe:2004b}. This may explain their extreme [\ion{O}{iii}] blueshifts \citep[e.g.,][]{Bian:2005} as well as the relative faintness of [\ion{O}{iii}] with respect to the broad H$\beta$ component, because the accretion-disk winds and associated NLR would be more compact leading to higher velocity outflows and higher gas densities prone to collisional de-excitation of forbidden lines. If strong \ion{Fe}{ii} emission is predominantly seen in AGN with narrower broad lines (smaller BH mass) as seen in Fig.~\ref{fig:CARS_population}, the \ion{Fe}{ii} line strength might be linked to such time evolution effects and decay with time. 

The relatively short timescale of $t_\mathrm{AGN}$ of less than a Myr across a large range of BH masses can provide an explanation for the lack of total star formation suppression observed for moderately luminous AGN samples \citep[e.g.,][]{Harrison:2012b,Shimizu:2017,Rosario:2018,Scholtz:2020} and for our CARS sample \citep{Smirnova-Pinchukova:2021}. For a large range in BH mass, the AGN lifetime would simply be too short to already have an impact on the total SFR of a galaxy with a characteristic time scale of several Myrs. For systems with more massive BHs, the AGN lifetime might be significantly longer and the time integrated energy input is much larger. Therefore, BH mass seems to be an important parameter in addition to the AGN luminosity in controlling the AGN feeding and feedback cycle of AGN, as also suggested by hydrodynamical simulations (e.g.~including CCA; \citealt{Gaspari:2020} for a review). This hypothesis needs to be further explored in future studies, which is more difficult for obscured AGN samples where the single-epoch BH mass method is not applicable.

\begin{acknowledgements}
We thank the anonymous referee for the constructive feedback that improved the quality and clarity of the manuscript.
The authors thank G. Worseck, K. Jahnke and H.-W. Rix for instructive discussions. BH is grateful for the financial support from the DFG grant GE625/17-1, DLR grant 50OR1911 and DAAD grant 57509925. The work of MS was supported in part by the University of Manitoba Faculty of Science Graduate Fellowship (Cangene Award), and by the University of Manitoba Graduate Enhancement of Tri-Council Stipends (GETS) program. JS is supported by the international Gemini Observatory, a program of NSF’s NOIRLab, which is managed by the Association of Universities for Research in Astronomy (AURA) under a cooperative agreement with the National Science Foundation, on behalf of the Gemini partnership of Argentina, Brazil, Canada, Chile, the Republic of Korea, and the United States of America. The Science, Technology and Facilities Council is acknowledged by JN for support through the Consolidated Grant Cosmology and Astrophysics at Portsmouth, ST/S000550/1. ISP acknowledges funding from the DLR grant 50OR2006. 
SMC and RM acknowledge support from the Australian Research Council through grant DP190102714.
VNB gratefully acknowledges assistance from a National Science Foundation (NSF) Research at Undergraduate Institutions (RUI) grant AST-1909297. We note that findings and conclusions do not necessarily represent views of the NSF. 
TAD acknowledges support from STFC grant ST/S00033X/1. TR is supported by the Science and Technology Facilities Council (STFC) through grant ST/R504725/1.  MK acknowledges support by DFG grant KR 3338/4-1.
MG acknowledges partial support by NASA Chandra GO8-19104X/GO9-20114X and HST GO-15890.020-A grants. 
MPT acknowledges financial support from the State Agency for Research of the Spanish MCIU through the ``Center of Excellence Severo Ochoa'' award to the Instituto de Astrofísica de Andalucía (SEV-2017-0709) and through grant PGC2018-098915-B-C21 (MCI/AEI/FEDER, UE).
GRT acknowledges support from NASA through grant numbers HST-GO-15411-.002-A, HST-GO-15440.002-A, and HST-GO-16173.001-A from the Space Telescope Science Institute, which is operated by AURA, Inc., under NASA contract NAS 5-26555.

%% IFU data acknowledgements
Based on observations collected at the European Southern Observatory under ESO program(s) 083.B-0801(A), 094.B-0345(A), 095.B-0015(A), 099.B-0242(B), and 099.B-0249(A). 
Based on observations collected at the Centro Astronómico Hispano-Alemán (CAHA) at Calar Alto, operated jointly by Junta de Andalucía and Consejo Superior de Investigaciones Científicas (IAA-CSIC).

%% Pan-STARRS acknowledgements
The Pan-STARRS1 Surveys (PS1) and the PS1 public science archive have been made possible through contributions by the Institute for Astronomy, the University of Hawaii, the Pan-STARRS Project Office, the Max-Planck Society and its participating institutes, the Max Planck Institute for Astronomy, Heidelberg and the Max Planck Institute for Extraterrestrial Physics, Garching, The Johns Hopkins University, Durham University, the University of Edinburgh, the Queen's University Belfast, the Harvard-Smithsonian Center for Astrophysics, the Las Cumbres Observatory Global Telescope Network Incorporated, the National Central University of Taiwan, the Space Telescope Science Institute, the National Aeronautics and Space Administration under Grant No. NNX08AR22G issued through the Planetary Science Division of the NASA Science Mission Directorate, the National Science Foundation Grant No. AST-1238877, the University of Maryland, Eotvos Lorand University (ELTE), the Los Alamos National Laboratory, and the Gordon and Betty Moore Foundation.

%% SkyMapper acknowledgements
The national facility capability for SkyMapper has been funded through ARC LIEF grant LE130100104 from the Australian Research Council, awarded to the University of Sydney, the Australian National University, Swinburne University of Technology, the University of Queensland, the University of Western Australia, the University of Melbourne, Curtin University of Technology, Monash University and the Australian Astronomical Observatory. SkyMapper is owned and operated by The Australian National University's Research School of Astronomy and Astrophysics. The survey data were processed and provided by the SkyMapper Team at ANU. The SkyMapper node of the All-Sky Virtual Observatory (ASVO) is hosted at the National Computational Infrastructure (NCI). Development and support the SkyMapper node of the ASVO has been funded in part by Astronomy Australia Limited (AAL) and the Australian Government through the Commonwealth's Education Investment Fund (EIF) and National Collaborative Research Infrastructure Strategy (NCRIS), particularly the National eResearch Collaboration Tools and Resources (NeCTAR) and the Australian National Data Service Projects (ANDS).

\end{acknowledgements}

%-------------------------------------------------------------------

\bibliographystyle{aa}
\bibliography{references}

\begin{thebibliography}{317}
\expandafter\ifx\csname natexlab\endcsname\relax\def\natexlab#1{#1}\fi

\bibitem[{{Abazajian} {et~al.}(2009){Abazajian}, {Adelman-McCarthy},
  {Ag{\"u}eros}, {Allam}, {Allende Prieto}, {An}, {Anderson}, {Anderson},
  {Annis}, {Bahcall}, \& et~al.}]{Abazajian:2009}
{Abazajian}, K.~N., {Adelman-McCarthy}, J.~K., {Ag{\"u}eros}, M.~A., {et~al.}
  2009, \apjs, 182, 543

\bibitem[{{Alonso} {et~al.}(2013){Alonso}, {Coldwell}, \&
  {Lambas}}]{Alonso:2013}
{Alonso}, M.~S., {Coldwell}, G., \& {Lambas}, D.~G. 2013, \aap, 549, A141

\bibitem[{{Alonso-Herrero} {et~al.}(2019){Alonso-Herrero},
  {Garc{\'\i}a-Burillo}, {Pereira-Santaella}, {Davies}, {Combes},
  {Vestergaard}, {Raimundo}, {Bunker}, {D{\'\i}az-Santos}, {Gandhi},
  {Garc{\'\i}a-Bernete}, {Hicks}, {H{\"o}nig}, {Hunt}, {Imanishi}, {Izumi},
  {Levenson}, {Maciejewski}, {Packham}, {Ramos Almeida}, {Ricci}, {Rigopoulou},
  {Roche}, {Rosario}, {Schartmann}, {Usero}, \& {Ward}}]{Alonso-Herrero:2019}
{Alonso-Herrero}, A., {Garc{\'\i}a-Burillo}, S., {Pereira-Santaella}, M.,
  {et~al.} 2019, \aap, 628, A65

\bibitem[{{Angl{\'e}s-Alc{\'a}zar} {et~al.}(2017){Angl{\'e}s-Alc{\'a}zar},
  {Faucher-Gigu{\`e}re}, {Quataert}, {Hopkins}, {Feldmann}, {Torrey}, {Wetzel},
  \& {Kere{\v{s}}}}]{Angles-Alcazar:2017}
{Angl{\'e}s-Alc{\'a}zar}, D., {Faucher-Gigu{\`e}re}, C.-A., {Quataert}, E.,
  {et~al.} 2017, \mnras, 472, L109

\bibitem[{{Antonucci}(1993)}]{Antonucci:1993}
{Antonucci}, R. 1993, \araa, 31, 473

\bibitem[{{Auchettl} {et~al.}(2017){Auchettl}, {Guillochon}, \&
  {Ramirez-Ruiz}}]{Auchettl:2017}
{Auchettl}, K., {Guillochon}, J., \& {Ramirez-Ruiz}, E. 2017, \apj, 838, 149

\bibitem[{{Azadi} {et~al.}(2015){Azadi}, {Aird}, {Coil}, {Moustakas}, {Mendez},
  {Blanton}, {Cool}, {Eisenstein}, {Wong}, \& {Zhu}}]{Azadi:2015}
{Azadi}, M., {Aird}, J., {Coil}, A.~L., {et~al.} 2015, \apj, 806, 187

\bibitem[{{Ba{\~n}ados} {et~al.}(2018){Ba{\~n}ados}, {Venemans},
  {Mazzucchelli}, {Farina}, {Walter}, {Wang}, {Decarli}, {Stern}, {Fan},
  {Davies}, {Hennawi}, {Simcoe}, {Turner}, {Rix}, {Yang}, {Kelson}, {Rudie}, \&
  {Winters}}]{Banados:2018}
{Ba{\~n}ados}, E., {Venemans}, B.~P., {Mazzucchelli}, C., {et~al.} 2018, \nat,
  553, 473

\bibitem[{{Bacon} {et~al.}(2010){Bacon}, {Accardo}, {Adjali}, {Anwand},
  {Bauer}, {Biswas}, {Blaizot}, {Boudon}, {Brau-Nogue}, {Brinchmann},
  {Caillier}, {Capoani}, {Carollo}, {Contini}, {Couderc}, {Daguis{\'e}},
  {Deiries}, {Delabre}, {Dreizler}, {Dubois}, {Dupieux}, {Dupuy}, {Emsellem},
  {Fechner}, {Fleischmann}, {Fran{\c c}ois}, {Gallou}, {Gharsa}, {Glindemann},
  {Gojak}, {Guiderdoni}, {Hansali}, {Hahn}, {Jarno}, {Kelz}, {Koehler},
  {Kosmalski}, {Laurent}, {Le Floch}, {Lilly}, {Lizon}, {Loupias}, {Manescau},
  {Monstein}, {Nicklas}, {Olaya}, {Pares}, {Pasquini}, {P{\'e}contal-Rousset},
  {Pell{\'o}}, {Petit}, {Popow}, {Reiss}, {Remillieux}, {Renault}, {Roth},
  {Rupprecht}, {Serre}, {Schaye}, {Soucail}, {Steinmetz}, {Streicher}, {Stuik},
  {Valentin}, {Vernet}, {Weilbacher}, {Wisotzki}, \& {Yerle}}]{Bacon:2010}
{Bacon}, R., {Accardo}, M., {Adjali}, L., {et~al.} 2010, SPIE Conf. Ser., 7735,
  8

\bibitem[{{Bacon} {et~al.}(2014){Bacon}, {Vernet}, {Borisova}, {Bouch{\'e}},
  {Brinchmann}, {Carollo}, {Carton}, {Caruana}, {Cerda}, {Contini}, {Franx},
  {Girard}, {Guerou}, {Haddad}, {Hau}, {Herenz}, {Herrera}, \&
  {Husemann}}]{Bacon:2014a}
{Bacon}, R., {Vernet}, J., {Borisova}, E., {et~al.} 2014, The Messenger, 157,
  13

\bibitem[{{Bai} {et~al.}(2017){Bai}, {Zhang}, {Yuan}, {Li}, {Chen}, {Lei},
  {Yang}, {Dong}, {Wang}, \& {Zhao}}]{Bai:2017}
{Bai}, Z.-R., {Zhang}, H.-T., {Yuan}, H.-L., {et~al.} 2017, Research in
  Astronomy and Astrophysics, 17, 091

\bibitem[{{Bajtlik} {et~al.}(1988){Bajtlik}, {Duncan}, \&
  {Ostriker}}]{Bajtlik:1988}
{Bajtlik}, S., {Duncan}, R.~C., \& {Ostriker}, J.~P. 1988, \apj, 327, 570

\bibitem[{{Baldwin} {et~al.}(1981){Baldwin}, {Phillips}, \&
  {Terlevich}}]{Baldwin:1981}
{Baldwin}, J.~A., {Phillips}, M.~M., \& {Terlevich}, R. 1981, \pasp, 93, 5

\bibitem[{{Balmaverde} {et~al.}(2021){Balmaverde}, {Capetti}, {Marconi},
  {Venturi}, {Chiaberge}, {Baldi}, {Baum}, {Gilli}, {Grandi}, {Meyer}, {Miley},
  {O'Dea}, {Sparks}, {Torresi}, \& {Tremblay}}]{Balmverde:2021}
{Balmaverde}, B., {Capetti}, A., {Marconi}, A., {et~al.} 2021, \aap, 645, A12

\bibitem[{{Barai} {et~al.}(2016){Barai}, {Murante}, {Borgani}, {Gaspari},
  {Granato}, {Monaco}, \& {Ragone-Figueroa}}]{Barai:2016}
{Barai}, P., {Murante}, G., {Borgani}, S., {et~al.} 2016, \mnras, 461, 1548

\bibitem[{{Baum} {et~al.}(1988){Baum}, {Heckman}, {Bridle}, {van Breugel}, \&
  {Miley}}]{Baum:1988}
{Baum}, S.~A., {Heckman}, T.~M., {Bridle}, A., {van Breugel}, W.~J.~M., \&
  {Miley}, G.~K. 1988, \apjs, 68, 643

\bibitem[{{Becker} {et~al.}(1995){Becker}, {White}, \& {Helfand}}]{Becker:1995}
{Becker}, R.~H., {White}, R.~L., \& {Helfand}, D.~J. 1995, \apj, 450, 559

\bibitem[{{Bennert} {et~al.}(2002){Bennert}, {Falcke}, {Schulz}, {Wilson}, \&
  {Wills}}]{Bennert:2002}
{Bennert}, N., {Falcke}, H., {Schulz}, H., {Wilson}, A.~S., \& {Wills}, B.~J.
  2002, \apjl, 574, L105

\bibitem[{{Bennert} {et~al.}(2006){Bennert}, {Jungwiert}, {Komossa}, {Haas}, \&
  {Chini}}]{Bennert:2006a}
{Bennert}, N., {Jungwiert}, B., {Komossa}, S., {Haas}, M., \& {Chini}, R. 2006,
  \aap, 456, 953

\bibitem[{{Bennert} {et~al.}(2018){Bennert}, {Loveland}, {Donohue}, {Cosens},
  {Lewis}, {Komossa}, {Treu}, {Malkan}, {Milgram}, {Flatland}, {Auger}, {Park},
  \& {Lazarova}}]{Bennert:2018}
{Bennert}, V.~N., {Loveland}, D., {Donohue}, E., {et~al.} 2018, \mnras, 481,
  138

\bibitem[{Bennert {et~al.}(2015)Bennert, Treu, Auger, Cosens, Park, Rosen,
  Harris, Malkan, \& Woo}]{Bennert:2015}
Bennert, V.~N., Treu, T., Auger, M.~W., {et~al.} 2015, \apj, 809, 20

\bibitem[{{Bentz} {et~al.}(2013){Bentz}, {Denney}, {Grier}, {Barth},
  {Peterson}, {Vestergaard}, {Bennert}, {Canalizo}, {De Rosa}, {Filippenko},
  {Gates}, {Greene}, {Li}, {Malkan}, {Pogge}, {Stern}, {Treu}, \&
  {Woo}}]{Bentz:2013}
{Bentz}, M.~C., {Denney}, K.~D., {Grier}, C.~J., {et~al.} 2013, \apj, 767, 149

\bibitem[{{Bentz} {et~al.}(2009{\natexlab{a}}){Bentz}, {Peterson}, {Pogge}, \&
  {Vestergaard}}]{Bentz:2009b}
{Bentz}, M.~C., {Peterson}, B.~M., {Pogge}, R.~W., \& {Vestergaard}, M.
  2009{\natexlab{a}}, \apjl, 694, L166

\bibitem[{{Bentz} {et~al.}(2009{\natexlab{b}}){Bentz}, {Walsh}, {Barth},
  {Baliber}, {Bennert}, {Canalizo}, {Filippenko}, {Ganeshalingam}, {Gates},
  {Greene}, {Hidas}, {Hiner}, {Lee}, {Li}, {Malkan}, {Minezaki}, {Sakata},
  {Serduke}, {Silverman}, {Steele}, {Stern}, {Street}, {Thornton}, {Treu},
  {Wang}, {Woo}, \& {Yoshii}}]{Bentz:2009c}
{Bentz}, M.~C., {Walsh}, J.~L., {Barth}, A.~J., {et~al.} 2009{\natexlab{b}},
  \apj, 705, 199

\bibitem[{{Bertram} {et~al.}(2007){Bertram}, {Eckart}, {Fischer}, {Zuther},
  {Straubmeier}, {Wisotzki}, \& {Krips}}]{Bertram:2007}
{Bertram}, T., {Eckart}, A., {Fischer}, S., {et~al.} 2007, \aap, 470, 571

\bibitem[{{Best} {et~al.}(2005){Best}, {Kauffmann}, {Heckman}, {Brinchmann},
  {Charlot}, {Ivezi{\'c}}, \& {White}}]{Best:2005}
{Best}, P.~N., {Kauffmann}, G., {Heckman}, T.~M., {et~al.} 2005, \mnras, 362,
  25

\bibitem[{{Bian} {et~al.}(2005){Bian}, {Yuan}, \& {Zhao}}]{Bian:2005}
{Bian}, W., {Yuan}, Q., \& {Zhao}, Y. 2005, \mnras, 364, 187

\bibitem[{{Bian} \& {Zhao}(2003)}]{Bian:2003}
{Bian}, W. \& {Zhao}, Y. 2003, \mnras, 343, 164

\bibitem[{{Biava} {et~al.}(2021){Biava}, {Brienza}, {Bonafede}, {Gitti},
  {Bonnassieux}, {Harwood}, {Edge}, {Riseley}, \& {Vantyghem}}]{Biava:2021}
{Biava}, N., {Brienza}, M., {Bonafede}, A., {et~al.} 2021, \aap, 650, A170

\bibitem[{{Binette} {et~al.}(1994){Binette}, {Magris}, {Stasi{\'n}ska}, \&
  {Bruzual}}]{Binette:1994}
{Binette}, L., {Magris}, C.~G., {Stasi{\'n}ska}, G., \& {Bruzual}, A.~G. 1994,
  \aap, 292, 13

\bibitem[{{B{\^\i}rzan} {et~al.}(2004){B{\^\i}rzan}, {Rafferty}, {McNamara},
  {Wise}, \& {Nulsen}}]{Birzan:2004}
{B{\^\i}rzan}, L., {Rafferty}, D.~A., {McNamara}, B.~R., {Wise}, M.~W., \&
  {Nulsen}, P.~E.~J. 2004, \apj, 607, 800

\bibitem[{{Bischetti} {et~al.}(2021){Bischetti}, {Feruglio}, {Piconcelli},
  {Duras}, {P{\'e}rez-Torres}, {Herrero}, {Venturi}, {Carniani}, {Bruni},
  {Gavignaud}, {Testa}, {Bongiorno}, {Brusa}, {Circosta}, {Cresci},
  {D'Odorico}, {Maiolino}, {Marconi}, {Mingozzi}, {Pappalardo}, {Perna},
  {Traianou}, {Travascio}, {Vietri}, {Zappacosta}, \& {Fiore}}]{Bischetti:2021}
{Bischetti}, M., {Feruglio}, C., {Piconcelli}, E., {et~al.} 2021, \aap, 645,
  A33

\bibitem[{{Bischetti} {et~al.}(2017){Bischetti}, {Piconcelli}, {Vietri},
  {Bongiorno}, {Fiore}, {Sani}, {Marconi}, {Duras}, {Zappacosta}, {Brusa},
  {Comastri}, {Cresci}, {Feruglio}, {Giallongo}, {La Franca}, {Mainieri},
  {Mannucci}, {Martocchia}, {Ricci}, {Schneider}, {Testa}, \&
  {Vignali}}]{Bischetti:2017}
{Bischetti}, M., {Piconcelli}, E., {Vietri}, G., {et~al.} 2017, \aap, 598, A122

\bibitem[{{Bittner} {et~al.}(2019){Bittner}, {Falc{\'o}n-Barroso}, {Nedelchev},
  {Dorta}, {Gadotti}, {Sarzi}, {Molaeinezhad}, {Iodice}, {Rosado-Belza}, {de
  Lorenzo-C{\'a}ceres}, {Fragkoudi}, {Gal{\'a}n-de Anta}, {Husemann},
  {M{\'e}ndez-Abreu}, {Neumann}, {Pinna}, {Querejeta},
  {S{\'a}nchez-Bl{\'a}zquez}, \& {Seidel}}]{Bittner:2019}
{Bittner}, A., {Falc{\'o}n-Barroso}, J., {Nedelchev}, B., {et~al.} 2019, \aap,
  628, A117

\bibitem[{{Bolton} \& {Haehnelt}(2007)}]{Bolton:2007}
{Bolton}, J.~S. \& {Haehnelt}, M.~G. 2007, \mnras, 374, 493

\bibitem[{{Borisova} {et~al.}(2016){Borisova}, {Cantalupo}, {Lilly}, {Marino},
  {Gallego}, {Bacon}, {Blaizot}, {Bouch{\'e}}, {Brinchmann}, {Carollo},
  {Caruana}, {Finley}, {Herenz}, {Richard}, {Schaye}, {Straka}, {Turner},
  {Urrutia}, {Verhamme}, \& {Wisotzki}}]{Borisova:2016}
{Borisova}, E., {Cantalupo}, S., {Lilly}, S.~J., {et~al.} 2016, \apj, 831, 39

\bibitem[{{Boroson} \& {Green}(1992)}]{Boroson:1992}
{Boroson}, T.~A. \& {Green}, R.~F. 1992, \apjs, 80, 109

\bibitem[{Bradley {et~al.}(2019)Bradley, Sipőcz, Robitaille, Tollerud, {Zé
  Vinícius}, Deil, Barbary, Wilson, Busko, Günther, Cara, Conseil,
  Droettboom, {Azalee Bostroem}, {E. M. Bray}, Bratholm, {P. L. Lim}, Craig,
  Barentsen, Pascual, Donath, Greco, Perren, Kerzendorf, Val-Borro, Dencheva,
  Ferreira, Souchereau, D'Eugenio, \& Weaver}]{Bradley:2019}
Bradley, L., Sipőcz, B., Robitaille, T., {et~al.} 2019, astropy/photutils:
  v0.7.2

\bibitem[{{Br{\"u}ggen} \& {Kaiser}(2002)}]{Brueggen:2002}
{Br{\"u}ggen}, M. \& {Kaiser}, C.~R. 2002, \nat, 418, 301

\bibitem[{{Busch} {et~al.}(2016){Busch}, {Fazeli}, {Eckart}, {Valencia-S.},
  {Smaji{\'c}}, {Moser}, {Scharw{\"a}chter}, {Dierkes}, \&
  {Fischer}}]{Busch:2016}
{Busch}, G., {Fazeli}, N., {Eckart}, A., {et~al.} 2016, \aap, 587, A138

\bibitem[{{Busch} {et~al.}(2014){Busch}, {Zuther}, {Valencia-S.}, {Moser},
  {Fischer}, {Eckart}, {Scharw{\"a}chter}, {Gadotti}, \&
  {Wisotzki}}]{Busch:2014}
{Busch}, G., {Zuther}, J., {Valencia-S.}, M., {et~al.} 2014, \aap, 561, A140

\bibitem[{{Buta} {et~al.}(2015){Buta}, {Sheth}, {Athanassoula}, {Bosma},
  {Knapen}, {Laurikainen}, {Salo}, {Elmegreen}, {Ho}, {Zaritsky}, {Courtois},
  {Hinz}, {Mu{\~n}oz-Mateos}, {Kim}, {Regan}, {Gadotti}, {Gil de Paz}, {Laine},
  {Men{\'e}ndez-Delmestre}, {Comer{\'o}n}, {Erroz Ferrer}, {Seibert},
  {Mizusawa}, {Holwerda}, \& {Madore}}]{Buta:2015}
{Buta}, R.~J., {Sheth}, K., {Athanassoula}, E., {et~al.} 2015, \apjs, 217, 32

\bibitem[{{Caglar} {et~al.}(2020){Caglar}, {Burtscher}, {Brandl}, {Brinchmann},
  {Davies}, {Hicks}, {Koss}, {Lin}, {Maciejewski}, {M{\"u}ller-S{\'a}nchez},
  {Riffel}, {Riffel}, {Rosario}, {Schartmann}, {Schnorr-M{\"u}ller}, {Shimizu},
  {Storchi-Bergmann}, {Veilleux}, {Orban de Xivry}, \& {Bennert}}]{Caglar:2020}
{Caglar}, T., {Burtscher}, L., {Brandl}, B., {et~al.} 2020, \aap, 634, A114

\bibitem[{{Cantalupo} {et~al.}(2019){Cantalupo}, {Pezzulli}, {Lilly}, {Marino},
  {Gallego}, {Schaye}, {Bacon}, {Feltre}, {Kollatschny}, {Nanayakkara},
  {Richard}, {Wendt}, {Wisotzki}, \& {Prochaska}}]{Cantalupo:2019}
{Cantalupo}, S., {Pezzulli}, G., {Lilly}, S.~J., {et~al.} 2019, \mnras, 483,
  5188

\bibitem[{{Caplar} {et~al.}(2017){Caplar}, {Lilly}, \&
  {Trakhtenbrot}}]{Caplar:2017}
{Caplar}, N., {Lilly}, S.~J., \& {Trakhtenbrot}, B. 2017, \apj, 834, 111

\bibitem[{{Cappellari} \& {Emsellem}(2004)}]{Cappellari:2004}
{Cappellari}, M. \& {Emsellem}, E. 2004, \pasp, 116, 138

\bibitem[{{Cardelli} {et~al.}(1989){Cardelli}, {Clayton}, \&
  {Mathis}}]{Cardelli:1989}
{Cardelli}, J.~A., {Clayton}, G.~C., \& {Mathis}, J.~S. 1989, \apj, 345, 245

\bibitem[{{Carniani} {et~al.}(2015){Carniani}, {Marconi}, {Maiolino},
  {Balmaverde}, {Brusa}, {Cano-D{\'{\i}}az}, {Cicone}, {Comastri}, {Cresci},
  {Fiore}, {Feruglio}, {La Franca}, {Mainieri}, {Mannucci}, {Nagao}, {Netzer},
  {Piconcelli}, {Risaliti}, {Schneider}, \& {Shemmer}}]{Carniani:2015}
{Carniani}, S., {Marconi}, A., {Maiolino}, R., {et~al.} 2015, \aap, 580, A102

\bibitem[{{Chambers} {et~al.}(2016){Chambers}, {Magnier}, {Metcalfe},
  {Flewelling}, {Huber}, {Waters}, {Denneau}, {Draper}, {Farrow}, {Finkbeiner},
  {Holmberg}, {Koppenhoefer}, {Price}, {Saglia}, {Schlafly}, {Smartt},
  {Sweeney}, {Wainscoat}, {Burgett}, {Grav}, {Heasley}, {Hodapp}, {Jedicke},
  {Kaiser}, {Kudritzki}, {Luppino}, {Lupton}, {Monet}, {Morgan}, {Onaka},
  {Stubbs}, {Tonry}, {Banados}, {Bell}, {Bender}, {Bernard}, {Botticella},
  {Casertano}, {Chastel}, {Chen}, {Chen}, {Cole}, {Deacon}, {Frenk},
  {Fitzsimmons}, {Gezari}, {Goessl}, {Goggia}, {Goldman}, {Grebel}, {Hambly},
  {Hasinger}, {Heavens}, {Heckman}, {Henderson}, {Henning}, {Holman}, {Hopp},
  {Ip}, {Isani}, {Keyes}, {Koekemoer}, {Kotak}, {Long}, {Lucey}, {Liu},
  {Martin}, {McLean}, {Morganson}, {Murphy}, {Nieto-Santisteban}, {Norberg},
  {Peacock}, {Pier}, {Postman}, {Primak}, {Rae}, {Rest}, {Riess}, {Riffeser},
  {Rix}, {Roser}, {Schilbach}, {Schultz}, {Scolnic}, {Szalay}, {Seitz},
  {Shiao}, {Small}, {Smith}, {Soderblom}, {Taylor}, {Thakar}, {Thiel},
  {Thilker}, {Urata}, {Valenti}, {Walter}, {Watters}, {Werner}, {White},
  {Wood-Vasey}, \& {Wyse}}]{Chambers:2016}
{Chambers}, K.~C., {Magnier}, E.~A., {Metcalfe}, N., {et~al.} 2016, ArXiv
  e-prints [\eprint[arXiv]{1612.05560}]

\bibitem[{{Chen} {et~al.}(2013){Chen}, {Hickox}, {Alberts}, {Brodwin}, {Jones},
  {Murray}, {Alexander}, {Assef}, {Brown}, {Dey}, {Forman}, {Gorjian},
  {Goulding}, {Le Floc'h}, {Jannuzi}, {Mullaney}, \& {Pope}}]{Chen:2013}
{Chen}, C.-T.~J., {Hickox}, R.~C., {Alberts}, S., {et~al.} 2013, \apj, 773, 3

\bibitem[{{Chen} {et~al.}(2019{\natexlab{a}}){Chen}, {Shi}, {Dempsey}, {Law},
  {Chen}, {Yan}, {Bing}, {Rembold}, {Li}, {Yu}, {Riffel}, {Brownstein}, \&
  {Riffel}}]{Chen:2019b}
{Chen}, J., {Shi}, Y., {Dempsey}, R., {et~al.} 2019{\natexlab{a}}, \mnras, 489,
  855

\bibitem[{{Chen} {et~al.}(2019{\natexlab{b}}){Chen}, {Akiyama}, {Noda},
  {Abdurro'uf}, {Yamamura}, {Kawaguchi}, {Kokubo}, \& {Ichikawa}}]{Chen:2019}
{Chen}, X., {Akiyama}, M., {Noda}, H., {et~al.} 2019{\natexlab{b}}, \pasj, 71,
  29

\bibitem[{{Christensen} {et~al.}(2006){Christensen}, {Jahnke}, {Wisotzki}, \&
  {S{\'a}nchez}}]{Christensen:2006}
{Christensen}, L., {Jahnke}, K., {Wisotzki}, L., \& {S{\'a}nchez}, S.~F. 2006,
  \aap, 459, 717

\bibitem[{{Cicone} {et~al.}(2018){Cicone}, {Brusa}, {Ramos Almeida}, {Cresci},
  {Husemann}, \& {Mainieri}}]{Cicone:2018}
{Cicone}, C., {Brusa}, M., {Ramos Almeida}, C., {et~al.} 2018, Nature
  Astronomy, 2, 176

\bibitem[{{Cid Fernandes} {et~al.}(2011){Cid Fernandes}, {Stasi{\'n}ska},
  {Mateus}, \& {Vale Asari}}]{CidFernandes:2011}
{Cid Fernandes}, R., {Stasi{\'n}ska}, G., {Mateus}, A., \& {Vale Asari}, N.
  2011, \mnras, 413, 1687

\bibitem[{{Circosta} {et~al.}(2021){Circosta}, {Mainieri}, {Lamperti},
  {Padovani}, {Bischetti}, {Harrison}, {Kakkad}, {Zanella}, {Vietri},
  {Lanzuisi}, {Salvato}, {Brusa}, {Carniani}, {Cicone}, {Cresci}, {Feruglio},
  {Husemann}, {Mannucci}, {Marconi}, {Perna}, {Piconcelli}, {Puglisi},
  {Saintonge}, {Schramm}, {Vignali}, \& {Zappacosta}}]{Circosta:2021}
{Circosta}, C., {Mainieri}, V., {Lamperti}, I., {et~al.} 2021, \aap, 646, A96

\bibitem[{{Circosta} {et~al.}(2018){Circosta}, {Mainieri}, {Padovani},
  {Lanzuisi}, {Salvato}, {Harrison}, {Kakkad}, {Puglisi}, {Vietri}, \&
  {Zamorani}}]{Circosta:2018}
{Circosta}, C., {Mainieri}, V., {Padovani}, P., {et~al.} 2018, \aap, 620, A82

\bibitem[{{Cisternas} {et~al.}(2011){Cisternas}, {Jahnke}, {Inskip},
  {Kartaltepe}, {Koekemoer}, {Lisker}, {Robaina}, {Scodeggio}, {Sheth},
  {Trump}, {Andrae}, \& {Miyaji}}]{Cisternas:2011}
{Cisternas}, M., {Jahnke}, K., {Inskip}, K.~J., {et~al.} 2011, \apj, 726, 57

\bibitem[{{Cisternas} {et~al.}(2015){Cisternas}, {Sheth}, {Salvato}, {Knapen},
  {Civano}, \& {Santini}}]{Cisternas:2015}
{Cisternas}, M., {Sheth}, K., {Salvato}, M., {et~al.} 2015, \apj, 802, 137

\bibitem[{{Coffey} {et~al.}(2019){Coffey}, {Salvato}, {Merloni}, {Boller},
  {Nandra}, {Dwelly}, {Comparat}, {Schulze}, {Del Moro}, \&
  {Schneider}}]{Coffey:2019}
{Coffey}, D., {Salvato}, M., {Merloni}, A., {et~al.} 2019, \aap, 625, A123

\bibitem[{{Cohen} {et~al.}(1986){Cohen}, {Puetter}, {Rudy}, {Ake}, \&
  {Foltz}}]{Cohen:1986}
{Cohen}, R.~D., {Puetter}, R.~C., {Rudy}, R.~J., {Ake}, T.~B., \& {Foltz},
  C.~B. 1986, \apj, 311, 135

\bibitem[{{Collin} {et~al.}(2006){Collin}, {Kawaguchi}, {Peterson}, \&
  {Vestergaard}}]{Collin:2006}
{Collin}, S., {Kawaguchi}, T., {Peterson}, B.~M., \& {Vestergaard}, M. 2006,
  \aap, 456, 75

\bibitem[{{Combes} {et~al.}(2013){Combes}, {Garc{\'{\i}}a-Burillo}, {Casasola},
  {Hunt}, {Krips}, {Baker}, {Boone}, {Eckart}, {Marquez}, {Neri}, {Schinnerer},
  \& {Tacconi}}]{Combes:2013}
{Combes}, F., {Garc{\'{\i}}a-Burillo}, S., {Casasola}, V., {et~al.} 2013, \aap,
  558, A124

\bibitem[{{Condon} {et~al.}(1998){Condon}, {Cotton}, {Greisen}, {Yin},
  {Perley}, {Taylor}, \& {Broderick}}]{Condon:1998}
{Condon}, J.~J., {Cotton}, W.~D., {Greisen}, E.~W., {et~al.} 1998, \aj, 115,
  1693

\bibitem[{{Crain} {et~al.}(2015){Crain}, {Schaye}, {Bower}, {Furlong},
  {Schaller}, {Theuns}, {Dalla Vecchia}, {Frenk}, {McCarthy}, {Helly},
  {Jenkins}, {Rosas-Guevara}, {White}, \& {Trayford}}]{Crain:2015}
{Crain}, R.~A., {Schaye}, J., {Bower}, R.~G., {et~al.} 2015, \mnras, 450, 1937

\bibitem[{{Cresci} {et~al.}(2015){Cresci}, {Mainieri}, {Brusa}, {Marconi},
  {Perna}, {Mannucci}, {Piconcelli}, {Maiolino}, {Feruglio}, {Fiore},
  {Bongiorno}, {Lanzuisi}, {Merloni}, {Schramm}, {Silverman}, \&
  {Civano}}]{Cresci:2015}
{Cresci}, G., {Mainieri}, V., {Brusa}, M., {et~al.} 2015, \apj, 799, 82

\bibitem[{{Croom} {et~al.}(2021){Croom}, {Owers}, {Scott}, {Poetrodjojo},
  {Groves}, {van de Sande}, {Barone}, {Cortese}, {D'Eugenio}, {Bland-Hawthorn},
  {Bryant}, {Oh}, {Brough}, {Agostino}, {Casura}, {Catinella}, {Colless},
  {Cecil}, {Davies}, {Drinkwater}, {Driver}, {Ferreras}, {Foster},
  {Fraser-McKelvie}, {Lawrence}, {Leslie}, {Liske}, {L{\'o}pez-S{\'a}nchez},
  {Lorente}, {McElroy}, {Medling}, {Obreschkow}, {Richards}, {Sharp}, {Sweet},
  {Taranu}, {Taylor}, {Tescari}, {Thomas}, {Tocknell}, \&
  {Vaughan}}]{Croom:2021}
{Croom}, S.~M., {Owers}, M.~S., {Scott}, N., {et~al.} 2021, \mnras, 505, 991

\bibitem[{{Croom} {et~al.}(2004){Croom}, {Smith}, {Boyle}, {Shanks}, {Miller},
  {Outram}, \& {Loaring}}]{Croom:2004}
{Croom}, S.~M., {Smith}, R.~J., {Boyle}, B.~J., {et~al.} 2004, \mnras, 349,
  1397

\bibitem[{{Czerny} {et~al.}(2009){Czerny}, {Siemiginowska}, {Janiuk},
  {Nikiel-Wroczy{\'n}ski}, \& {Stawarz}}]{Czerny:2009}
{Czerny}, B., {Siemiginowska}, A., {Janiuk}, A., {Nikiel-Wroczy{\'n}ski}, B.,
  \& {Stawarz}, {\L}. 2009, \apj, 698, 840

\bibitem[{{Davies} {et~al.}(2016){Davies}, {Groves}, {Kewley}, {Dopita},
  {Hampton}, {Shastri}, {Scharw{\"a}chter}, {Sutherland}, {Kharb}, {Bhatt},
  {Jin}, {Banfield}, {Zaw}, {James}, {Juneau}, \& {Srivastava}}]{Davies:2016}
{Davies}, R.~L., {Groves}, B., {Kewley}, L.~J., {et~al.} 2016, \mnras, 462,
  1616

\bibitem[{{Davies} {et~al.}(2014){Davies}, {Kewley}, {Ho}, \&
  {Dopita}}]{Davies:2014b}
{Davies}, R.~L., {Kewley}, L.~J., {Ho}, I.-T., \& {Dopita}, M.~A. 2014, \mnras,
  444, 3961

\bibitem[{{Dempsey} \& {Zakamska}(2018)}]{Dempsey:2018}
{Dempsey}, R. \& {Zakamska}, N.~L. 2018, \mnras, 477, 4615

\bibitem[{{Denney} {et~al.}(2014){Denney}, {De Rosa}, {Croxall}, {Gupta},
  {Bentz}, {Fausnaugh}, {Grier}, {Martini}, {Mathur}, {Peterson}, {Pogge}, \&
  {Shappee}}]{Denney:2014}
{Denney}, K.~D., {De Rosa}, G., {Croxall}, K., {et~al.} 2014, \apj, 796, 134

\bibitem[{{Denney} {et~al.}(2009){Denney}, {Peterson}, {Dietrich},
  {Vestergaard}, \& {Bentz}}]{Denney:2009}
{Denney}, K.~D., {Peterson}, B.~M., {Dietrich}, M., {Vestergaard}, M., \&
  {Bentz}, M.~C. 2009, \apj, 692, 246

\bibitem[{{Di Matteo} {et~al.}(2005){Di Matteo}, {Springel}, \&
  {Hernquist}}]{Matteo:2005}
{Di Matteo}, T., {Springel}, V., \& {Hernquist}, L. 2005, \nat, 433, 604

\bibitem[{{Diamond-Stanic} {et~al.}(2012){Diamond-Stanic}, {Moustakas},
  {Tremonti}, {Coil}, {Hickox}, {Robaina}, {Rudnick}, \&
  {Sell}}]{Diamond-Stanic:2012}
{Diamond-Stanic}, A.~M., {Moustakas}, J., {Tremonti}, C.~A., {et~al.} 2012,
  \apjl, 755, L26

\bibitem[{{Drake} {et~al.}(2019){Drake}, {Farina}, {Neeleman}, {Walter},
  {Venemans}, {Banados}, {Mazzucchelli}, \& {Decarli}}]{Drake:2019}
{Drake}, A.~B., {Farina}, E.~P., {Neeleman}, M., {et~al.} 2019, \apj, 881, 131

\bibitem[{{Dwelly} {et~al.}(2017){Dwelly}, {Salvato}, {Merloni}, {Brusa},
  {Buchner}, {Anderson}, {Boller}, {Brandt}, {Budav{\'a}ri}, {Clerc}, {Coffey},
  {Del Moro}, {Georgakakis}, {Green}, {Jin}, {Menzel}, {Myers}, {Nandra},
  {Nichol}, {Ridl}, {Schwope}, \& {Simm}}]{Dwelly:2017}
{Dwelly}, T., {Salvato}, M., {Merloni}, A., {et~al.} 2017, \mnras, 469, 1065

\bibitem[{{Eilers} {et~al.}(2017){Eilers}, {Davies}, {Hennawi}, {Prochaska},
  {Luki{\'c}}, \& {Mazzucchelli}}]{Eilers:2017}
{Eilers}, A.-C., {Davies}, F.~B., {Hennawi}, J.~F., {et~al.} 2017, \apj, 840,
  24

\bibitem[{{Elvis} {et~al.}(2009){Elvis}, {Civano}, {Vignali}, {Puccetti},
  {Fiore}, {Cappelluti}, {Aldcroft}, {Fruscione}, {Zamorani}, {Comastri},
  {Brusa}, {Gilli}, {Miyaji}, {Damiani}, {Koekemoer}, {Finoguenov}, {Brunner},
  {Urry}, {Silverman}, {Mainieri}, {Hasinger}, {Griffiths}, {Carollo}, {Hao},
  {Guzzo}, {Blain}, {Calzetti}, {Carilli}, {Capak}, {Ettori}, {Fabbiano},
  {Impey}, {Lilly}, {Mobasher}, {Rich}, {Salvato}, {Sanders}, {Schinnerer},
  {Scoville}, {Shopbell}, {Taylor}, {Taniguchi}, \& {Volonteri}}]{Elvis:2009}
{Elvis}, M., {Civano}, F., {Vignali}, C., {et~al.} 2009, \apjs, 184, 158

\bibitem[{{Erroz-Ferrer} {et~al.}(2019){Erroz-Ferrer}, {Carollo}, {den Brok},
  {Onodera}, {Brinchmann}, {Marino}, {Monreal-Ibero}, {Schaye}, {Woo},
  {Cibinel}, {Debattista}, {Inami}, {Maseda}, {Richard}, {Tacchella}, \&
  {Wisotzki}}]{Erroz-Ferrer:2019}
{Erroz-Ferrer}, S., {Carollo}, C.~M., {den Brok}, M., {et~al.} 2019, \mnras,
  484, 5009

\bibitem[{{Erwin}(2018)}]{Erwin:2018}
{Erwin}, P. 2018, \mnras, 474, 5372

\bibitem[{{Evans} {et~al.}(2006){Evans}, {Solomon}, {Tacconi}, {Vavilkin}, \&
  {Downes}}]{Evans:2006}
{Evans}, A.~S., {Solomon}, P.~M., {Tacconi}, L.~J., {Vavilkin}, T., \&
  {Downes}, D. 2006, \aj, 132, 2398

\bibitem[{{Fabian}(2012)}]{Fabian:2012}
{Fabian}, A.~C. 2012, \araa, 50, 455

\bibitem[{{Fabian} {et~al.}(2006{\natexlab{a}}){Fabian}, {Celotti}, \&
  {Erlund}}]{Fabian:2006}
{Fabian}, A.~C., {Celotti}, A., \& {Erlund}, M.~C. 2006{\natexlab{a}}, \mnras,
  373, L16

\bibitem[{{Fabian} {et~al.}(2006{\natexlab{b}}){Fabian}, {Sanders}, {Taylor},
  {Allen}, {Crawford}, {Johnstone}, \& {Iwasawa}}]{Fabian:2006b}
{Fabian}, A.~C., {Sanders}, J.~S., {Taylor}, G.~B., {et~al.}
  2006{\natexlab{b}}, \mnras, 366, 417

\bibitem[{{Feruglio} {et~al.}(2010){Feruglio}, {Maiolino}, {Piconcelli},
  {Menci}, {Aussel}, {Lamastra}, \& {Fiore}}]{Feruglio:2010}
{Feruglio}, C., {Maiolino}, R., {Piconcelli}, E., {et~al.} 2010, \aap, 518,
  L155

\bibitem[{{Fiore} {et~al.}(2017){Fiore}, {Feruglio}, {Shankar}, {Bischetti},
  {Bongiorno}, {Brusa}, {Carniani}, {Cicone}, {Duras}, {Lamastra}, {Mainieri},
  {Marconi}, {Menci}, {Maiolino}, {Piconcelli}, {Vietri}, \&
  {Zappacosta}}]{Fiore:2017}
{Fiore}, F., {Feruglio}, C., {Shankar}, F., {et~al.} 2017, \aap, 601, A143

\bibitem[{{Fischer} {et~al.}(2006){Fischer}, {Iserlohe}, {Zuther}, {Bertram},
  {Straubmeier}, {Sch{\"o}del}, \& {Eckart}}]{Fischer:2006}
{Fischer}, S., {Iserlohe}, C., {Zuther}, J., {et~al.} 2006, \aap, 452, 827

\bibitem[{{Fischer} {et~al.}(2018){Fischer}, {Kraemer}, {Schmitt}, {Longo
  Micchi}, {Crenshaw}, {Revalski}, {Vestergaard}, {Elvis}, {Gaskell}, {Hamann},
  {Ho}, {Hutchings}, {Mushotzky}, {Netzer}, {Storchi-Bergmann}, {Straughn},
  {Turner}, \& {Ward}}]{Fischer:2018}
{Fischer}, T.~C., {Kraemer}, S.~B., {Schmitt}, H.~R., {et~al.} 2018, \apj, 856,
  102

\bibitem[{{Foreman-Mackey} {et~al.}(2013){Foreman-Mackey}, {Hogg}, {Lang}, \&
  {Goodman}}]{Foreman-Mackey:2013}
{Foreman-Mackey}, D., {Hogg}, D.~W., {Lang}, D., \& {Goodman}, J. 2013, \pasp,
  125, 306

\bibitem[{{Forman} {et~al.}(2007){Forman}, {Jones}, {Churazov}, {Markevitch},
  {Nulsen}, {Vikhlinin}, {Begelman}, {B{\"o}hringer}, {Eilek}, {Heinz},
  {Kraft}, {Owen}, \& {Pahre}}]{Forman:2007}
{Forman}, W., {Jones}, C., {Churazov}, E., {et~al.} 2007, \apj, 665, 1057

\bibitem[{{Fruchter} \& {Hook}(2002)}]{Fruchter:2002}
{Fruchter}, A.~S. \& {Hook}, R.~N. 2002, \pasp, 114, 144

\bibitem[{{Fu} \& {Stockton}(2008)}]{Fu:2008}
{Fu}, H. \& {Stockton}, A. 2008, \apj, 677, 79

\bibitem[{{Fu} \& {Stockton}(2009)}]{Fu:2009}
{Fu}, H. \& {Stockton}, A. 2009, \apj, 690, 953

\bibitem[{{Gadotti}(2008)}]{Gadotti:2008}
{Gadotti}, D.~A. 2008, \mnras, 384, 420

\bibitem[{{Garc{\'\i}a-Burillo} {et~al.}(2005){Garc{\'\i}a-Burillo}, {Combes},
  {Schinnerer}, {Boone}, \& {Hunt}}]{Garcia-Burillo:2005}
{Garc{\'\i}a-Burillo}, S., {Combes}, F., {Schinnerer}, E., {Boone}, F., \&
  {Hunt}, L.~K. 2005, \aap, 441, 1011

\bibitem[{{Gaspari} {et~al.}(2012){Gaspari}, {Brighenti}, \&
  {Temi}}]{Gaspari:2012}
{Gaspari}, M., {Brighenti}, F., \& {Temi}, P. 2012, \mnras, 424, 190

\bibitem[{{Gaspari} {et~al.}(2019){Gaspari}, {Eckert}, {Ettori}, {Tozzi},
  {Bassini}, {Rasia}, {Brighenti}, {Sun}, {Borgani}, {Johnson}, {Tremblay},
  {Stone}, {Temi}, {Yang}, {Tombesi}, \& {Cappi}}]{Gaspari:2019}
{Gaspari}, M., {Eckert}, D., {Ettori}, S., {et~al.} 2019, \apj, 884, 169

\bibitem[{{Gaspari} {et~al.}(2018){Gaspari}, {McDonald}, {Hamer}, {Brighenti},
  {Temi}, {Gendron-Marsolais}, {Hlavacek-Larrondo}, {Edge}, {Werner}, {Tozzi},
  {Sun}, {Stone}, {Tremblay}, {Hogan}, {Eckert}, {Ettori}, {Yu}, {Biffi}, \&
  {Planelles}}]{Gaspari:2018}
{Gaspari}, M., {McDonald}, M., {Hamer}, S.~L., {et~al.} 2018, \apj, 854, 167

\bibitem[{{Gaspari} {et~al.}(2017){Gaspari}, {Temi}, \&
  {Brighenti}}]{Gaspari:2017}
{Gaspari}, M., {Temi}, P., \& {Brighenti}, F. 2017, \mnras, 466, 677

\bibitem[{{Gaspari} {et~al.}(2020){Gaspari}, {Tombesi}, \&
  {Cappi}}]{Gaspari:2020}
{Gaspari}, M., {Tombesi}, F., \& {Cappi}, M. 2020, Nature Astronomy, 4, 10

\bibitem[{{Gezari} {et~al.}(2009){Gezari}, {Heckman}, {Cenko}, {Eracleous},
  {Forster}, {Gon{\c{c}}alves}, {Martin}, {Morrissey}, {Neff}, {Seibert},
  {Schiminovich}, \& {Wyder}}]{Gezari:2009}
{Gezari}, S., {Heckman}, T., {Cenko}, S.~B., {et~al.} 2009, \apj, 698, 1367

\bibitem[{{Gordon} {et~al.}(2021){Gordon}, {Boyce}, {O'Dea}, {Rudnick},
  {Andernach}, {Vantyghem}, {Baum}, {Bui}, {Dionyssiou}, {Safi-Harb}, \&
  {Sander}}]{Gordon:2021}
{Gordon}, Y.~A., {Boyce}, M.~M., {O'Dea}, C.~P., {et~al.} 2021, \apjs, 255, 30

\bibitem[{{Graham} {et~al.}(2017){Graham}, {Djorgovski}, {Drake}, {Stern},
  {Mahabal}, {Glikman}, {Larson}, \& {Christensen}}]{Graham:2017}
{Graham}, M.~J., {Djorgovski}, S.~G., {Drake}, A.~J., {et~al.} 2017, \mnras,
  470, 4112

\bibitem[{{Granato} {et~al.}(2004){Granato}, {De Zotti}, {Silva}, {Bressan}, \&
  {Danese}}]{Granato:2004}
{Granato}, G.~L., {De Zotti}, G., {Silva}, L., {Bressan}, A., \& {Danese}, L.
  2004, \apj, 600, 580

\bibitem[{{Greene} \& {Ho}(2005{\natexlab{a}})}]{Greene:2005b}
{Greene}, J.~E. \& {Ho}, L.~C. 2005{\natexlab{a}}, \apj, 627, 721

\bibitem[{{Greene} \& {Ho}(2005{\natexlab{b}})}]{Greene:2005}
{Greene}, J.~E. \& {Ho}, L.~C. 2005{\natexlab{b}}, \apj, 630, 122

\bibitem[{{Greene} \& {Ho}(2007)}]{Greene:2007}
{Greene}, J.~E. \& {Ho}, L.~C. 2007, \apj, 670, 92

\bibitem[{{Greene} {et~al.}(2014){Greene}, {Pooley}, {Zakamska}, {Comerford},
  \& {Sun}}]{Greene:2014}
{Greene}, J.~E., {Pooley}, D., {Zakamska}, N.~L., {Comerford}, J.~M., \& {Sun},
  A.-L. 2014, \apj, 788, 54

\bibitem[{{Greene} {et~al.}(2011){Greene}, {Zakamska}, {Ho}, \&
  {Barth}}]{Greene:2011}
{Greene}, J.~E., {Zakamska}, N.~L., {Ho}, L.~C., \& {Barth}, A.~J. 2011, \apj,
  732, 9

\bibitem[{{Greene} {et~al.}(2012){Greene}, {Zakamska}, \&
  {Smith}}]{Greene:2012}
{Greene}, J.~E., {Zakamska}, N.~L., \& {Smith}, P.~S. 2012, \apj, 746, 86

\bibitem[{{Grier} {et~al.}(2017){Grier}, {Pancoast}, {Barth}, {Fausnaugh},
  {Brewer}, {Treu}, \& {Peterson}}]{Grier:2017}
{Grier}, C.~J., {Pancoast}, A., {Barth}, A.~J., {et~al.} 2017, \apj, 849, 146

\bibitem[{{Grier} {et~al.}(2013){Grier}, {Peterson}, {Horne}, {Bentz}, {Pogge},
  {Denney}, {De Rosa}, {Martini}, {Kochanek}, {Zu}, {Shappee}, {Siverd},
  {Beatty}, {Sergeev}, {Kaspi}, {Araya Salvo}, {Bird}, {Bord}, {Borman}, {Che},
  {Chen}, {Cohen}, {Dietrich}, {Doroshenko}, {Efimov}, {Free}, {Ginsburg},
  {Henderson}, {King}, {Mogren}, {Molina}, {Mosquera}, {Nazarov}, {Okhmat},
  {Pejcha}, {Rafter}, {Shields}, {Skowron}, {Szczygiel}, {Valluri}, \& {van
  Saders}}]{Grier:2013b}
{Grier}, C.~J., {Peterson}, B.~M., {Horne}, K., {et~al.} 2013, \apj, 764, 47

\bibitem[{{Grupe}(2004)}]{Grupe:2004b}
{Grupe}, D. 2004, \aj, 127, 1799

\bibitem[{{Haan} {et~al.}(2009){Haan}, {Schinnerer}, {Emsellem},
  {Garc{\'\i}a-Burillo}, {Combes}, {Mundell}, \& {Rix}}]{Haan:2009}
{Haan}, S., {Schinnerer}, E., {Emsellem}, E., {et~al.} 2009, \apj, 692, 1623

\bibitem[{{Hainline} {et~al.}(2013){Hainline}, {Hickox}, {Greene}, {Myers}, \&
  {Zakamska}}]{Hainline:2013}
{Hainline}, K.~N., {Hickox}, R., {Greene}, J.~E., {Myers}, A.~D., \&
  {Zakamska}, N.~L. 2013, \apj, 774, 145

\bibitem[{{Hainline} {et~al.}(2014){Hainline}, {Hickox}, {Greene}, {Myers},
  {Zakamska}, {Liu}, \& {Liu}}]{Hainline:2014}
{Hainline}, K.~N., {Hickox}, R.~C., {Greene}, J.~E., {et~al.} 2014, \apj, 787,
  65

\bibitem[{{Hardcastle} {et~al.}(1998){Hardcastle}, {Alexander}, {Pooley}, \&
  {Riley}}]{Hardcastle:1998}
{Hardcastle}, M.~J., {Alexander}, P., {Pooley}, G.~G., \& {Riley}, J.~M. 1998,
  \mnras, 296, 445

\bibitem[{{H{\"a}ring} \& {Rix}(2004)}]{Haering:2004}
{H{\"a}ring}, N. \& {Rix}, H.-W. 2004, \apjl, 604, L89

\bibitem[{{Harrison}(2017)}]{Harrison:2017}
{Harrison}, C.~M. 2017, Nature Astronomy, 1, 0165

\bibitem[{{Harrison} {et~al.}(2012){Harrison}, {Alexander}, {Mullaney},
  {Altieri}, {Coia}, {Charmandaris}, {Daddi}, {Dannerbauer}, {Dasyra}, {Del
  Moro}, {Dickinson}, {Hickox}, {Ivison}, {Kartaltepe}, {Le Floc'h}, {Leiton},
  {Magnelli}, {Popesso}, {Rovilos}, {Rosario}, \& {Swinbank}}]{Harrison:2012b}
{Harrison}, C.~M., {Alexander}, D.~M., {Mullaney}, J.~R., {et~al.} 2012, \apjl,
  760, L15

\bibitem[{{Harrison} {et~al.}(2016){Harrison}, {Alexander}, {Mullaney},
  {Stott}, {Swinbank}, {Arumugam}, {Bauer}, {Bower}, {Bunker}, \&
  {Sharples}}]{Harrison:2016b}
{Harrison}, C.~M., {Alexander}, D.~M., {Mullaney}, J.~R., {et~al.} 2016,
  \mnras, 456, 1195

\bibitem[{{Harrison} {et~al.}(2014){Harrison}, {Alexander}, {Mullaney}, \&
  {Swinbank}}]{Harrison:2014}
{Harrison}, C.~M., {Alexander}, D.~M., {Mullaney}, J.~R., \& {Swinbank}, A.~M.
  2014, \mnras, 441, 3306

\bibitem[{{Harrison} {et~al.}(2018){Harrison}, {Costa}, {Tadhunter},
  {Fl{\"u}tsch}, {Kakkad}, {Perna}, \& {Vietri}}]{Harrison:2018}
{Harrison}, C.~M., {Costa}, T., {Tadhunter}, C.~N., {et~al.} 2018, Nature
  Astronomy, 2, 198

\bibitem[{{Harrison} {et~al.}(2015){Harrison}, {Thomson}, {Alexander}, {Bauer},
  {Edge}, {Hogan}, {Mullaney}, \& {Swinbank}}]{Harrison:2015}
{Harrison}, C.~M., {Thomson}, A.~P., {Alexander}, D.~M., {et~al.} 2015, \apj,
  800, 45

\bibitem[{{Heckman} {et~al.}(1991){Heckman}, {Lehnert}, {Miley}, \& {van
  Breugel}}]{Heckman:1991}
{Heckman}, T.~M., {Lehnert}, M.~D., {Miley}, G.~K., \& {van Breugel}, W. 1991,
  \apj, 381, 373

\bibitem[{{Herenz} {et~al.}(2015){Herenz}, {Wisotzki}, {Roth}, \&
  {Anders}}]{Herenz:2015}
{Herenz}, E.~C., {Wisotzki}, L., {Roth}, M., \& {Anders}, F. 2015, \aap, 576,
  A115

\bibitem[{{Hickox} {et~al.}(2014){Hickox}, {Mullaney}, {Alexander}, {Chen},
  {Civano}, {Goulding}, \& {Hainline}}]{Hickox:2014}
{Hickox}, R.~C., {Mullaney}, J.~R., {Alexander}, D.~M., {et~al.} 2014, \apj,
  782, 9

\bibitem[{{Ho} {et~al.}(2016){Ho}, {Medling}, {Bland-Hawthorn}, {Groves},
  {Kewley}, {Kobayashi}, {Dopita}, {Leslie}, {Sharp}, {Allen}, {Bourne},
  {Bryant}, {Cortese}, {Croom}, {Dunne}, {Fogarty}, {Goodwin}, {Green},
  {Konstantopoulos}, {Lawrence}, {Lorente}, {Owers}, {Richards}, {Sweet},
  {Tescari}, \& {Valiante}}]{Ho:2016}
{Ho}, I.-T., {Medling}, A.~M., {Bland-Hawthorn}, J., {et~al.} 2016, \mnras,
  457, 1257

\bibitem[{{Holt} {et~al.}(2008){Holt}, {Tadhunter}, \& {Morganti}}]{Holt:2008}
{Holt}, J., {Tadhunter}, C.~N., \& {Morganti}, R. 2008, \mnras, 387, 639

\bibitem[{{Hopkins} {et~al.}(2008){Hopkins}, {Hernquist}, {Cox}, \& {Kere{\v
  s}}}]{Hopkins:2008a}
{Hopkins}, P.~F., {Hernquist}, L., {Cox}, T.~J., \& {Kere{\v s}}, D. 2008,
  \apjs, 175, 356

\bibitem[{{Husemann} {et~al.}(2019{\natexlab{a}}){Husemann}, {Bennert},
  {Jahnke}, {Davis}, {Woo}, {Scharw{\"a}chter}, {Schulze}, {Gaspari}, \&
  {Zwaan}}]{Husemann:2019b}
{Husemann}, B., {Bennert}, V.~N., {Jahnke}, K., {et~al.} 2019{\natexlab{a}},
  \apj, 879, 75

\bibitem[{{Husemann} {et~al.}(2017{\natexlab{a}}){Husemann}, {Davis}, {Jahnke},
  {Dannerbauer}, {Urrutia}, \& {Hodge}}]{Husemann:2017}
{Husemann}, B., {Davis}, T.~A., {Jahnke}, K., {et~al.} 2017{\natexlab{a}},
  \mnras, 470, 1570

\bibitem[{{Husemann} \& {Harrison}(2018)}]{Husemann:2018b}
{Husemann}, B. \& {Harrison}, C.~M. 2018, Nature Astronomy, 2, 196

\bibitem[{{Husemann} {et~al.}(2013{\natexlab{a}}){Husemann}, {Jahnke},
  {S{\'a}nchez}, {Barrado}, {Bekerait*error*{\.e}}, {Bomans},
  {Castillo-Morales}, {Catal{\'a}n-Torrecilla}, {Cid Fernandes},
  {Falc{\'o}n-Barroso}, {Garc{\'{\i}}a-Benito}, {Gonz{\'a}lez Delgado},
  {Iglesias-P{\'a}ramo}, {Johnson}, {Kupko}, {L{\'o}pez-Fernandez},
  {Lyubenova}, {Marino}, {Mast}, {Miskolczi}, {Monreal-Ibero}, {Gil de Paz},
  {P{\'e}rez}, {P{\'e}rez}, {Rosales-Ortega}, {Ruiz-Lara}, {Schilling}, {van de
  Ven}, {Walcher}, {Alves}, {de Amorim}, {Backsmann}, {Barrera-Ballesteros},
  {Bland-Hawthorn}, {Cortijo}, {Dettmar}, {Demleitner}, {D{\'{\i}}az}, {Enke},
  {Florido}, {Flores}, {Galbany}, {Gallazzi}, {Garc{\'{\i}}a-Lorenzo}, {Gomes},
  {Gruel}, {Haines}, {Holmes}, {Jungwiert}, {Kalinova}, {Kehrig}, {Kennicutt},
  {Klar}, {Lehnert}, {L{\'o}pez-S{\'a}nchez}, {de Lorenzo-C{\'a}ceres},
  {M{\'a}rmol-Queralt{\'o}}, {M{\'a}rquez}, {Mendez-Abreu}, {Moll{\'a}}, {del
  Olmo}, {Meidt}, {Papaderos}, {Puschnig}, {Quirrenbach}, {Roth},
  {S{\'a}nchez-Bl{\'a}zquez}, {Spekkens}, {Singh}, {Stanishev}, {Trager},
  {Vilchez}, {Wild}, {Wisotzki}, {Zibetti}, \& {Ziegler}}]{Husemann:2013b}
{Husemann}, B., {Jahnke}, K., {S{\'a}nchez}, S.~F., {et~al.}
  2013{\natexlab{a}}, \aap, 549, A87

\bibitem[{{Husemann} {et~al.}(2014){Husemann}, {Jahnke}, {S{\'a}nchez},
  {Wisotzki}, {Nugroho}, {Kupko}, \& {Schramm}}]{Husemann:2014}
{Husemann}, B., {Jahnke}, K., {S{\'a}nchez}, S.~F., {et~al.} 2014, \mnras, 443,
  755

\bibitem[{{Husemann} {et~al.}(2019{\natexlab{b}}){Husemann},
  {Scharw{\"a}chter}, {Davis}, {P{\'e}rez-Torres}, {Smirnova-Pinchukova},
  {Tremblay}, {Krumpe}, {Combes}, {Baum}, \& {Busch}}]{Husemann:2019a}
{Husemann}, B., {Scharw{\"a}chter}, J., {Davis}, T.~A., {et~al.}
  2019{\natexlab{b}}, \aap, 627, A53

\bibitem[{{Husemann} {et~al.}(2017{\natexlab{b}}){Husemann}, {Tremblay},
  {Davis}, {Busch}, {McElroy}, {Neumann}, {Urrutia}, {Krumpe},
  {Scharw{\"a}chter}, {Powell}, {Perez-Torres}, \& {The CARS
  Team}}]{Husemann:2017b}
{Husemann}, B., {Tremblay}, G., {Davis}, T., {et~al.} 2017{\natexlab{b}}, The
  Messenger, 169, 42

\bibitem[{{Husemann} {et~al.}(2016){Husemann}, {Urrutia}, {Tremblay}, {Krumpe},
  {Dexter}, {Busch}, {Combes}, {Croom}, {Davis}, {Eckart}, {McElroy},
  {Perez-Torres}, {Powell}, \& {Scharw{\"a}chter}}]{Husemann:2016b}
{Husemann}, B., {Urrutia}, T., {Tremblay}, G.~R., {et~al.} 2016, \aap, 593, L9

\bibitem[{{Husemann} {et~al.}(2008){Husemann}, {Wisotzki}, {S{\'a}nchez}, \&
  {Jahnke}}]{Husemann:2008}
{Husemann}, B., {Wisotzki}, L., {S{\'a}nchez}, S.~F., \& {Jahnke}, K. 2008,
  \aap, 488, 145

\bibitem[{{Husemann} {et~al.}(2013{\natexlab{b}}){Husemann}, {Wisotzki},
  {S{\'a}nchez}, \& {Jahnke}}]{Husemann:2013a}
{Husemann}, B., {Wisotzki}, L., {S{\'a}nchez}, S.~F., \& {Jahnke}, K.
  2013{\natexlab{b}}, \aap, 549, A43

\bibitem[{{Jahnke} {et~al.}(2004{\natexlab{a}}){Jahnke}, {Kuhlbrodt}, \&
  {Wisotzki}}]{Jahnke:2004b}
{Jahnke}, K., {Kuhlbrodt}, B., \& {Wisotzki}, L. 2004{\natexlab{a}}, \mnras,
  352, 399

\bibitem[{{Jahnke} {et~al.}(2004{\natexlab{b}}){Jahnke}, {Wisotzki},
  {S{\'a}nchez}, {Christensen}, {Becker}, {Kelz}, \& {Roth}}]{Jahnke:2004}
{Jahnke}, K., {Wisotzki}, L., {S{\'a}nchez}, S.~F., {et~al.}
  2004{\natexlab{b}}, AN, 325, 128

\bibitem[{{Jarvis} {et~al.}(2021){Jarvis}, {Harrison}, {Mainieri}, {Alexander},
  {Arrigoni Battaia}, {Calistro Rivera}, {Circosta}, {Costa}, {De Breuck},
  {Edge}, {Girdhar}, {Kakkad}, {Kharb}, {Lansbury}, {Molyneux}, {Mukherjee},
  {Mullaney}, {Farina}, {Silpa}, {Thomson}, \& {Ward}}]{Jarvis:2021}
{Jarvis}, M.~E., {Harrison}, C.~M., {Mainieri}, V., {et~al.} 2021, \mnras, 503,
  1780

\bibitem[{{Jarvis} {et~al.}(2019){Jarvis}, {Harrison}, {Thomson}, {Circosta},
  {Mainieri}, {Alexander}, {Edge}, {Lansbury}, {Molyneux}, \&
  {Mullaney}}]{Jarvis:2019}
{Jarvis}, M.~E., {Harrison}, C.~M., {Thomson}, A.~P., {et~al.} 2019, \mnras,
  485, 2710

\bibitem[{Jedrzejewski(1987)}]{Jedrzejewski:1987}
Jedrzejewski, R.~I. 1987, \mnras, 226, 747

\bibitem[{{Jun} {et~al.}(2015){Jun}, {Im}, {Lee}, {Ohyama}, {Woo}, {Fan},
  {Goto}, {Kim}, {Kim}, {Kim}, {Lee}, {Nakagawa}, {Pearson}, \&
  {Serjeant}}]{Jun:2015}
{Jun}, H.~D., {Im}, M., {Lee}, H.~M., {et~al.} 2015, \apj, 806, 109

\bibitem[{{Kakkad} {et~al.}(2018){Kakkad}, {Groves}, {Dopita}, {Thomas},
  {Davies}, {Mainieri}, {Kharb}, {Scharw{\"a}chter}, {Hampton}, \&
  {Ho}}]{Kakkad:2018}
{Kakkad}, D., {Groves}, B., {Dopita}, M., {et~al.} 2018, \aap, 618, A6

\bibitem[{{Kakkad} {et~al.}(2017){Kakkad}, {Mainieri}, {Brusa}, {Padovani},
  {Carniani}, {Feruglio}, {Sargent}, {Husemann}, {Bongiorno}, {Bonzini},
  {Piconcelli}, {Silverman}, \& {Rujopakarn}}]{Kakkad:2017}
{Kakkad}, D., {Mainieri}, V., {Brusa}, M., {et~al.} 2017, \mnras, 468, 4205

\bibitem[{{Kakkad} {et~al.}(2020){Kakkad}, {Mainieri}, {Vietri}, {Carniani},
  {Harrison}, {Perna}, {Scholtz}, {Circosta}, {Cresci}, {Husemann},
  {Bischetti}, {Feruglio}, {Fiore}, {Marconi}, {Padovani}, {Brusa}, {Cicone},
  {Comastri}, {Lanzuisi}, {Mannucci}, {Menci}, {Netzer}, {Piconcelli},
  {Puglisi}, {Salvato}, {Schramm}, {Silverman}, {Vignali}, {Zamorani}, \&
  {Zappacosta}}]{Kakkad:2020}
{Kakkad}, D., {Mainieri}, V., {Vietri}, G., {et~al.} 2020, \aap, 642, A147

\bibitem[{{Kang} \& {Woo}(2018)}]{Kang:2018}
{Kang}, D. \& {Woo}, J.-H. 2018, \apj, 864, 124

\bibitem[{{Kaspi} {et~al.}(2000){Kaspi}, {Smith}, {Netzer}, {Maoz}, {Jannuzi},
  \& {Giveon}}]{Kaspi:2000}
{Kaspi}, S., {Smith}, P.~S., {Netzer}, H., {et~al.} 2000, \apj, 533, 631

\bibitem[{{Kauffmann} {et~al.}(2003){Kauffmann}, {Heckman}, {Tremonti},
  {Brinchmann}, {Charlot}, {White}, {Ridgway}, {Brinkmann}, {Fukugita}, {Hall},
  {Ivezi{\'c}}, {Richards}, \& {Schneider}}]{Kauffmann:2003}
{Kauffmann}, G., {Heckman}, T.~M., {Tremonti}, C., {et~al.} 2003, \mnras, 346,
  1055

\bibitem[{{Keel} {et~al.}(2012){Keel}, {Chojnowski}, {Bennert}, {Schawinski},
  {Lintott}, {Lynn}, {Pancoast}, {Harris}, {Nierenberg}, {Sonnenfeld}, \&
  {Proctor}}]{Keel:2012}
{Keel}, W.~C., {Chojnowski}, S.~D., {Bennert}, V.~N., {et~al.} 2012, \mnras,
  420, 878

\bibitem[{{Keel} {et~al.}(2017){Keel}, {Lintott}, {Maksym}, {Bennert},
  {Chojnowski}, {Moiseev}, {Smirnova}, {Schawinski}, {Sartori}, {Urry},
  {Pancoast}, {Schirmer}, {Scott}, {Showley}, \& {Flatland}}]{Keel:2017}
{Keel}, W.~C., {Lintott}, C.~J., {Maksym}, W.~P., {et~al.} 2017, \apj, 835, 256

\bibitem[{{Kelly} \& {Shen}(2013)}]{Kelly:2013}
{Kelly}, B.~C. \& {Shen}, Y. 2013, \apj, 764, 45

\bibitem[{{Kelz} {et~al.}(2006){Kelz}, {Verheijen}, {Roth}, {Bauer}, {Becker},
  {Paschke}, {Popow}, {S{\'a}nchez}, \& {Laux}}]{Kelz:2006}
{Kelz}, A., {Verheijen}, M.~A.~W., {Roth}, M.~M., {et~al.} 2006, \pasp, 118,
  129

\bibitem[{{Kewley} {et~al.}(2001){Kewley}, {Dopita}, {Sutherland}, {Heisler},
  \& {Trevena}}]{Kewley:2001}
{Kewley}, L.~J., {Dopita}, M.~A., {Sutherland}, R.~S., {Heisler}, C.~A., \&
  {Trevena}, J. 2001, \apj, 556, 121

\bibitem[{{Kewley} {et~al.}(2006){Kewley}, {Groves}, {Kauffmann}, \&
  {Heckman}}]{Kewley:2006}
{Kewley}, L.~J., {Groves}, B., {Kauffmann}, G., \& {Heckman}, T. 2006, \mnras,
  372, 961

\bibitem[{{Khrykin} {et~al.}(2021){Khrykin}, {Hennawi}, {Worseck}, \&
  {Davies}}]{Khrykin:2021}
{Khrykin}, I.~S., {Hennawi}, J.~F., {Worseck}, G., \& {Davies}, F.~B. 2021,
  \mnras, 505, 649

\bibitem[{{Kollmeier} {et~al.}(2006){Kollmeier}, {Onken}, {Kochanek}, {Gould},
  {Weinberg}, {Dietrich}, {Cool}, {Dey}, {Eisenstein}, {Jannuzi}, {Le Floc'h},
  \& {Stern}}]{Kollmeier:2006}
{Kollmeier}, J.~A., {Onken}, C.~A., {Kochanek}, C.~S., {et~al.} 2006, \apj,
  648, 128

\bibitem[{{K{\"o}nig} {et~al.}(2009){K{\"o}nig}, {Eckart},
  {Garc{\'{\i}}a-Mar{\'{\i}}n}, \& {Huchtmeier}}]{Koenig:2009}
{K{\"o}nig}, S., {Eckart}, A., {Garc{\'{\i}}a-Mar{\'{\i}}n}, M., \&
  {Huchtmeier}, W.~K. 2009, \aap, 507, 757

\bibitem[{{Kormendy} \& {Ho}(2013)}]{Kormendy:2013}
{Kormendy}, J. \& {Ho}, L.~C. 2013, \araa, 51, 511

\bibitem[{{Koss} {et~al.}(2017){Koss}, {Trakhtenbrot}, {Ricci}, {Lamperti},
  {Oh}, {Berney}, {Schawinski}, {Balokovi{\'c}}, {Baronchelli}, {Crenshaw},
  {Fischer}, {Gehrels}, {Harrison}, {Hashimoto}, {Hogg}, {Ichikawa}, {Masetti},
  {Mushotzky}, {Sartori}, {Stern}, {Treister}, {Ueda}, {Veilleux}, \&
  {Winter}}]{Koss:2017}
{Koss}, M., {Trakhtenbrot}, B., {Ricci}, C., {et~al.} 2017, \apj, 850, 74

\bibitem[{{Koss} {et~al.}(2021){Koss}, {Strittmatter}, {Lamperti}, {Shimizu},
  {Trakhtenbrot}, {Saintonge}, {Treister}, {Cicone}, {Mushotzky}, {Oh},
  {Ricci}, {Stern}, {Ananna}, {Bauer}, {Privon}, {B{\"a}r}, {De Breuck},
  {Harrison}, {Ichikawa}, {Powell}, {Rosario}, {Sanders}, {Schawinski}, {Shao},
  {Megan Urry}, \& {Veilleux}}]{Koss:2021}
{Koss}, M.~J., {Strittmatter}, B., {Lamperti}, I., {et~al.} 2021, \apjs, 252,
  29

\bibitem[{{Kova{\v c}evi{\'c}} {et~al.}(2010){Kova{\v c}evi{\'c}},
  {Popovi{\'c}}, \& {Dimitrijevi{\'c}}}]{Kovacevic:2010}
{Kova{\v c}evi{\'c}}, J., {Popovi{\'c}}, L.~{\v C}., \& {Dimitrijevi{\'c}},
  M.~S. 2010, \apjs, 189, 15

\bibitem[{{Krumpe} {et~al.}(2017){Krumpe}, {Husemann}, {Tremblay}, {Urrutia},
  {Powell}, {Davis}, {Scharw{\"a}chter}, {Dexter}, {Busch}, {Combes}, {Croom},
  {Eckart}, {McElroy}, {Perez-Torres}, \& {Leung}}]{Krumpe:2017}
{Krumpe}, M., {Husemann}, B., {Tremblay}, G.~R., {et~al.} 2017, \aap, 607, L9

\bibitem[{{Krumpe} {et~al.}(2015){Krumpe}, {Miyaji}, {Husemann}, {Fanidakis},
  {Coil}, \& {Aceves}}]{Krumpe:2015}
{Krumpe}, M., {Miyaji}, T., {Husemann}, B., {et~al.} 2015, \apj, 815, 21

\bibitem[{{Kurtz} \& {Mink}(2000)}]{Kurtz:2000}
{Kurtz}, M.~J. \& {Mink}, D.~J. 2000, \apjl, 533, L183

\bibitem[{{Lacerda} {et~al.}(2018){Lacerda}, {Cid Fernandes}, {Couto},
  {Stasi{\'n}ska}, {Garc{\'\i}a-Benito}, {Vale Asari}, {P{\'e}rez},
  {Gonz{\'a}lez Delgado}, {S{\'a}nchez}, \& {de Amorim}}]{Lacerda:2018}
{Lacerda}, E.~A.~D., {Cid Fernandes}, R., {Couto}, G.~S., {et~al.} 2018,
  \mnras, 474, 3727

\bibitem[{{Lacy} {et~al.}(2020){Lacy}, {Baum}, {Chandler}, {Chatterjee},
  {Clarke}, {Deustua}, {English}, {Farnes}, {Gaensler}, {Gugliucci},
  {Hallinan}, {Kent}, {Kimball}, {Law}, {Lazio}, {Marvil}, {Mao}, {Medlin},
  {Mooley}, {Murphy}, {Myers}, {Osten}, {Richards}, {Rosolowsky}, {Rudnick},
  {Schinzel}, {Sivakoff}, {Sjouwerman}, {Taylor}, {White}, {Wrobel},
  {Andernach}, {Beasley}, {Berger}, {Bhatnager}, {Birkinshaw}, {Bower},
  {Brandt}, {Brown}, {Burke-Spolaor}, {Butler}, {Comerford}, {Demorest}, {Fu},
  {Giacintucci}, {Golap}, {G{\"u}th}, {Hales}, {Hiriart}, {Hodge}, {Horesh},
  {Ivezi{\'c}}, {Jarvis}, {Kamble}, {Kassim}, {Liu}, {Loinard}, {Lyons},
  {Masters}, {Mezcua}, {Moellenbrock}, {Mroczkowski}, {Nyland}, {O'Dea},
  {O'Sullivan}, {Peters}, {Radford}, {Rao}, {Robnett}, {Salcido}, {Shen},
  {Sobotka}, {Witz}, {Vaccari}, {van Weeren}, {Vargas}, {Williams}, \&
  {Yoon}}]{Lacy:2020}
{Lacy}, M., {Baum}, S.~A., {Chandler}, C.~J., {et~al.} 2020, \pasp, 132, 035001

\bibitem[{{Laha} {et~al.}(2018){Laha}, {Ghosh}, {Guainazzi}, \&
  {Markowitz}}]{Laha:2018}
{Laha}, S., {Ghosh}, R., {Guainazzi}, M., \& {Markowitz}, A.~G. 2018, \mnras,
  480, 1522

\bibitem[{{Lansbury} {et~al.}(2018){Lansbury}, {Jarvis}, {Harrison},
  {Alexander}, {Del Moro}, {Edge}, {Mullaney}, \& {Thomson}}]{Lansbury:2018}
{Lansbury}, G.~B., {Jarvis}, M.~E., {Harrison}, C.~M., {et~al.} 2018, \apjl,
  856, L1

\bibitem[{{Lanzuisi} {et~al.}(2014){Lanzuisi}, {Ponti}, {Salvato}, {Hasinger},
  {Cappelluti}, {Bongiorno}, {Brusa}, {Lusso}, {Nandra}, {Merloni},
  {Silverman}, {Trump}, {Vignali}, {Comastri}, {Gilli}, {Schramm},
  {Steinhardt}, {Sanders}, {Kartaltepe}, {Rosario}, \&
  {Trakhtenbrot}}]{Lanzuisi:2014}
{Lanzuisi}, G., {Ponti}, G., {Salvato}, M., {et~al.} 2014, \apj, 781, 105

\bibitem[{{Le F{\`e}vre} {et~al.}(2013){Le F{\`e}vre}, {Cassata}, {Cucciati},
  {Garilli}, {Ilbert}, {Le Brun}, {Maccagni}, {Moreau}, {Scodeggio}, {Tresse},
  {Zamorani}, {Adami}, {Arnouts}, {Bardelli}, {Bolzonella}, {Bondi},
  {Bongiorno}, {Bottini}, {Cappi}, {Charlot}, {Ciliegi}, {Contini}, {de la
  Torre}, {Foucaud}, {Franzetti}, {Gavignaud}, {Guzzo}, {Iovino}, {Lemaux},
  {L{\'o}pez-Sanjuan}, {McCracken}, {Marano}, {Marinoni}, {Mazure}, {Mellier},
  {Merighi}, {Merluzzi}, {Paltani}, {Pell{\`o}}, {Pollo}, {Pozzetti},
  {Scaramella}, {Tasca}, {Vergani}, {Vettolani}, {Zanichelli}, \&
  {Zucca}}]{LeFevre:2013}
{Le F{\`e}vre}, O., {Cassata}, P., {Cucciati}, O., {et~al.} 2013, \aap, 559,
  A14

\bibitem[{{Le F{\'e}vre} {et~al.}(2003){Le F{\'e}vre}, {Saisse}, {Mancini},
  {Brau-Nogue}, {Caputi}, {Castinel}, {D'Odorico}, {Garilli}, {Kissler-Patig},
  {Lucuix}, {Mancini}, {Pauget}, {Sciarretta}, {Scodeggio}, {Tresse}, \&
  {Vettolani}}]{LeFevre:2003}
{Le F{\'e}vre}, O., {Saisse}, M., {Mancini}, D., {et~al.} 2003, in in Proc.
  SPIE, ed. {M.~Iye \& A.~F.~M.~Moorwood}, Vol. 4841, 1671--1681

\bibitem[{{Levy} {et~al.}(2019){Levy}, {Bolatto}, {S{\'a}nchez}, {Blitz},
  {Colombo}, {Kalinova}, {L{\'o}pez-Cob{\'a}}, {Ostriker}, {Teuben}, {Utomo},
  {Vogel}, \& {Wong}}]{Levy:2019}
{Levy}, R.~C., {Bolatto}, A.~D., {S{\'a}nchez}, S.~F., {et~al.} 2019, \apj,
  882, 84

\bibitem[{{Lintott} {et~al.}(2009){Lintott}, {Schawinski}, {Keel}, {van Arkel},
  {Bennert}, {Edmondson}, {Thomas}, {Smith}, {Herbert}, {Jarvis}, {Virani},
  {Andreescu}, {Bamford}, {Land}, {Murray}, {Nichol}, {Raddick}, {Slosar},
  {Szalay}, \& {Vandenberg}}]{Lintott:2009}
{Lintott}, C.~J., {Schawinski}, K., {Keel}, W., {et~al.} 2009, \mnras, 399, 129

\bibitem[{{Liu} {et~al.}(2014){Liu}, {Zakamska}, \& {Greene}}]{Liu:2014}
{Liu}, G., {Zakamska}, N.~L., \& {Greene}, J.~E. 2014, \mnras, 442, 1303

\bibitem[{{Liu} {et~al.}(2013{\natexlab{a}}){Liu}, {Zakamska}, {Greene},
  {Nesvadba}, \& {Liu}}]{Liu:2013}
{Liu}, G., {Zakamska}, N.~L., {Greene}, J.~E., {Nesvadba}, N.~P.~H., \& {Liu},
  X. 2013{\natexlab{a}}, \mnras, 430, 2327

\bibitem[{{Liu} {et~al.}(2013{\natexlab{b}}){Liu}, {Zakamska}, {Greene},
  {Nesvadba}, \& {Liu}}]{Liu:2013b}
{Liu}, G., {Zakamska}, N.~L., {Greene}, J.~E., {Nesvadba}, N.~P.~H., \& {Liu},
  X. 2013{\natexlab{b}}, \mnras, 436, 2576

\bibitem[{{L{\'o}pez-Cob{\'a}} {et~al.}(2019){L{\'o}pez-Cob{\'a}},
  {S{\'a}nchez}, {Bland-Hawthorn}, {Moiseev}, {Cruz-Gonz{\'a}lez},
  {Garc{\'\i}a-Benito}, {Barrera-Ballesteros}, \& {Galbany}}]{Lopez-Coba:2019}
{L{\'o}pez-Cob{\'a}}, C., {S{\'a}nchez}, S.~F., {Bland-Hawthorn}, J., {et~al.}
  2019, \mnras, 482, 4032

\bibitem[{{Lutz} {et~al.}(2008){Lutz}, {Sturm}, {Tacconi}, {Valiante},
  {Schweitzer}, {Netzer}, {Maiolino}, {Andreani}, {Shemmer}, \&
  {Veilleux}}]{Lutz:2008}
{Lutz}, D., {Sturm}, E., {Tacconi}, L.~J., {et~al.} 2008, \apj, 684, 853

\bibitem[{{MacLeod} {et~al.}(2016){MacLeod}, {Ross}, {Lawrence}, {Goad},
  {Horne}, {Burgett}, {Chambers}, {Flewelling}, {Hodapp}, {Kaiser}, {Magnier},
  {Wainscoat}, \& {Waters}}]{MacLeod:2016}
{MacLeod}, C.~L., {Ross}, N.~P., {Lawrence}, A., {et~al.} 2016, \mnras, 457,
  389

\bibitem[{{Magnier} {et~al.}(2020){Magnier}, {Schlafly}, {Finkbeiner}, {Tonry},
  {Goldman}, {R{\"o}ser}, {Schilbach}, {Casertano}, {Chambers}, {Flewelling},
  {Huber}, {Price}, {Sweeney}, {Waters}, {Denneau}, {Draper}, {Hodapp},
  {Jedicke}, {Kaiser}, {Kudritzki}, {Metcalfe}, {Stubbs}, \&
  {Wainscoat}}]{Magnier:2020}
{Magnier}, E.~A., {Schlafly}, E.~F., {Finkbeiner}, D.~P., {et~al.} 2020, \apjs,
  251, 6

\bibitem[{{Magorrian} {et~al.}(1998){Magorrian}, {Tremaine}, {Richstone},
  {Bender}, {Bower}, {Dressler}, {Faber}, {Gebhardt}, {Green}, {Grillmair},
  {Kormendy}, \& {Lauer}}]{Magorrian:1998}
{Magorrian}, J., {Tremaine}, S., {Richstone}, D., {et~al.} 1998, \aj, 115, 2285

\bibitem[{{Marconi} \& {Hunt}(2003)}]{Marconi:2003}
{Marconi}, A. \& {Hunt}, L.~K. 2003, \apjl, 589, L21

\bibitem[{{Marian} {et~al.}(2019){Marian}, {Jahnke}, {Mechtley}, {Cohen},
  {Husemann}, {Jones}, {Koekemoer}, {Schulze}, {van der Wel}, {Villforth}, \&
  {Windhorst}}]{Marian:2019}
{Marian}, V., {Jahnke}, K., {Mechtley}, M., {et~al.} 2019, \apj, 882, 141

\bibitem[{{Masters} {et~al.}(2011){Masters}, {Nichol}, {Hoyle}, {Lintott},
  {Bamford}, {Edmondson}, {Fortson}, {Keel}, {Schawinski}, {Smith}, \&
  {Thomas}}]{Masters:2011}
{Masters}, K.~L., {Nichol}, R.~C., {Hoyle}, B., {et~al.} 2011, \mnras, 411,
  2026

\bibitem[{{Mathur}(2000)}]{Mathur:2000}
{Mathur}, S. 2000, \nar, 44, 469

\bibitem[{{Matsuoka} {et~al.}(2007){Matsuoka}, {Oyabu}, {Tsuzuki}, \&
  {Kawara}}]{Matsuoka:2007}
{Matsuoka}, Y., {Oyabu}, S., {Tsuzuki}, Y., \& {Kawara}, K. 2007, \apj, 663,
  781

\bibitem[{{McConnell} {et~al.}(2011){McConnell}, {Ma}, {Gebhardt}, {Wright},
  {Murphy}, {Lauer}, {Graham}, \& {Richstone}}]{McConnell:2011}
{McConnell}, N.~J., {Ma}, C.-P., {Gebhardt}, K., {et~al.} 2011, \nat, 480, 215

\bibitem[{{McElroy} {et~al.}(2015){McElroy}, {Croom}, {Pracy}, {Sharp}, {Ho},
  \& {Medling}}]{McElroy:2015}
{McElroy}, R., {Croom}, S.~M., {Pracy}, M., {et~al.} 2015, \mnras, 446, 2186

\bibitem[{{McElroy} {et~al.}(2016){McElroy}, {Husemann}, {Croom}, {Davis},
  {Bennert}, {Busch}, {Combes}, {Eckart}, {Perez-Torres}, {Powell},
  {Scharwaechter}, {Tremblay}, \& {Urrutia}}]{McElroy:2016}
{McElroy}, R., {Husemann}, B., {Croom}, S.~M., {et~al.} 2016, \aap, 593, L8

\bibitem[{{McNamara} {et~al.}(2000){McNamara}, {Wise}, {Nulsen}, {David},
  {Sarazin}, {Bautz}, {Markevitch}, {Vikhlinin}, {Forman}, {Jones}, \&
  {Harris}}]{McNamara:2000}
{McNamara}, B.~R., {Wise}, M., {Nulsen}, P.~E.~J., {et~al.} 2000, \apjl, 534,
  L135

\bibitem[{{Mechtley} {et~al.}(2016){Mechtley}, {Jahnke}, {Windhorst}, {Andrae},
  {Cisternas}, {Cohen}, {Hewlett}, {Koekemoer}, {Schramm}, {Schulze},
  {Silverman}, {Villforth}, {van der Wel}, \& {Wisotzki}}]{Mechtley:2016}
{Mechtley}, M., {Jahnke}, K., {Windhorst}, R.~A., {et~al.} 2016, \apj, 830, 156

\bibitem[{{Mehrgan} {et~al.}(2019){Mehrgan}, {Thomas}, {Saglia}, {Mazzalay},
  {Erwin}, {Bender}, {Kluge}, \& {Fabricius}}]{Mehrgan:2019}
{Mehrgan}, K., {Thomas}, J., {Saglia}, R., {et~al.} 2019, \apj, 887, 195

\bibitem[{{Merloni} {et~al.}(2014){Merloni}, {Bongiorno}, {Brusa}, {Iwasawa},
  {Mainieri}, {Magnelli}, {Salvato}, {Berta}, {Cappelluti}, {Comastri},
  {Fiore}, {Gilli}, {Koekemoer}, {Le Floc'h}, {Lusso}, {Lutz}, {Miyaji},
  {Pozzi}, {Riguccini}, {Rosario}, {Silverman}, {Symeonidis}, {Treister},
  {Vignali}, \& {Zamorani}}]{Merloni:2014}
{Merloni}, A., {Bongiorno}, A., {Brusa}, M., {et~al.} 2014, \mnras, 437, 3550

\bibitem[{{Merloni} {et~al.}(2015){Merloni}, {Dwelly}, {Salvato},
  {Georgakakis}, {Greiner}, {Krumpe}, {Nandra}, {Ponti}, \&
  {Rau}}]{Merloni:2015}
{Merloni}, A., {Dwelly}, T., {Salvato}, M., {et~al.} 2015, \mnras, 452, 69

\bibitem[{{Mingozzi} {et~al.}(2019){Mingozzi}, {Cresci}, {Venturi}, {Marconi},
  {Mannucci}, {Perna}, {Belfiore}, {Carniani}, {Balmaverde}, {Brusa}, {Cicone},
  {Feruglio}, {Gallazzi}, {Mainieri}, {Maiolino}, {Nagao}, {Nardini}, {Sani},
  {Tozzi}, \& {Zibetti}}]{Mingozzi:2019}
{Mingozzi}, M., {Cresci}, G., {Venturi}, G., {et~al.} 2019, \aap, 622, A146

\bibitem[{{Moffat}(1969)}]{Moffat:1969}
{Moffat}, A.~F.~J. 1969, \aap, 3, 455

\bibitem[{{Morganti} {et~al.}(2005){Morganti}, {Tadhunter}, \&
  {Oosterloo}}]{Morganti:2005}
{Morganti}, R., {Tadhunter}, C.~N., \& {Oosterloo}, T.~A. 2005, \aap, 444, L9

\bibitem[{{Moser} {et~al.}(2012){Moser}, {Zuther}, {Busch}, {Valencia-S.}, \&
  {Eckart}}]{Moser:2012}
{Moser}, L., {Zuther}, J., {Busch}, G., {Valencia-S.}, M., \& {Eckart}, A.
  2012, in Proceedings of Nuclei of Seyfert galaxies and QSOs - Central engine
  conditions of star formation (Seyfert 2012). 6-8 November, 2012.
  Max-Planck-Insitut f{\"u}r Radioastronomie (MPIfR), Bonn, Germany, id.69, 69

\bibitem[{{Mullaney} {et~al.}(2013){Mullaney}, {Alexander}, {Fine}, {Goulding},
  {Harrison}, \& {Hickox}}]{Mullaney:2013}
{Mullaney}, J.~R., {Alexander}, D.~M., {Fine}, S., {et~al.} 2013, \mnras, 433,
  622

\bibitem[{{Mullaney} {et~al.}(2012){Mullaney}, {Pannella}, {Daddi},
  {Alexander}, {Elbaz}, {Hickox}, {Bournaud}, {Altieri}, {Aussel}, {Coia},
  {Dannerbauer}, {Dasyra}, {Dickinson}, {Hwang}, {Kartaltepe}, {Leiton},
  {Magdis}, {Magnelli}, {Popesso}, {Valtchanov}, {Bauer}, {Brandt}, {Del Moro},
  {Hanish}, {Ivison}, {Juneau}, {Luo}, {Lutz}, {Sargent}, {Scott}, \&
  {Xue}}]{Mullaney:2012b}
{Mullaney}, J.~R., {Pannella}, M., {Daddi}, E., {et~al.} 2012, \mnras, 419, 95

\bibitem[{{Nelson} {et~al.}(2019){Nelson}, {Pillepich}, {Springel}, {Pakmor},
  {Weinberger}, {Genel}, {Torrey}, {Vogelsberger}, {Marinacci}, \&
  {Hernquist}}]{Nelson:2019}
{Nelson}, D., {Pillepich}, A., {Springel}, V., {et~al.} 2019, \mnras, 490, 3234

\bibitem[{{Nesvadba} {et~al.}(2008){Nesvadba}, {Lehnert}, {De Breuck},
  {Gilbert}, \& {van Breugel}}]{Nesvadba:2008}
{Nesvadba}, N.~P.~H., {Lehnert}, M.~D., {De Breuck}, C., {Gilbert}, A.~M., \&
  {van Breugel}, W. 2008, \aap, 491, 407

\bibitem[{{Netzer}(2019)}]{Netzer:2019}
{Netzer}, H. 2019, \mnras, 488, 5185

\bibitem[{{Neumann} {et~al.}(2019){Neumann}, {Gadotti}, {Wisotzki}, {Husemann},
  {Busch}, {Combes}, {Croom}, {Davis}, {Gaspari}, {Krumpe}, {P{\'e}rez-Torres},
  {Scharw{\"a}chter}, {Smirnova-Pinchukova}, {Tremblay}, \&
  {Urrutia}}]{Neumann:2019}
{Neumann}, J., {Gadotti}, D.~A., {Wisotzki}, L., {et~al.} 2019, \aap, 627, A26

\bibitem[{{Noda} \& {Done}(2018)}]{Noda:2018}
{Noda}, H. \& {Done}, C. 2018, \mnras, 480, 3898

\bibitem[{{O'Neill} {et~al.}(2005){O'Neill}, {Nandra}, {Papadakis}, \&
  {Turner}}]{O'Neill:2005}
{O'Neill}, P.~M., {Nandra}, K., {Papadakis}, I.~E., \& {Turner}, T.~J. 2005,
  \mnras, 358, 1405

\bibitem[{{Onken} {et~al.}(2004){Onken}, {Ferrarese}, {Merritt}, {Peterson},
  {Pogge}, {Vestergaard}, \& {Wandel}}]{Onken:2004}
{Onken}, C.~A., {Ferrarese}, L., {Merritt}, D., {et~al.} 2004, \apj, 615, 645

\bibitem[{{Onken} {et~al.}(2019){Onken}, {Wolf}, {Bessell}, {Chang}, {Da
  Costa}, {Luvaul}, {Mackey}, {Schmidt}, \& {Shao}}]{Onken:2019}
{Onken}, C.~A., {Wolf}, C., {Bessell}, M.~S., {et~al.} 2019, \pasa, 36, e033

\bibitem[{{Osterbrock}(1977)}]{Osterbrock:1977}
{Osterbrock}, D.~E. 1977, \apj, 215, 733

\bibitem[{{Padovani} {et~al.}(2017){Padovani}, {Alexander}, {Assef}, {De
  Marco}, {Giommi}, {Hickox}, {Richards}, {Smol{\v{c}}i{\'c}},
  {Hatziminaoglou}, {Mainieri}, \& {Salvato}}]{Padovani:2017}
{Padovani}, P., {Alexander}, D.~M., {Assef}, R.~J., {et~al.} 2017, \aapr, 25, 2

\bibitem[{{Pancoast} {et~al.}(2014){Pancoast}, {Brewer}, {Treu}, {Park},
  {Barth}, {Bentz}, \& {Woo}}]{Pancoast:2014}
{Pancoast}, A., {Brewer}, B.~J., {Treu}, T., {et~al.} 2014, \mnras, 445, 3073

\bibitem[{{Park} {et~al.}(2012){Park}, {Woo}, {Treu}, {Barth}, {Bentz},
  {Bennert}, {Canalizo}, {Filippenko}, {Gates}, {Greene}, {Malkan}, \&
  {Walsh}}]{Park:2012}
{Park}, D., {Woo}, J.-H., {Treu}, T., {et~al.} 2012, \apj, 747, 30

\bibitem[{{Peng} {et~al.}(2010{\natexlab{a}}){Peng}, {Ho}, {Impey}, \&
  {Rix}}]{Peng:2010}
{Peng}, C.~Y., {Ho}, L.~C., {Impey}, C.~D., \& {Rix}, H. 2010{\natexlab{a}},
  \aj, 139, 2097

\bibitem[{{Peng} {et~al.}(2010{\natexlab{b}}){Peng}, {Lilly}, {Kova{\v c}},
  {Bolzonella}, {Pozzetti}, {Renzini}, {Zamorani}, {Ilbert}, {Knobel},
  {Iovino}, {Maier}, {Cucciati}, {Tasca}, {Carollo}, {Silverman}, {Kampczyk},
  {de Ravel}, {Sanders}, {Scoville}, {Contini}, {Mainieri}, {Scodeggio},
  {Kneib}, {Le F{\`e}vre}, {Bardelli}, {Bongiorno}, {Caputi}, {Coppa}, {de la
  Torre}, {Franzetti}, {Garilli}, {Lamareille}, {Le Borgne}, {Le Brun},
  {Mignoli}, {Perez Montero}, {Pello}, {Ricciardelli}, {Tanaka}, {Tresse},
  {Vergani}, {Welikala}, {Zucca}, {Oesch}, {Abbas}, {Barnes}, {Bordoloi},
  {Bottini}, {Cappi}, {Cassata}, {Cimatti}, {Fumana}, {Hasinger}, {Koekemoer},
  {Leauthaud}, {Maccagni}, {Marinoni}, {McCracken}, {Memeo}, {Meneux}, {Nair},
  {Porciani}, {Presotto}, \& {Scaramella}}]{Peng:2010b}
{Peng}, Y.-j., {Lilly}, S.~J., {Kova{\v c}}, K., {et~al.} 2010{\natexlab{b}},
  \apj, 721, 193

\bibitem[{{Peterson} {et~al.}(2004){Peterson}, {Ferrarese}, {Gilbert}, {Kaspi},
  {Malkan}, {Maoz}, {Merritt}, {Netzer}, {Onken}, {Pogge}, {Vestergaard}, \&
  {Wandel}}]{Peterson:2004}
{Peterson}, B.~M., {Ferrarese}, L., {Gilbert}, K.~M., {et~al.} 2004, \apj, 613,
  682

\bibitem[{{Peterson} \& {Wandel}(2000)}]{Peterson:2000}
{Peterson}, B.~M. \& {Wandel}, A. 2000, \apjl, 540, L13

\bibitem[{{Pierre} {et~al.}(2016){Pierre}, {Pacaud}, {Adami}, {Alis},
  {Altieri}, {Baran}, {Benoist}, {Birkinshaw}, {Bongiorno}, {Bremer}, {Brusa},
  {Butler}, {Ciliegi}, {Chiappetti}, {Clerc}, {Corasaniti}, {Coupon}, {De
  Breuck}, {Democles}, {Desai}, {Delhaize}, {Devriendt}, {Dubois}, {Eckert},
  {Elyiv}, {Ettori}, {Evrard}, {Faccioli}, {Farahi}, {Ferrari}, {Finet},
  {Fotopoulou}, {Fourmanoit}, {Gandhi}, {Gastaldello}, {Gastaud},
  {Georgantopoulos}, {Giles}, {Guennou}, {Guglielmo}, {Horellou}, {Husband},
  {Huynh}, {Iovino}, {Kilbinger}, {Koulouridis}, {Lavoie}, {Le Brun}, {Le
  Fevre}, {Lidman}, {Lieu}, {Lin}, {Mantz}, {Maughan}, {Maurogordato},
  {McCarthy}, {McGee}, {Melin}, {Melnyk}, {Menanteau}, {Novak}, {Paltani},
  {Plionis}, {Poggianti}, {Pomarede}, {Pompei}, {Ponman}, {Ramos-Ceja},
  {Ranalli}, {Rapetti}, {Raychaudury}, {Reiprich}, {Rottgering}, {Rozo},
  {Rykoff}, {Sadibekova}, {Santos}, {Sauvageot}, {Schimd}, {Sereno}, {Smith},
  {Smol{\v{c}}i{\'c}}, {Snowden}, {Spergel}, {Stanford}, {Surdej}, {Valageas},
  {Valotti}, {Valtchanov}, {Vignali}, {Willis}, \& {Ziparo}}]{Pierre:2016}
{Pierre}, M., {Pacaud}, F., {Adami}, C., {et~al.} 2016, \aap, 592, A1

\bibitem[{{Powell} {et~al.}(2018){Powell}, {Husemann}, {Tremblay}, {Krumpe},
  {Urrutia}, {Baum}, {Busch}, {Combes}, {Croom}, {Davis}, {Eckart}, {O'Dea},
  {P{\'e}rez-Torres}, {Scharw{\"a}chter}, {Smirnova-Pinchukova}, \&
  {Urry}}]{Powell:2018}
{Powell}, M.~C., {Husemann}, B., {Tremblay}, G.~R., {et~al.} 2018, \aap, 618,
  A27

\bibitem[{{Predehl} {et~al.}(2021){Predehl}, {Andritschke}, {Arefiev},
  {Babyshkin}, {Batanov}, {Becker}, {B{\"o}hringer}, {Bogomolov}, {Boller},
  {Borm}, {Bornemann}, {Br{\"a}uninger}, {Br{\"u}ggen}, {Brunner}, {Brusa},
  {Bulbul}, {Buntov}, {Burwitz}, {Burkert}, {Clerc}, {Churazov}, {Coutinho},
  {Dauser}, {Dennerl}, {Doroshenko}, {Eder}, {Emberger}, {Eraerds},
  {Finoguenov}, {Freyberg}, {Friedrich}, {Friedrich}, {F{\"u}rmetz},
  {Georgakakis}, {Gilfanov}, {Granato}, {Grossberger}, {Gueguen}, {Gureev},
  {Haberl}, {H{\"a}lker}, {Hartner}, {Hasinger}, {Huber}, {Ji}, {Kienlin},
  {Kink}, {Korotkov}, {Kreykenbohm}, {Lamer}, {Lomakin}, {Lapshov}, {Liu},
  {Maitra}, {Meidinger}, {Menz}, {Merloni}, {Mernik}, {Mican}, {Mohr},
  {M{\"u}ller}, {Nandra}, {Nazarov}, {Pacaud}, {Pavlinsky}, {Perinati},
  {Pfeffermann}, {Pietschner}, {Ramos-Ceja}, {Rau}, {Reiffers}, {Reiprich},
  {Robrade}, {Salvato}, {Sanders}, {Santangelo}, {Sasaki}, {Scheuerle},
  {Schmid}, {Schmitt}, {Schwope}, {Shirshakov}, {Steinmetz}, {Stewart},
  {Str{\"u}der}, {Sunyaev}, {Tenzer}, {Tiedemann}, {Tr{\"u}mper}, {Voron},
  {Weber}, {Wilms}, \& {Yaroshenko}}]{Predehl:2021}
{Predehl}, P., {Andritschke}, R., {Arefiev}, V., {et~al.} 2021, \aap, 647, A1

\bibitem[{{Racine}(1996)}]{Racine:1996}
{Racine}, R. 1996, \pasp, 108, 699

\bibitem[{{Raimundo} {et~al.}(2019){Raimundo}, {Vestergaard}, {Koay},
  {Lawther}, {Casasola}, \& {Peterson}}]{Raimundo:2019}
{Raimundo}, S.~I., {Vestergaard}, M., {Koay}, J.~Y., {et~al.} 2019, \mnras,
  486, 123

\bibitem[{{Richards} {et~al.}(2006){Richards}, {Lacy}, {Storrie-Lombardi},
  {Hall}, {Gallagher}, {Hines}, {Fan}, {Papovich}, {Vanden Berk}, {Trammell},
  {Schneider}, {Vestergaard}, {York}, {Jester}, {Anderson}, {Budav{\'a}ri}, \&
  {Szalay}}]{Richards:2006}
{Richards}, G.~T., {Lacy}, M., {Storrie-Lombardi}, L.~J., {et~al.} 2006, \apjs,
  166, 470

\bibitem[{{Richardson} {et~al.}(2014){Richardson}, {Allen}, {Baldwin},
  {Hewett}, \& {Ferland}}]{Richardson:2014}
{Richardson}, C.~T., {Allen}, J.~T., {Baldwin}, J.~A., {Hewett}, P.~C., \&
  {Ferland}, G.~J. 2014, \mnras, 437, 2376

\bibitem[{{Rosario} {et~al.}(2018){Rosario}, {Burtscher}, {Davies}, {Koss},
  {Ricci}, {Lutz}, {Riffel}, {Alexander}, {Genzel}, {Hicks}, {Lin},
  {Maciejewski}, {M{\"u}ller-S{\'a}nchez}, {Orban de Xivry}, {Riffel},
  {Schartmann}, {Schawinski}, {Schnorr-M{\"u}ller}, {Saintonge}, {Shimizu},
  {Sternberg}, {Storchi-Bergmann}, {Sturm}, {Tacconi}, {Treister}, \&
  {Veilleux}}]{Rosario:2018}
{Rosario}, D.~J., {Burtscher}, L., {Davies}, R.~I., {et~al.} 2018, \mnras, 473,
  5658

\bibitem[{{Rosario} {et~al.}(2013){Rosario}, {Trakhtenbrot}, {Lutz}, {Netzer},
  {Trump}, {Silverman}, {Schramm}, {Lusso}, {Berta}, {Bongiorno}, {Brusa},
  {F{\"o}rster-Schreiber}, {Genzel}, {Lilly}, {Magnelli}, {Mainieri},
  {Maiolino}, {Merloni}, {Mignoli}, {Nordon}, {Popesso}, {Salvato}, {Santini},
  {Tacconi}, \& {Zamorani}}]{Rosario:2013}
{Rosario}, D.~J., {Trakhtenbrot}, B., {Lutz}, D., {et~al.} 2013, \aap, 560, A72

\bibitem[{{Rose} {et~al.}(2019){Rose}, {Edge}, {Combes}, {Gaspari}, {Hamer},
  {Nesvadba}, {Peck}, {Sarazin}, {Tremblay}, {Baum}, {Bremer}, {McNamara},
  {O'Dea}, {Oonk}, {Russell}, {Salom{\'e}}, {Donahue}, {Fabian}, {Ferland},
  {Mittal}, \& {Vantyghem}}]{Rose:2019}
{Rose}, T., {Edge}, A.~C., {Combes}, F., {et~al.} 2019, \mnras, 489, 349

\bibitem[{{Roth} {et~al.}(2005){Roth}, {Kelz}, {Fechner}, {Hahn}, {Bauer},
  {Becker}, {B{\"o}hm}, {Christensen}, {Dionies}, {Paschke}, {Popow}, {Wolter},
  {Schmoll}, {Laux}, \& {Altmann}}]{Roth:2005}
{Roth}, M.~M., {Kelz}, A., {Fechner}, T., {et~al.} 2005, \pasp, 117, 620

\bibitem[{{Ruan} {et~al.}(2016){Ruan}, {Anderson}, {Cales}, {Eracleous},
  {Green}, {Morganson}, {Runnoe}, {Shen}, {Wilkinson}, {Blanton}, {Dwelly},
  {Georgakakis}, {Greene}, {LaMassa}, {Merloni}, \& {Schneider}}]{Ruan:2016}
{Ruan}, J.~J., {Anderson}, S.~F., {Cales}, S.~L., {et~al.} 2016, \apj, 826, 188

\bibitem[{{Runnoe} {et~al.}(2012){Runnoe}, {Brotherton}, \&
  {Shang}}]{Runnoe:2012b}
{Runnoe}, J.~C., {Brotherton}, M.~S., \& {Shang}, Z. 2012, \mnras, 422, 478

\bibitem[{{S{\'a}nchez} {et~al.}(2012){S{\'a}nchez}, {Kennicutt}, {Gil de Paz},
  {van de Ven}, {V{\'{\i}}lchez}, {Wisotzki}, {Walcher}, {Mast}, {Aguerri},
  {Albiol-P{\'e}rez}, {Alonso-Herrero}, {Alves}, {Bakos}, {Bart{\'a}kov{\'a}},
  {Bland-Hawthorn}, {Boselli}, {Bomans}, {Castillo-Morales}, {Cortijo-Ferrero},
  {de Lorenzo-C{\'a}ceres}, {Del Olmo}, {Dettmar}, {D{\'{\i}}az}, {Ellis},
  {Falc{\'o}n-Barroso}, {Flores}, {Gallazzi}, {Garc{\'{\i}}a-Lorenzo},
  {Gonz{\'a}lez Delgado}, {Gruel}, {Haines}, {Hao}, {Husemann},
  {Igl{\'e}sias-P{\'a}ramo}, {Jahnke}, {Johnson}, {Jungwiert}, {Kalinova},
  {Kehrig}, {Kupko}, {L{\'o}pez-S{\'a}nchez}, {Lyubenova}, {Marino},
  {M{\'a}rmol-Queralt{\'o}}, {M{\'a}rquez}, {Masegosa}, {Meidt},
  {Mendez-Abreu}, {Monreal-Ibero}, {Montijo}, {Mour{\~a}o}, {Palacios-Navarro},
  {Papaderos}, {Pasquali}, {Peletier}, {P{\'e}rez}, {P{\'e}rez}, {Quirrenbach},
  {Rela{\~n}o}, {Rosales-Ortega}, {Roth}, {Ruiz-Lara},
  {S{\'a}nchez-Bl{\'a}zquez}, {Sengupta}, {Singh}, {Stanishev}, {Trager},
  {Vazdekis}, {Viironen}, {Wild}, {Zibetti}, \& {Ziegler}}]{Sanchez:2012a}
{S{\'a}nchez}, S.~F., {Kennicutt}, R.~C., {Gil de Paz}, A., {et~al.} 2012,
  \aap, 538, A8

\bibitem[{{Santini} {et~al.}(2012){Santini}, {Rosario}, {Shao}, {Lutz},
  {Maiolino}, {Alexander}, {Altieri}, {Andreani}, {Aussel}, {Bauer}, {Berta},
  {Bongiovanni}, {Brandt}, {Brusa}, {Cepa}, {Cimatti}, {Daddi}, {Elbaz},
  {Fontana}, {F{\"o}rster Schreiber}, {Genzel}, {Grazian}, {Le Floc'h},
  {Magnelli}, {Mainieri}, {Nordon}, {P{\'e}rez Garcia}, {Poglitsch}, {Popesso},
  {Pozzi}, {Riguccini}, {Rodighiero}, {Salvato}, {Sanchez-Portal}, {Sturm},
  {Tacconi}, {Valtchanov}, \& {Wuyts}}]{Santini:2012}
{Santini}, P., {Rosario}, D.~J., {Shao}, L., {et~al.} 2012, \aap, 540, A109

\bibitem[{{Santoro} {et~al.}(2020){Santoro}, {Tadhunter}, {Baron}, {Morganti},
  \& {Holt}}]{Santoro:2020}
{Santoro}, F., {Tadhunter}, C., {Baron}, D., {Morganti}, R., \& {Holt}, J.
  2020, \aap, 644, A54

\bibitem[{{Schawinski} {et~al.}(2015){Schawinski}, {Koss}, {Berney}, \&
  {Sartori}}]{Schawinski:2015}
{Schawinski}, K., {Koss}, M., {Berney}, S., \& {Sartori}, L.~F. 2015, \mnras,
  451, 2517

\bibitem[{{Schawinski} {et~al.}(2012){Schawinski}, {Simmons}, {Urry},
  {Treister}, \& {Glikman}}]{Schawinski:2012}
{Schawinski}, K., {Simmons}, B.~D., {Urry}, C.~M., {Treister}, E., \&
  {Glikman}, E. 2012, \mnras, 425, L61

\bibitem[{{Schlafly} \& {Finkbeiner}(2011)}]{Schlafly:2011}
{Schlafly}, E.~F. \& {Finkbeiner}, D.~P. 2011, \apj, 737, 103

\bibitem[{{Schlegel} {et~al.}(1998){Schlegel}, {Finkbeiner}, \&
  {Davis}}]{Schlegel:1998}
{Schlegel}, D.~J., {Finkbeiner}, D.~P., \& {Davis}, M. 1998, \apj, 500, 525

\bibitem[{{Schmidt} {et~al.}(2017){Schmidt}, {Worseck}, {Hennawi}, {Prochaska},
  \& {Crighton}}]{Schmidt:2017}
{Schmidt}, T.~M., {Worseck}, G., {Hennawi}, J.~F., {Prochaska}, J.~X., \&
  {Crighton}, N.~H.~M. 2017, \apj, 847, 81

\bibitem[{{Schmitt} {et~al.}(2003){Schmitt}, {Donley}, {Antonucci},
  {Hutchings}, {Kinney}, \& {Pringle}}]{Schmitt:2003b}
{Schmitt}, H.~R., {Donley}, J.~L., {Antonucci}, R.~R.~J., {et~al.} 2003, \apj,
  597, 768

\bibitem[{{Scholtz} {et~al.}(2020){Scholtz}, {Harrison}, {Rosario},
  {Alexander}, {Chen}, {Kakkad}, {Mainieri}, {Tiley}, {Turner}, {Cirasuolo},
  {Sharples}, \& {Stach}}]{Scholtz:2020}
{Scholtz}, J., {Harrison}, C.~M., {Rosario}, D.~J., {et~al.} 2020, \mnras, 492,
  3194

\bibitem[{{Schulze} {et~al.}(2015){Schulze}, {Bongiorno}, {Gavignaud},
  {Schramm}, {Silverman}, {Merloni}, {Zamorani}, {Hirschmann}, {Mainieri},
  {Wisotzki}, {Shankar}, {Fiore}, {Koekemoer}, \& {Temporin}}]{Schulze:2015}
{Schulze}, A., {Bongiorno}, A., {Gavignaud}, I., {et~al.} 2015, \mnras, 447,
  2085

\bibitem[{{Schulze} \& {Wisotzki}(2010)}]{Schulze:2010}
{Schulze}, A. \& {Wisotzki}, L. 2010, \aap, 516, A87+

\bibitem[{{Schulze} {et~al.}(2009){Schulze}, {Wisotzki}, \&
  {Husemann}}]{Schulze:2009}
{Schulze}, A., {Wisotzki}, L., \& {Husemann}, B. 2009, \aap, 507, 781

\bibitem[{{Schweizer} {et~al.}(2013){Schweizer}, {Seitzer}, {Kelson},
  {Villanueva}, \& {Walth}}]{Schweizer:2013}
{Schweizer}, F., {Seitzer}, P., {Kelson}, D.~D., {Villanueva}, E.~V., \&
  {Walth}, G.~L. 2013, \apj, 773, 148

\bibitem[{{Scodeggio} {et~al.}(2018){Scodeggio}, {Guzzo}, {Garilli}, {Granett},
  {Bolzonella}, {de la Torre}, {Abbas}, {Adami}, {Arnouts}, {Bottini}, {Cappi},
  {Coupon}, {Cucciati}, {Davidzon}, {Franzetti}, {Fritz}, {Iovino}, {Krywult},
  {Le Brun}, {Le F{\`e}vre}, {Maccagni}, {Ma{\l}ek}, {Marchetti}, {Marulli},
  {Polletta}, {Pollo}, {Tasca}, {Tojeiro}, {Vergani}, {Zanichelli}, {Bel},
  {Branchini}, {De Lucia}, {Ilbert}, {McCracken}, {Moutard}, {Peacock},
  {Zamorani}, {Burden}, {Fumana}, {Jullo}, {Marinoni}, {Mellier}, {Moscardini},
  \& {Percival}}]{Scodeggio:2018}
{Scodeggio}, M., {Guzzo}, L., {Garilli}, B., {et~al.} 2018, \aap, 609, A84

\bibitem[{{Scoville} {et~al.}(2003){Scoville}, {Frayer}, {Schinnerer}, \&
  {Christopher}}]{Scoville:2003}
{Scoville}, N.~Z., {Frayer}, D.~T., {Schinnerer}, E., \& {Christopher}, M.
  2003, \apjl, 585, L105

\bibitem[{{Shangguan} {et~al.}(2020){Shangguan}, {Ho}, {Bauer}, {Wang}, \&
  {Treister}}]{Shangguan:2020}
{Shangguan}, J., {Ho}, L.~C., {Bauer}, F.~E., {Wang}, R., \& {Treister}, E.
  2020, \apjs, 247, 15

\bibitem[{{Shangguan} {et~al.}(2018){Shangguan}, {Ho}, \&
  {Xie}}]{Shangguan:2018}
{Shangguan}, J., {Ho}, L.~C., \& {Xie}, Y. 2018, \apj, 854, 158

\bibitem[{{Shao} {et~al.}(2010){Shao}, {Lutz}, {Nordon}, {Maiolino},
  {Alexander}, {Altieri}, {Andreani}, {Aussel}, {Bauer}, {Berta},
  {Bongiovanni}, {Brandt}, {Brusa}, {Cava}, {Cepa}, {Cimatti}, {Daddi},
  {Dominguez-Sanchez}, {Elbaz}, {F{\"o}rster Schreiber}, {Geis}, {Genzel},
  {Grazian}, {Gruppioni}, {Magdis}, {Magnelli}, {Mainieri}, {P{\'e}rez
  Garc{\'\i}a}, {Poglitsch}, {Popesso}, {Pozzi}, {Riguccini}, {Rodighiero},
  {Rovilos}, {Saintonge}, {Salvato}, {Sanchez Portal}, {Santini}, {Sturm},
  {Tacconi}, {Valtchanov}, {Wetzstein}, \& {Wieprecht}}]{Shao:2010}
{Shao}, L., {Lutz}, D., {Nordon}, R., {et~al.} 2010, \aap, 518, L26

\bibitem[{{Sharp} \& {Birchall}(2010)}]{Sharp:2010}
{Sharp}, R. \& {Birchall}, M.~N. 2010, \pasa, 27, 91

\bibitem[{{Shen} {et~al.}(2008){Shen}, {Greene}, {Strauss}, {Richards}, \&
  {Schneider}}]{Shen:2008b}
{Shen}, Y., {Greene}, J.~E., {Strauss}, M.~A., {Richards}, G.~T., \&
  {Schneider}, D.~P. 2008, \apj, 680, 169

\bibitem[{{Shen} \& {Liu}(2012)}]{Shen:2012}
{Shen}, Y. \& {Liu}, X. 2012, \apj, 753, 125

\bibitem[{{Shimizu} {et~al.}(2015){Shimizu}, {Mushotzky}, {Mel{\'e}ndez},
  {Koss}, \& {Rosario}}]{Shimizu:2015}
{Shimizu}, T.~T., {Mushotzky}, R.~F., {Mel{\'e}ndez}, M., {Koss}, M., \&
  {Rosario}, D.~J. 2015, \mnras, 452, 1841

\bibitem[{{Shimizu} {et~al.}(2017){Shimizu}, {Mushotzky}, {Mel{\'e}ndez},
  {Koss}, {Barger}, \& {Cowie}}]{Shimizu:2017}
{Shimizu}, T.~T., {Mushotzky}, R.~F., {Mel{\'e}ndez}, M., {et~al.} 2017,
  \mnras, 466, 3161

\bibitem[{{Shimwell} {et~al.}(2019){Shimwell}, {Tasse}, {Hardcastle}, {Mechev},
  {Williams}, {Best}, {R{\"o}ttgering}, {Callingham}, {Dijkema}, {de Gasperin},
  {Hoang}, {Hugo}, {Mirmont}, {Oonk}, {Prandoni}, {Rafferty}, {Sabater},
  {Smirnov}, {van Weeren}, {White}, {Atemkeng}, {Bester}, {Bonnassieux},
  {Br{\"u}ggen}, {Brunetti}, {Chy{\.z}y}, {Cochrane}, {Conway}, {Croston},
  {Danezi}, {Duncan}, {Haverkorn}, {Heald}, {Iacobelli}, {Intema}, {Jackson},
  {Jamrozy}, {Jarvis}, {Lakhoo}, {Mevius}, {Miley}, {Morabito}, {Morganti},
  {Nisbet}, {Orr{\'u}}, {Perkins}, {Pizzo}, {Schrijvers}, {Smith}, {Vermeulen},
  {Wise}, {Alegre}, {Bacon}, {van Bemmel}, {Beswick}, {Bonafede}, {Botteon},
  {Bourke}, {Brienza}, {Calistro Rivera}, {Cassano}, {Clarke}, {Conselice},
  {Dettmar}, {Drabent}, {Dumba}, {Emig}, {En{\ss}lin}, {Ferrari}, {Garrett},
  {G{\'e}nova-Santos}, {Goyal}, {G{\"u}rkan}, {Hale}, {Harwood}, {Heesen},
  {Hoeft}, {Horellou}, {Jackson}, {Kokotanekov}, {Kondapally},
  {Kunert-Bajraszewska}, {Mahatma}, {Mahony}, {Mandal}, {McKean}, {Merloni},
  {Mingo}, {Miskolczi}, {Mooney}, {Nikiel-Wroczy{\'n}ski}, {O'Sullivan},
  {Quinn}, {Reich}, {Roskowi{\'n}ski}, {Rowlinson}, {Savini}, {Saxena},
  {Schwarz}, {Shulevski}, {Sridhar}, {Stacey}, {Urquhart}, {van der Wiel},
  {Varenius}, {Webster}, \& {Wilber}}]{Shimwell:2019}
{Shimwell}, T.~W., {Tasse}, C., {Hardcastle}, M.~J., {et~al.} 2019, \aap, 622,
  A1

\bibitem[{{Silk} \& {Rees}(1998)}]{Silk:1998}
{Silk}, J. \& {Rees}, M.~J. 1998, \aap, 331, L1

\bibitem[{{Singh} {et~al.}(2013){Singh}, {van de Ven}, {Jahnke}, {Lyubenova},
  {Falc{\'o}n-Barroso}, {Alves}, {Cid Fernandes}, {Galbany},
  {Garc{\'{\i}}a-Benito}, {Husemann}, {Kennicutt}, {Marino}, {M{\'a}rquez},
  {Masegosa}, {Mast}, {Pasquali}, {S{\'a}nchez}, {Walcher}, {Wild}, {Wisotzki},
  \& {Ziegler}}]{Singh:2013}
{Singh}, R., {van de Ven}, G., {Jahnke}, K., {et~al.} 2013, \aap, 558, A43

\bibitem[{{Singha} {et~al.}(2021){Singha}, {Husemann}, {Urrutia}, {O'Dea},
  {Scharw{\"a}chter}, {Gaspari}, {Combes}, {Nevin}, {Terrazas},
  {P{\'e}rez-Torres}, {Rose}, {Davis}, {Tremblay}, {Neumann},
  {Smirnova-Pinchukova}, \& {Baum}}]{Singha:2021}
{Singha}, M., {Husemann}, B., {Urrutia}, T., {et~al.} 2021, arXiv e-prints,
  arXiv:2111.10418

\bibitem[{{Smirnova-Pinchukova} {et~al.}(2019){Smirnova-Pinchukova},
  {Husemann}, {Busch}, {Appleton}, {Bethermin}, {Combes}, {Croom}, {Davis},
  {Fischer}, {Gaspari}, {Groves}, {Klein}, {O'Dea}, {P{\'e}rez-Torres},
  {Scharw{\"a}chter}, {Singha}, {Tremblay}, \&
  {Urrutia}}]{Smirnova-Pinchukova:2019}
{Smirnova-Pinchukova}, I., {Husemann}, B., {Busch}, G., {et~al.} 2019, \aap,
  626, L3

\bibitem[{{Smirnova-Pinchukova} {et~al.}(2021){Smirnova-Pinchukova},
  {Husemann}, {Davis}, {Smith}, {Singha}, {Tremblay}, {Klessen}, {Powell},
  {Connor}, {Baum}, {Combes}, {Croom}, {Gaspari}, {Neumann}, {O'Dea},
  {P{\'e}rez-Torres}, {Rosario}, {Rose}, {Scharw{\"a}chter}, \&
  {Winkel}}]{Smirnova-Pinchukova:2021}
{Smirnova-Pinchukova}, I., {Husemann}, B., {Davis}, T.~A., {et~al.} 2021, arXiv
  e-prints, arXiv:2111.10419

\bibitem[{{Smol{\v{c}}i{\'c}} {et~al.}(2017){Smol{\v{c}}i{\'c}}, {Novak},
  {Bondi}, {Ciliegi}, {Mooley}, {Schinnerer}, {Zamorani}, {Navarrete},
  {Bourke}, {Karim}, {Vardoulaki}, {Leslie}, {Delhaize}, {Carilli}, {Myers},
  {Baran}, {Delvecchio}, {Miettinen}, {Banfield}, {Balokovi{\'c}}, {Bertoldi},
  {Capak}, {Frail}, {Hallinan}, {Hao}, {Herrera Ruiz}, {Horesh}, {Ilbert},
  {Intema}, {Jeli{\'c}}, {Kl{\"o}ckner}, {Krpan}, {Kulkarni}, {McCracken},
  {Laigle}, {Middleberg}, {Murphy}, {Sargent}, {Scoville}, \&
  {Sheth}}]{Smolcic:2017}
{Smol{\v{c}}i{\'c}}, V., {Novak}, M., {Bondi}, M., {et~al.} 2017, \aap, 602, A1

\bibitem[{{Soltan}(1982)}]{Soltan:1982}
{Soltan}, A. 1982, \mnras, 200, 115

\bibitem[{{Somerville} {et~al.}(2008){Somerville}, {Hopkins}, {Cox},
  {Robertson}, \& {Hernquist}}]{Somerville:2008}
{Somerville}, R.~S., {Hopkins}, P.~F., {Cox}, T.~J., {Robertson}, B.~E., \&
  {Hernquist}, L. 2008, \mnras, 391, 481

\bibitem[{{Soto} {et~al.}(2016){Soto}, {Lilly}, {Bacon}, {Richard}, \&
  {Conseil}}]{Soto:2016}
{Soto}, K.~T., {Lilly}, S.~J., {Bacon}, R., {Richard}, J., \& {Conseil}, S.
  2016, \mnras, 458, 3210

\bibitem[{{Springel} {et~al.}(2005){Springel}, {Di Matteo}, \&
  {Hernquist}}]{Springel:2005}
{Springel}, V., {Di Matteo}, T., \& {Hernquist}, L. 2005, \apjl, 620, L79

\bibitem[{{Stasi{\'n}ska} {et~al.}(2008){Stasi{\'n}ska}, {Vale Asari}, {Cid
  Fernandes}, {Gomes}, {Schlickmann}, {Mateus}, {Schoenell}, \&
  {Sodr{\'e}}}]{Stasinska:2008}
{Stasi{\'n}ska}, G., {Vale Asari}, N., {Cid Fernandes}, R., {et~al.} 2008,
  \mnras, 391, L29

\bibitem[{{Steinborn} {et~al.}(2015){Steinborn}, {Dolag}, {Hirschmann},
  {Prieto}, \& {Remus}}]{Steinborn:2015}
{Steinborn}, L.~K., {Dolag}, K., {Hirschmann}, M., {Prieto}, M.~A., \& {Remus},
  R.-S. 2015, \mnras, 448, 1504

\bibitem[{{Stern} \& {Laor}(2012)}]{Stern:2012a}
{Stern}, J. \& {Laor}, A. 2012, \mnras, 423, 600

\bibitem[{{Stockton} \& {MacKenty}(1983)}]{Stockton:1983}
{Stockton}, A. \& {MacKenty}, J.~W. 1983, \nat, 305, 678

\bibitem[{{Stockton} \& {MacKenty}(1987)}]{Stockton:1987}
{Stockton}, A. \& {MacKenty}, J.~W. 1987, \apj, 316, 584

\bibitem[{{Storey} \& {Zeippen}(2000)}]{Storey:2000}
{Storey}, P.~J. \& {Zeippen}, C.~J. 2000, \mnras, 312, 813

\bibitem[{{Suberlak} {et~al.}(2021){Suberlak}, {Ivezi{\'c}}, \&
  {MacLeod}}]{Suberlak:2021}
{Suberlak}, K.~L., {Ivezi{\'c}}, {\v{Z}}., \& {MacLeod}, C. 2021, \apj, 907, 96

\bibitem[{{Sulentic} {et~al.}(2002){Sulentic}, {Marziani}, {Zamanov}, {Bachev},
  {Calvani}, \& {Dultzin-Hacyan}}]{Sulentic:2002}
{Sulentic}, J.~W., {Marziani}, P., {Zamanov}, R., {et~al.} 2002, \apjl, 566,
  L71

\bibitem[{{Sulentic} {et~al.}(2000){Sulentic}, {Zwitter}, {Marziani}, \&
  {Dultzin-Hacyan}}]{Sulentic:2000}
{Sulentic}, J.~W., {Zwitter}, T., {Marziani}, P., \& {Dultzin-Hacyan}, D. 2000,
  \apjl, 536, L5

\bibitem[{{Sun} {et~al.}(2017){Sun}, {Greene}, \& {Zakamska}}]{Sun:2017}
{Sun}, A.-L., {Greene}, J.~E., \& {Zakamska}, N.~L. 2017, \apj, 835, 222

\bibitem[{{Sun} {et~al.}(2018){Sun}, {Greene}, {Zakamska}, {Goulding},
  {Strauss}, {Huang}, {Johnson}, {Kawaguchi}, {Matsuoka}, {Marsteller},
  {Nagao}, \& {Toba}}]{Sun:2018}
{Sun}, A.-L., {Greene}, J.~E., {Zakamska}, N.~L., {et~al.} 2018, \mnras, 480,
  2302

\bibitem[{{Trujillo} {et~al.}(2001){Trujillo}, {Aguerri}, {Cepa}, \&
  {Guti{\'e}rrez}}]{Trujillo:2001}
{Trujillo}, I., {Aguerri}, J.~A.~L., {Cepa}, J., \& {Guti{\'e}rrez}, C.~M.
  2001, \mnras, 328, 977

\bibitem[{{Unger} {et~al.}(1987){Unger}, {Pedlar}, {Axon}, {Whittle}, {Meurs},
  \& {Ward}}]{Unger:1987}
{Unger}, S.~W., {Pedlar}, A., {Axon}, D.~J., {et~al.} 1987, \mnras, 228, 671

\bibitem[{{Urry} \& {Padovani}(1995)}]{Urry:1995}
{Urry}, C.~M. \& {Padovani}, P. 1995, \pasp, 107, 803

\bibitem[{{Valdes} {et~al.}(2004){Valdes}, {Gupta}, {Rose}, {Singh}, \&
  {Bell}}]{Valdes:2004}
{Valdes}, F., {Gupta}, R., {Rose}, J.~A., {Singh}, H.~P., \& {Bell}, D.~J.
  2004, \apjs, 152, 251

\bibitem[{{van den Bosch}(2016)}]{Bosch:2016}
{van den Bosch}, R. C.~E. 2016, \apj, 831, 134

\bibitem[{{Veilleux} \& {Osterbrock}(1987)}]{Veilleux:1987}
{Veilleux}, S. \& {Osterbrock}, D.~E. 1987, \apjs, 63, 295

\bibitem[{{V{\'e}ron-Cetty} {et~al.}(2004){V{\'e}ron-Cetty}, {Joly}, \&
  {V{\'e}ron}}]{Veron-Cetty:2004}
{V{\'e}ron-Cetty}, M., {Joly}, M., \& {V{\'e}ron}, P. 2004, \aap, 417, 515

\bibitem[{{Vestergaard}(2002)}]{Vestergaard:2002}
{Vestergaard}, M. 2002, \apj, 571, 733

\bibitem[{{Vestergaard} \& {Peterson}(2006)}]{Vestergaard:2006}
{Vestergaard}, M. \& {Peterson}, B.~M. 2006, \apj, 641, 689

\bibitem[{{Vietri} {et~al.}(2020){Vietri}, {Mainieri}, {Kakkad}, {Netzer},
  {Perna}, {Circosta}, {Harrison}, {Zappacosta}, {Husemann}, {Padovani},
  {Bischetti}, {Bongiorno}, {Brusa}, {Carniani}, {Cicone}, {Comastri},
  {Cresci}, {Feruglio}, {Fiore}, {Lanzuisi}, {Mannucci}, {Marconi},
  {Piconcelli}, {Puglisi}, {Salvato}, {Schramm}, {Schulze}, {Scholtz},
  {Vignali}, \& {Zamorani}}]{Vietri:2020}
{Vietri}, G., {Mainieri}, V., {Kakkad}, D., {et~al.} 2020, \aap, 644, A175

\bibitem[{{Vietri} {et~al.}(2018){Vietri}, {Piconcelli}, {Bischetti}, {Duras},
  {Martocchia}, {Bongiorno}, {Marconi}, {Zappacosta}, {Bisogni}, {Bruni},
  {Brusa}, {Comastri}, {Cresci}, {Feruglio}, {Giallongo}, {La Franca},
  {Mainieri}, {Mannucci}, {Ricci}, {Sani}, {Testa}, {Tombesi}, {Vignali}, \&
  {Fiore}}]{Vietri:2018}
{Vietri}, G., {Piconcelli}, E., {Bischetti}, M., {et~al.} 2018, \aap, 617, A81

\bibitem[{{Villar-Mart{\'{\i}}n} {et~al.}(2016){Villar-Mart{\'{\i}}n},
  {Arribas}, {Emonts}, {Humphrey}, {Tadhunter}, {Bessiere}, {Cabrera Lavers},
  \& {Ramos Almeida}}]{Villar-Martin:2016}
{Villar-Mart{\'{\i}}n}, M., {Arribas}, S., {Emonts}, B., {et~al.} 2016, \mnras,
  460, 130

\bibitem[{{Villar-Mart{\'{\i}}n} {et~al.}(2018){Villar-Mart{\'{\i}}n},
  {Cabrera-Lavers}, {Humphrey}, {Silva}, {Ramos Almeida}, {Piqueras-L{\'o}pez},
  \& {Emonts}}]{Villar-Martin:2018}
{Villar-Mart{\'{\i}}n}, M., {Cabrera-Lavers}, A., {Humphrey}, A., {et~al.}
  2018, \mnras, 474, 2302

\bibitem[{{Villar-Mart{\'{\i}}n} {et~al.}(2013){Villar-Mart{\'{\i}}n},
  {Rodr{\'{\i}}guez}, {Drouart}, {Emonts}, {Colina}, {Humphrey}, {Garc{\'{\i}}a
  Burillo}, {Graci{\'a} Carpio}, {Planesas}, {P{\'e}rez Torres}, \&
  {Arribas}}]{Villar-Martin:2013}
{Villar-Mart{\'{\i}}n}, M., {Rodr{\'{\i}}guez}, M., {Drouart}, G., {et~al.}
  2013, \mnras, 434, 978

\bibitem[{{Villar-Martin} {et~al.}(1997){Villar-Martin}, {Tadhunter}, \&
  {Clark}}]{Villar-Martin:1997}
{Villar-Martin}, M., {Tadhunter}, C., \& {Clark}, N. 1997, \aap, 323, 21

\bibitem[{{Villforth} {et~al.}(2017){Villforth}, {Hamilton}, {Pawlik},
  {Hewlett}, {Rowlands}, {Herbst}, {Shankar}, {Fontana}, {Hamann}, {Koekemoer},
  {Pforr}, {Trump}, \& {Wuyts}}]{Villforth:2017}
{Villforth}, C., {Hamilton}, T., {Pawlik}, M.~M., {et~al.} 2017, \mnras, 466,
  812

\bibitem[{{Vink} {et~al.}(2006){Vink}, {Snellen}, {Mack}, \&
  {Schilizzi}}]{Vink:2006}
{Vink}, J., {Snellen}, I., {Mack}, K.-H., \& {Schilizzi}, R. 2006, \mnras, 367,
  928

\bibitem[{{Vogelsberger} {et~al.}(2014){Vogelsberger}, {Genel}, {Springel},
  {Torrey}, {Sijacki}, {Xu}, {Snyder}, {Bird}, {Nelson}, \&
  {Hernquist}}]{Vogelsberger:2014}
{Vogelsberger}, M., {Genel}, S., {Springel}, V., {et~al.} 2014, \nat, 509, 177

\bibitem[{{Voges} {et~al.}(1999){Voges}, {Aschenbach}, {Boller},
  {Br{\"a}uninger}, {Briel}, {Burkert}, {Dennerl}, {Englhauser}, {Gruber},
  {Haberl}, {Hartner}, {Hasinger}, {K{\"u}rster}, {Pfeffermann}, {Pietsch},
  {Predehl}, {Rosso}, {Schmitt}, {Tr{\"u}mper}, \& {Zimmermann}}]{Voges:1999}
{Voges}, W., {Aschenbach}, B., {Boller}, T., {et~al.} 1999, \aap, 349, 389

\bibitem[{{Walcher} {et~al.}(2015){Walcher}, {Coelho}, {Gallazzi}, {Bruzual},
  {Charlot}, \& {Chiappini}}]{Walcher:2015}
{Walcher}, C.~J., {Coelho}, P.~R.~T., {Gallazzi}, A., {et~al.} 2015, \aap, 582,
  A46

\bibitem[{{Weaver} {et~al.}(2018){Weaver}, {Husemann}, {Kuntschner},
  {Mart{\'{\i}}n-Navarro}, {Bournaud}, {Duc}, {Emsellem}, {Krajnovi{\'c}},
  {Lyubenova}, \& {McDermid}}]{Weaver:2018}
{Weaver}, J., {Husemann}, B., {Kuntschner}, H., {et~al.} 2018, \aap, 614, A32

\bibitem[{{Weilbacher} {et~al.}(2020){Weilbacher}, {Palsa}, {Streicher},
  {Bacon}, {Urrutia}, {Wisotzki}, {Conseil}, {Husemann}, {Jarno}, {Kelz},
  {P{\'e}contal-Rousset}, {Richard}, {Roth}, {Selman}, \&
  {Vernet}}]{Weilbacher:2020}
{Weilbacher}, P.~M., {Palsa}, R., {Streicher}, O., {et~al.} 2020, \aap, 641,
  A28

\bibitem[{{Weilbacher} {et~al.}(2012){Weilbacher}, {Streicher}, {Urrutia},
  {Jarno}, {P{\'e}contal-Rousset}, {Bacon}, \& {B{\"o}hm}}]{Weilbacher:2012}
{Weilbacher}, P.~M., {Streicher}, O., {Urrutia}, T., {et~al.} 2012, SPIE Conf.
  Ser., 8451

\bibitem[{{Weilbacher} {et~al.}(2014){Weilbacher}, {Streicher}, {Urrutia},
  {P{\'e}contal-Rousset}, {Jarno}, \& {Bacon}}]{Weilbacher:2014}
{Weilbacher}, P.~M., {Streicher}, O., {Urrutia}, T., {et~al.} 2014, in
  Astronomical Society of the Pacific Conference Series, Vol. 485, Astronomical
  Data Analysis Software and Systems XXIII, ed. N.~{Manset} \& P.~{Forshay},
  451

\bibitem[{{Weinberger} {et~al.}(2017){Weinberger}, {Springel}, {Hernquist},
  {Pillepich}, {Marinacci}, {Pakmor}, {Nelson}, {Genel}, {Vogelsberger},
  {Naiman}, \& {Torrey}}]{Weinberger:2017}
{Weinberger}, R., {Springel}, V., {Hernquist}, L., {et~al.} 2017, \mnras, 465,
  3291

\bibitem[{{Westfall} {et~al.}(2019){Westfall}, {Cappellari}, {Bershady},
  {Bundy}, {Belfiore}, {Ji}, {Law}, {Schaefer}, {Shetty}, {Tremonti}, {Yan},
  {Andrews}, {Brownstein}, {Cherinka}, {Coccato}, {Drory}, {Maraston},
  {Parikh}, {S{\'a}nchez-Gallego}, {Thomas}, {Weijmans}, {Barrera-Ballesteros},
  {Du}, {Goddard}, {Li}, {Masters}, {Ibarra Medel}, {S{\'a}nchez}, {Yang},
  {Zheng}, \& {Zhou}}]{Westfall:2019}
{Westfall}, K.~B., {Cappellari}, M., {Bershady}, M.~A., {et~al.} 2019, \aj,
  158, 231

\bibitem[{{Williams} {et~al.}(2018){Williams}, {Pancoast}, {Treu}, {Brewer},
  {Barth}, {Bennert}, {Buehler}, {Canalizo}, {Cenko}, {Clubb}, {Cooper},
  {Filippenko}, {Gates}, {Hoenig}, {Joner}, {Kandrashoff}, {Laney}, {Lazarova},
  {Li}, {Malkan}, {Rex}, {Silverman}, {Tollerud}, {Walsh}, \&
  {Woo}}]{Williams:2018}
{Williams}, P.~R., {Pancoast}, A., {Treu}, T., {et~al.} 2018, \apj, 866, 75

\bibitem[{{Wisotzki} {et~al.}(2000){Wisotzki}, {Christlieb}, {Bade},
  {Beckmann}, {K{\"o}hler}, {Vanelle}, \& {Reimers}}]{Wisotzki:2000}
{Wisotzki}, L., {Christlieb}, N., {Bade}, N., {et~al.} 2000, \aap, 358, 77

\bibitem[{{Woo} {et~al.}(2016){Woo}, {Bae}, {Son}, \& {Karouzos}}]{Woo:2016}
{Woo}, J.-H., {Bae}, H.-J., {Son}, D., \& {Karouzos}, M. 2016, \apj, 817, 108

\bibitem[{{Woo} {et~al.}(2013){Woo}, {Schulze}, {Park}, {Kang}, {Kim}, \&
  {Riechers}}]{Woo:2013}
{Woo}, J.-H., {Schulze}, A., {Park}, D., {et~al.} 2013, \apj, 772, 49

\bibitem[{{Woo} {et~al.}(2015){Woo}, {Yoon}, {Park}, {Park}, \&
  {Kim}}]{Woo:2015}
{Woo}, J.-H., {Yoon}, Y., {Park}, S., {Park}, D., \& {Kim}, S.~C. 2015, \apj,
  801, 38

\bibitem[{{Worseck} {et~al.}(2021){Worseck}, {Khrykin}, {Hennawi}, {Prochaska},
  \& {Farina}}]{Worseck:2021}
{Worseck}, G., {Khrykin}, I.~S., {Hennawi}, J.~F., {Prochaska}, J.~X., \&
  {Farina}, E.~P. 2021, \mnras, 505, 5084

\bibitem[{{Xue} {et~al.}(2011){Xue}, {Luo}, {Brandt}, {Bauer}, {Lehmer},
  {Broos}, {Schneider}, {Alexander}, {Brusa}, {Comastri}, {Fabian}, {Gilli},
  {Hasinger}, {Hornschemeier}, {Koekemoer}, {Liu}, {Mainieri}, {Paolillo},
  {Rafferty}, {Rosati}, {Shemmer}, {Silverman}, {Smail}, {Tozzi}, \&
  {Vignali}}]{Xue:2011}
{Xue}, Y.~Q., {Luo}, B., {Brandt}, W.~N., {et~al.} 2011, \apjs, 195, 10

\bibitem[{{Yang} {et~al.}(2019){Yang}, {Gaspari}, \& {Marlow}}]{Yang:2019}
{Yang}, H. Y.~K., {Gaspari}, M., \& {Marlow}, C. 2019, \apj, 871, 6

\bibitem[{{York} {et~al.}(2000){York}, {Adelman}, {Anderson}, {Anderson},
  {Annis}, {Bahcall}, {Bakken}, {Barkhouser}, {Bastian}, {Berman}, {Boroski},
  {Bracker}, {Briegel}, {Briggs}, {Brinkmann}, {Brunner}, {Burles}, {Carey},
  {Carr}, {Castander}, {Chen}, {Colestock}, {Connolly}, {Crocker}, {Csabai},
  {Czarapata}, {Davis}, {Doi}, {Dombeck}, {Eisenstein}, {Ellman}, {Elms},
  {Evans}, {Fan}, {Federwitz}, {Fiscelli}, {Friedman}, {Frieman}, {Fukugita},
  {Gillespie}, {Gunn}, {Gurbani}, {de Haas}, {Haldeman}, {Harris}, {Hayes},
  {Heckman}, {Hennessy}, {Hindsley}, {Holm}, {Holmgren}, {Huang}, {Hull},
  {Husby}, {Ichikawa}, {Ichikawa}, {Ivezi{\'c}}, {Kent}, {Kim}, {Kinney},
  {Klaene}, {Kleinman}, {Kleinman}, {Knapp}, {Korienek}, {Kron}, {Kunszt},
  {Lamb}, {Lee}, {Leger}, {Limmongkol}, {Lindenmeyer}, {Long}, {Loomis},
  {Loveday}, {Lucinio}, {Lupton}, {MacKinnon}, {Mannery}, {Mantsch}, {Margon},
  {McGehee}, {McKay}, {Meiksin}, {Merelli}, {Monet}, {Munn}, {Narayanan},
  {Nash}, {Neilsen}, {Neswold}, {Newberg}, {Nichol}, {Nicinski}, {Nonino},
  {Okada}, {Okamura}, {Ostriker}, {Owen}, {Pauls}, {Peoples}, {Peterson},
  {Petravick}, {Pier}, {Pope}, {Pordes}, {Prosapio}, {Rechenmacher}, {Quinn},
  {Richards}, {Richmond}, {Rivetta}, {Rockosi}, {Ruthmansdorfer}, {Sandford},
  {Schlegel}, {Schneider}, {Sekiguchi}, {Sergey}, {Shimasaku}, {Siegmund},
  {Smee}, {Smith}, {Snedden}, {Stone}, {Stoughton}, {Strauss}, {Stubbs},
  {SubbaRao}, {Szalay}, {Szapudi}, {Szokoly}, {Thakar}, {Tremonti}, {Tucker},
  {Uomoto}, {Vanden Berk}, {Vogeley}, {Waddell}, {Wang}, {Watanabe},
  {Weinberg}, {Yanny}, \& {Yasuda}}]{York:2000}
{York}, D.~G., {Adelman}, J., {Anderson}, Jr., J.~E., {et~al.} 2000, \aj, 120,
  1579

\bibitem[{{Zhuang} {et~al.}(2021){Zhuang}, {Ho}, \& {Shangguan}}]{Zhuang:2021}
{Zhuang}, M.-Y., {Ho}, L.~C., \& {Shangguan}, J. 2021, \apj, 906, 38

\end{thebibliography}

\begin{appendix}
 \section{Bayesian model for AGN lifetime}\label{apx:pdfs}
In order to statistically infer the average AGN lifetime $\langle t_\mathrm{AGN}\rangle$ from an observed AGN sample, we follow the basic prescription developed by \citet{Khrykin:2021}. They use the \ion{He}{ii} proximity zone sizes of about 20 $z\sim3$ luminous AGN to determine $\langle t_\mathrm{AGN}\rangle$. Instead of the proximity zones, we use the size of the ENLR as an alternative proxy for the time an AGN has been active during the current episode, the ``on-time'' ($t_\mathrm{on}$) where $t_{on}\in [0;t_\mathrm{AGN}]$. Only $t_\mathrm{on}$ can be measured and we need to use a predictive model to infer $\langle t_\mathrm{AGN}\rangle$ from the whole sample as $t_\mathrm{AGN}$ cannot be directly inferred for a given AGN by the nature of the problem.

\begin{figure}
 \resizebox{\hsize}{!}{\includegraphics{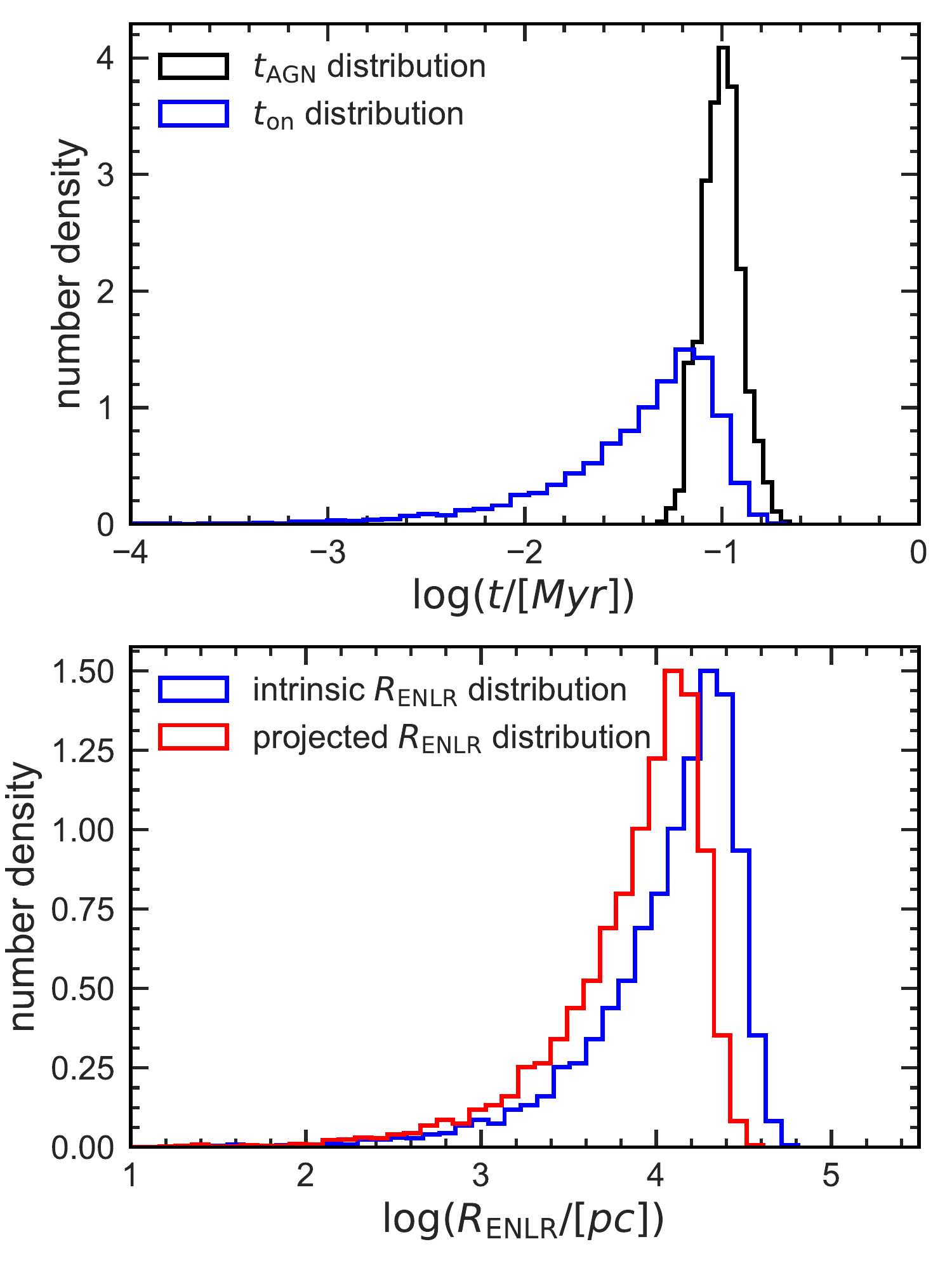}}
 \caption{Example on the PDF changes from a log-normal $t_\mathrm{AGN}$ distribution to corresponding projected $R_\mathrm{ENLR}$ distribution. For the input log-normal distribution we use $\mu=\log(10^5/[\mathrm{yr}])$ and $\sigma=0.1$\,dex from the larger parameter grid.}\label{fig:pdf_dist}
\end{figure}

Following \citet{Khrykin:2021} we start by assuming a log-normal distribution function for the AGN lifetime as 
\begin{equation}
 p(t_\mathrm{AGN}) = \frac{1}{t_\mathrm{AGN}}\frac{\log(\mathrm{e})}{\sigma\sqrt{2\pi}}\times\exp\left[-\frac{(\log(t_\mathrm{AGN}/[\mathrm{Myr}])-\mu)^2}{2\sigma^2}\right] \quad ,
\end{equation}
where $\mu=\langle \log(t_\mathrm{AGN}/\mathrm{Myr}])\rangle$ and $\sigma=\sigma_{\log t_\mathrm{AGN}}$. In order to apply Bayesian inference we need to determine the likelihood function $\mathcal{L}_\mathrm{AGN}(R_{\mathrm{ENLR},i}|\mu,\sigma)$ for each AGN in the sample. We start with the construction of a parameter grid for $\mu$ and $\sigma$ in the range of $\mu=[-5,1.5]$ with a step size of $\Delta\mu=0.1$ and $\sigma=[0.01,1.0]$ with $\Delta\sigma=0.05$. For each of those grid points we perform the following sequence of steps to determine the likelihood function for each parameter combination:\\
(1) A sample of 1000 AGN lifetimes $\log(t_\mathrm{AGN}/[\mathrm{Myr}])$ is randomly drawn from a log-normal distribution function.\\
(2) For each of those 1000 values of $t_\mathrm{AGN}$ we draw 10 values of $t_\mathrm{on}$ adopting a uniform distribution with $p(t_\mathrm{on}) = \frac{1}{t_\mathrm{AGN}}$ between $0$ and $t_\mathrm{AGN}$. This takes into account the random sampling of AGN observations during their overall lifetime.\\
(3) We convert the drawn $t_\mathrm{on}$ values into an intrinsic ENLR size $R_\mathrm{ENLR}$ with the light speed.\\
(4) Adopting an intrinsic half-opening angle $\theta_\mathrm{ENLR}=60\degr$ for the ENLR ionization cones, the probability density function (PDF) $p$ for the cone inclination $i_\mathrm{ENLR}$ is given by the ratio of the surface area of a spherical ring $\mathrm{d}A_\mathrm{ring}(i) = 2\pi r^2\sin(i) \mathrm{d}i$ to the total area of the cone surface $A_\mathrm{cone} = 2\pi r^2 (1-\cos(\theta))$:
\begin{equation}
p(i_\mathrm{ENLR})=\begin{cases}\frac{\sin(i_\mathrm{ENLR})}{1-\cos(\theta_\mathrm{ENLR})} & 0 < i_\mathrm{ENLR} < \theta_\mathrm{ENLR}\\
                                 0  & i_\mathrm{ENLR} > \theta_\mathrm{ENLR}
                   \end{cases}\quad .
 \end{equation}
 This PDF implies a mean inclination angle of $\langle i_\mathrm{ENLR}\rangle=39\degr$, which we apply to convert intrinsic $R_\mathrm{ENLR}$ sizes to projected $R_\mathrm{ENLR,proj}$ sizes. We basically assume that the large opening angle of the cones are removing any dependence on object-to-object variations and apply only a global projection.\\
 (5) The continuous PDF of $p(R_\mathrm{ENLR,proj})$ is determined by applying a kernel density estimation (KDE) to the 10\,000 drawn values of $R_\mathrm{ENLR,proj}$. This way the impact of the random sampling of $t_\mathrm{on}$ is effeciently marginalized in the combined PDF.
 
 We provide an example of all steps in Fig.~\ref{fig:pdf_dist} for a single parameter set of $\mu$ and $\sigma$. The likelihood $\mathcal{L}_\mathrm{AGN}(R_\mathrm{ENLR}|\mu,\sigma)$ can be obtained by evaluating the PDF of $p(R_\mathrm{ENLR,proj})$ for a specific parameter set of $\mu$ and $\sigma$. Here, we use a simple linear interpolation of the PDF between our precomputed grid points. Since our data imply that $\mu$ is a function of $\log(M_\mathrm{BH})$, we assume a linear relation $\mu = m\log(M_\mathrm{BH}) + b$, which simply change the likelihood function to $\mathcal{L}_\mathrm{AGN}(R_\mathrm{ENLR},\log(M_\mathrm{BH})|m,b,\sigma)$.
 
 The joint likelihood function for our entire sample can then be computed as 
 \begin{equation}
  \mathcal{L}_\mathrm{joint} = \prod_{0}^{N_\mathrm{AGN}}\mathcal{L}_\mathrm{AGN}(R_\mathrm{ENLR,i},\log(M_\mathrm{BH,i})|m,b,\sigma)
 \end{equation}
where $N_\mathrm{AGN}=32$ is the number of AGN observed with MUSE excluding non-AGN and CLAGN sources as well as HE~0021$-$1810 due to its low data quality. As described in the main text we use the MCMC to sample $\mathcal{L}_\mathrm{joint}$  to infer the posterior distribution function for  $m$, $b$  and $\sigma$ from our observed data.

\section{Stellar and emission line maps}\label{apx:maps_all}
 \begin{figure*}
 \includegraphics[width=0.85\textwidth]{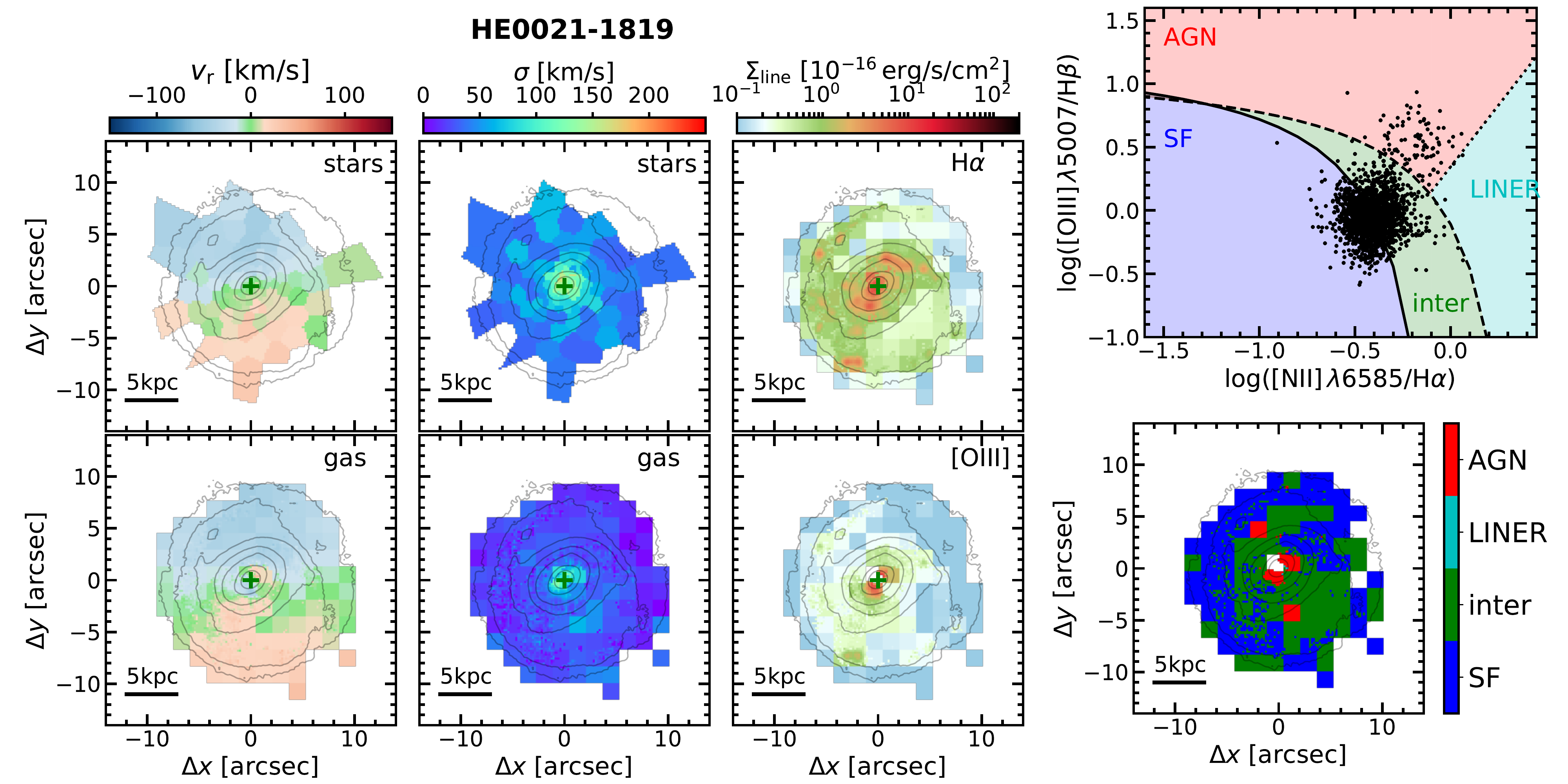}\\
 \includegraphics[width=0.85\textwidth]{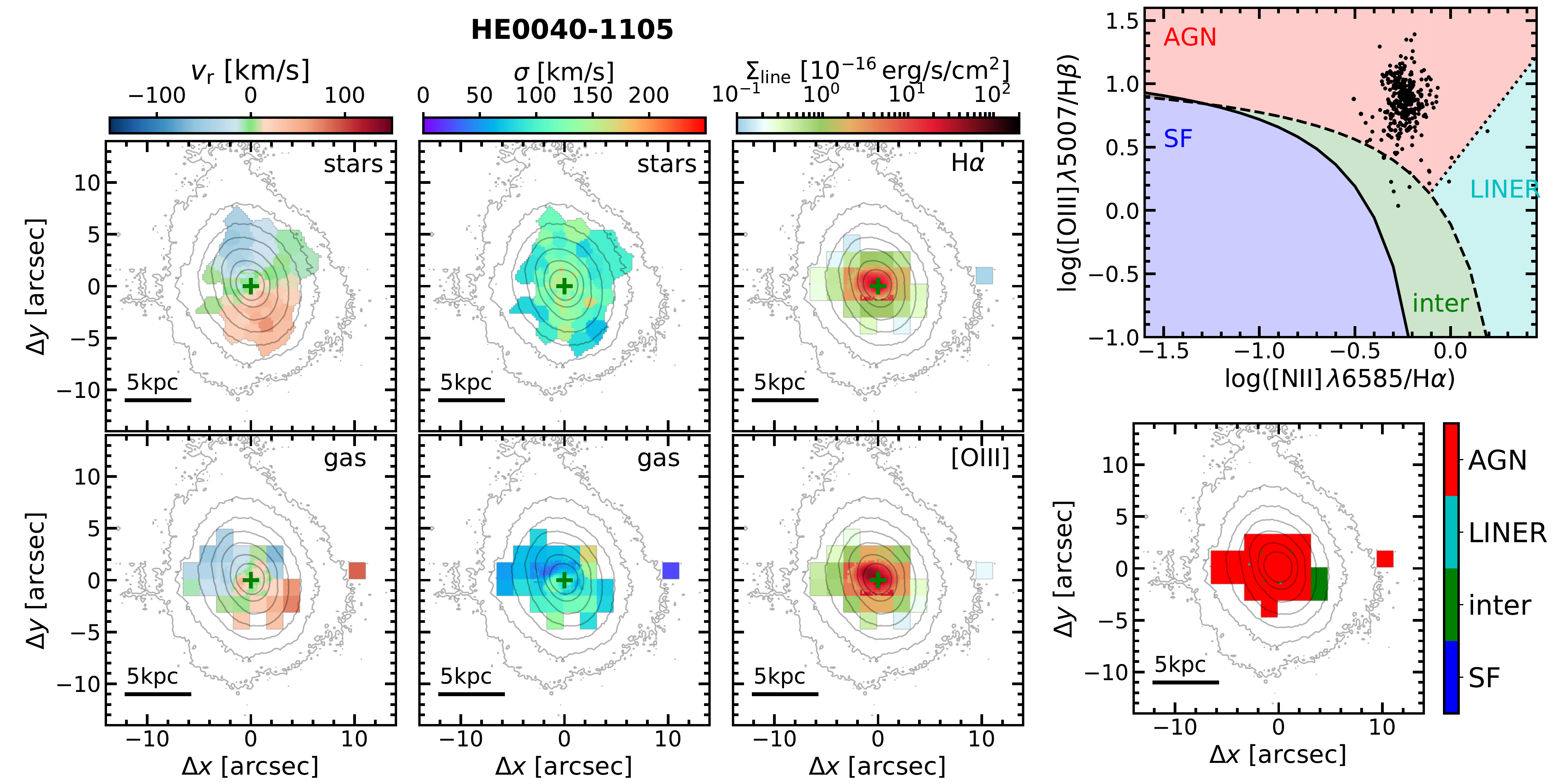}\\
 \includegraphics[width=0.85\textwidth]{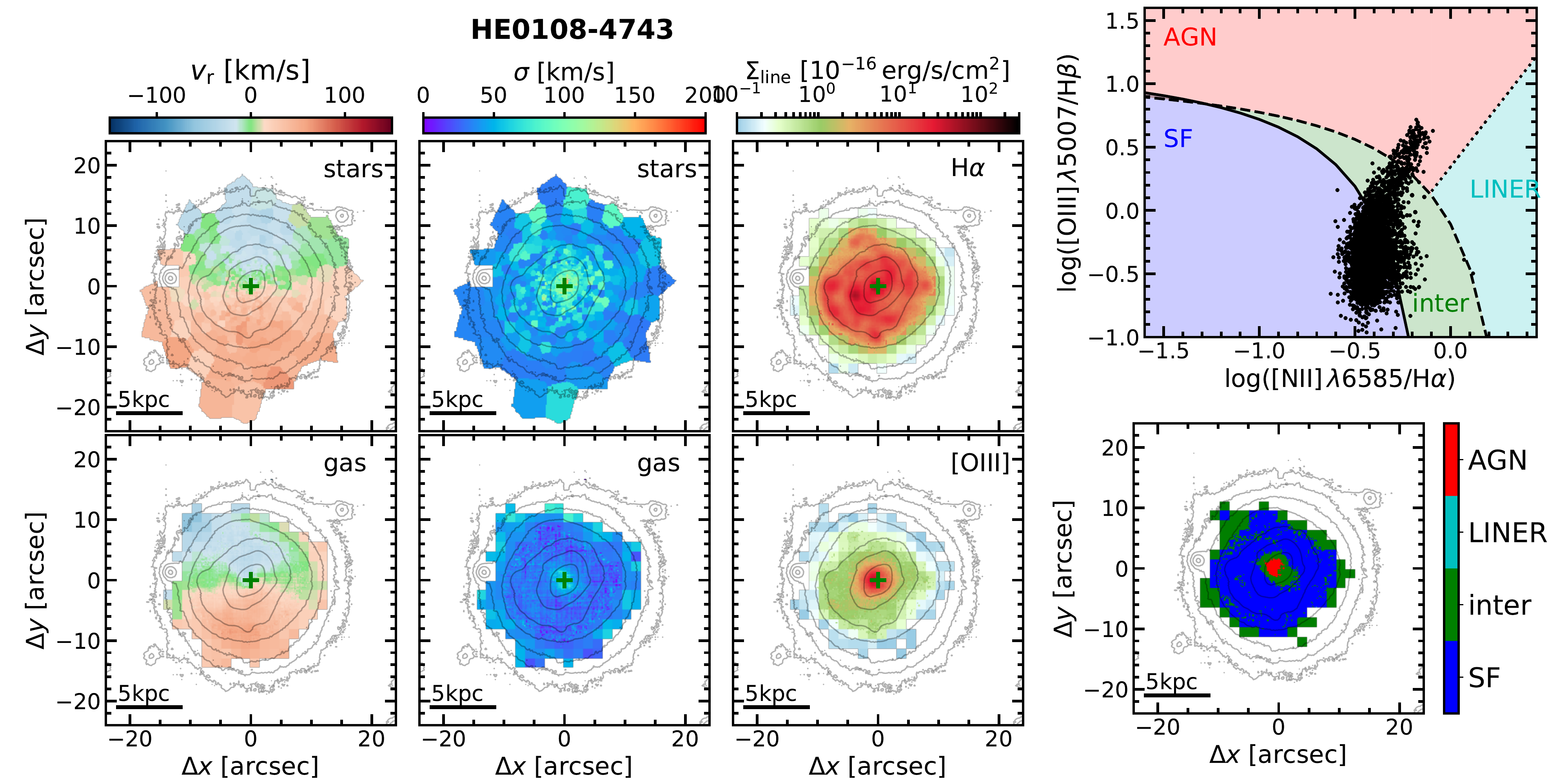}
 \caption{Same as in Fig.~\ref{fig:PyParadise_results}, but for the full sample.}\label{fig:full_maps}
 \end{figure*}
  \addtocounter{figure}{-1}
  
 \begin{figure*}
 \includegraphics[width=0.85\textwidth]{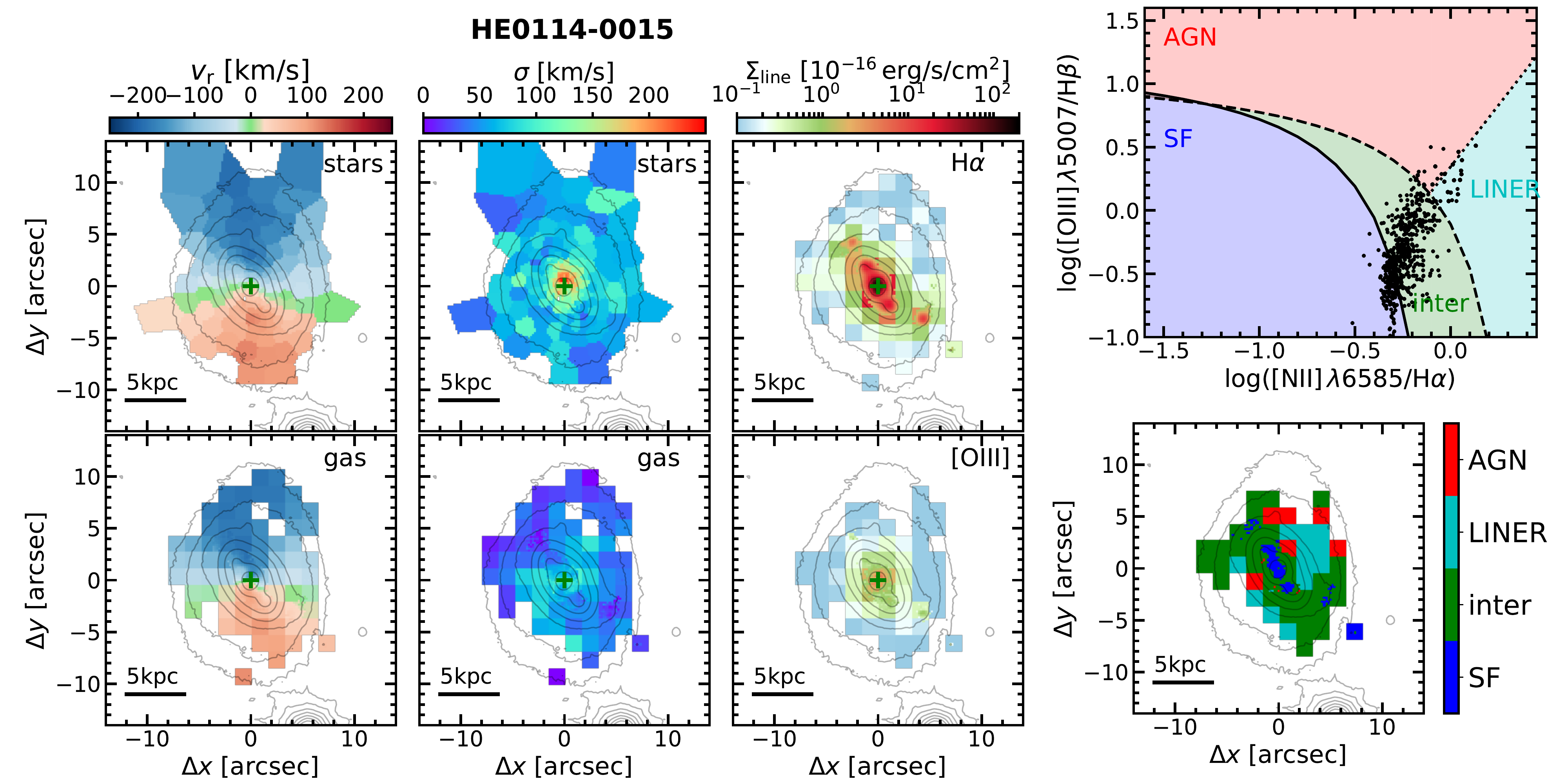}\\
 \includegraphics[width=0.85\textwidth]{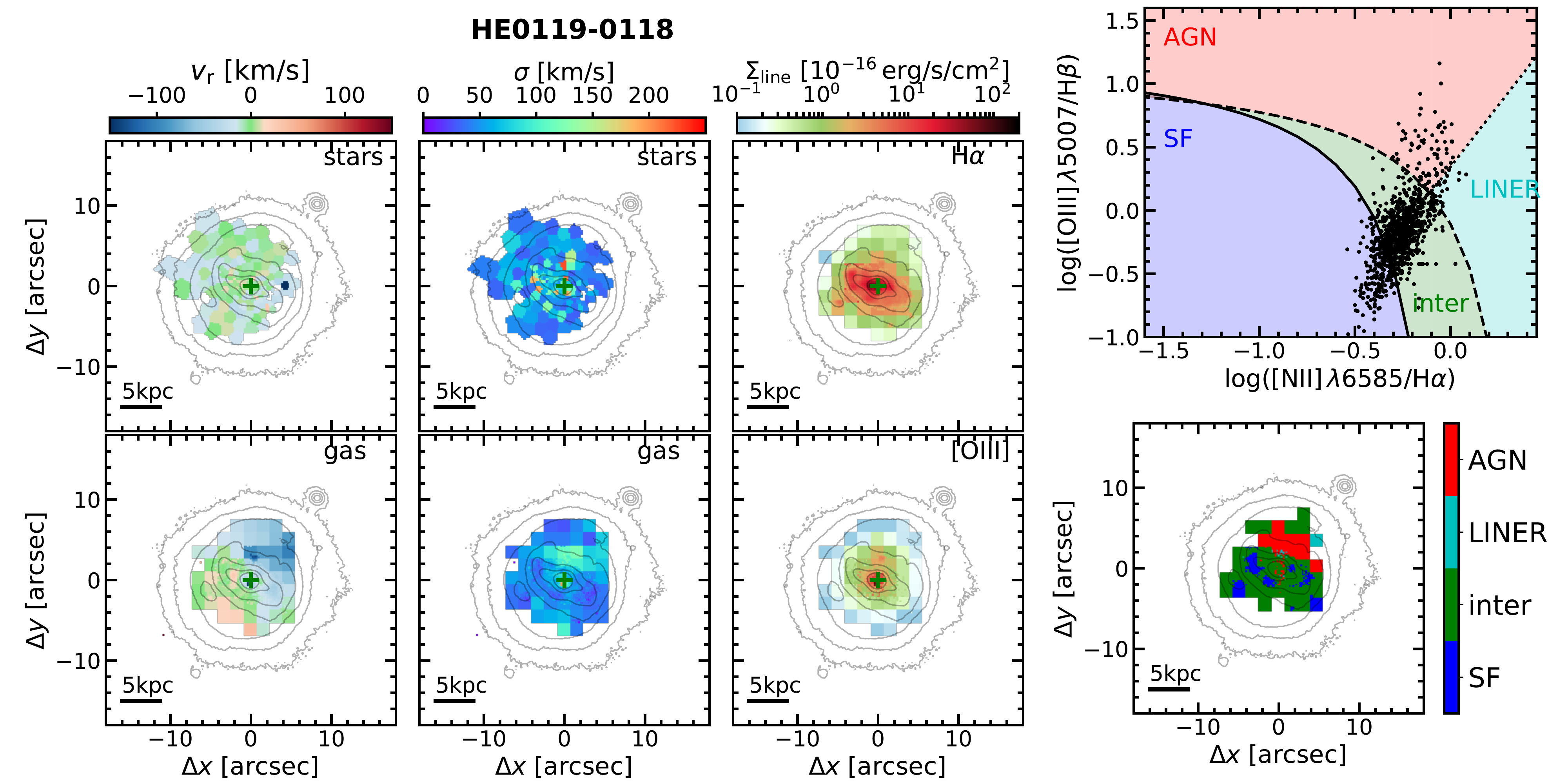}\\
 \includegraphics[width=0.85\textwidth]{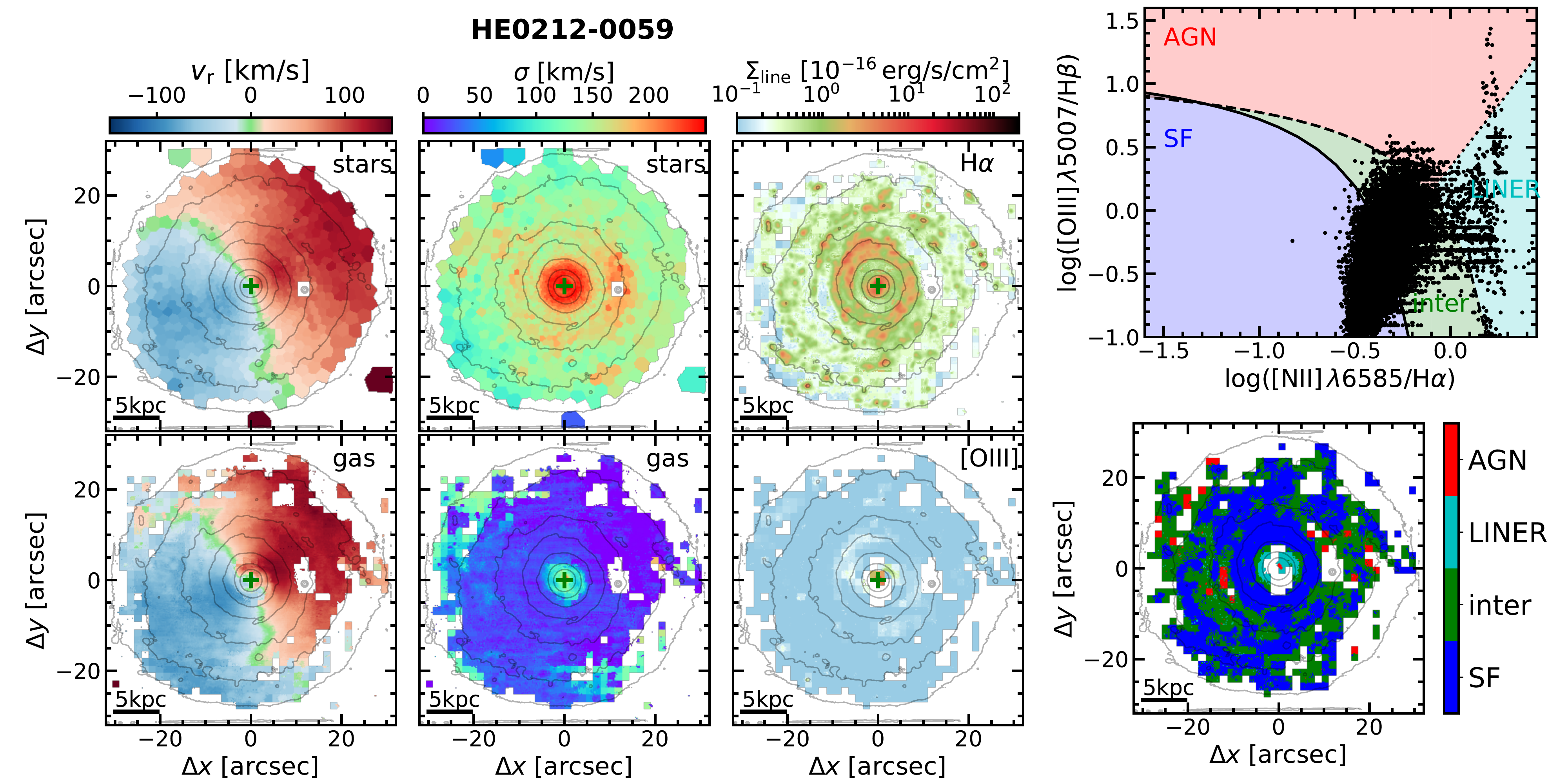}
 \caption{continued.}
 \end{figure*}
 \addtocounter{figure}{-1}
 
 \begin{figure*}
 \includegraphics[width=0.85\textwidth]{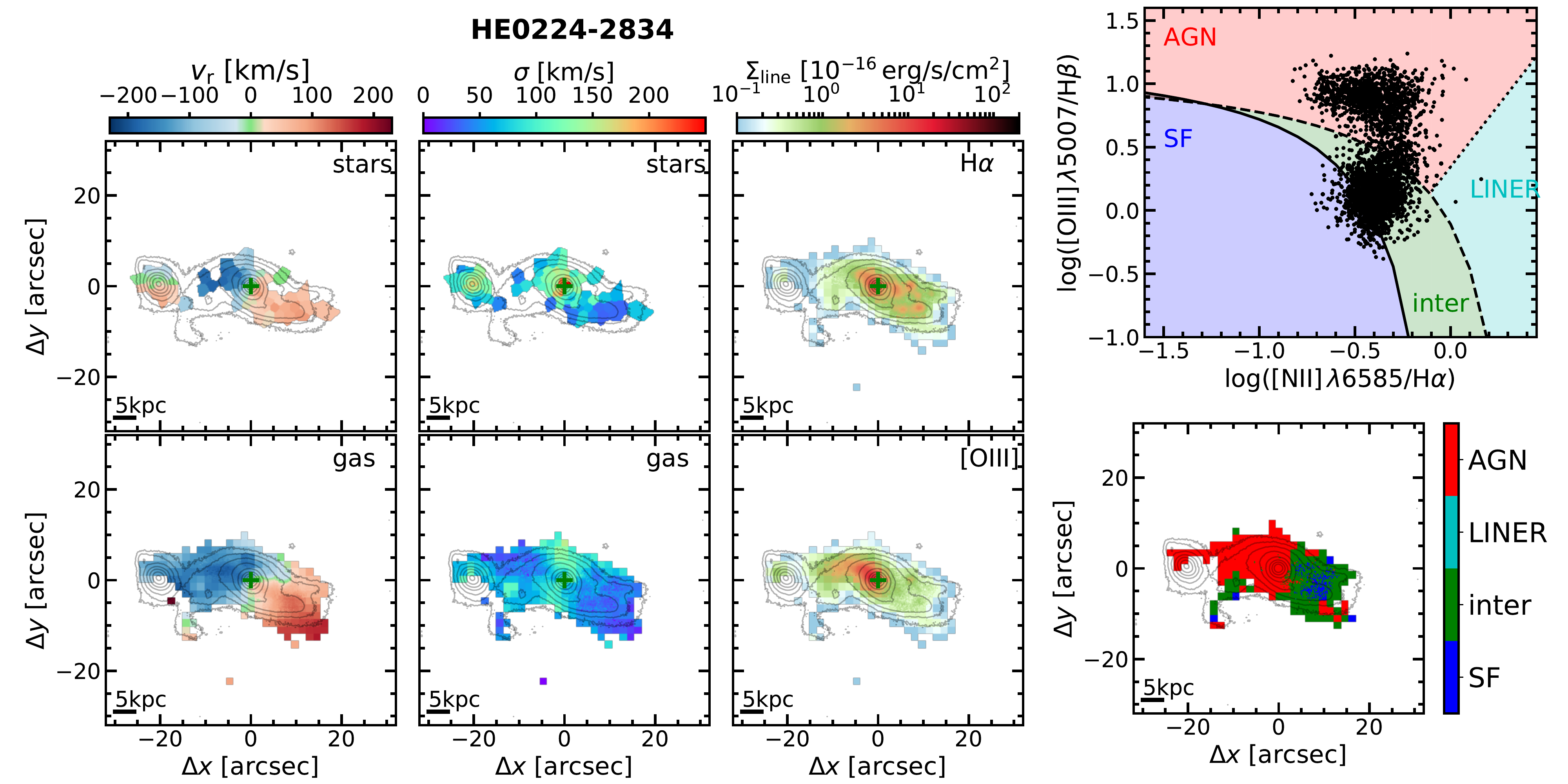}\\
 \includegraphics[width=0.85\textwidth]{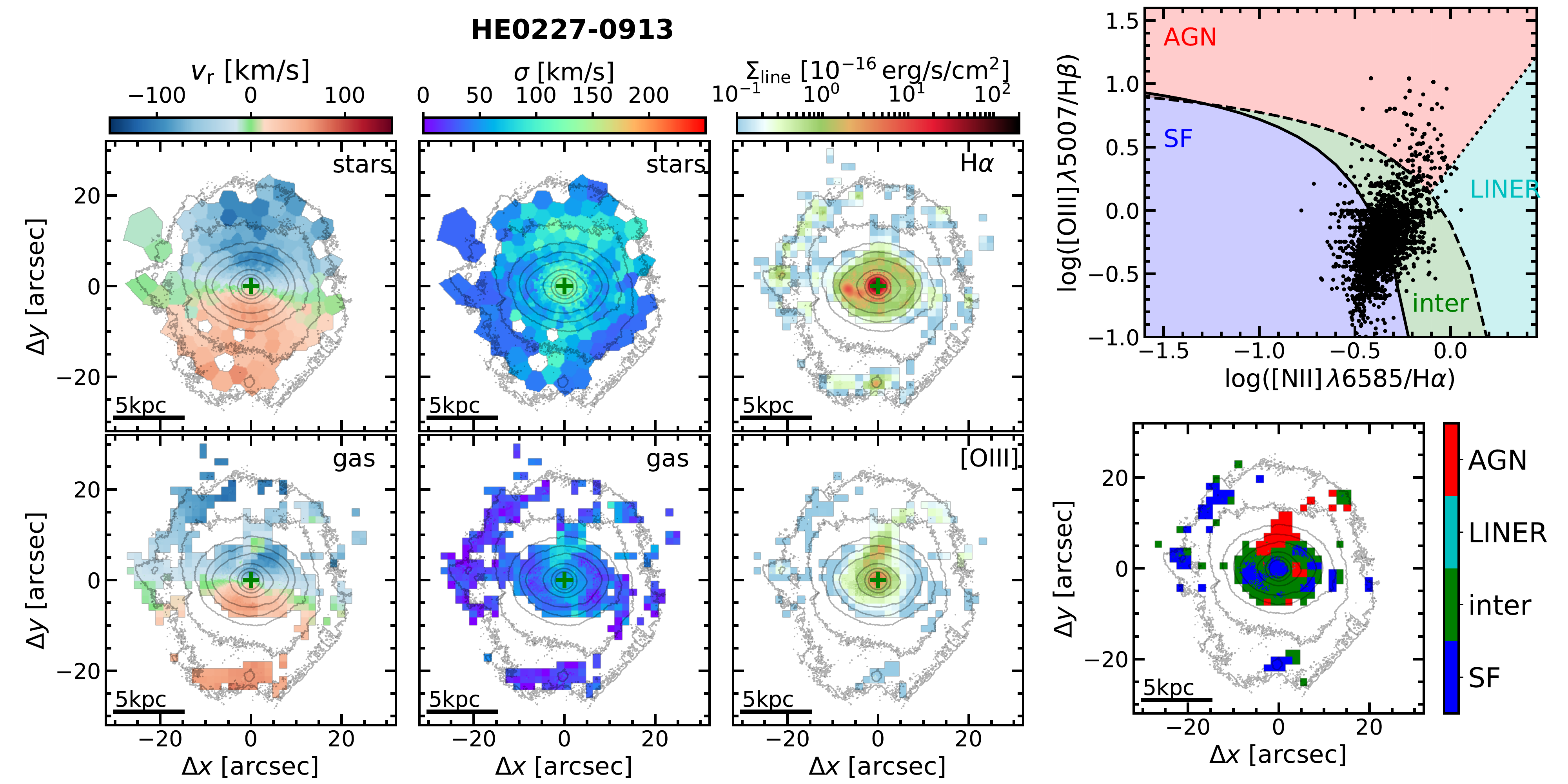}\\
 \includegraphics[width=0.85\textwidth]{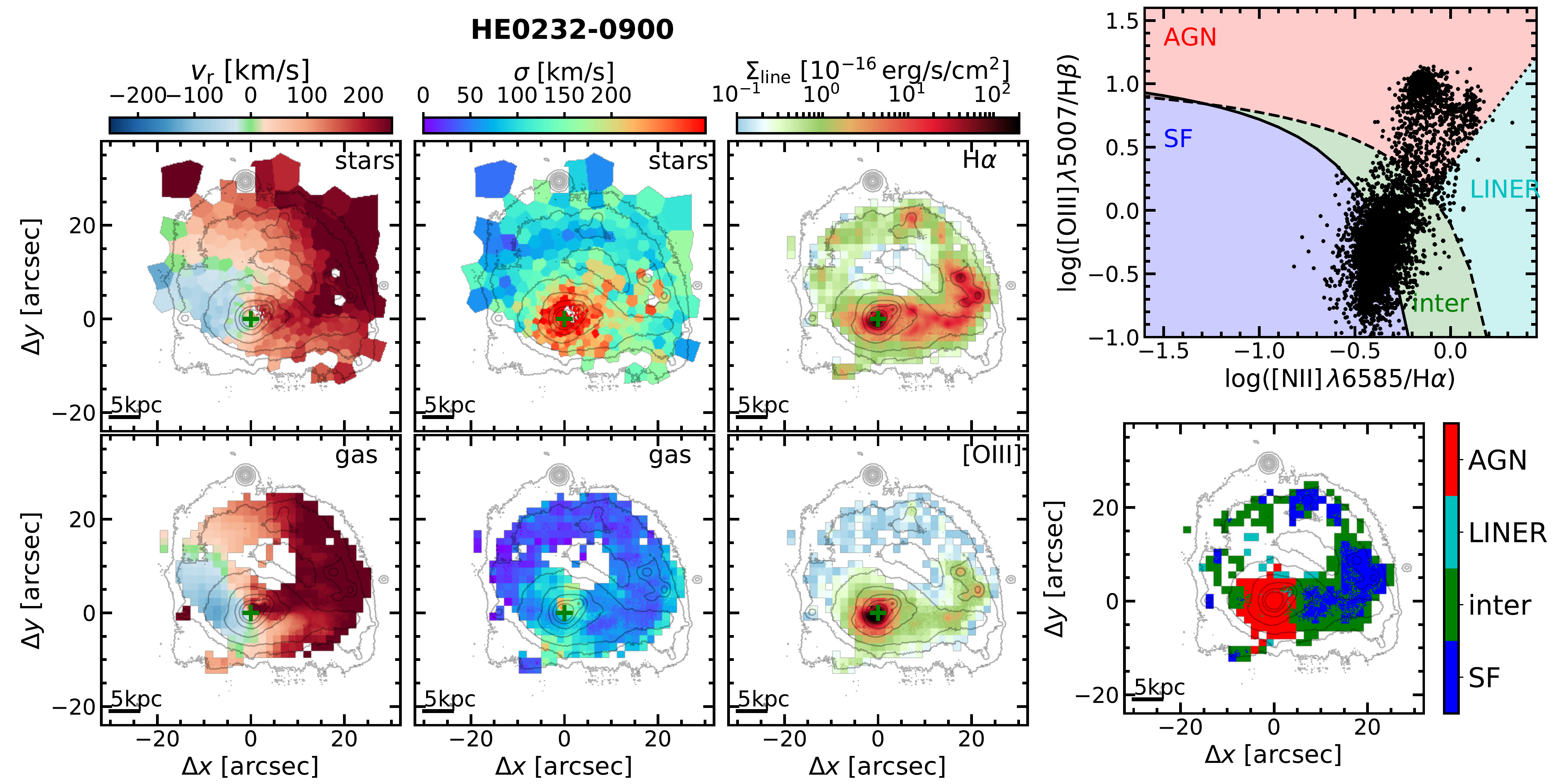}
 \caption{continued.}
 \end{figure*}
 \addtocounter{figure}{-1}
 
 \begin{figure*}
 \includegraphics[width=0.85\textwidth]{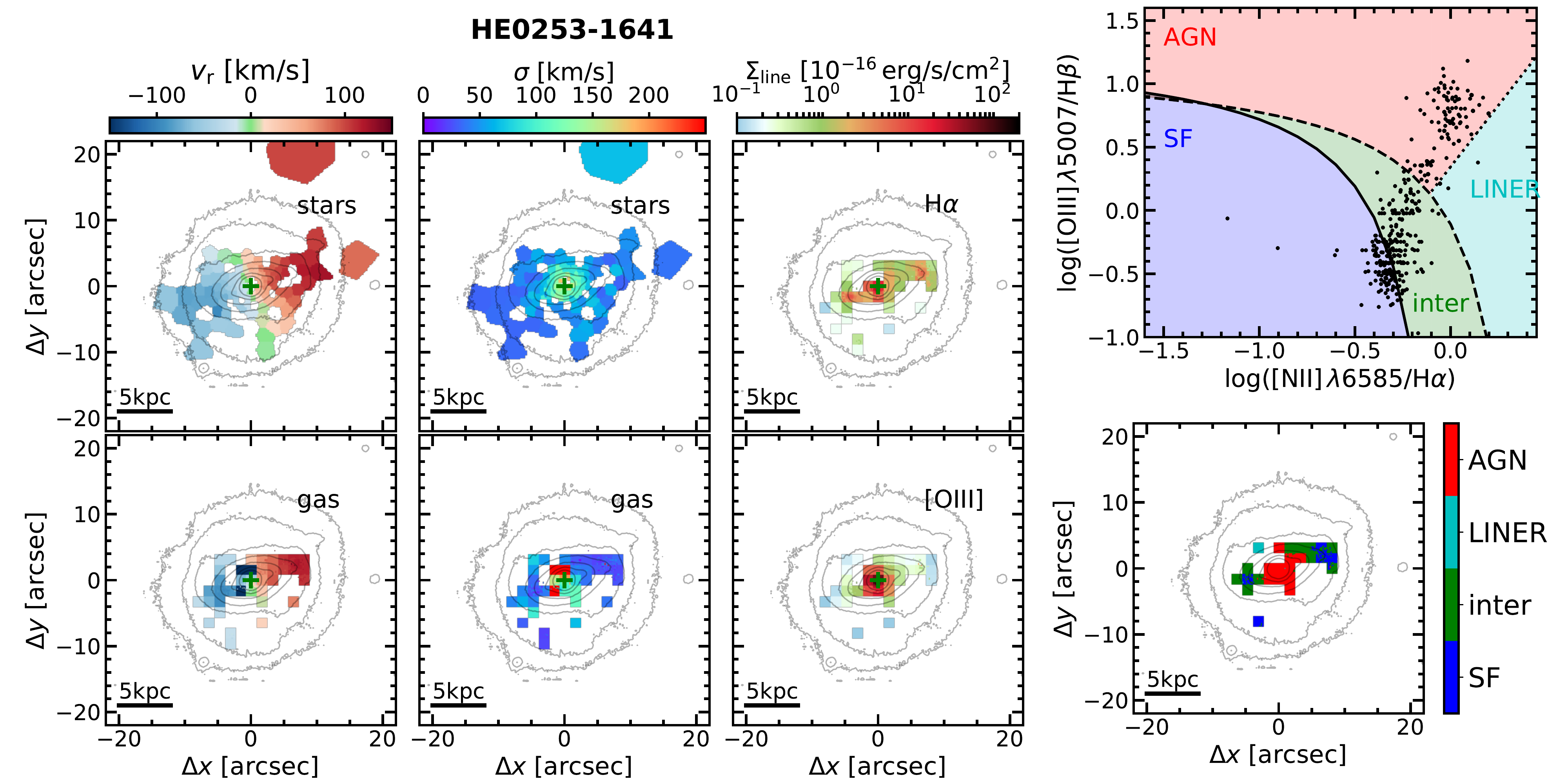}\\
 \includegraphics[width=0.85\textwidth]{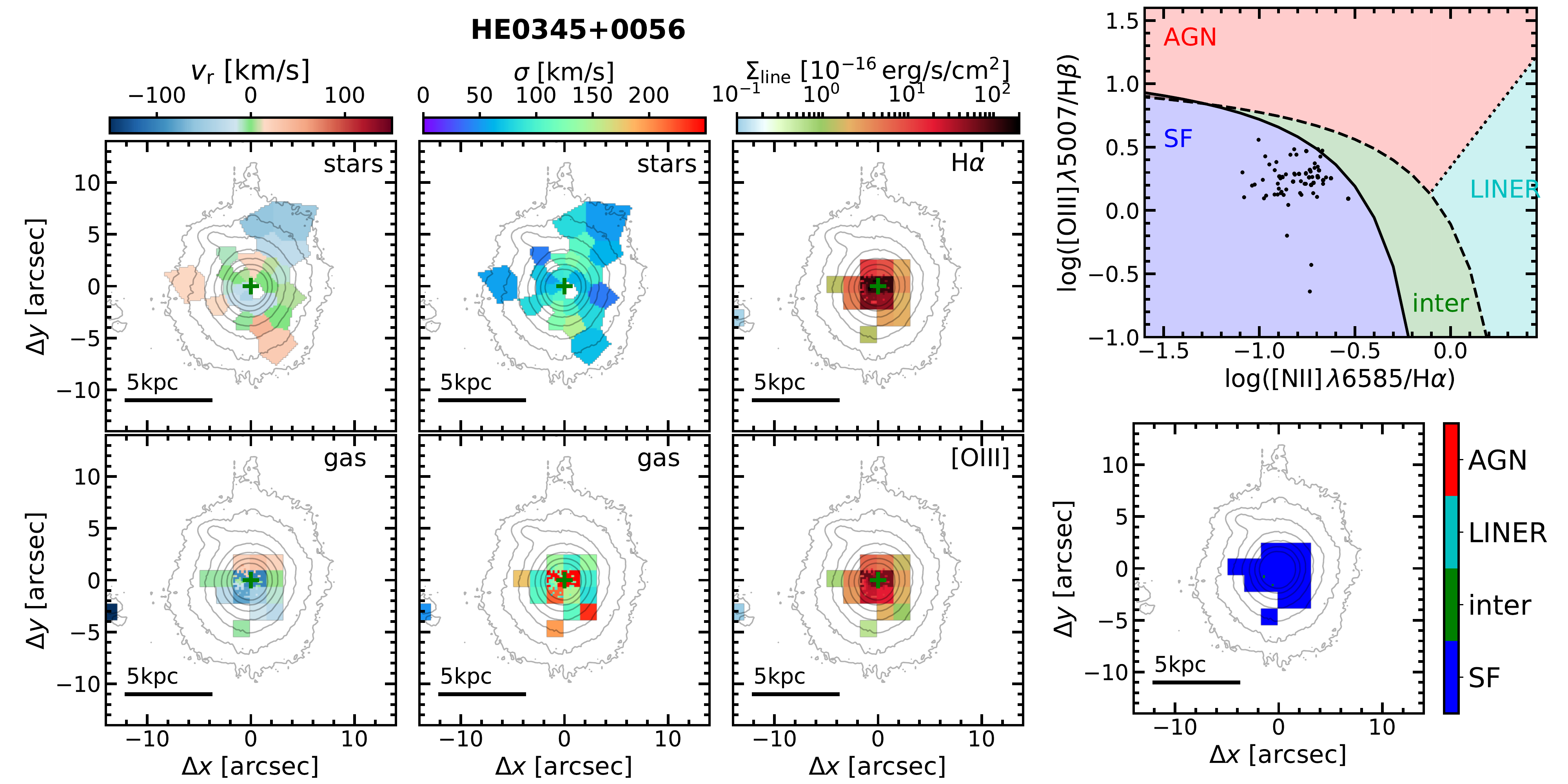}\\
 \includegraphics[width=0.85\textwidth]{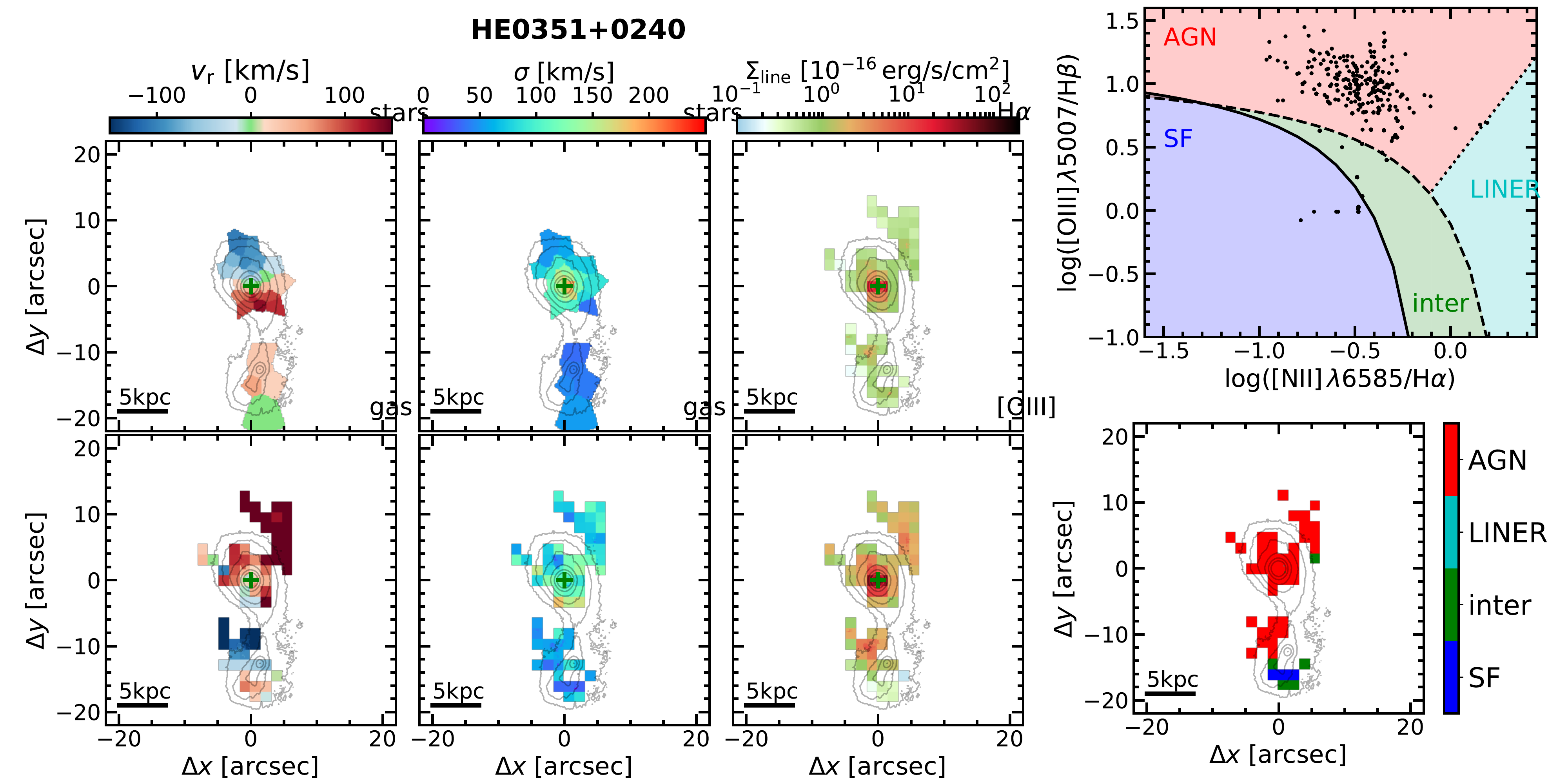}
 \caption{continued.}
 \end{figure*}
 \addtocounter{figure}{-1}
 
 \begin{figure*}
  \includegraphics[width=0.85\textwidth]{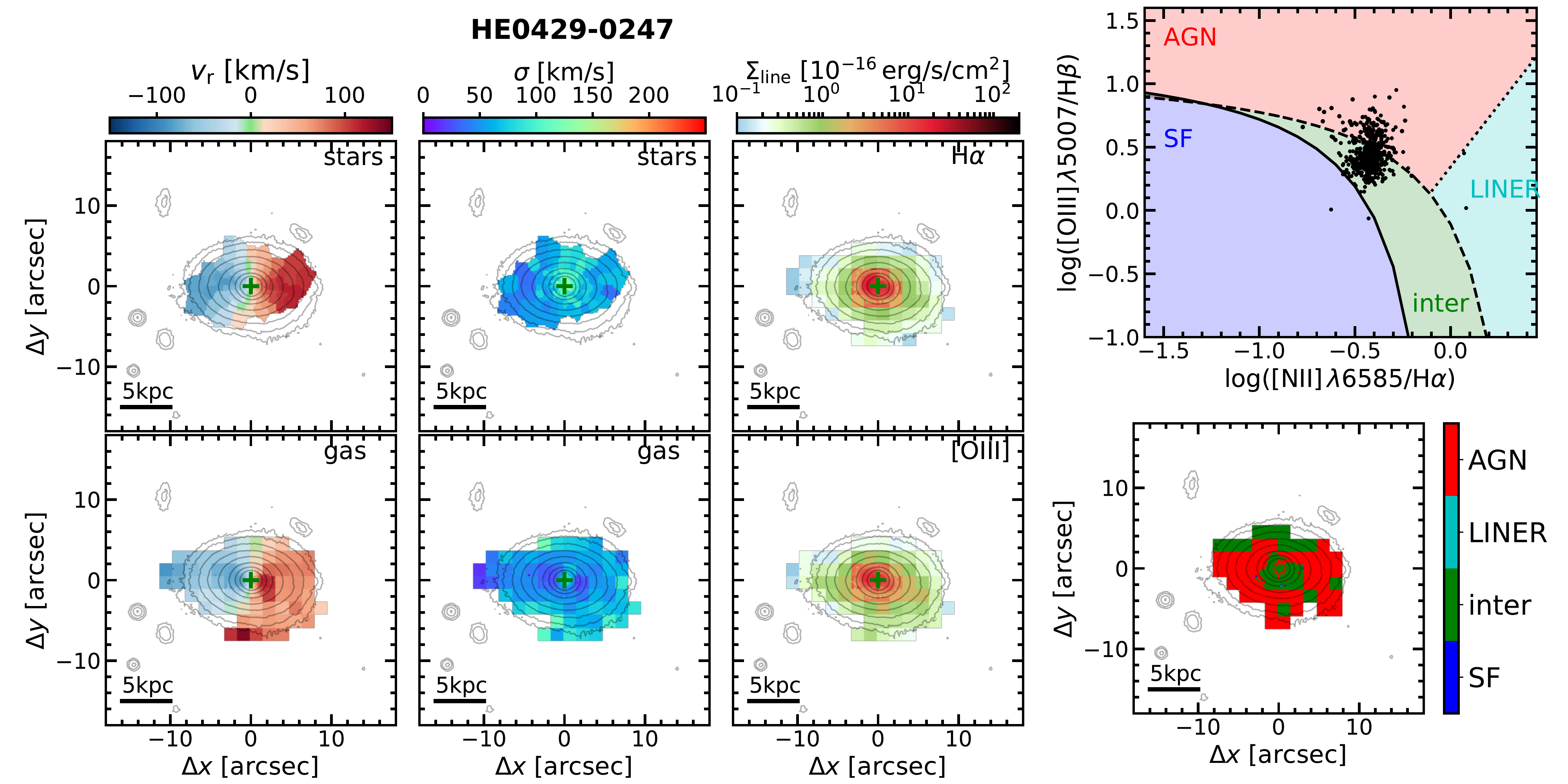}\\
  \includegraphics[width=0.85\textwidth]{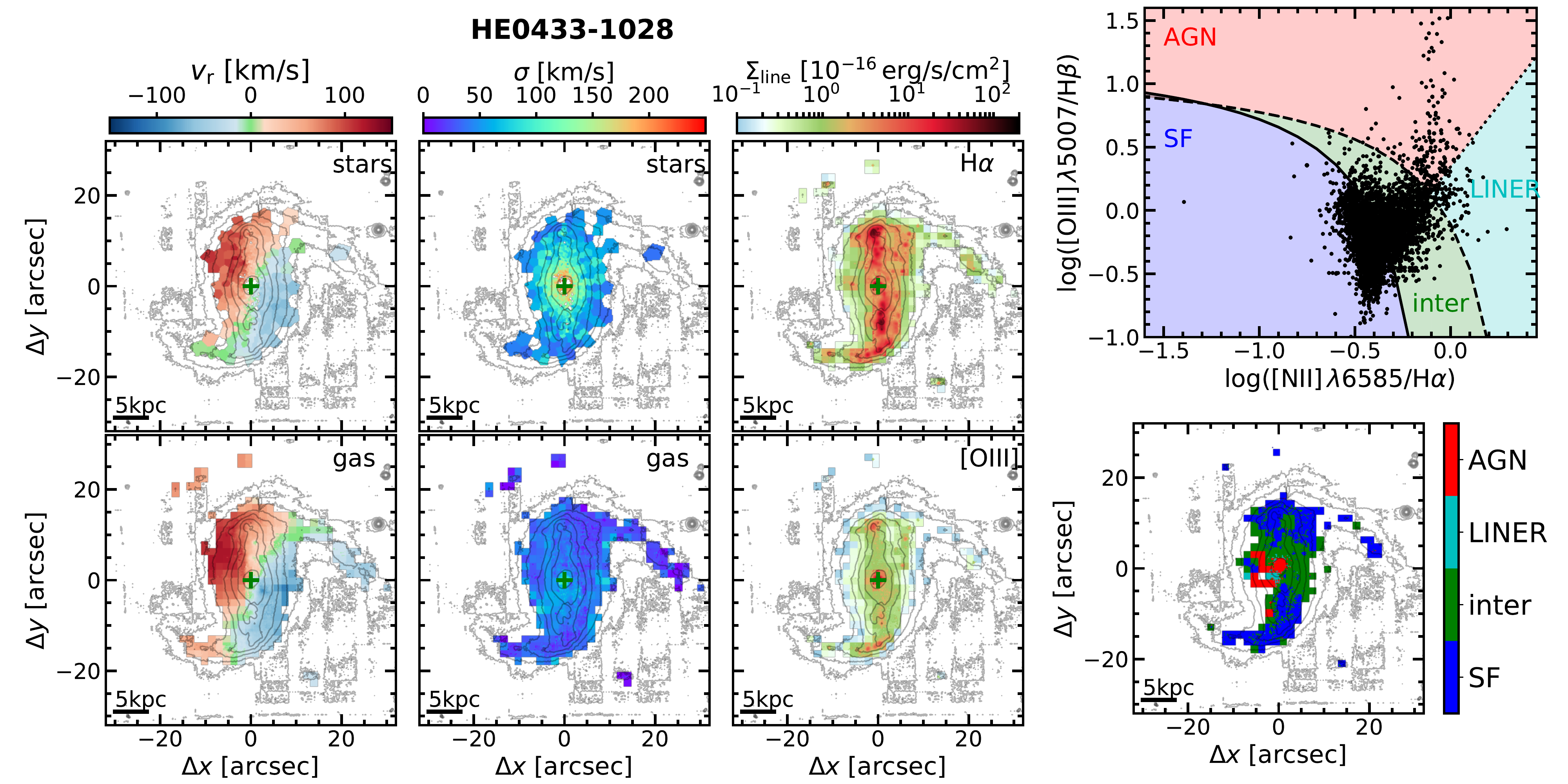}\\
  \includegraphics[width=0.85\textwidth]{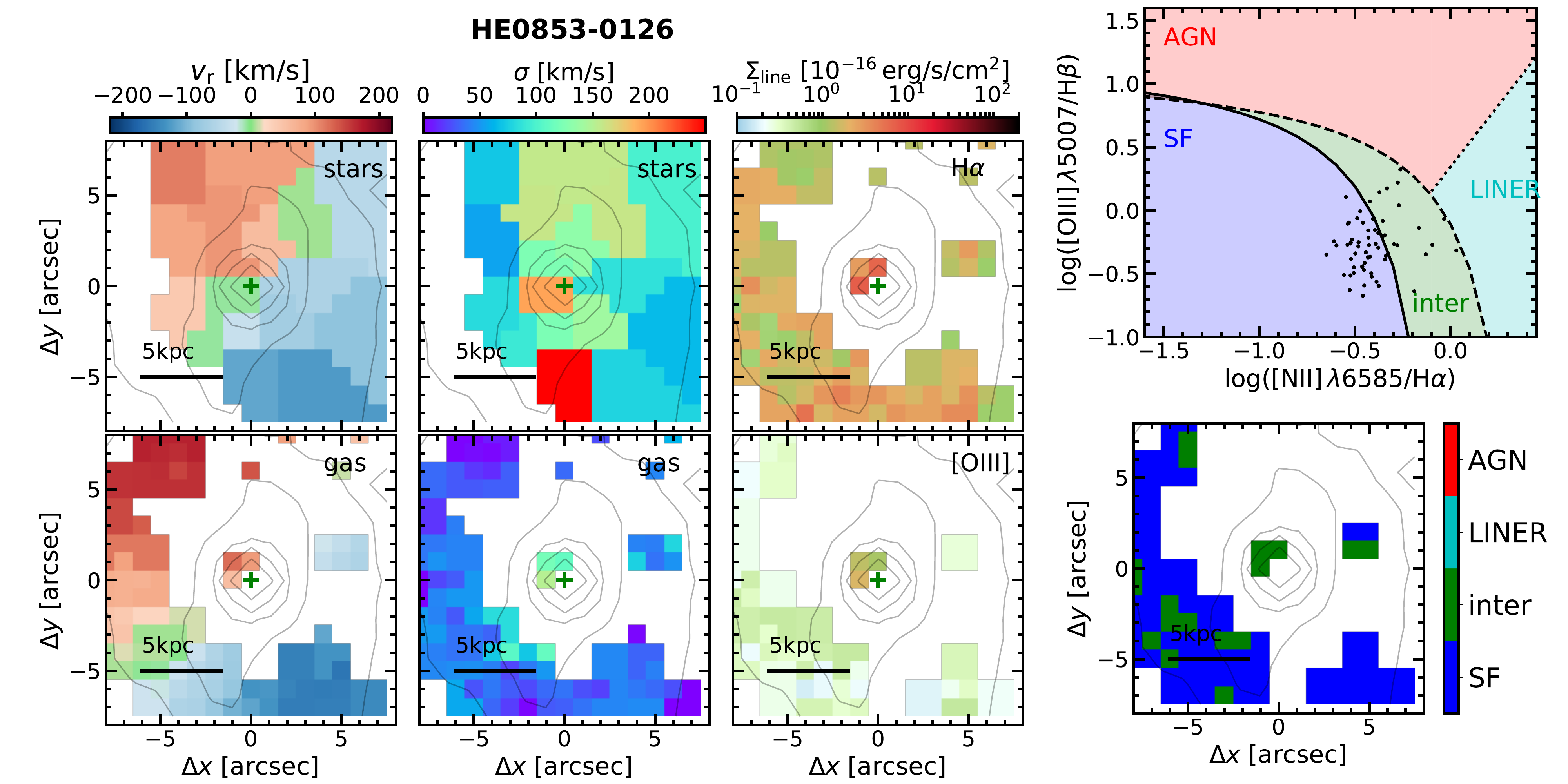}
  \caption{continued.}
 \end{figure*}
 \addtocounter{figure}{-1}
 
 \begin{figure*}
  \includegraphics[width=0.85\textwidth]{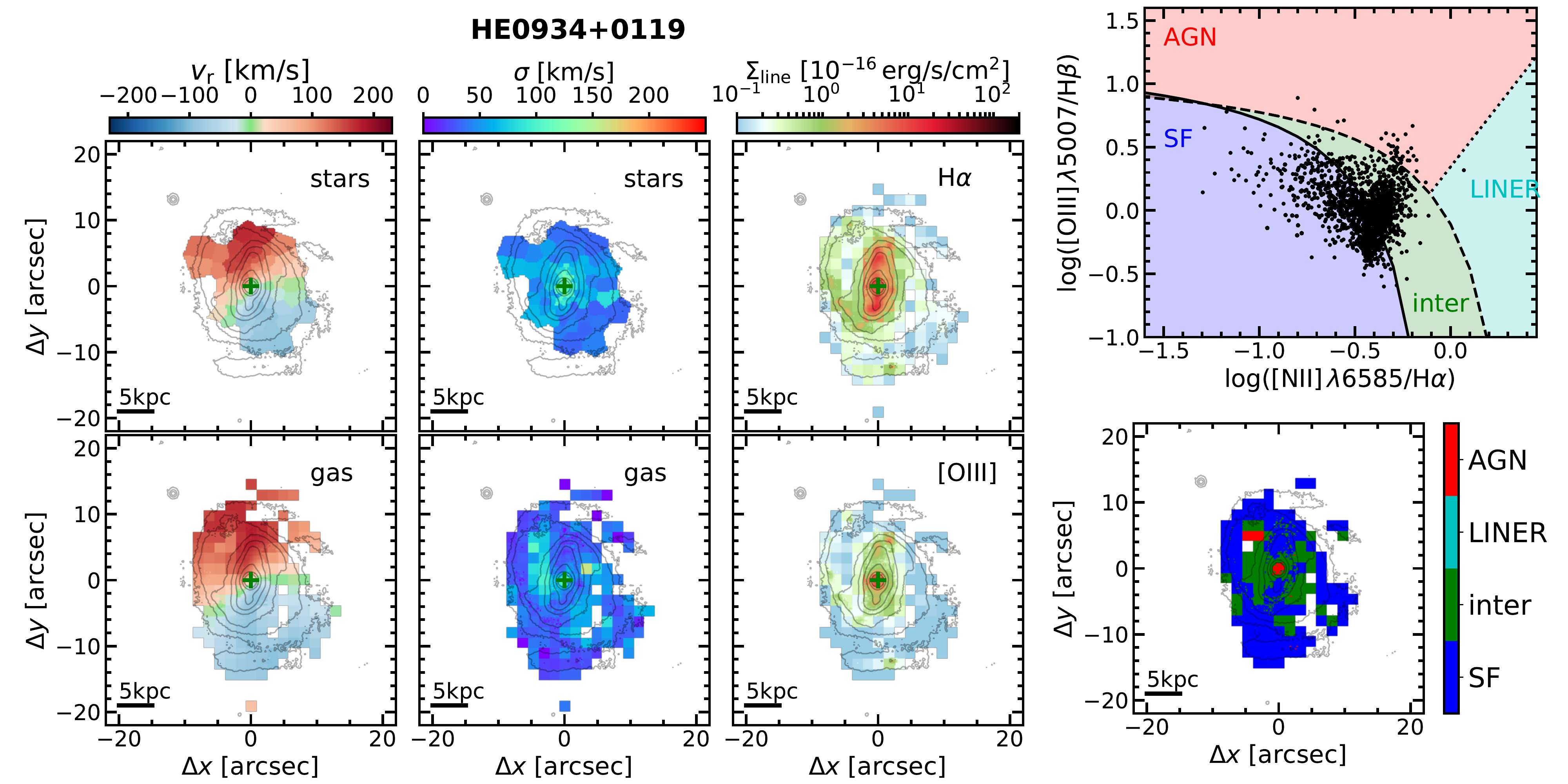}\\
  \includegraphics[width=0.85\textwidth]{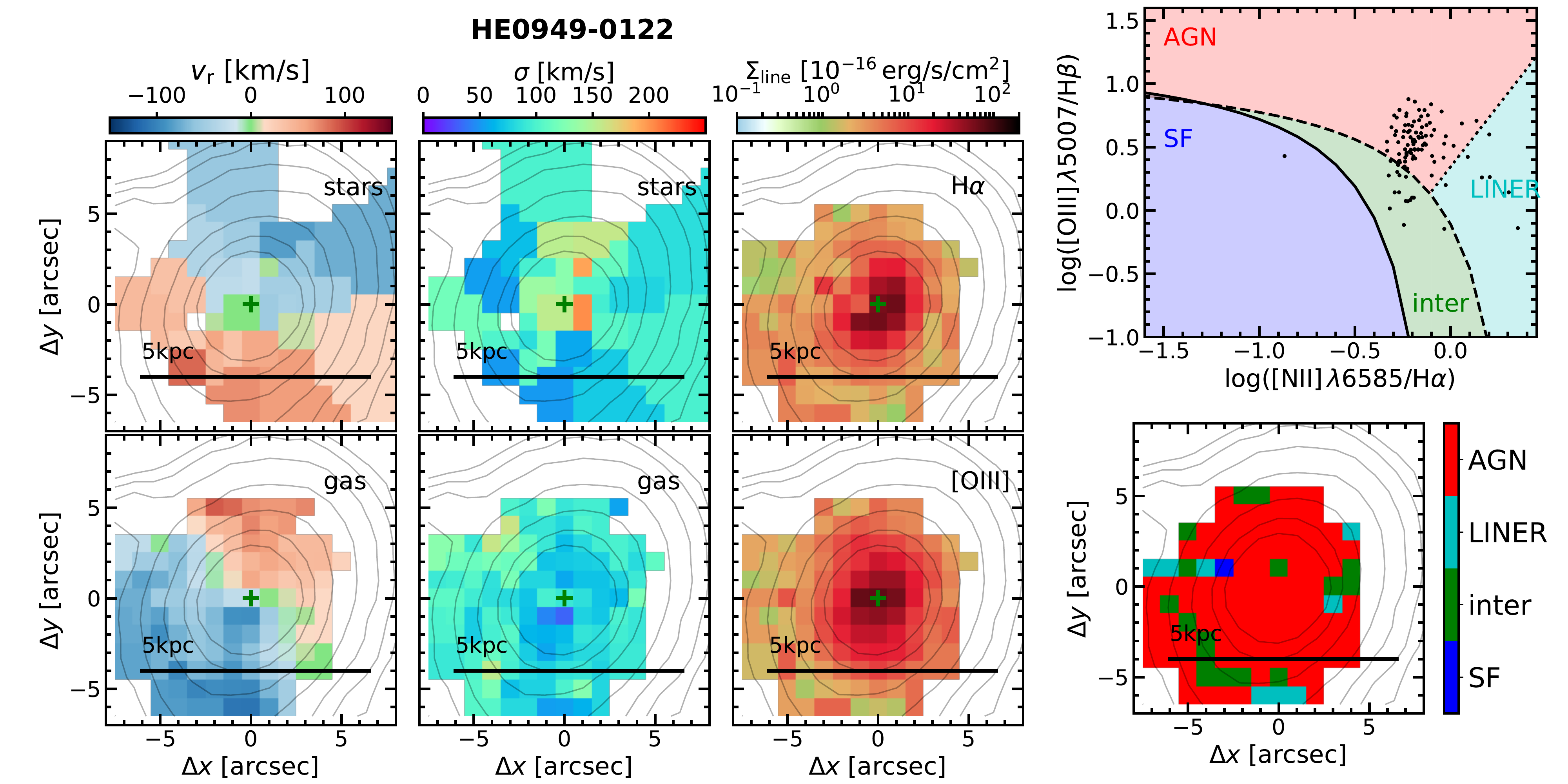}\\
  \includegraphics[width=0.85\textwidth]{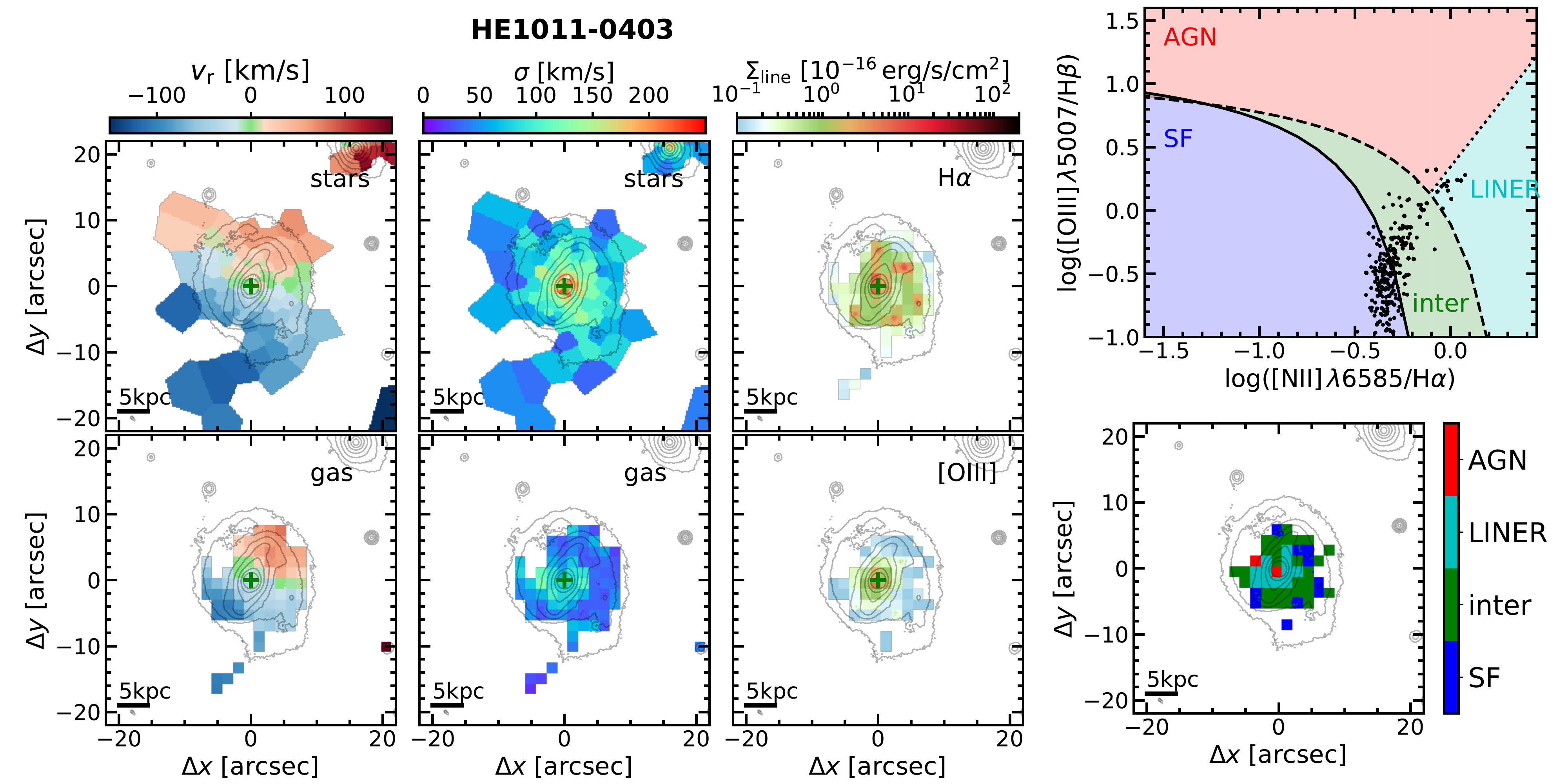}
  \caption{continued.}
 \end{figure*}
 \addtocounter{figure}{-1}
 
 \begin{figure*}
  \includegraphics[width=0.85\textwidth]{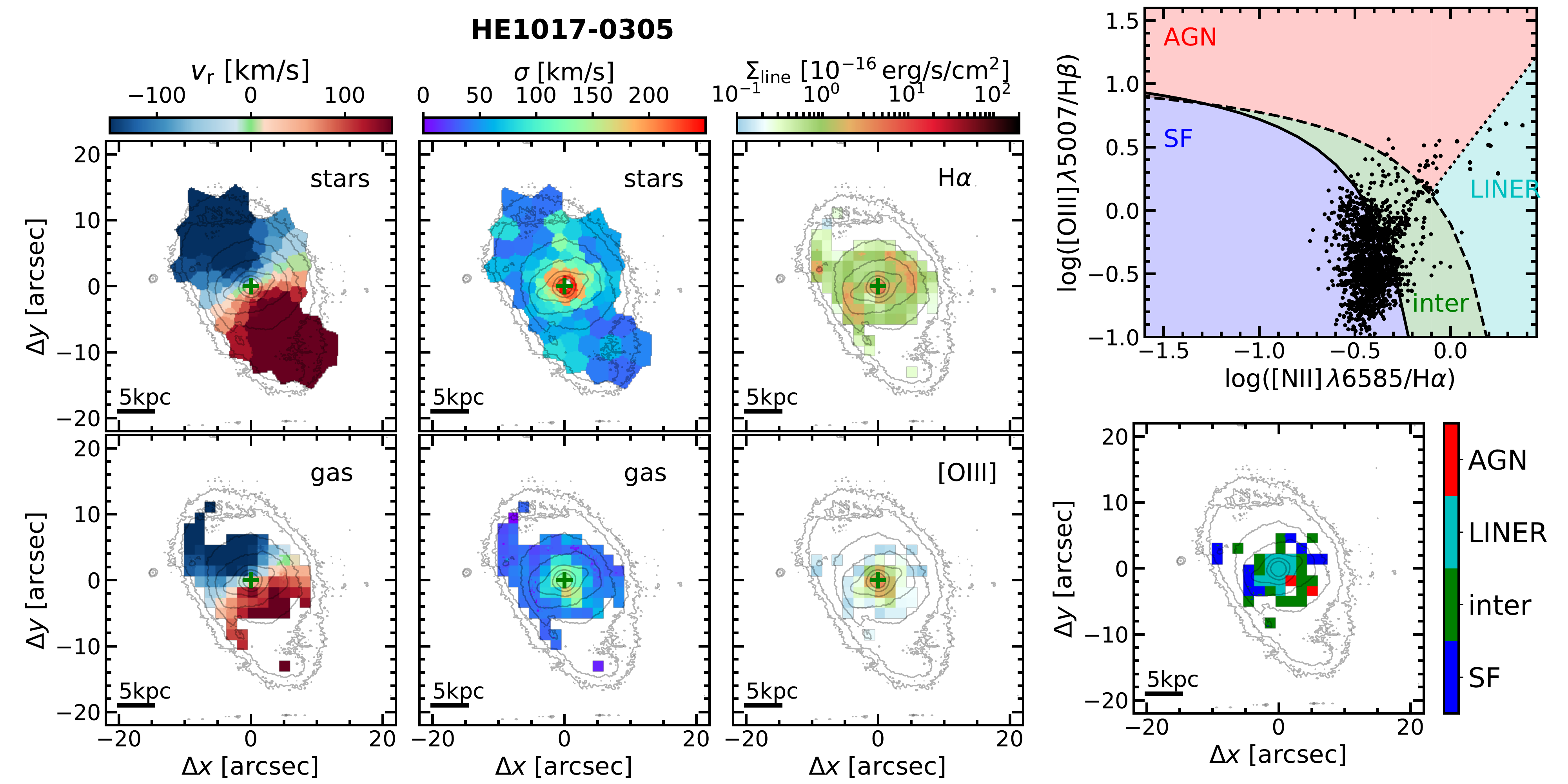}\\
  \includegraphics[width=0.85\textwidth]{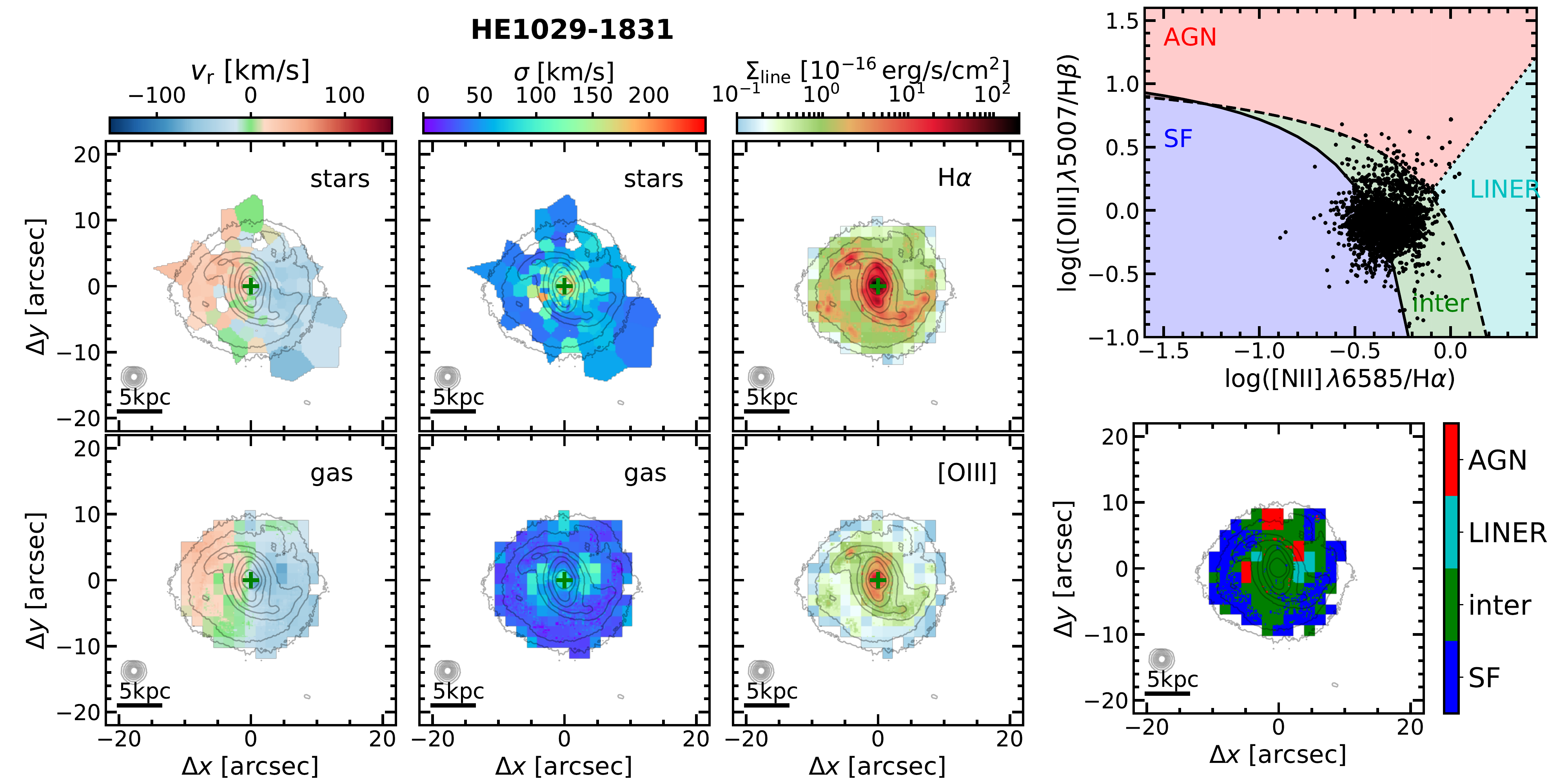}\\
  \includegraphics[width=0.85\textwidth]{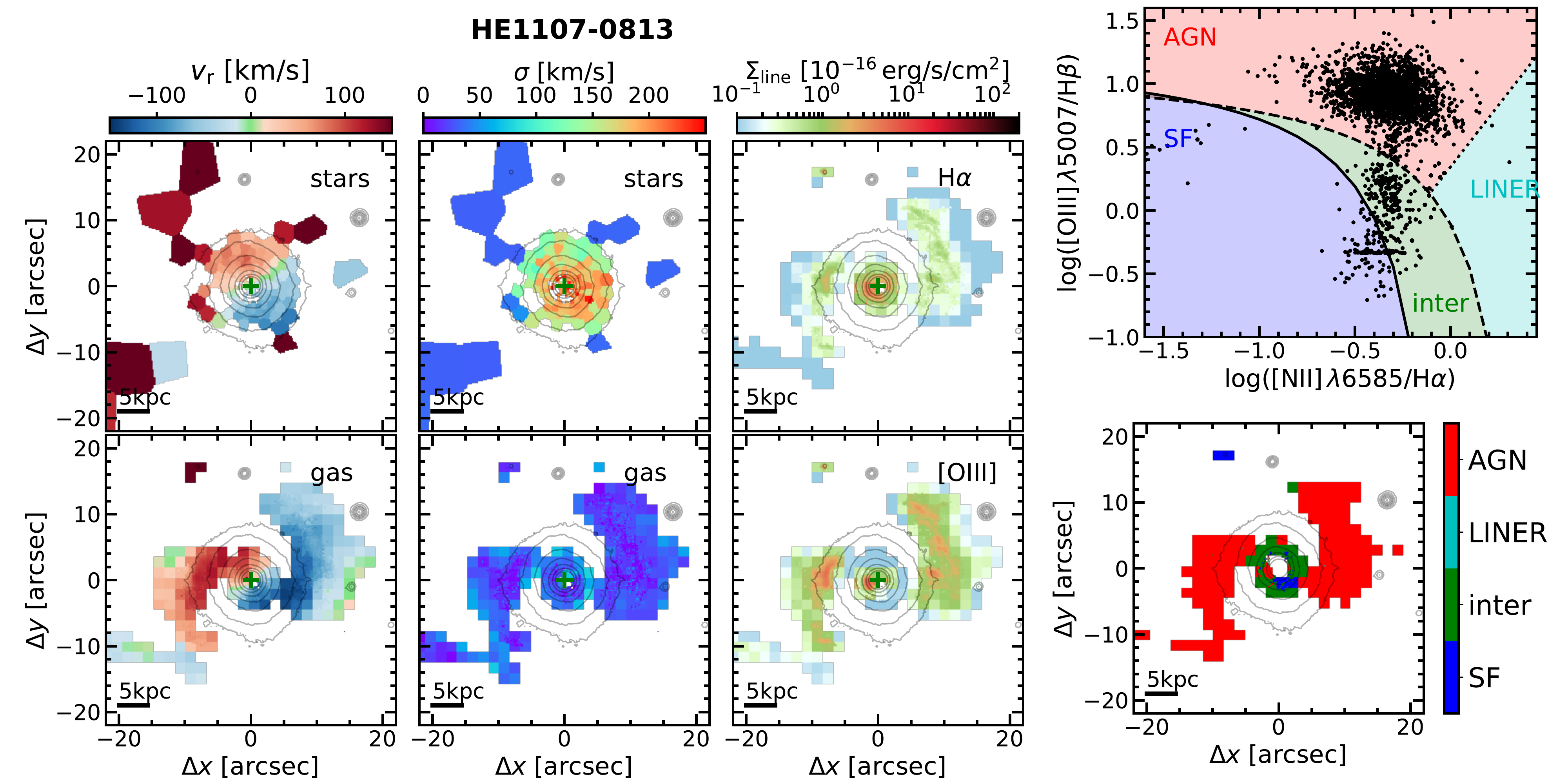}
  \caption{continued.}
 \end{figure*}
 \addtocounter{figure}{-1}
 
 \begin{figure*}
  \includegraphics[width=0.85\textwidth]{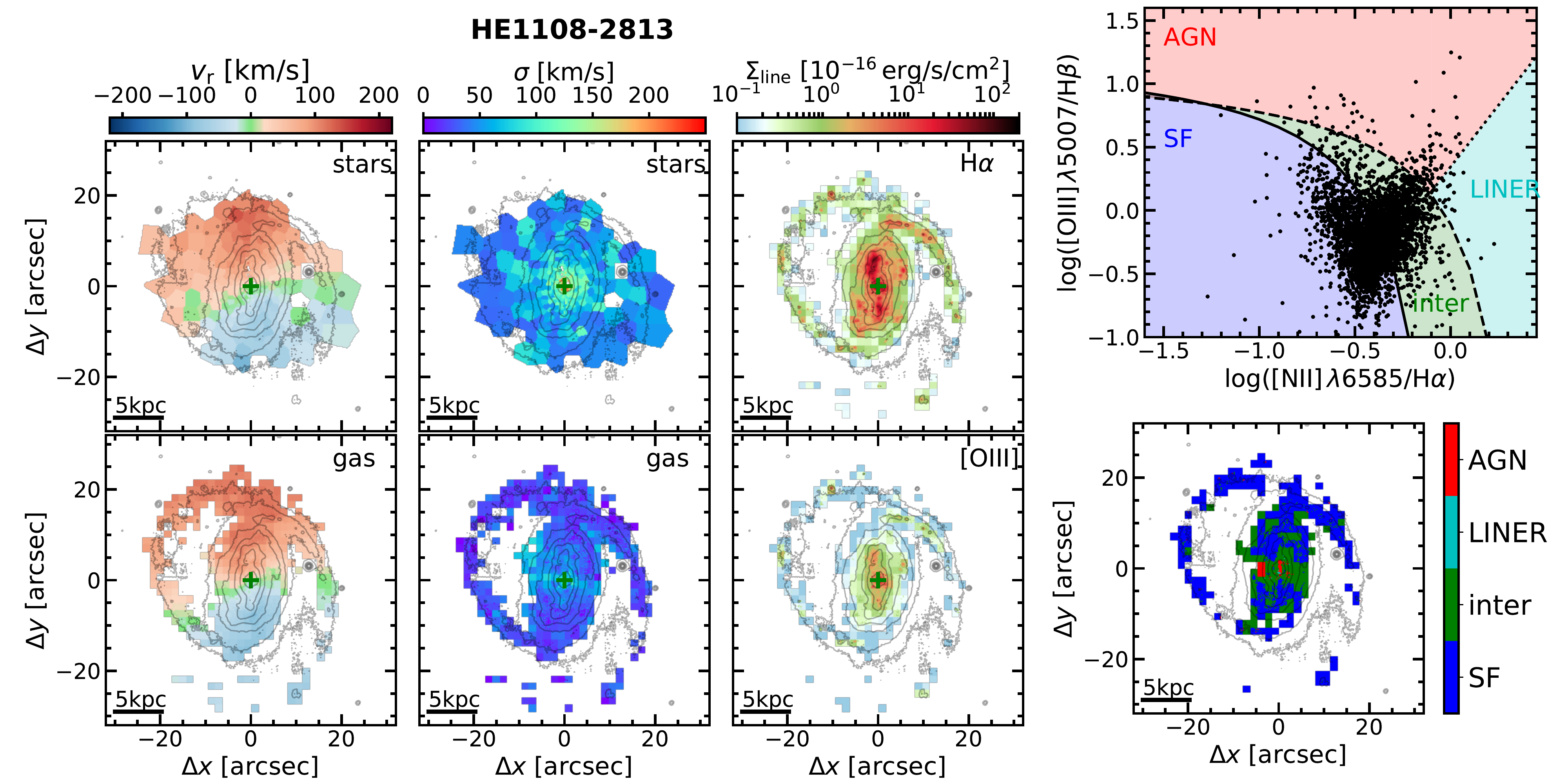}\\
  \includegraphics[width=0.85\textwidth]{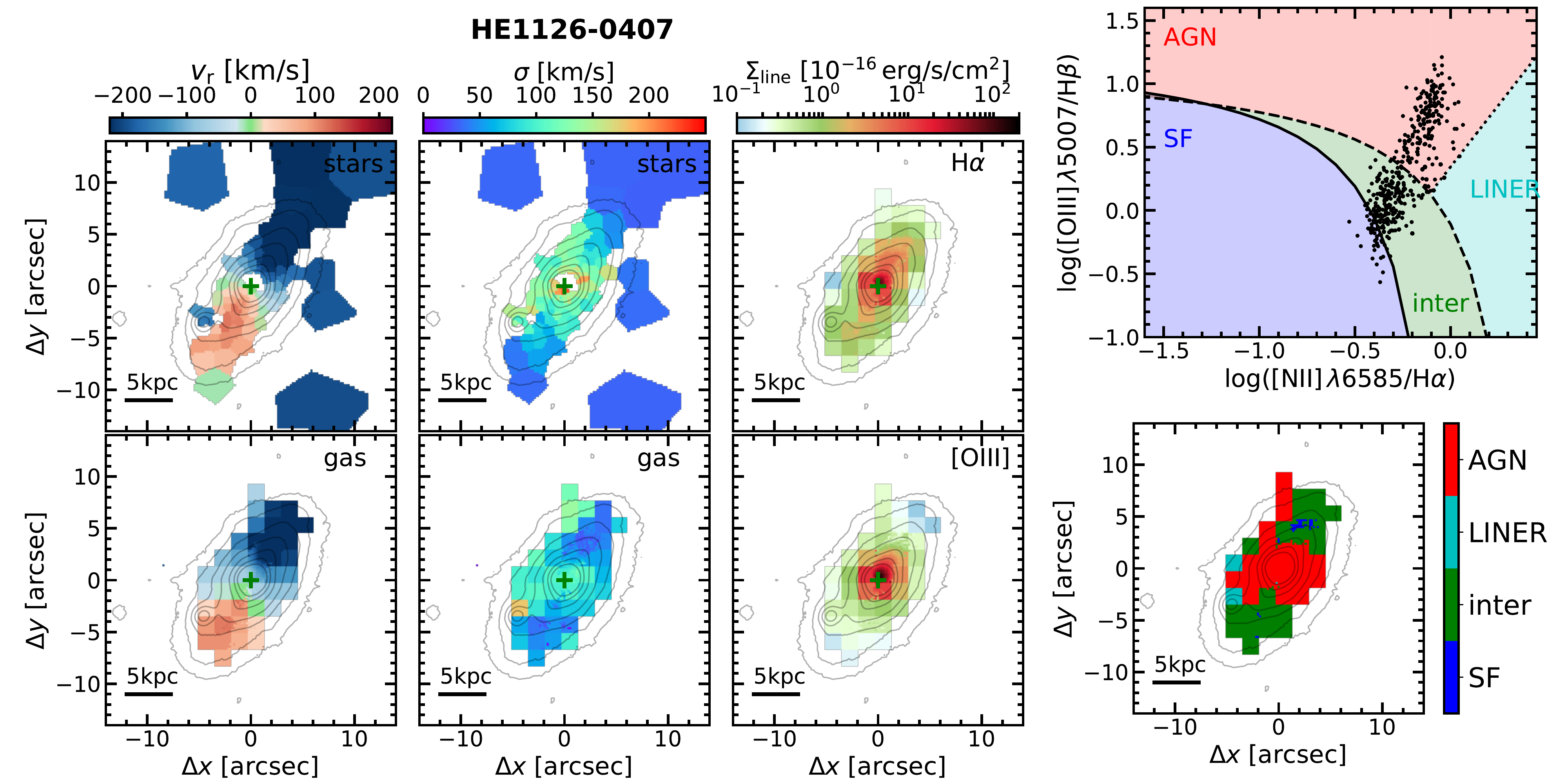}\\
  \includegraphics[width=0.85\textwidth]{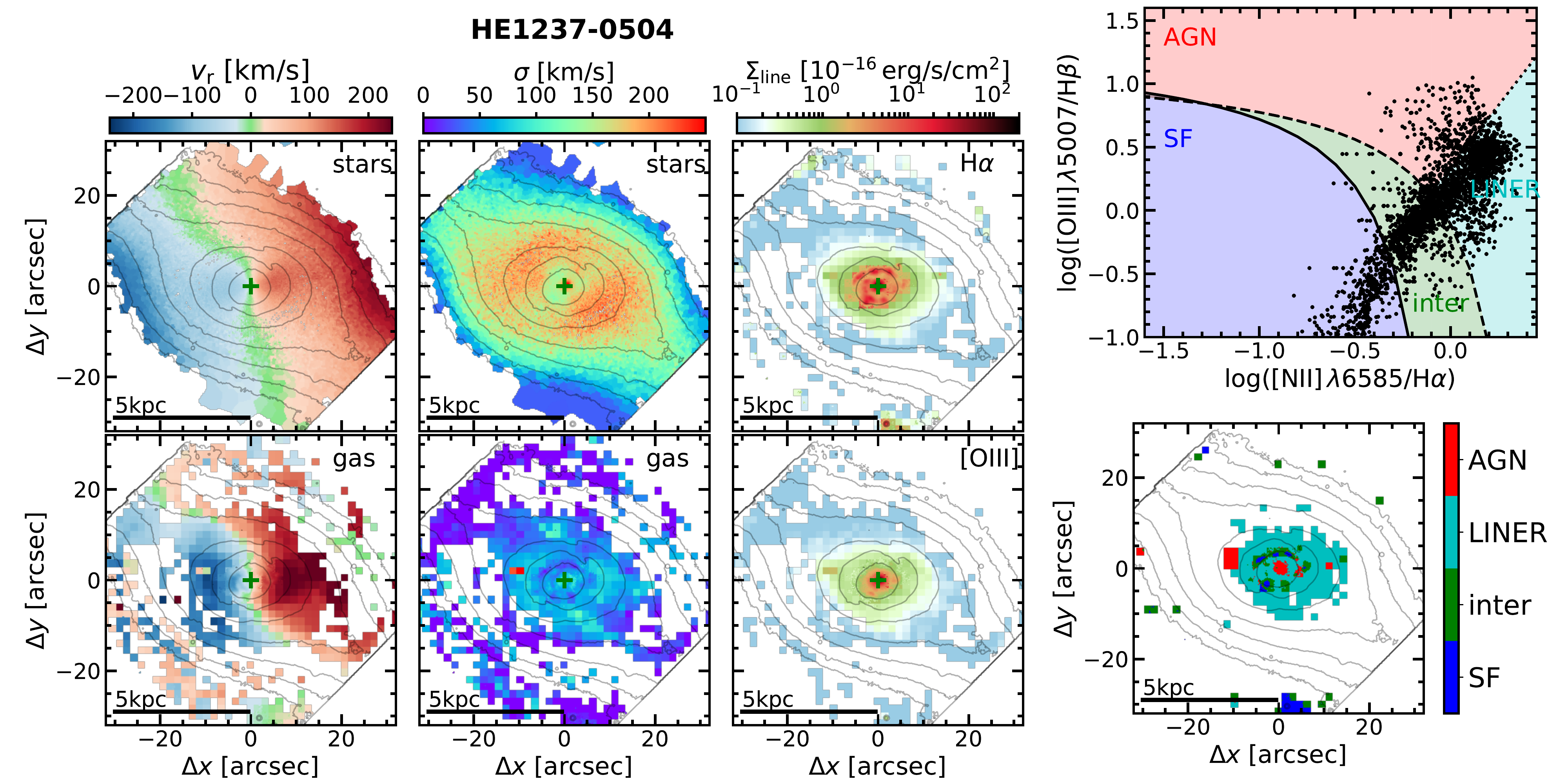}
  \caption{continued.}
 \end{figure*}
 \addtocounter{figure}{-1}
 
 \begin{figure*}
  \includegraphics[width=0.85\textwidth]{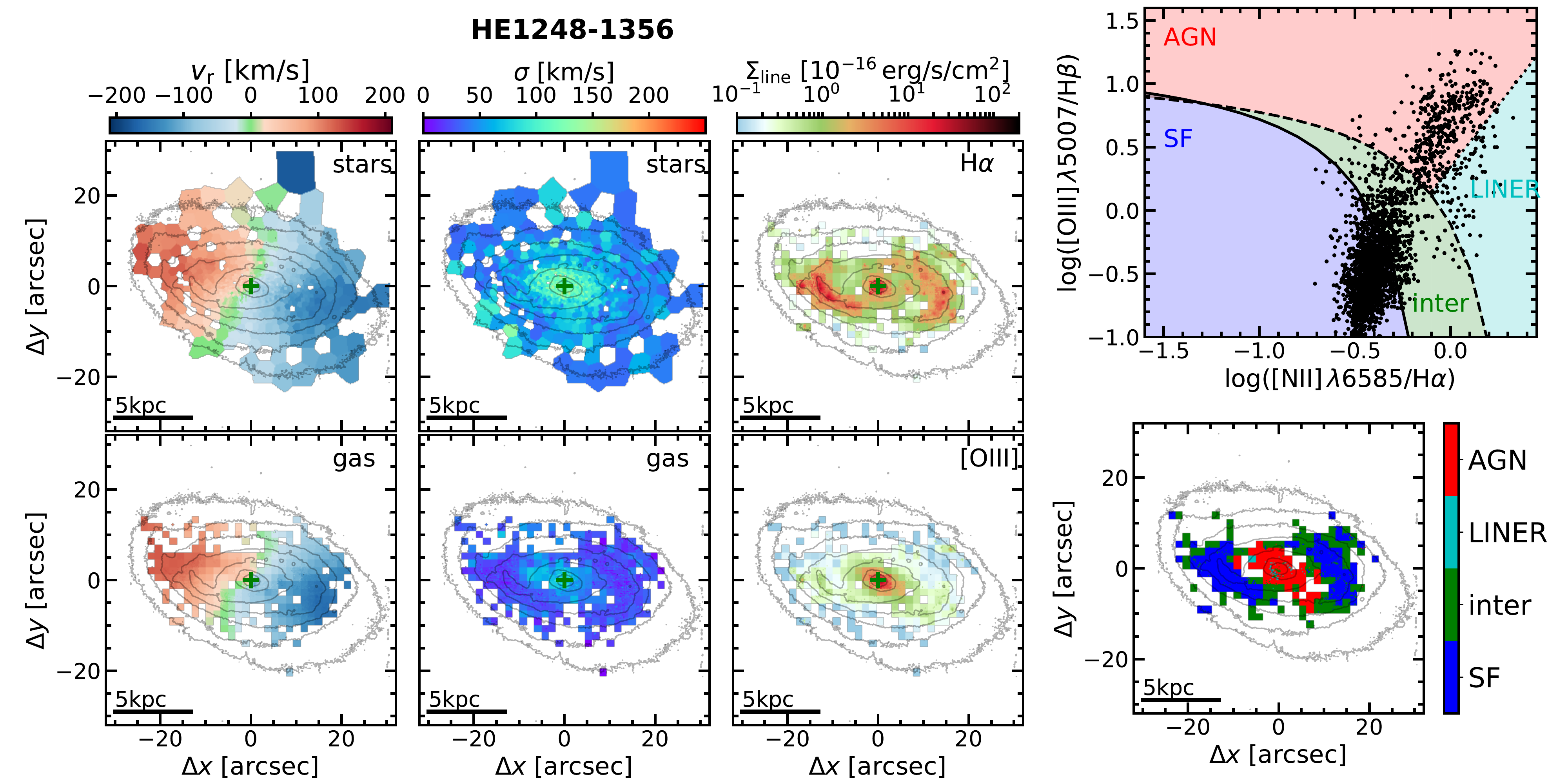}\\
  \includegraphics[width=0.85\textwidth]{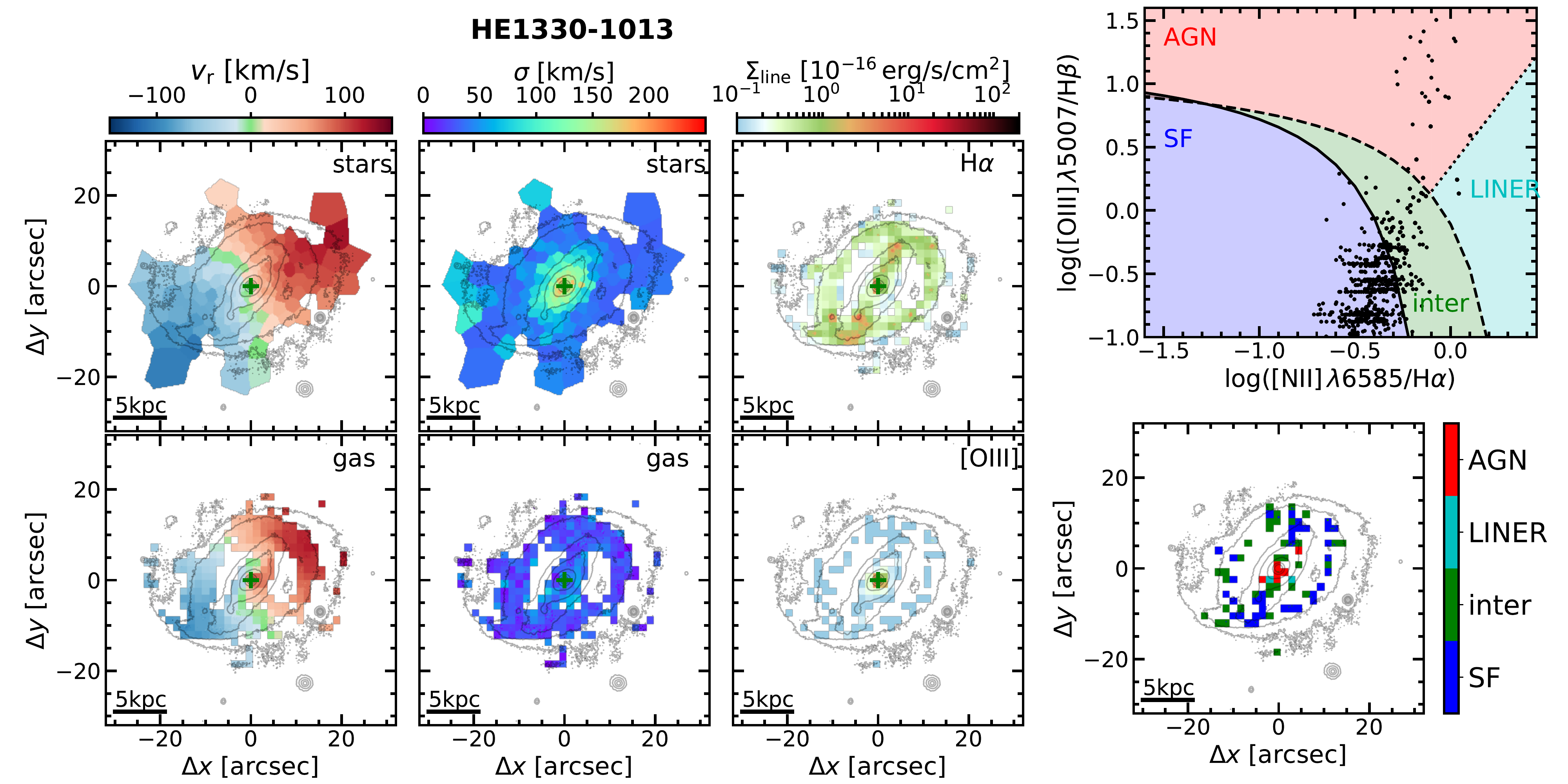}\\
  \includegraphics[width=0.85\textwidth]{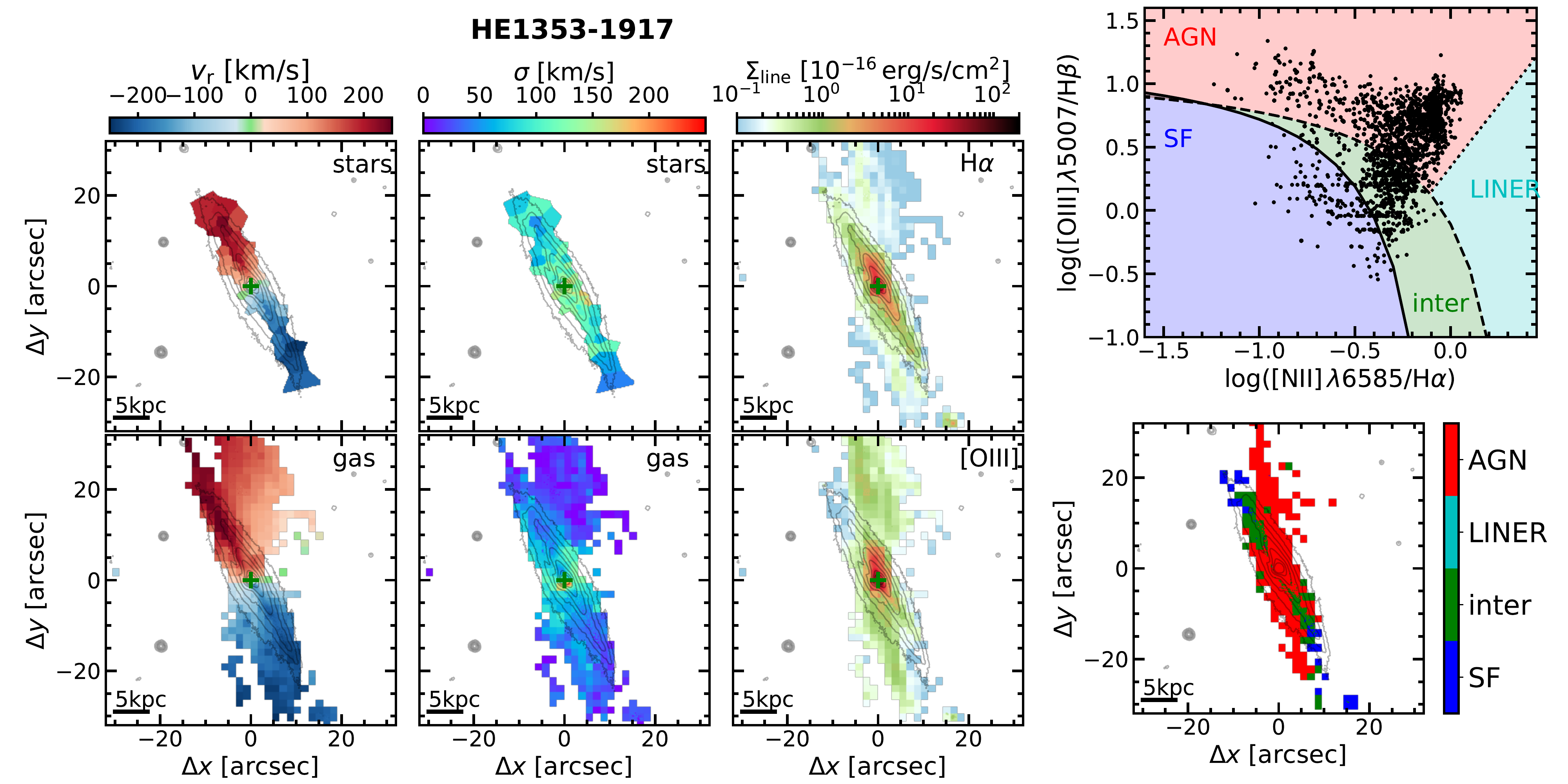}
  \caption{continued.}
 \end{figure*}
 \addtocounter{figure}{-1}
 
 \begin{figure*}
  \includegraphics[width=0.85\textwidth]{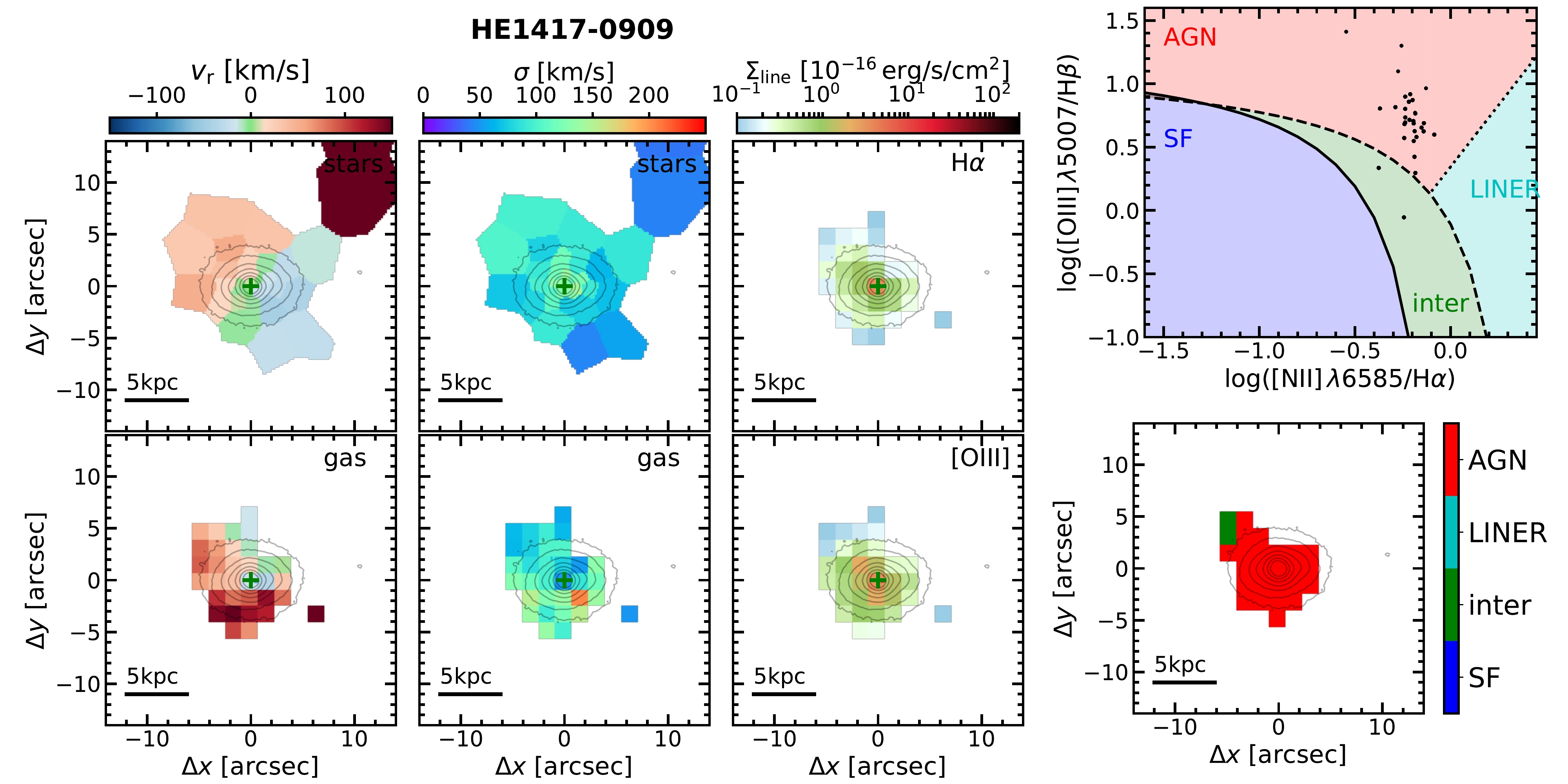}\\
  \includegraphics[width=0.85\textwidth]{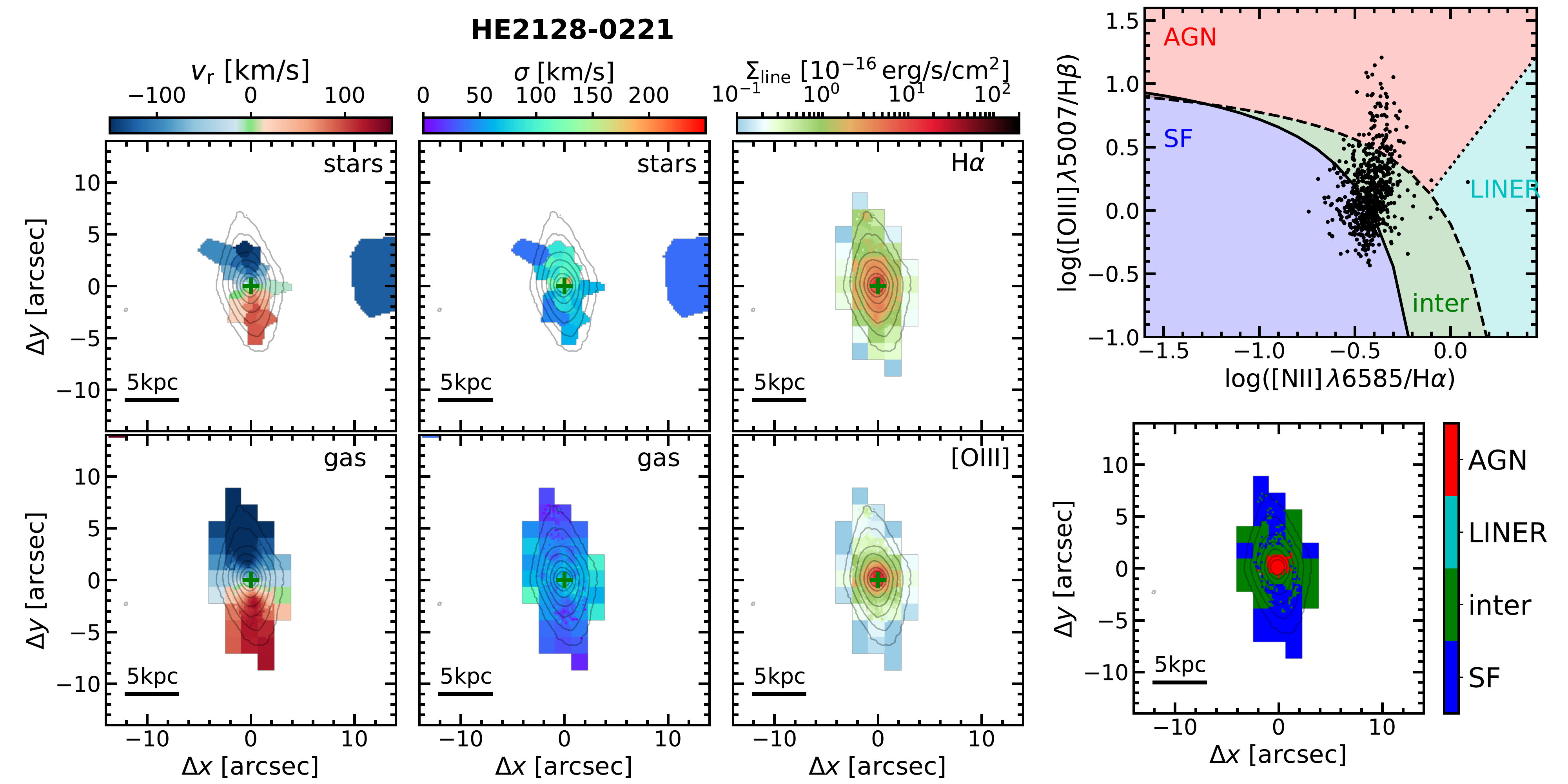}\\
  \includegraphics[width=0.85\textwidth]{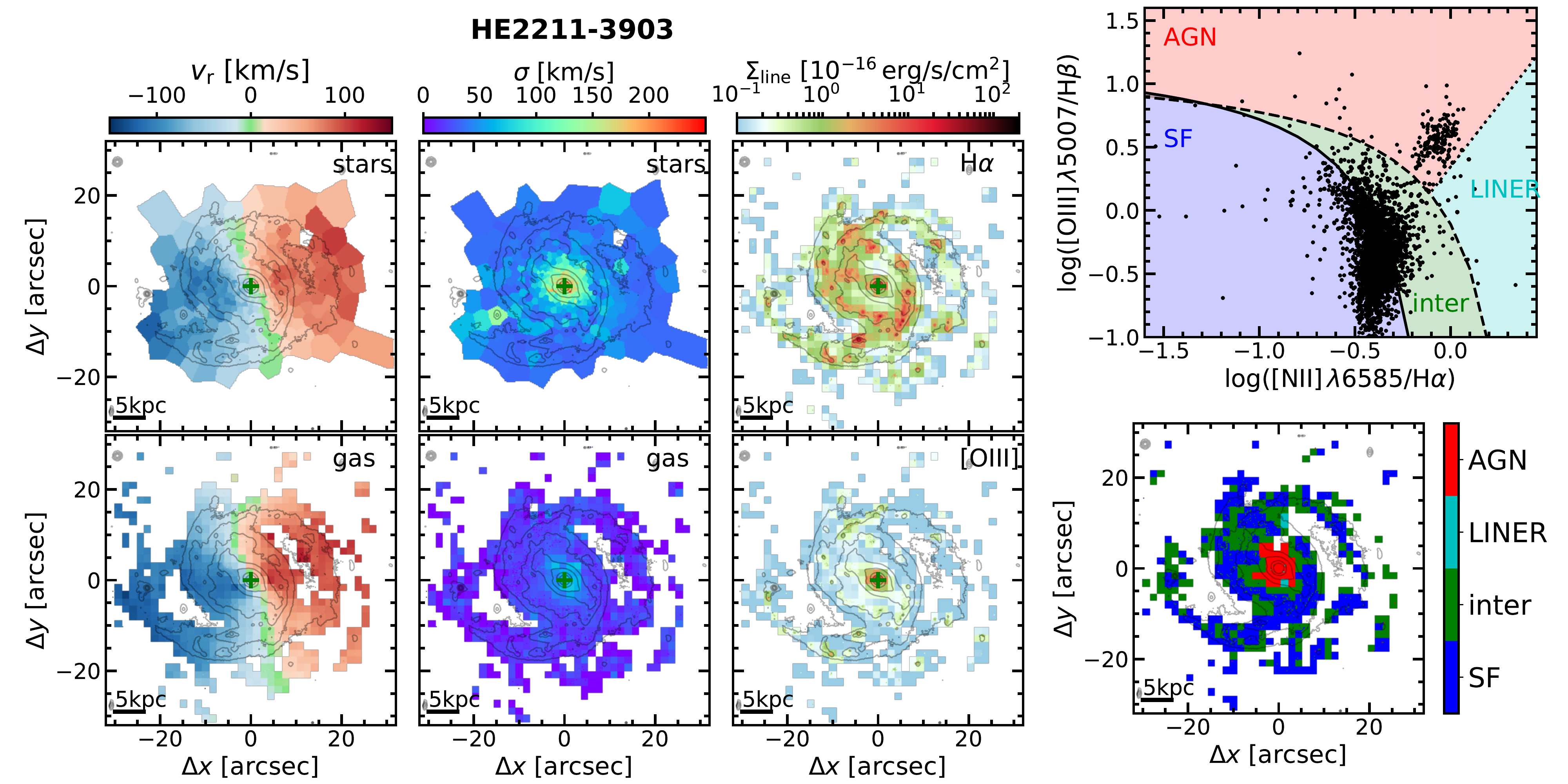}
  \caption{continued.}
 \end{figure*}
 \addtocounter{figure}{-1}
 
 \begin{figure*}
  \includegraphics[width=0.85\textwidth]{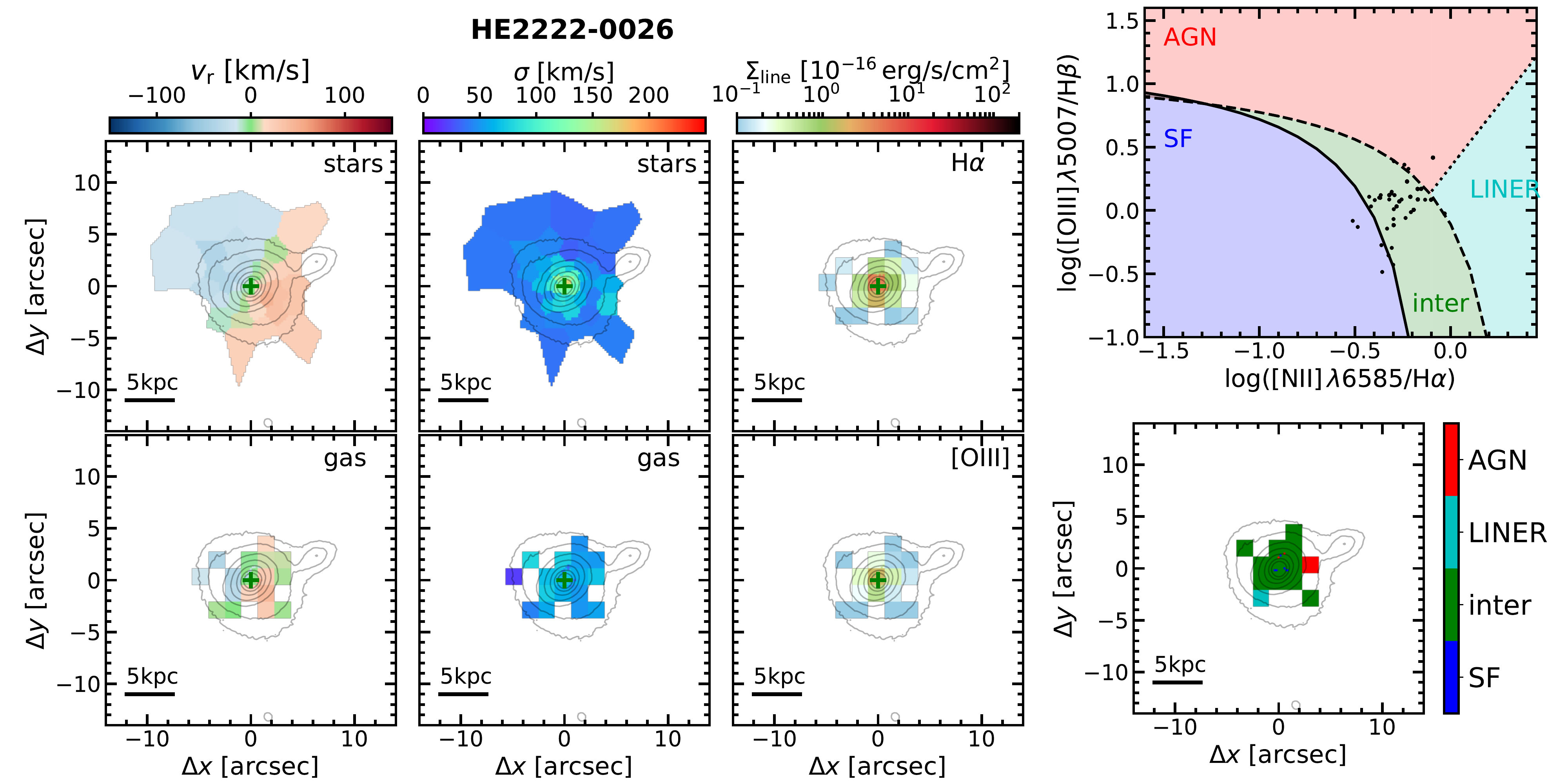}\\
  \includegraphics[width=0.85\textwidth]{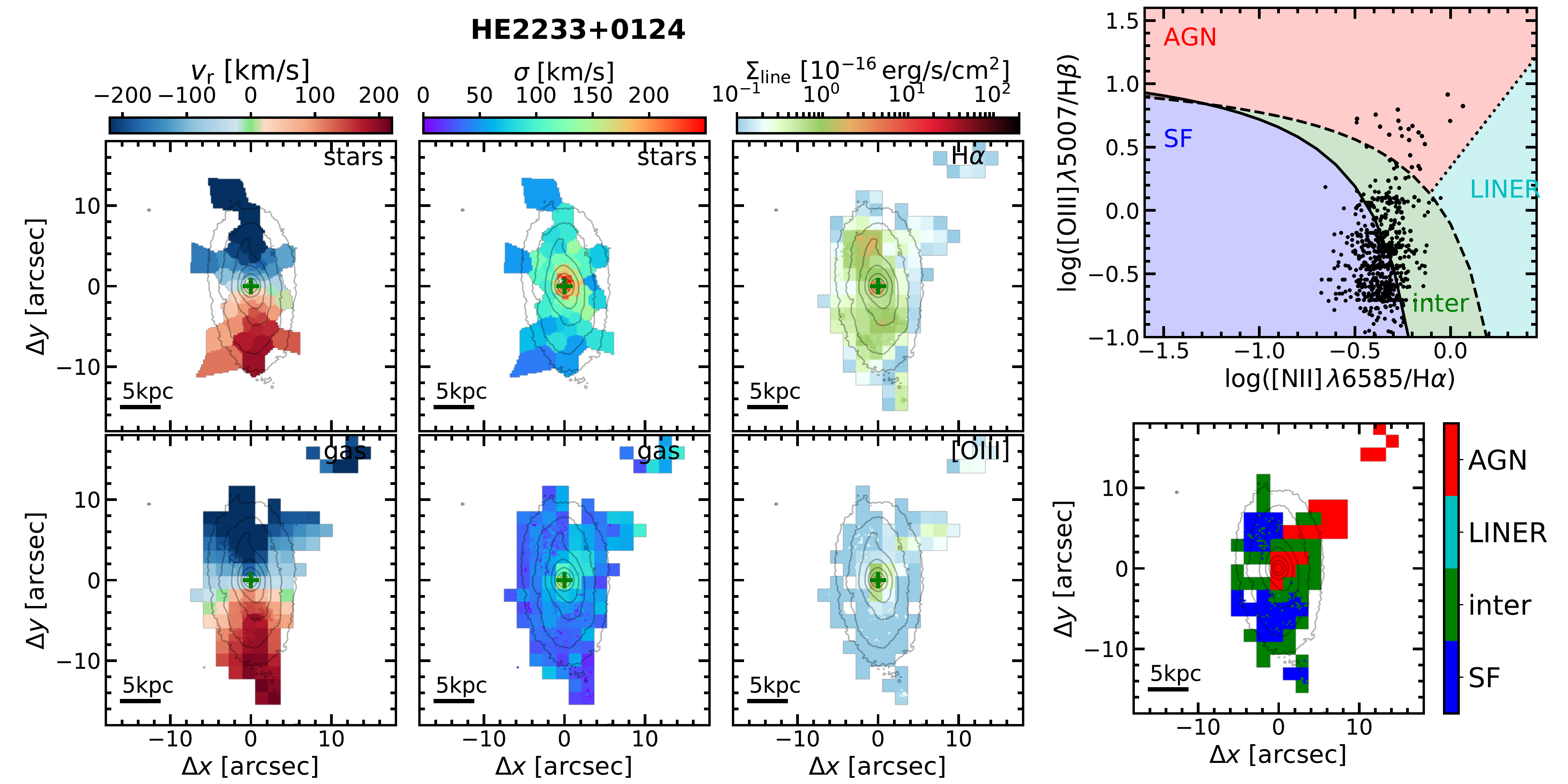}\\
  \includegraphics[width=0.85\textwidth]{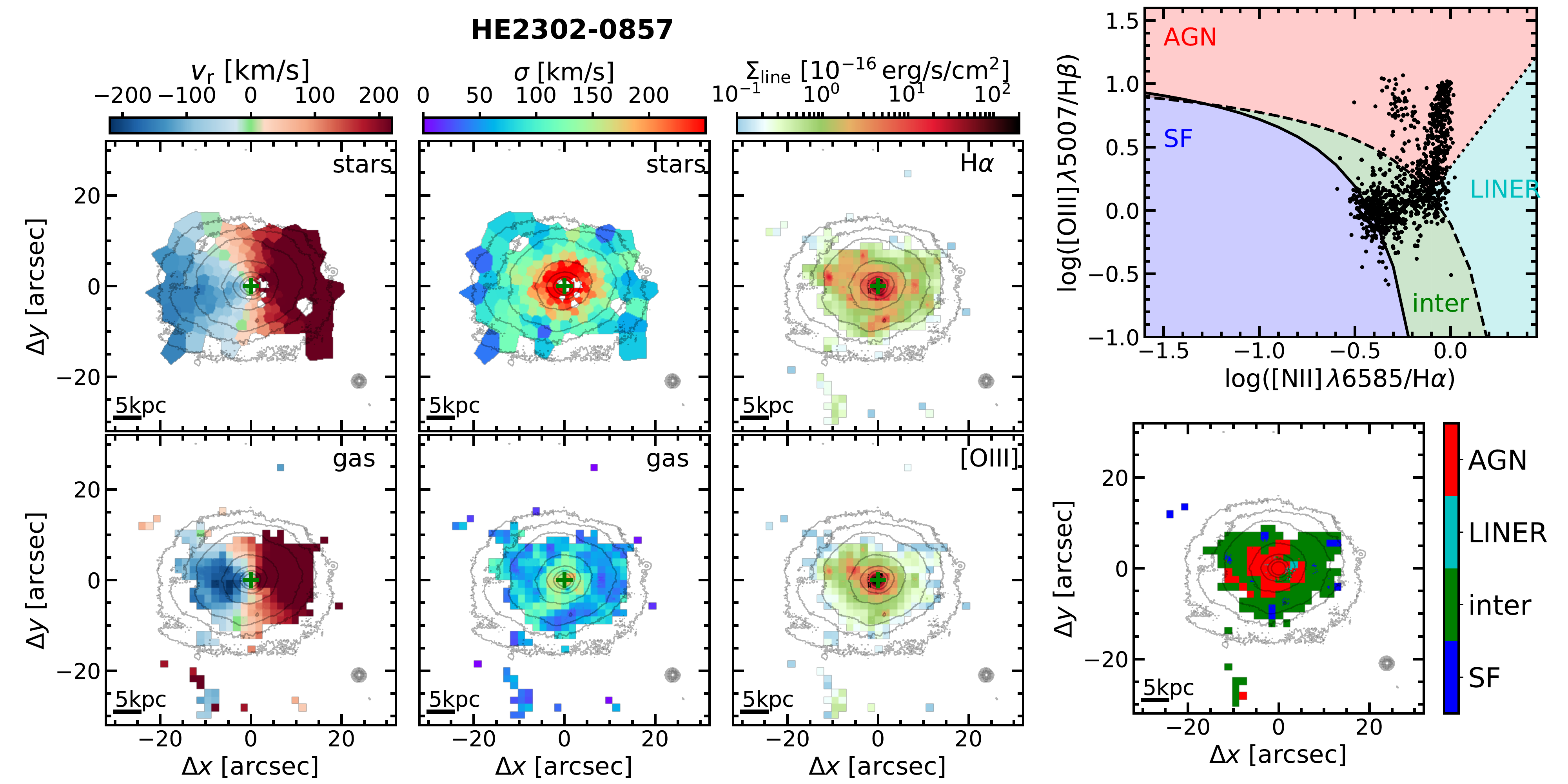}
  \caption{continued.}
 \end{figure*}
\end{appendix}
\end{document}